\documentclass[aip,
               preprint,
               amsmath,
               superscriptaddress,
               floatfix,
               nobibnotes,
               ]{revtex4-2}

\usepackage{amssymb}
\usepackage{mathtools}
\usepackage[mathscr]{eucal}
\usepackage[dvips]{graphicx}
\usepackage{color}
\usepackage[utf8]{inputenc}
\usepackage[flushleft]{caption}
\usepackage{subfigure}
\usepackage{rotating}
\usepackage{txfonts}
\usepackage{booktabs}
\usepackage[unicode]{hyperref}
\usepackage{bm}
\DeclareMathAlphabet{\mathpzc}{OT1}{pzc}{m}{it}

\def\figfoot{Quan-Mol-Geom}
\def\mathbi#1{\textbf{\em #1}}

\def\mi{\textrm{i}}

\makeatletter

\makeatletter

\newcommand{\figcaption}[2]{
    \noindent {\bf Figure \ref{#1}:} #2
    \vspace{1cm}}

\begin{document}
\newtheorem{algo}{Algorithm}
\newtheorem{theo}{Theorem}
\newtheorem{defi}{Definition}
\newtheorem{post}{Postulate}
\newtheorem{prop}{Proposition}
\newtheorem{lemma}{Lemma}
\newtheorem{coro}{Corollary}[theo]
\newtheorem{corop}{Corollary}[prop]
\newtheorem{remark}{Remark}
\newtheorem{assu}{Assumption}

%--------------------------------------------------
% Title, authors, and address
%--------------------------------------------------

\title{Variational Framework for Quantum Molecular Reaction Dynamics}

\author{Xingyu Zhang}
 \affiliation{Theoretical Chemistry,
              Department of Chemistry, 
              Northwestern Polytechnical University,
              West Youyi Road 127, 710072 Xi'an,
              China}

\author{Jinke Yu}
 \affiliation{Theoretical Chemistry,
              Department of Chemistry, 
              Northwestern Polytechnical University,
              West Youyi Road 127, 710072 Xi'an,
              China}

\author{Weijia Guo}
 \affiliation{Theoretical Chemistry,
              Department of Chemistry, 
              Northwestern Polytechnical University,
              West Youyi Road 127, 710072 Xi'an,
              China}
                                          
\author{Qingyong Meng*}
\thanks{To whom correspondence should be addressed}
 \email{qingyong.meng@nwpu.edu.cn}
  \affiliation{Theoretical Chemistry,
               Department of Chemistry, 
               Northwestern Polytechnical University,
               West Youyi Road 127, 710072 Xi'an,
               China}

\date{\today}

%----------------------------------------------------------
% Abstract
%----------------------------------------------------------

\begin{abstract}

\noindent {\bf Abstract}:
This work proposes a variational framework of quantum molecular reaction
dynamics based on the principles of (1) least action, (2) equivalence,
and (3) local Lorentz invariance. By these three postulates, all
propositions can be derived through the variational principle, aiming to
``see'' how a chemical reaction occurs in a general hyperbolic spacetime
by solving the Schr{\"o}dinger equation. Unifying the theories of chemical
reaction, a theory of chemistry on a general hyperbolic curved spacetime
is discussed using the geometric language. After describing preliminaries,
we deeply discuss the relation between space metric and the kinetic energy
operator, function approximation techniques for building the potential
energy surface, electromagnetic interactions in curved spacetime, and
electronic structure. Consequently, we discuss the nuclear Hamiltonian
in nonzero curvature spacetime, implying possibility to introduce gauge
fields through the space curvature. For quantum reaction dynamics, we
generally discuss the single-particle approximation (SPA) and then give
powerful ansatzes to solve both electronic and nuclear Schr{\"o}dinger
equation by the variational principle. To connect theory with experiment,
we summarize quantum scattering and resonance theories for chemical
reactions in hyperbolic curved spacetime. Finally, discussions from
optimization insight are made, including postulates of this work, quantum computation for quantum molecular dynamics,
generative artificial intelligence, role of perturbation, and Markov
process in optimization.
\\~~\\ 
{\bf Keywords}: {\it Geometric Theory}; {\it Quantum Dynamics}; {\it Molecular
Reaction}; {\it Hamiltonian Operator}; {\it Variational Principle}

\end{abstract}
\maketitle
\tableofcontents
\clearpage

%--------------------------------------------------------------
% Introduction
%------------------------------------------------------------
\section{Introduction\label{sec:intro}}

By solving the Schr{\"o}dinger equation (SE) of a molecular system, one
can directly ``see'' how a chemical reaction occurs
\cite{zha25:20397,son24:597,son22:11128,men22:16415,men21:2702,lar24:e2306881,sza12:108,goi18:e1341,li20:9951,sve23:e1666}.
To this end, both electronic structure and quantum molecular dynamics
must be performed to separately solve electronic and nuclear SEs,
respectively, under the Born-Oppenheimer approximation (BOA)
\cite{yar96:985,yar96:10456,yar98:971,yar01:6277,jua07:044317,bou08:124322,jua08:211101,alt08:214117,ced04:3,bae06:boa,ced13:224110}.
Over the past decades these aspects, especially those based on a
variational framework
\cite{zha25:20397,son24:597,son22:11128,men22:16415,men21:2702,lar24:e2306881,sza12:108,goi18:e1341,li20:9951,sve23:e1666},
have been extensively explored and discussed yielding a wide range of
algorithms. In general, most of these algorithms employ the single-particle
approximation (SPA), where perturbation theory for corrections might be
further adopted by some of them. Core idea of the SPA is to repeatedly
combine several coupled degrees of freedom (DOFs) into a single mode
making the system defined hierarchically. The ansatz is then expanded
by products of single-particle terms. For instance, function-based
variational methods for electronic structure \cite{sza12:108,goi18:e1341,li20:9951,sve23:e1666},
such as Hartree-Fock (HF) and multiconfigurational self-consistent field
(MCSCF) together with truncated configuration interaction (CI) and its
multi-reference version (called MRCI), expand the electronic wave function
through products of single-electron functions, called molecular orbitals
(MOs). For another example, method to propagate nuclear wavepacket adopts
the ansatz expanded hierarchically by products of single-particle functions
(SPFs). Typical example of such method is the multilayer multiconfiguration
time-dependent Hartree (ML-MCTDH) method \cite{wan03:1289,man08:164116,ven11:044135},
together with its two-layer form, called the MCTDH method \cite{bec00:1,mey03:251,wor08:569,mey12:351}.
In addition to the above function-based methods, for one- or two-dimensional
quantum lattices, quantum states of either electrons or nuclei can be
represented by the matrix product states (MPSs) and then the hierarchical
framework becomes the tensor network (TN) or tree tensor network (TTN).
One of typical examples is the density matrix renormalization group
(DMRG) in either time-independent or time-dependent fashion, denoted
by TI-DMRG or TD-DMRG. Due to the characteristic of parameter redundancy,
MPSs were found to the mathematical structure of a principal fiber bundle
\cite{hae14:021902,hae11:070601,van19:7}. Despite this, the mathematical
principles underlying their power require more profound elucidation.
In this work, extending previously reported variational framework for
wavepacket propagation \cite{zha25:20397}, we provide a unified
description of variational framework moleuclar reaction dynamics by
fiber-bundle theory, differential geometry, geometric quantum mechanics,
and so on, to deeply explore variational framework of chemical reaction,
as collected in Table \ref{tab:summ-var-gauge}.

Although quantum molecular reaction dynamics is the main purpose of this
work, we would like to give a general comment on other associated fields.
It is well known that electronic structure, together with molecular
dynamics and statistical ensemble model, forms the core of computational
chemistry. Under the BOA, solving the SE of electrons in molecules,
electronic structure methods provide energetic properties of individual
molecules. By moleuclar dynamics, one solves the nuclear Newton or
Schr{\"o}dinger equation on the basis of electronic energy that has been
solved by electronic structure, providing dynamics of the molecular system.
Statistical-mechanical simulation provides macroscopic thermodynamics
and kinetics quantities by simulating ensemble motions in phase space
based on the molecular dynamics. The above three branches form a
form a hierarchical and deeply interconnected framework of computational
chemistry that bridges electronic motion, atomic motion, and macroscopic
properties. Together, they constitute the engine of modern computational
chemistry, enabling predictions ranging from single-molecule dynamics
to material properties under realistic environment. One may be also
interested in the fact that the boundaries among the three branches are
becoming blurred due to developments of modern computational techniques,
such as data-driven and/or agent-based artificial intelligence (AI) and
quantum computing techniques. However, the foundational logic remains
making these branches will continue to serve as the conceptual framework
of computational chemistry for the future to come. In this work, as
given above, we focus on variational framework of quantum molecular
reaction dynamics, which is one of sub-branches in molecular dynamics.
Its classical counterpart is quasi-classical trajectory simulation,
which is beyond the scope of the present work.

Although the present topic of quantum molecular dynamics has been
comprehensively surveyed by numerous authors, we have chosen to unify
the theory using variational principles and geometric language, so as
to open a door to new ideas and algorithms. It must be mentioned that
the idea of geometric language \cite{dor01:19,hae11:070601,hae14:021902,van19:7}
for quantum dynamics is not new. It plays an important role in various
fields. Taking variable separation as the first example, one can
separates parameters from the system of interest by treating the
separated parameters as an effective background, called parameter
space, when the change in these parameters differs greatly from that
in focus. It is often called adiabatic approximation \cite{ber84:45}
leading to an additional phase, called the Berry phase \cite{ber84:45}
in the wave function of the remaining part. Over the past decades, the
Berry phase was widely discussed \cite{ber84:45} and was found to be
the holonomy in a Hermitian line bundle since the adiabatic theorem
naturally defines a connection in such a bundle. The BOA is a typical
adiabatic approximations
\cite{yar96:985,yar96:10456,yar98:971,yar01:6277,jua07:044317,bou08:124322,jua08:211101,alt08:214117,ced04:3,bae06:boa,ced13:224110},
where the nuclear coordinates are separated from the molecular wave
function as parameters leading the Berry phase in the electronic wave
function. Taking the time-dependent variational principle (TDVP) as
the second example, it was applied to MPSs for efficiently simulating
infinite one-dimensional quantum lattices \cite{hae11:070601}. The TDVP
was found to be optimal without Trotter error preserving all symmetries
and conservation laws and performing with low computational complexity
\cite{hae11:070601}. In this context, the total space or bundle space
corresponds to the parameter space of tensors associated to every
physical site. The base manifold is embedded in Hilbert space and can
be given the structure of a K{\"a}hler manifold by inducing the Hilbert
space metric \cite{dor01:19,hae11:070601,hae14:021902,van19:7}. The main
interest for variation framework is the tangent space to the base manifold.
By lifting the tangent vectors to the tangent space of the bundle space
using a principal bundle connection, one can define an inverse metric
and introduce differential geometric concepts, such as parallel transport
related to the Levi-Civita connection and the Riemann curvature tensor
\cite{dor01:19}. More generally, a complete theory of quantum mechanics
can be formulated in which the usual projective Hilbert space is replaced
by the smaller manifold.

In addition to dynamic system itself, variational principles and thus
target-oriented optimizations are employed in constructing the nuclear
Hamiltonian, which is a prerequisite of quantum dynamics calculations.
For instance, construction of a potential energy surface (PES) is one
of prerequisites for studying molecular dynamics. It requires function
approximation through a set of electronic energies \cite{son22:1983}.
With the fast development of computational techniques, machine learning
(ML) or artificial intelligence (AI) has become an increasingly popular
method \cite{son22:1983}, such as neural network (NN) based on generalized
linear regression (GLR) and Gaussian process regression (GPR) based on
kernel-model regression (KMR). The standard training of AI models
typically performs optimization in a Euclidean space of model parameters,
relying on gradients of the loss function with respect to the model
parameters \cite{son22:1983} and facing challenges of convergence to
local minima and of vanishing or exploding gradients. In contrast,
recognizing that the model parameter space possesses a non-Euclidean
space, Riemannian optimization may offer a more stable and reliable
process in fitting the PES, together with enhanced adaptability to
complex data structures \cite{fei25:123}, if one can overcome the
following open challenges. First, the energy database is often
geometrically complex implying further database-oriented development,
such as the NN fitting approaches based on permutation invariant
polynomial (PIP) and its fundamental invariant (FI) version. Second,
model-oriented approach should be considered to explore its
generalizability, computational efficiency, and interpretability, such
as the NN fitting approaches with sum-of-products (SOP) form.
Third, manifold-oriented geometric optimization may be a direction
worthy of consideration to find an appropriate manifold for
construction of PES \cite{fei25:123}. The second application of variational
principles to construction of the nuclear Hamiltonian is to derive the
kinetic energy operator (KEO) from a given coordinate set. Recently, a
general dependence of the KEO on metric of the configuration space and
on metric of the spacetime, in which the reaction occurs, was discussed
\cite{zha26:1613} using the $[3+1]$ decomposition of the 4D spacetime.
For the primitive case, the reference frames are fixed on the Eulerian
observers \cite{zha26:1613}, implying that the absolute space and
universal time viewpoint of Galileo are adopted. Thus, this choice
allows us to explore the molecular dynamics for a generally curved
spacetime.

In general, this work provides a unified theory of quantum molecular
reaction dynamics through variational principles, as shown in Table
\ref{tab:summ-var-gauge}. First, the present unified theory provides
a geometric description on the SPA providing powful ansatzes to solve
the Schr{\"o}dinger equation by variational principle. Previous
comparison \cite{zha25:20397} of the function-based methods with those
in the form of TN/TTN provides a unified framework for ansatz. Second,
we turn to construction of the nuclear Hamiltonian operator. The
present unified theory provides general expression of the KEO from
the viewpoint of geometric characteristics of configuration space.
Using curvilinear coordinates, the configuration space is typically
curved. Its curvature depends on the coordinate frame and the curvature
of the spacetime where the reaction occurs \cite{zha25:20397,zha26:1613}.
Deriving the KEO necessitates incorporating the variation of kinetic
energy to properly account for the motion along geodesics in configuration
space \cite{zha25:20397,zha26:1613}. Third, variable separation leads
to the Berry phase effects. The present theory provides a geometric
interpretation of the Berry phase effects and clarifies its role in
reaction dynamics. Based on previous discussions \cite{zha25:20397},
we further discuss application of gauge theory to molecular reaction
dynamics in this work, particularly the possibility of designing quantum
simulators for quantum computing on molecular reaction dynamics. Fourth,
in practical optimization within the variational principle, stochastic
algorithms are often employed to improve initialization sampling, escape
local minima, enhance convergence efficiency, and so on. In addition,
due to the inherent randomness in some algorithms (say stochastic gradient
descent and random initialization) and hardware (say floating-point
rounding), the stochastic characteristics of practical optimization
might be significant. Therefore, in this work we discuss the numerical
accuracy of optimization from the perspective of stochastic process,
aiming to deepen our understanding of the practical accuracy limits
encountered.

It is worth noting that molecular reaction dynamics is only observed
by crossed molecular-beam experiments
\cite{her60:1870,lee69:1402,bus79:34,pol81:1837,lee87:793,sch94:1345,qui06:1440,yan07:433,zha10:12782,don10:1501,xia11:440,yan12:1650,yan14:3106,yan15:60,yua18:1289,che20:9202,xie20:767},
which provide essentially quantum scattering processes between reactant
state (the ``in'' state) and product state (the ``out'' state). In crossed
molecular-beam experiments
\cite{her60:1870,lee69:1402,bus79:34,pol81:1837,lee87:793,sch94:1345,qui06:1440,yan07:433,zha10:12782,don10:1501,xia11:440,yan12:1650,yan14:3106,yan15:60,yua18:1289,che20:9202,xie20:767},
the molecular beams consist of reactant molecules that are effectively
non-interacting, which arises from the fact that it is difficult or
impossible to observe a single molecular scattering process by
experiments. Thus, the reactant molecules in each molecular beam are
approximately at the same quantum state state, which is one of eigenstates
of the molecular Hamiltonian operator in a body-fixed (BF) coordinate
frame. Unlike the above theory of wavepacket propagation, quantum
scattering plays the time-independent role in providing differential
scattering cross-section (DCS) and detailed state-to-state mechanism.
Keeping this idea in the mind, eigenvalue problems at the reactant and
product regions are required to solve. It should be stressed that
``in'' and ``out'' states do not inhabit two different Hilbert spaces.
Therefore, the collision of the ``in'' and ``out'' states can be
considered as a transition from the ``in'' state to the ``out'' state.
The rationale for considering scattering and introducing so-called
scattering matrix between the ``in'' and ``out'' states is not only
that they have proved useful but also that the scattering formalism
makes it easy to choose interaction. Therefore, the quantum scattering
theory is often referred as time-independent dynamics theory, in contrast
to the time-dependent wavepacket propagation dynamics. In this work,
although we focus on framework of time-dependent dynamics theory, the
time-independent one together with theory of resonance are also
incorporated into the present framework for completeness. These
contents especially focus on reaction scattering by the viewpoint
of geometry, including eigenstates, potential scattering, and
scattering resonances.

The paper is organized as follows. Section \ref{sec:theory} describes
preliminaries of the present work. Section \ref{sec:hamiltonian} describes
construction techniques of the nuclear Hamiltonian. Sections \ref{sec:dynamics}
and \ref{sec:scatt-res} include main contents of the geometric theories
of wavepacket propagation and quantum scattering, respectively. Section
\ref{sec:perspec} discusses the present theory from the viewpoint of
optimization. Section \ref{sec:con} concludes with a summary and outlook.
Finally, unless otherwise specified, Einstein summation
convention, natural units, and mass-weighted coordiantes are always
used for simple.

%----------------------------------------------------
% Theory
%----------------------------------------------------
\section{Preliminaries\label{sec:theory}}

In this section, preliminaries of the this work are given, as collected
in Table \ref{tab:theory-comp}. These contents include two of three
postulates of this work, the mountain pass theorem, and electronic
structure theory. Section \ref{sec:math-frame} describes the principle of
least action, which is the first postulate of this work. Section
\ref{sec:mpt-math} describes the mountain pass theorem which states
that there exists a critical point between two minima on a smooth
hypersurface. Section \ref{sec:elec-struc} concisely describes
electronic structure theory (or called quantum chemistry), providing
the energetic foundation of molecular dynamics. Although it is not
the goal of this work, we describe electronic structure theory in
Section \ref{sec:elec-struc} simply as an example of the variational
principle. Moreover, since electronic structure theory constitutes a
significant branch of theoretical and computational chemistry, a deep
description is required to maintain the completeness of the this work.
Finally, in Section \ref{sec:keo-metric} we discuss the principle of
equivalence and then show the relation between the Laplacian operator
and variational principle. It is the second postulate of this work and
a foundation of KEO derivation.

%-------------------------------------------------------
% Mathematics
%-------------------------------------------------------
\subsection{Principle of Least Action\label{sec:math-frame}}

\begin{figure}
 \centering
  \includegraphics[width=18cm]{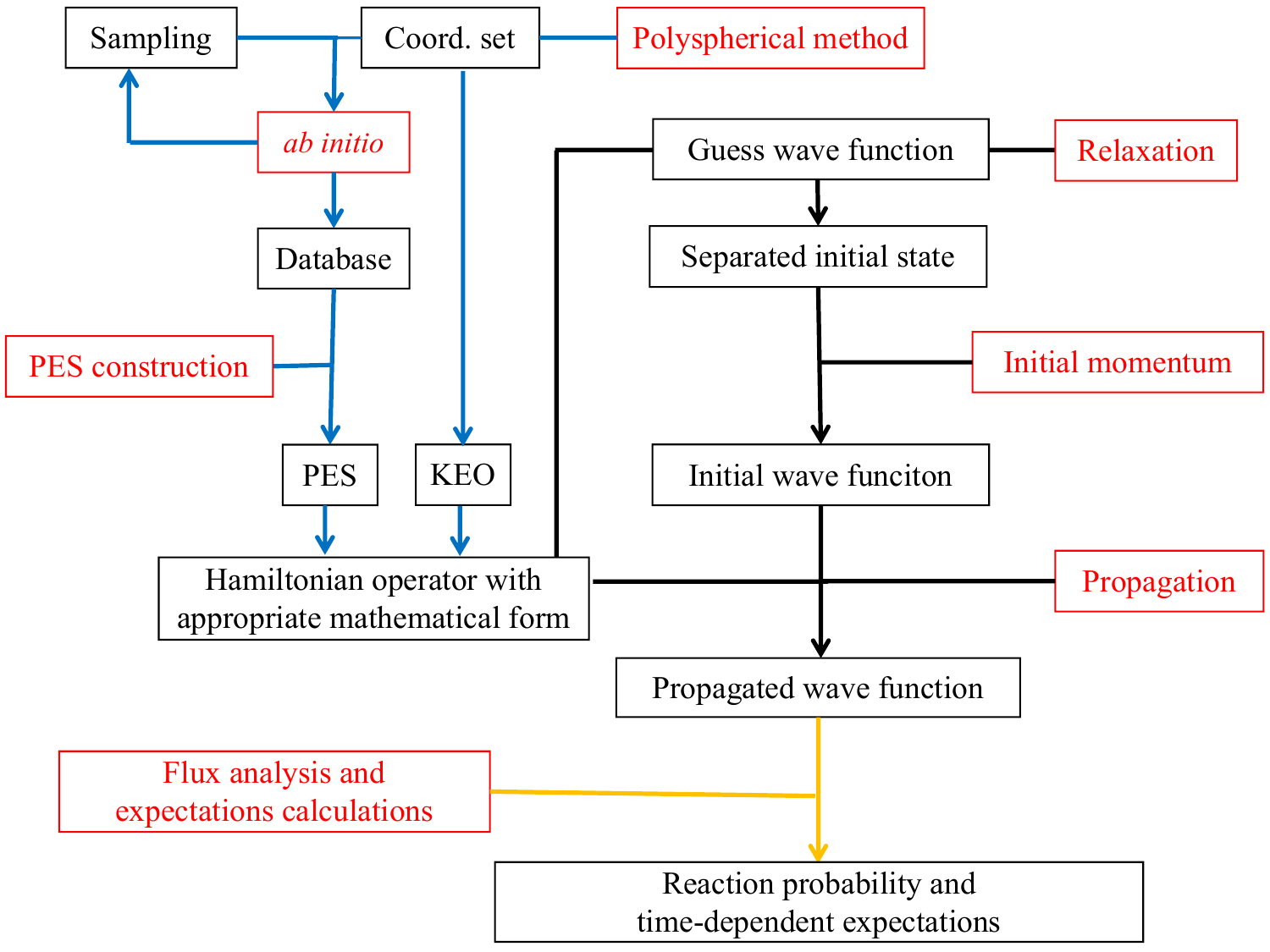}
   \caption{\figfoot}
    \label{fig:framework}
     \end{figure}

Illustrated in Figure \ref{fig:framework} are typical framework for
quantum molecular dynamics \cite{zha25:20397}, in which the variational
principle plays a fundamental role. To introduce this point, this
section presents unified geometric descriptions on the principle of
least (or stationary) action as one of postulates of the this work. 
\begin{post}[principle of stationary action]
Among all possible paths a system could take to go from one event to
another event, the path that satisfies the equation of motion (EOM) is
the one that makes the action stationary. 
\label{post:pls}
\end{post}
By providing a quantitative constraint or law on motion, that is the
EOM, the principle of least action is one of the fundamental postulates
of dynamics. In 2020, Hackl and Guaita and co-workers \cite{hac20:048}
presented a systematic geometric framework based on variational
families for real or imaginary time evolution and excitation spectra
on the K{\"a}hler manifold, where multiplication by the imaginary unit
preserves the tangent spaces, together with the non-K{\"a}hler manifold,
which has already been encountered occasionally. Moreover, it should
be noted that the so-called system mentioned in Postulate \ref{post:pls}
may or may not be a realistic system. It could be a molecular system
or other mathematical structures, such as variational target in constructing
the KEO and PES terms. In this context, all of steps shown in Figure
\ref{fig:framework} can be unified through the variational principles
for various target based on different manifold.
\begin{remark}
If the system in Postulate \ref{post:pls} is chosen to be a realistic
system, then the principle of least action become one of postulates of
dynamics. In this context, by the principle of least action, one can
derive the EOM of the realistic system, either Newton equation or
Schr{\"o}dinger equation, depending on how the action is defined.
\label{rem:pla-varia}
\end{remark}
\begin{remark}
A K{\"a}hler manifold is a special type of smooth manifold that
simultaneously carries three compatible structures, including
a complex structure, a Riemannian metric, and a symplectic structure,
which must be mutually compatible in a precise sense.
\label{remark:kaehler-manifold}
\end{remark}

To simply understand physical insight of the principle of stationary
action, we deifne $d$-dimensional configuration space $X$ as a set of
atomic coordinates for a molecular system. According to Postulate
\ref{post:pls}, from time $t_0$ to $t_1$ this $d$-dimensional system
moves along a curve $x:[t_0,t_1]\to\mathbb{R}^d$, called a Euler-Lagrange
trajectory, where the EOM satisfies the principle of stationary action,
\begin{equation}
\delta S\Big\vert_{x(t_0)}^{x(t_1)}=\left.\delta\int_{t_0}^{t_1}
L\Big(X(t),\dot{X}(t)\Big)\mathrm{d}t\right\vert_{x(t_0)}^{x(t_1)}=0,
\label{eq:000-least-action}
\end{equation}
where $L(X(t),\dot{X}(t))$ and $S$ are Lagrangian and action, respectively,
and $\dot{X}(t)$ denotes the derivative with respect to $t$.
\begin{defi}
The Lagrangian $L:TX\to\mathbb{R}$ is a real-valued function defined
on the tangent bundle $TX$ of the configuration space $X$, providing the
states of the system.
\label{def:lagrangian}
\end{defi}
It should be mentioned that $t$ can be generally seen as a parameter
(may or may not time). If $t$ represents time, then it must be real.
Thus, Postulate \ref{post:pls} requires that the
Euler-Lagrange trajectory extremizes the action with fixed end points
on each finite interval $[t_0,t_1]$. By Equation \eqref{eq:000-least-action},
the Lagrange must satisfy the Euler-Lagrange equation
\begin{equation}
\frac{\mathrm{d}}{\mathrm{d}t}\partial_{\dot{X}}
L\Big(X(t),\dot{X}(t)\Big)-\partial_XL\Big(X(t),\dot{X}(t)\Big)=0,
\label{eq:001-least-action}
\end{equation}
providing the variational character of solutions of the Lagrangian
system. The Euler-Lagrange equation given by Equation \eqref{eq:001-least-action}
is one kind of the EOM of the Euler-Lagrange trajectory. Of couese, there
exists other EOM. Based on the $d$-dimensional configuration space $X$,
introducing its $2d$-dimensional phase space one can obtain the Hamiltonian
system from the Lagrangian system by the Legendre transform associated
with a Lagrangian $L$.
\begin{defi}
The Hamiltonian $H:T^*X\to\mathbb{R}$ is a real-valued function defined
on the cotangent bundle $T^*X$ of the configuration space $X$, providing
the states of the system.
\label{def:hamiltonian}
\end{defi}
The Legendre transform yields a symplectic structure on the cotangent
bundle $T^*X$, that is the phase space, and the corresponding Hamiltonian
$H:T^*X\to\mathbb{R}$. Setting $P$ to be the fiber variable, the Legendre
transform of Equation \eqref{eq:001-least-action} gives the Hamiltonian
equations
\begin{equation}
\dot{X}(t)=\partial_YH\Big(X(t),P(t)\Big),\quad
\dot{P}(t)=-\partial_XH\Big(X(t),P(t)\Big),
\label{eq:002-least-action}
\end{equation}
which are another kind of the EOM. Therefore, the above derivations of
the Euler-Lagrange and Hamiltonian equations are typical applications
of the principle of stationary action and the variational principle.
\begin{remark}
It is worth noting concepts of tangent bundle, cotangent bundle, and
fiber variable. Based on the configuration space $X$, called base space,
the tangent bundle $TX$ describes how the system is moving in $X$ and
thus contains all points in $X$ and all possible $\dot{X}$. The fiber
variable $P$ provides additional information for defining cotangent
bundle $T^*X$ from base space $X$. The cotangent bundle describes the
potential for system motion and thus the phase space $\{X,P\}$ containing
all points in $X$ and the fiber variable $P$.
\label{rem:contangent-fiber}
\end{remark}

Mathematically, there exist action minimizing measures with any given
asymptotic direction. Such measures can be described by a homological
rotation vector. Associating each rotation vector to the action of a
minimal measure, the minimal action functional is defined as
\begin{equation}
\alpha:H_1\big(X,\mathbb{R}\big)\to\mathbb{R},
\label{eq:003-least-action}
\end{equation}
where $H_1(X,\mathbb{R})$ is the first real homology group of the
base space $X$. By its definition in Equation \eqref{eq:003-least-action},
the minimal action $\alpha$ only partially describes the full dynamics,
but concentrates on a very special part of it, that is the homological
part of the dynamics, not the full details. 
\begin{defi}
The homological part of the dynamics ignores the fine details and captures
only the large-scale, net behavior that can be detected by topology.
\label{def:hopl-part-dyn-action}
\end{defi}
To understand this point,
considering $d$-dimensional torus $\mathbb{T}^d$ (that is, product of
$d$ copies of the circle) as a periodic base space and letting $\theta$
be the Liouville $1$-form in the cotangent bundle, we consider the
cotangent bundle $T^*X=T^*\mathbb{T}^d$ with its canonical symplectic
form $\omega_0=\mathrm{d}\theta$ and suppose a convex Hamiltonian
$H:T^*\mathbb{T}^d\to\mathbb{R}$. Since $H$ is $t$-independent, all
trajectories lie on fiberwise convex $(2d-1)$-dimensional hypersurfaces
$\Sigma$ with the symplectic form $\omega_0$ vanishing, called Lagrangian
submanifolds. Here, we should emphasize again that $t$ may or may not
time. By the pull-back of the Liouville form $\theta$ to the Lagrangian
submanifold, one can introduce a Liouville class that determines the
Lagrangian submanifold uniquely. Letting $U$ be a fiberwise convex
subset of $T^*\mathbb{T}^d$, every cohomology class can be represented
as the Liouville class of some Lagrangian submanifold in $U$ and
actually by a Lagrangian section contained in $U$. Thus, Lagrangian
sections actually belong to symplectic geometry. Further, the above
result allows symplectic descriptions of seemingly non-symplectic
objects. The stable norm of a Riemannian metric on $\mathbb{T}^d$ and
the minimal action of a convex Lagrangian both admit a symplectically
invariant description. In summary, the long-term behavior of a dynamical
system can be encoded in the geometry of Lagrangian submanifolds. This
encoding is powerful and unifies concepts from Riemannian geometry and
dynamics under a single symplectic framework.
\begin{remark}
If the parameter $t$ is time, we then connect the motion of a
particle moving on a torus with symplectic geometry which is the
mathematical language of Hamiltonian dynamics. We consider a periodic
configuration space $\mathbb{T}^d$, such as that of a bounded molecule
with $d$ vibrational modes. The canonical symplectic form $\omega_0$
ensures that time evolution is a canonical transformation. Since $H$
is time-independent, the energy is conserved implying that the molecular
motion is confined to a $(2d-1)$-dimensional surface $\Sigma$. The
symplectic form $\omega_0$ is a $2$-form that measures the area of a
parallelogram surface in phase space projected onto the position-momentum
plane. If $\omega_0=0$ on $\Sigma$, the $(2d-1)$-dimensional surface $\Sigma$
is maximally isotropic. Thus, $\Sigma$ becomes a ``sheet'' and cuts
through the phase space in a way where position and momentum are not
independent. The Liouville form $\theta$ can be integrated to define a
cohomology class, called Liouville class, which essentially stores the
quantization conditions. A Liouville class determines a Lagrangian
submanifold uniquely. Different Lagrangian submanifolds correspond to
different sets of actions. Thus, the system motions are all related to
finding a specific Lagrangian submanifold that minimizes a certain
quantity (say action). In other words, the minimal action $\alpha$
only describes the homological part of the dynamics, not the full
details.
\label{rem:pla-alpha-explain}
\end{remark}
\begin{remark}
We are interested in more closely looking the concept given by Definition
\ref{def:hopl-part-dyn-action}.
Among the paths of the system motion, the principle of least action determines
the most probable path, or called classical path, while the uncertainty
relationship dictates that position and momentum are not independent.
Consequently, the system explores a superposition of all possible
paths, and the interference between these paths gives rise to so-called
quantum phenomena. In this way, the system obeys quantum mechanics in
the path-integral formulation.
\label{rem:qm-path-in}
\end{remark}

With the path integral fashion of Remark \ref{rem:qm-path-in}, one can
derive the EOM for quantum phenomena, called the Schr{\"o}dinger equation,
through the principle of least action (Postulate \ref{post:pls}). The
Schr{\"o}dinger equation is another kind of EOM, distinct from Equations
\eqref{eq:001-least-action} and \eqref{eq:002-least-action}. As given
in Remarks \eqref{rem:pla-alpha-explain} and \ref{rem:qm-path-in}, the
system explores all possible paths in the space. The probability
amplitude for the transition $X(t)\leftarrow X(t_0)$, called
propagator, is given by the path integral
\begin{equation}
K\big(X(t),X(t_0)\big)=\int\mathcal{D}\big[X(t)\big]
\exp\Big(\mi S\big[X(t)\big]\Big),
\label{eq:path-int-000}
\end{equation}
where $\mathcal{D}[X(t)]$ denotes integration over all paths connecting
$X(t)$ and $X(t_0)$. By Equation \eqref{eq:path-int-000}, in a very
short time interval $\epsilon>0$, the propagator from $X(t)$ to $X(t+\epsilon)$
can be evaluated explicitly. Due to the local homogeneity and isotropy
of space together with the local homogeneity of time, the Lagrangian
is given by $2L(X,\dot{X})=m\dot{X}^2-2V(X)$, where $V(X)$ is external
field. According to Equation \eqref{eq:000-least-action}, then the
action over the short interval is $2S\approx m(X(t+\epsilon)-X(t))^2/\epsilon-2\epsilon
V((X(t)+X(t+\epsilon))/2)$. Substituting $S$ into Equation \eqref{eq:path-int-000},
one can obtain the propagator of interval $t\to t+\epsilon$. Integrating
the propagator of $t\to t+\epsilon$ against the wave function at $t$,
one can obtain the wave function at $t+\epsilon$,
\begin{align}
\Psi\big(X(t+\epsilon)\big)&=\int_{-\infty}^{\infty}K\big(X(t+\epsilon),X(t)\big)
\Psi\big(X(t)\big)\mathrm{d}X
\allowdisplaybreaks[4] \nonumber \\
&=\sqrt{\frac{m}{2\mi\pi\epsilon}}\int_{-\infty}^{\infty}
\exp\left\{\mi\left[\dfrac{m\xi^2}{2\epsilon}-\epsilon
V\left(X(t+\epsilon)-\frac{\xi}{2}\right)\right]\right\}
\Psi\big(X(t+\epsilon)-\xi\big)\mathrm{d}\xi,
\label{eq:path-int-002}
\end{align}
where $\xi=X(t+\epsilon)-X(t)$. For small $\epsilon$, the exponential
of Equation \eqref{eq:path-int-002} oscillates rapidly unless $\xi$ is
small enough. Expanding $\Psi(X(t+\epsilon)-\xi)$ and $V(X(t+\epsilon)-\xi/2)$
in powers of $\xi$, Equation \eqref{eq:path-int-002} further becomes
\begin{equation}
\Psi\big(X(t+\epsilon)\big)=\Psi\big(X(t)\big)-\mi\epsilon
\left(-\dfrac{1}{2m}\partial_X^2\Psi+V(X)\Psi\right)+
\mathcal{O}\big(\epsilon^2\big).
\label{eq:path-int-003}
\end{equation}
Rearranging Equation \eqref{eq:path-int-003} and taking the limit
$\epsilon\to0^+$, we have
\begin{equation}
\mi\partial_t\Psi=\left(-\dfrac{1}{2m}\partial_X^2+V(X)\right)\Psi=
\big(\hat{T}(\dot{X})+V(X)\big)\Psi=\hat{H}\Psi,
\label{eq:path-int-004}
\end{equation}
which is precisely the Schr{\"o}dinger equation. The operator $\hat{H}$
in Equation \eqref{eq:path-int-004} is called Hamiltonian operator
which is prerequisite of quantum dynamics. In this context, the
Schr{\"o}dinger equation is not a postulate but a consequence of
the path integral formulation and the principle of least action.
Solving the Schr{\"o}dinger equations for both electorns and nuclei
is the core purpose of chemical dynamics.
\begin{remark}
The expressions of Lagrangian and Hamiltonian are $L=T-V$ and $H=T+V$,
respectively, where $T$ originates from the kinetic energy of a particle
moving freely along geodesics and depends on the geometric properties
of space, while $V$ is the external potential, which depends on the
external field of the motion.
\label{rem:hamil-keo-pes}
\end{remark}
\begin{remark}
The expression $L(X,\dot{X})=m\dot{X}^2/2-V(X)$ is arisen from symmetry
properties of the spacetime through Galileo principle of equivalence.
We will turn to the principle of equivalence in a general fashion later.
\label{rem:hamil-keo-99-pes}
\end{remark}

%-------------------------------------
\subsection{Mountain Pass Theorem\label{sec:mpt-math}}

\begin{figure}[h!]
   \centering  
    \includegraphics[width=18cm]{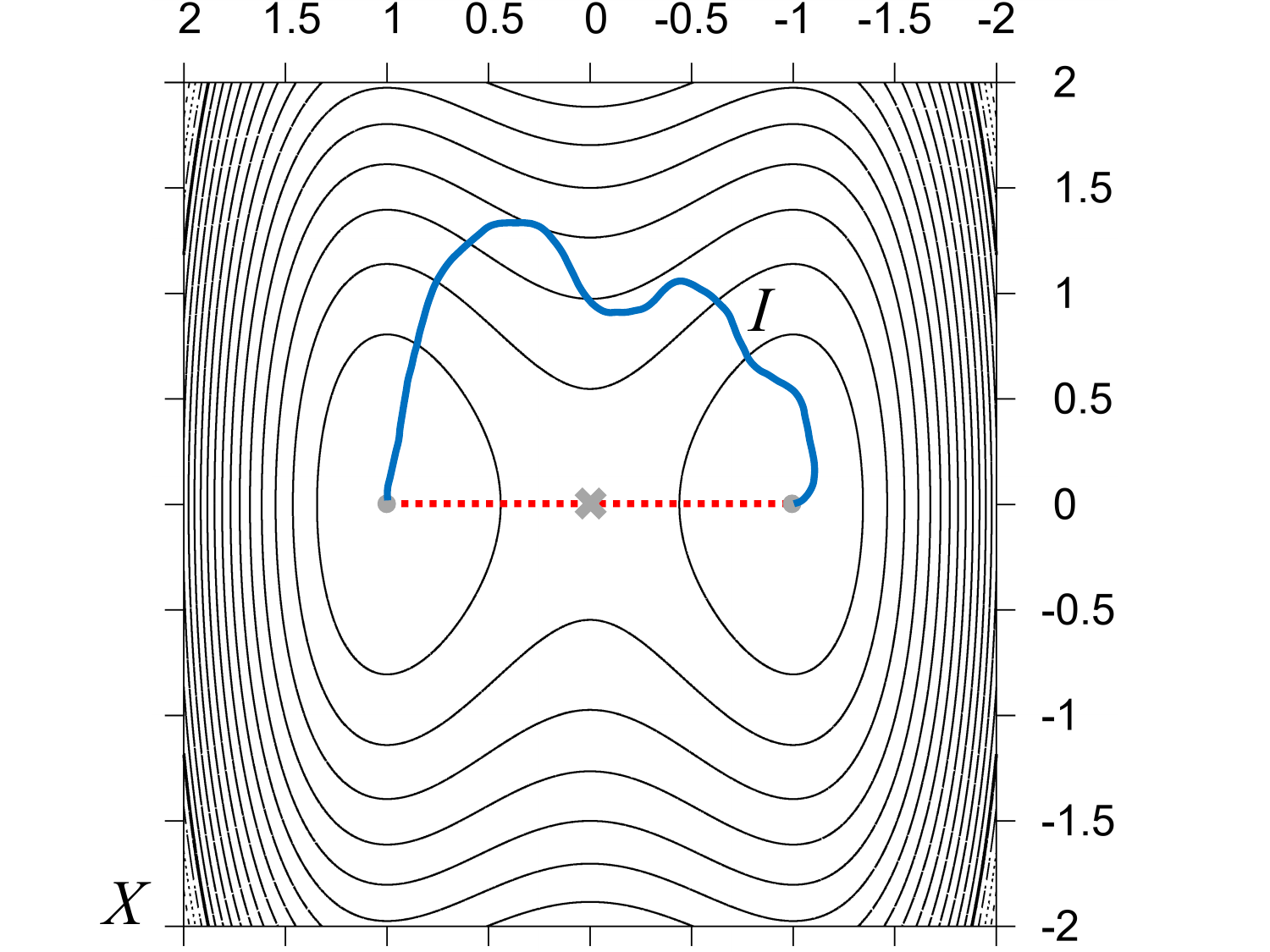}
     \caption{\figfoot}
      \label{fig:mountain-pass}
       \end{figure}

In Section \ref{sec:math-frame}, we have simply described the principle
of least action given in Postulate \ref{post:pls}, providing the importance
in deriving the EOM. In this section, we consider the variation itself
and describe the mountain pass theorem \cite{amb73:349} which has been
applied frequently to establish the existence of critical points for
functionals. Considering a real Banach space $X$ and a function
$I:X\to\mathbb{R}$ of class $C^1$ and setting
$I_k=\{x\in X\vert I(x)<k,\forall k\in\mathbb{R}\}$, let $Y$ be
a subset of $X$ and $k$ a positive number such that $I_k\cap Y\neq\emptyset$
and then define $\tilde{Y}=\{x\in X\vert\mathrm{dist}(x,Y)<1\}$.
\begin{remark}
The configuration space is a Banach manifold, which is locally modeled
on a real Banach space. A real Banach space serves as the local model
for the configuration space.
\label{rem:banach-space-conf}
\end{remark}
Pucci and Serrin \cite{puc85:142} have proved that, the set $I_k\cap\tilde{Y}$
contains a point $y$ such that either $I(y)<0$ or $\Vert I'(y)\Vert<2k$,
where $I'(y)$ is the derivative of function $I(X)$ at point $y$. Then,
Pucci and Serrin \cite{puc85:142} proved the mountain pass theorem.
\begin{prop}[mountain pass theorem]
Let function $I$ satisfy that (1) $\exists a,r,R\in\mathbb{R}$ such
that $0<r<R$ and $I(x)\geq a$ for $\forall x\in A=\{x\in X\vert
r<\Vert x\Vert<R\}$, (2) there are $I(0)\leq a$ and $I(e)\leq a$ for
some $e\in X$ with $\Vert e\Vert\geq R$, and (3) any sequence
$\{x_n\in X\}_{n=1}^{\infty}$ such that $\lim I(x_n)\geq a$ and
$\lim I'(x_n)=0$ possesses a convergent subsequence. There exists a
critical point $x_0\in X$, different from $0$ and $e$, with critical
value $I(x_0)=c\geq a$, in addition $x_0\in A$ when $c=a$.
\label{theo:mpt}
\end{prop}
In the case $c=a$, there are an infinite number of critical points in
set $A$ \cite{puc85:142}. Moreover, if the inequality $I(x)\geq a$ (in
condition (1)) is replaced by $I(x)>a$, then there is a critical point
with critical value $c>a$. Figure \ref{fig:mountain-pass} illustrates
the mountain pass theorem in Proposition \ref{theo:mpt} by a 2D smooth
function in configuration space $X$ as an example, where solid lines
are contour lines. Two distinct circle symbols represent two minima.
The symbol $\times$ represents the saddle point. The straight red
dotted line represents the shortest distance between two minima. The
curved blue line $I$ represents a general path.

To understand Proposition \ref{theo:mpt}, letting
$G=\{g\in C([0,1],X)\vert g(0)=0,\;g(1)=e\}$, for
$\forall g\in G$ we have $\max_{0\leq t\leq1}I\big(g(t)\big)\geq a$
and thus obtain
\begin{equation}
\inf_{g\in G}
\max_{0\leq t\leq1}I\big(g(t)\big)=c\geq a.
\label{eq:mpt-ts-opt-000}
\end{equation}
If $c>a$, a critical value of $I(x)$ is $c$ and corresponding critial
point $x_0$ must be different from $0$ and $e$. If $c=a$, we should
first define the annulus $A'=\{x\in X\vert r'<\Vert x\Vert<R'\}$ with
$0<r<r'<R'<R$. Considering the set $Y=\{x\in X\vert r'+1<\Vert x\Vert<R'-1\}$
together with a fixed number $k>0$. There exists some path $g\in G$
such that $I(g(t))<k$ for $\forall t\in[0,1]$. By continuity of
$t\to\Vert g(t)\Vert$ and the fact that $\Vert g(0)\Vert=0$ and
$\Vert g(e)\Vert>R'$, there exists a value $t_0\in[0,1]$ such that
$g(t_0)\in Y$. This implies that $g(t_0)\in I_k\cap Y\neq\emptyset$.
Therefore, since we have $\tilde{Y}=A'$ and $I(x)\geq0$ for $x\in A'$,
there exists a point $y\in I_k\cap A'$ such that $\Vert I'(y)\Vert<2k$.
Because $k>0$ is arbitrary, choosing the sequence $\{2^{-n}\}_{n=1}^{\infty}$
for $k$, one can find the existence of a sequence $\{x_n\}_{n=1}^{\infty}$
in $A'$ with $a\leq I(x_n)<2^{-n}$ and $\Vert I'(x_n)\Vert<2^{1-n}$.
By the condition (3), one can extract from $\{x_n\}_{n=1}^{\infty}$ a
subsequence converging to some $x_0\in A'\subset A$. By the regularity
of $I(x)$ one can have $I(x_0)=a$ and $I'(x_0)=0$. 
\begin{corop}
If condition (3) in Proposition \ref{theo:mpt} holds and function $I(x)$
has two different local minima, then $I(x)$ possesses a third critial
point.
\label{coro:mpt-coro}
\end{corop}
To see this result, changing variables the local minimum point with
higher value falls at the origin $0$. The other will then be denoted
by $e$, so that $\Vert e\Vert>0$ and $I(e)\leq I(0)$. We can find an
open ball $B_R$ with center $0$ and radius $R$ satisfying $0<R<\Vert
e\Vert$, such that $I(x)\geq I(0)$ with $\forall x\in B_R$. Choosing
$r=R/2$, the critical value at the third point is greater than or equal
to $I(0)$, that is as large as either local minimum.
\begin{remark}
From the viewpoint of chemistry, the mountain pass theorem provides a
mathematical foundation for the existence of transition states. Generally,
minima on the energy profile correspond to stable intermediates, and
the transformation between them defines a chemical reaction. Along the
minimum-energy path connecting these two intermediates, as illustrated
in Figure \ref{fig:mountain-pass}, the variation of the potential
functional, or free energy functional if statistical effects are
considered, must give rise to at least one barrier, which corresponds
to the transition state.
\label{rem:pass-mountain-th}
\end{remark}

Further, Pucci and Serrin \cite{puc85:142} extended the mountain pass
theorem to finite-dimensional generalization by additionally assuming
that (4) there exist two numbers $r>0$ and $a$ such that $I(x)\geq a$
for $\forall x\in S_R=\{x\in X\vert\Vert x\Vert=R\}$ and (5) there
exist $I(0)\leq a$ and $I(e)\leq a$ for some $e$ with $\Vert e\Vert>R$.
Under condition (3), Pucci and Serrin \cite{puc85:142} found that there
exists a critical point $x_0\in X$, different from $0$ and $e$, with
critical value $I(x_0)=c\geq a$. If $c=a$, the critical point can be
taken with $\Vert x_0\Vert=R$. Thus, one can find that there exists a
critical point in the closure of an annulus $A=\{x\in X\vert r<\Vert x\Vert<R\}$.
Since the distance from the boundary of $A$ to $S_R$ can be taken
arbitrarily small, a standard compactness argument shows that there is
a critical point $x_0\in S_R$ with $I(x_0)=0$. To understand this result,
one should turn to Equation \eqref{eq:mpt-ts-opt-000}. The case $c>a$
can be treated exactly as before. When $c=a$, if there is a point
$x_0\in S_R$ with $I'(x_0)=0$, we can directly obtain expected
results. Now, we should assume that $I'(x)\neq0$ for $\forall x\in S_R$.
Considering an arbitrary point $y\in S_R$, considtion (4) predicts either
$I(y)>a$ or $I(y)=a$. If $I(y)>a$, there exists an open pluri-interval
$J_y$ with center $y$ such that $I(x)>a$ for $\forall x\in J_y$. If
$I(y)=a$, the vector $I'(y)$ has zero projection onto the tangent space
to $S_R$ at $y$ implying that $I'(y)$ is orthogonal to $S_R$ at $y$.
Since $I'$ is continuous there is an open pluri-interval $J_y$ with
center $y$ such that $I(x)>a$ and $x\cdot I'(x)\neq0$ either for each
$x\in B_y=J_y\cap\{z\in X\vert\Vert z\Vert<R\}$ or for each
$x\in C_y=J_y\cap\{z\in X\vert\Vert z\Vert>R\}$. From the open covering
$\{J_y\vert y\in S_R\}$ of $S_R$ one can extract a finite subcovering
$\{J_{y_1},\cdots,J_{y_M}\}$. The set $\cup_{i=1}^MJ_{y_i}$ contains
an open annulus $A$ around $S_R$. Let us consider the set
$\tilde{Y}=\cup_{i=1}^M(Z_i\cap A)$, where $Z_i$ denotes $J_{y_i}$ if
$I(y_i)>a$ or $B_{y_i}$ or $C_{y_i}$ if $I(y_i)=a$ and $I'(y_i)$ is
the inner normal or the outer normal to $S_R$, respectively. If $Z_i$
and $Z_j$ are adjacent, then they cannot be of the type $B_{y_i}$ and
$C_{y_j}$ or $C_{y_i}$ and $B_{y_j}$, respectively. Evidently, for
$\forall x\in\tilde{Y}$ and moreover $S_R\subset\tilde{Y}$, we have
$I(x)>a$. Putting $Y_{\delta}=\{x\in\tilde{Y}\vert\mathrm{dist}(x,X
\setminus\tilde{Y})>\delta\}$, we may choose $\delta>0$ so small such
that $Y_{\delta}$ separates $0$ and $e$. Therefore, there exists a
critical point in the closure of the annulus $A$. Since the distance
from the boundary of $A$ to $S_R$ can be taken arbitrary small, there
is a critical point $x_0\in S_R$ with $I(x_0)=a$. Having the mountain
pass theorem \cite{amb73:349,puc85:142}, Pucci and Serrin \cite{puc87:115}
further shown that set of the critical points, called the critical set,
must have a well-defined structure. In particular, if the underlying
Banach space is infinite dimensional then either the critical set
contains a saddle point or the set of local minima intersects at least
two components of the set of saddle points. Pucci and Serrin
\cite{puc87:115} also established related conclusions for the finite
dimensional case.

The mountain pass theorem has become a tool for proving the existence
of critical points of nonlinear functionals. Such critical points are
solutions of semilinear partial differential equations (PDEs) or
periodic solutions of nonlinear Hamiltonian systems. Choi and McKenna
\cite{cho93:417} proposed an algorithm which is globally convergent and
which will always converge to a solution with the required mountain pass
property. Moreover, the mountain pass theorem and global homeomorphism
theorem are important tools for nonlinear problems in analysis \cite{kat94:189}.
By the mountain pass theorem, Katriel \cite{kat94:189} derived the
global homeomorphism theorems for certain topological and metric spaces,
generalizing known global homeomorphism theorems for mappings between
Banach spaces.
\begin{remark}
Mountain pass theorems and global homeomorphism theorems are among the
important tools for dealing with nonlinear problems in analysis. The
mountain pass theorems predict existence of critical points in nonlinear
functions. The global homeomorphism theorems predict conditions under
which a map is not just locally a homeomorphism, but globally a
homeomorphism over its entire domain.
\label{rem:mpt-non-linea-theo}
\end{remark}
With these results, Hill and Humphreys \cite{hil00:731} considered a
variational formulation for a set of coupled PDEs in the form
$\nabla^2u+f(u)=0$ and then proved the existence of at least one
nontrivial solution. We will see this by a special example in Section
\ref{sec:elec-struc}. Ruppen \cite{rup16:89} generalized the mountain
pass theorem and shown the existence of multiple critical points for
an even functional $f:\mathbb{H}\to\mathbb{R}$, where $\mathbb{H}$ is
the Hilbert space. Ruppen \cite{rup16:89} applied these results to a
Schr{\"o}dinger-type equation of the semi-linear eigenvalue problem.
Since the above mountain pass theorem for scalar functionals is a
foundation of the minimax methods in variational analysis, Molho and
co-workers \cite{bed11:569} extended it to the class of functions
$f:\mathbb{R}^n\to\mathbb{R}^m$ and proved the existence of a critical
point of $f$ as a solution of a minimax problem which consists of an
inner vector maximization problem and of an outer set-valued minimization
problem.

A perfect application of the minimax optimization is a two-player game,
yielding a foundation of generative artificial intelligence (GenAI)
whose capability can be further extended with tool use, memory, and
autonomous planning. Mathematically, the minimax optimization in GenAI
is a zero-sum game played between the generator and discriminator,
which are two competing maps on a data manifold $M\subset\mathbb{R}^d$.
The generator $G:\mathcal{Z}\to\mathbb{R}^d$ defines a parameterized
family of maps that push forward a simple latent distribution to a
generated distribution, where $\mathcal{Z}$ is the latent space
equipped with a simple prior distribution, typically a standard
Gaussian or uniform distribution. The discriminator $D:\mathbb{R}^d\to[0,1]$
acts as a trainable classifier that approximates the geometric distance
between the data distribution (probability measure) supported on $M$
and the generated distribution via the Wasserstein distance. With the
minimax loss function $L(G,D)$, the minimax game $\min_G\max_DL(G,D)$
seeks a saddle point at the Nash equilibrium. In this case, the generator
manifold coincides with $M$ up to a set of measure zero. while the
discriminator decision boundary becomes geometrically indistinguishable,
meaning that the push-forward measure matches the data distribution
exactly. The optimization landscape is thus governed by the curvature
and intrinsic dimension of $M$, implying that the success of minimax
optimization depends critically on both the Riemannian metric and the
topological complexity of $M$. In other words, the minimax optimization
is not merely an algorithm, but also a geometric procedure. Such procedure
gradually deforms support of the latent distribution until it isometrically
embeds into $M$, with the discriminator serving as a trainable gauge that
measures the geodesic discrepancy between the two probability measures
in the distribution space equipped with the Wasserstein metric.

For example, in a generative adversarial networks (GAN), the training
process is a minimax game, where the generator wants to minimize the
detective ability to give fake from real while the discriminator wants
to maximize its ability to catch fakes. The resulting GAN model corresponds
to a Nash equilibrium, which is a saddle point. For the second example,
the minimax optimization has significant and rapidly growing applications
in the development of AI agents for science, going far beyond its application
to game-playing problem. In general, by two primary ways the minimax
optimization has been applied to the development of AI agents. The
first way is strategic decision-making in games, which is the most
classic application, as described above for GenAI. In such minimax
algorithm, there are two AI models. The one model maximizes the target,
and the other one minimizes it. The other way is mult-agent reinforcement
learning to achieve high-performance. The minimax approaches are integral
to training AI agents to operate in complex and competitive environments
with other agents. This is a step beyond single-agent system, as the
environmental dynamics are influenced by other agents. To tackle more
complex problems, developers often have to combine minimax approaches
with other methods. As given above, unlike traditional AI models which
evaluate existing databases, GenAI models and AI agents learn the
underlying distribution of valid molecular structures and can generate
entirely new properties that satisfy multiple constraints simultaneously.
Integrating with numerical simulations, these models act as proposers
that iteratively refine candidates based on feedback from chemistry-based
evaluations by either calculations or experiments. This development
accelerates the design-experiment-test-analyze cycle, reducing both
time and cost for discovery.
\begin{remark}
Approaches based on minimax optmization are revolutionizing computational
chemistry by enabling the rapid exploration of chemical space and the
inverse design of molecules with targeted properties.
\label{rem:general-ai}
\end{remark}
\begin{remark}
As a specific class of GenAI, large language model (LLM) is an important
engine that powers modern AI agents for science, as it provides the
linguistic and reasoning backbone. In general, LLMs process natural
language, generate coherent text, and store vast amounts of world
knowledge encoded in their parameters.
\label{rem:general-ai-llm}
\end{remark}

%----------------------------------------
\subsection{Electronic Structure\label{sec:elec-struc}}

Now, we consider the electronic structure theory \cite{sza12:108,goi18:e1341,li20:9951,sve23:e1666}
through the variantional principle. This is the basis for building the
PES term \cite{zha25:20397}. For the molecular system in a static field
or in a field-free environment, the electronic Hamiltonian $\hat{H}_{\mathrm{ele}}$
is time-independence because the state of an electron is only determined
by the Coulomb interaction from the nuclei and other electrons. In this
case, the temporal coordinate is separable from the electronic SE indicating
that core idea of the electronic structure theory is to solve the eigenvalue
equation of $\hat{H}_{\mathrm{ele}}$, that is
$\hat{H}_{\mathrm{ele}}\vert\phi\rangle=\vert\phi\rangle\epsilon$.
\begin{defi}
If an operator $\hat{O}$ acting on a state vector $\vert\phi\rangle$
produces the vector itself multiplied by a constant $\epsilon$, then
$\vert\phi\rangle$ is an eigenvector of the operator $\hat{O}$ with an
eigenvalue $\epsilon$.
\label{def:eigenvalue-problem}
\end{defi}
According to Definition \ref{def:eigenvalue-problem}, solving the
eigenvalue equation is called an eigenvalue problem. Usually,
without loss of generality, we choose the eigenvectors of Hermitian
operators (with $\hat{O}^{\dagger}=\hat{O}$) to be normalized
$\langle\phi\vert\phi\rangle=1$.
\begin{prop}
The eigenvalues of a Herimitian operator are real, while its eigenvectors
are orthogonal.
\label{prop:eigen-hermi-00}
\end{prop}
Proof of Proposition \ref{prop:eigen-hermi-00} is direct if one is aware
of the fact that
$\epsilon^*\langle\phi\vert\phi\rangle=\langle\phi\vert\hat{O}^{\dagger}\vert\phi\rangle=
\langle\phi\vert\hat{O}\vert\phi\rangle=\langle\phi\vert\phi\rangle\epsilon$
for $\langle\phi\vert\phi\rangle\neq0$ and
$\langle\phi_i\vert\phi_j\rangle(\epsilon_i-\epsilon_j)=
\epsilon_i\langle\phi_i\vert\phi_j\rangle-\langle\phi_i\vert\phi_j\rangle\epsilon_j=
\langle\phi_i\vert\hat{O}^{\dagger}\vert\phi_j\rangle-
\langle\phi_i\vert\hat{O}\vert\phi_j\rangle=0$ means $\langle\phi_i\vert\phi_j\rangle=\delta_{ij}$.
Since $\hat{H}_{\mathrm{ele}}$ is a Herimitian operator, its eigenvectors
form a complete set. Moreover, Proposition \ref{prop:eigen-hermi-00}
indicates that the electronic eigenvectors are orthonormal. Thus,
transforming basis sets through some unitary transformation, one
can diagonalize a general representation of $\hat{H}_{\mathrm{ele}}$.
Such a diagonalization for matrix representation of $\hat{H}_{\mathrm{ele}}$
is the main purpose of the electronic structure theory. In practical
calculations, the dimension $N$ of the basis set is finite leading to
at most $N$ eigenvalues and eigenvectors. As $N\to\infty$, the basis
set becomes complete and the diagonalization become accuracy. Currently,
there exist numerous algorithms for diagonalizing matrix representation
of the Hermitian operator. Most of these algorithms were developed on
the basis of the varialtional principle.

Next, let us consider general variation method to solve the electronic
eigenvalue equation $\hat{H}_{\mathrm{ele}}\vert\phi\rangle=\vert\phi\rangle\epsilon$.
We are interested in approximately solving the electronic eigenvalue
equation because electronic problems cannot be solved exactly. Since
$\hat{H}_{\mathrm{ele}}$ is Hermitian, Proposition \ref{prop:eigen-hermi-00}
indicates that its exact eigenvalues can be given in the order
$\epsilon_1^{(\mathrm{exact})}\leq\epsilon_2^{(\mathrm{exact})}\leq
\cdots\leq\epsilon_i^{(\mathrm{exact})}\leq\cdots$. Ignoring ionization
and affinity, electrons in a molecular system always move within a
finite region. Moreover, the eigenvalues $\{\epsilon_i^{(\mathrm{exact})}\}$
are discrete and the eigenvectors are orthonormal
$\langle\phi_i\vert\phi_j\rangle=\delta_{ij}$ according to Proposition
\ref{prop:eigen-hermi-00}. Multiplying by a bra vector on the left,
the eigenvalue equation beocmes
$\langle\phi_j\vert\hat{H}_{\mathrm{ele}}\vert\phi_i\rangle=
\langle\phi_j\vert\phi_i\rangle\epsilon_i^{(\mathrm{exact})}=
\delta_{ji}\epsilon_i^{(\mathrm{exact})}$. Now, we can give
time-independent variational principle for solving the eigenvalue
equation of $\hat{H}_{\mathrm{ele}}$.
\begin{prop}[time-independent variational principle for ground state]
If a normalized state vector vanishes at infinity, then the expectation
of the Hamiltonian is an upper bound to the exact smallest eigenvalue.
\label{prop:vari-princi-000}
\end{prop}
In other words, if state vector satisfies $\langle\tilde{\phi}\vert\tilde{\phi}\rangle=1$
and appropriate boundary conditions, then Proposition \ref{prop:vari-princi-000}
indicates that $\langle\tilde{\phi}\vert\hat{H}_{\mathrm{ele}}\vert
\tilde{\phi}\rangle\geq\epsilon_1^{(\mathrm{exact})}$. In some cases,
the trial state is often referred as ansatz. The proof of Proposition
\ref{prop:vari-princi-000} is simple and direct. Since
$\hat{H}_{\mathrm{ele}}$ is Hermitian, its eigenvactor set
$\{\vert\phi_i\rangle\}$ is a complete and orthonormal. Taking
$\{\vert\phi_i\rangle\}$ as basis set, one can decompose
$\vert\tilde{\phi}\rangle$ and have expansion
\begin{align}
\Big\langle\tilde{\phi}\big\vert\hat{H}_{\mathrm{ele}}\big\vert\tilde{\phi}\Big\rangle&=
\sum_{ij}\Big\langle\tilde{\phi}\big\vert\phi_j\Big\rangle
\Big\langle\phi_j\big\vert\hat{H}_{\mathrm{ele}}\big\vert\phi_i\Big\rangle
\Big\langle\phi_i\big\vert\tilde{\phi}\Big\rangle=
%\sum_{ij}\Big\langle\tilde{\phi}\big\vert\phi_j\Big\rangle
%\delta_{ji}\epsilon_i
%\Big\langle\phi_i\big\vert\tilde{\phi}\Big\rangle=
\sum_i\big\vert\Big\langle\phi_i\big\vert\tilde{\phi}
\Big\rangle\big\vert^2\epsilon_i^{(\mathrm{exact})}
\allowdisplaybreaks[4] \nonumber \\
&\geq
\epsilon_1^{(\mathrm{exact})}\sum_i\big\vert\Big\langle\phi_i
\big\vert\tilde{\phi}\Big\rangle\big\vert^2=
\epsilon_1^{(\mathrm{exact})}\Big\langle\tilde{\phi}\big\vert\tilde{\phi}\Big\rangle=
\epsilon_1^{(\mathrm{exact})},
\label{eq:decomp-relation-000}
\end{align}
where the orthonormal and complete conditions for basis set. Then,
Equation \eqref{eq:decomp-relation-000} directly leads to Proposition
\ref{prop:vari-princi-000}. Therefore, the essence of variational
electronic structure methods is the reasonable formulation of trial
state from which these methods derive their name \cite{sza12:108,goi18:e1341,li20:9951,sve23:e1666}.
Furthermore, since electrons are spin-$1/2$ fermions, the trial functions
must be antisymmetric with respect to the interchange of any two electrons,
say the Slater determinant. Some of famous methbods, such as HF, MCSCF,
CI, MRCI, the coupled cluster (CC) theory and its multi-reference version
(denoted by MRCC), employ trail functions that are linearly decomposed by
a set of Slater determinants \cite{zha25:20397,sza12:108,goi18:e1341,li20:9951,sve23:e1666}.

To easily understand Proposition \ref{prop:vari-princi-000}, we now
consider a computationally inefficient but a direct procedure, called
linear variational approach or Rayleigh-Ritz variation method
\cite{sza12:108,goi18:e1341,li20:9951,sve23:e1666} to solve the eigenvalue
equation of $\hat{H}_{\mathrm{ele}}$ by finding the roots of the secular
determinant. It is the simplest variational approach.
\begin{prop}
By the linear variational approach defined by an $N$-dimensional basis
set, solving the secular determinant once approximately gives upper
bounds to the eigenvalues of the ground state and $N-1$ excited states
and $N$ mutually orthogonal eigenfunctions.
\label{prop:ratz-vari-prin-0}
\end{prop}
With an $N$-dimensional basis set, the eigenvalue equation
$\hat{H}_{\mathrm{ele}}\vert\phi\rangle=\vert\phi\rangle\epsilon$ can
be represented by matrix expression
$(\mathbf{H}_{\mathrm{ele}}-\mathbf{I}\epsilon)\boldsymbol{\phi}=\mathbf{0}$,
where $\mathbf{H}_{\mathrm{ele}}$ and $\mathbf{I}$ are $N\times N$
Hermitian and unity matrices, respectively, while $\boldsymbol{\phi}$
and $\mathbf{0}$ are $N\times1$ representation matrix of $\vert\phi\rangle$
and null matrix, respectively. Since $(\mathbf{H}_{\mathrm{ele}}-\mathbf{I}\epsilon)^{-1}$
does not exist when its determinant is equal to zero, the eigenvalue
equation can have a nontrivial solution $\boldsymbol{\phi}\neq\mathbf{0}$
only when $\det(\mathbf{H}_{\mathrm{ele}}-\mathbf{I}\epsilon)=0$. This
is called the secular determinant which is a polynomial of degree $N$
in $\epsilon$. Thus, solving the secular determinant, one can have $N$
roots $\{\epsilon_i\}_{i=1}^N$, namely the eigenvalues of the matrix
$\mathbf{H}_{\mathrm{ele}}$. Once we have got the eigenvalues we can
obtain the corresponding eigenvectors by substituting each $\epsilon_i$
into the eigenvalue equation and solving the resulting equation for the
$i$-th eigenvector $\boldsymbol{\phi}_i$ within a multiplicative constant.
Then, we should finally normalize $\boldsymbol{\phi}_i$. According to
Proposition \ref{prop:eigen-hermi-00}, since $\hat{H}_{\mathrm{ele}}$
and then $\mathbf{H}_{\mathrm{ele}}$ are Hermitian, the eigenvalues are
real and the eigenvectors are orthogonal. In this way, we diagonalize
$\mathbf{H}_{\mathrm{ele}}$ to the matrix $\boldsymbol{\epsilon}=
\mathbf{U}^{\dagger}\mathbf{H}_{\mathrm{ele}}\mathbf{U}=\mathbf{U}^{-1}\mathbf{H}_{\mathrm{ele}}\mathbf{U}$
with the $i$-th diagonal element $\epsilon_i$, where $\mathbf{U}=
(\boldsymbol{\phi}_1\;\cdots\;\boldsymbol{\phi}_N)$ is an $N\times N$
unitary matrix. If we restrict the above description in the coordiante
representation, then the state vector is represented by the wave
function. Using a set of $N$ functions as basis set, one can similarly
find the eigenvalue equation in the matrix form. Solving it produces
$N$ eigenvalues and corresponding eigenfunctions. This is the core
idea of linear variational approach.
\begin{remark}
H{\"u}ckel molecular orbital theory is the simplest example of the
linear variational approach for describing the $\pi$-electrons in
conjugated hydrocarbons, such as ethylene, benzene, butadiene, and so
on, where only the $\pi$-orbitals are explicitly considered as basis
set.
\label{huckel-mo-theo}
\end{remark}

Next, we extend Proposition \ref{prop:vari-princi-000} to the
Hylleraas-Undheim and MacDonald theorem for optimizing the eigenvalues
and corresponding eigenvectors of exicted states.
\begin{prop}[Hylleraas-Undheim-MacDonald theorem]
Letting exact eigenvalues of $\hat{H}_{\mathrm{ele}}$ be in order
$\epsilon_1^{(\mathrm{exact})}\leq\epsilon_2^{(\mathrm{exact})}\leq
\cdots\leq\epsilon_i^{(\mathrm{exact})}\leq\cdots$, the linear
variational approach spanned by an $N$-dimensional basis set gives
$N$ eigenvalues and predicts that the $k$-th eigenvalue is an upper
bound to the $k$-th exact eigenvalue.
\label{prop:uhm-theo-var}
\end{prop}
To prove Proposition \ref{prop:uhm-theo-var} in the $N$-dimensional
subspace $\mathbb{V}_N$ spanned by fixed basis set, the $k$-th eigenvalue should be
recognized as both a maximum and a minimum of the expectation value
under certain conditions, so that it lies between two exact eigenvalues.
According to Proposition \ref{theo:mpt}, there exists the $k$-th
eigenvalue in the minimax form
\begin{equation}
\epsilon_k=\min_{\mathbb{U}\subset\mathbb{V}_N,\:\dim\mathbb{U}=k}\;
\max_{\vert\phi\rangle\in\mathbb{U}}
\dfrac{\big\langle\phi\vert\hat{H}_{\mathrm{ele}}\vert\phi\big\rangle}
{\big\langle\phi\vert\phi\big\rangle}\geq
\min_{\mathbb{U}\subset\mathbb{H},\:\dim\mathbb{U}=k}\;
\max_{\vert\phi\rangle\in\mathbb{U}}
\dfrac{\big\langle\phi\vert\hat{H}_{\mathrm{ele}}\vert\phi\big\rangle}
{\big\langle\phi\vert\phi\big\rangle}=\epsilon_k^{(\mathrm{exact})},
\label{eq:minmax-theo-varis-0}
\end{equation}
where $\mathbb{U}$ is a $k$-dimensional subspace of $\mathbb{V}_N$.
Since the set of $k$-dimensional subspaces in $\mathbb{V}_N$ is a
subset of those in the Hilbert space $\mathbb{H}$, the minimum over
a subset is greater than or equal to the minimum over the whole set
(also called monotonicity of infimum). Thus, one can obtain inequality
in Equation \eqref{eq:minmax-theo-varis-0} and then Proposition
\ref{prop:uhm-theo-var}. Now, based on Propositions \ref{prop:ratz-vari-prin-0}
and \ref{prop:uhm-theo-var}, we can have the following result.
\begin{corop}
If $N$- and $(N+1)$-dimensional basis sets are employed, the resulting
eigenvalues $\{\epsilon_i^{(N)}\}_{i=1}^N$ and $\{\epsilon_i^{(N+1)}\}_{i=1}^{N+1}$
satisfy $\cdots\leq\epsilon_k^{(\mathrm{exact})}\leq\epsilon_k^{(N+1)}\leq
\epsilon_k^{(N)}\leq\epsilon_{k+1}^{(\mathrm{exact})}\leq\epsilon_{k+1}^{(N+1)}\leq\cdots$.
\label{coro:vari-ppp-01}
\end{corop}
To prove Corollary \ref{coro:vari-ppp-01}, we also adopt the minimax
relations as adopted in proving Proposition \ref{prop:uhm-theo-var}.
Introducing one more basis into $\mathbb{V}_N$, we obtain an $(N+1)$-dimensional
sapce $\mathbb{V}_{N+1}$, implying $\mathbb{V}_N\subset\mathbb{V}_{N+1}$
so that any $k$-dimensional optimization in $\mathbb{V}_N$ is a restricted
one in $\mathbb{V}_{N+1}$. The minimum for $\epsilon_k^{(N)}$ is taken
over $\mathbb{V}_N$, which is a subset of all $\mathbb{V}_{N+1}$. Due
to monotonicity of infimum, similar to Equation \eqref{eq:minmax-theo-varis-0},
we can have inequality
\begin{equation}
\epsilon_k^{(N+1)}=\min_{\mathbb{U}\subset\mathbb{V}_{N+1},\:\dim\mathbb{U}=k}\;
\max_{\vert\phi\rangle\in\mathbb{U}}
\dfrac{\big\langle\phi\vert\hat{H}_{\mathrm{ele}}\vert\phi\big\rangle}
{\big\langle\phi\vert\phi\big\rangle}\leq
\min_{\mathbb{U}\subset\mathbb{V}_{N},\:\dim\mathbb{U}=k}\;
\max_{\vert\phi\rangle\in\mathbb{U}}
\dfrac{\big\langle\phi\vert\hat{H}_{\mathrm{ele}}\vert\phi\big\rangle}
{\big\langle\phi\vert\phi\big\rangle}=\epsilon_k^{(N)}.
\label{eq:minmax-theo-varis-1}
\end{equation}
Obviously, Equation \eqref{eq:minmax-theo-varis-1} gives main result of
Corollary \ref{coro:vari-ppp-01}. According to the order condition of
eigenvalues, one can further obtain the order given in Corollary \ref{coro:vari-ppp-01}.
Moreover, Equation \eqref{eq:minmax-theo-varis-1} indicates the
variational convergence theorem \cite{sza12:108,goi18:e1341,li20:9951,sve23:e1666},
which states that as the basis set increases toward completeness the
eigenvalues converge monotonically from above to the exact eigenvalues,
namely $\lim_{N\to\infty}\epsilon_k^{(N)}=\epsilon_k^{(\mathrm{exact})}$.

According to Corollary \ref{coro:vari-ppp-01}, the guidelines for
choosing an appropriate basis set in electronic structure calculations
are conceptually clear. Adding more basis functions, especially polarization,
diffuse, and higher angular momentum functions, often systematically
improves accuracy up to a point. Despite this, realistic calculations
for real molecular systems are usually nasty and tricky. In this context,
choice of basis set is as much an art as it is a science 
\cite{sza12:108,goi18:e1341,li20:9951,sve23:e1666}. At least five points
should be noted on this issue. First, the monotonicity given by Corollary
\ref{coro:vari-ppp-01} is guaranteed only by a nested sequence of basis
sets, rather than a non-nested sequence. Moreover the so-called complete
basis set (CBS) limit is an aymptote but not a finish choice. One should
always inspect convergence with respect to size of basis set. Second,
in very large basis set, especially with diffuse function, basis functions
become linearly dependent mainly because of ill-conditioned overlap matrix.
To improve results in this case, one should employ stable diagonalization
algorithms, such as Cholesky decomposition with thresholding, and/or remove
near-linearly dependent functions (if possible). Third, since large basis
set leads to expensive calculations, a practical balance between accuracy
and cost must be carefully inspected. Fourth, for computing the interaction
between two monomers, basis functins on one fragment can artificially lower
the energy of the other fragment leading to overbinding, called basis set
superposition error (BSSE). It can be estimated by computing each monomer
with the full dimer basis set, called the counterpoise correction. Although
BSSE decreases with systematic basis set enlargement and vanishes in the
CBS limit, residual BSSE may persist for finite basis sets, in particular
if basis set improvement is unbalanced or incomplete. Finally, to consider
excited states, basis functions are required to describe nodal surfaces.
Moreover, the calculated $k$-th root may correspond to a higher state if
the basis set is too small, called root flipping. In this context, one
should always inspect
the nature of the excited state by checking natural orbitals to ensure the
correct of state assignment. In summary, the art of choosing basis set in
practical calculations must balance computational accuracy, size consistency,
and computational affordability. Although the variational principle
guarantees that larger and nested basis sets converge to the exact result,
practical calculations require careful and property-specific choice and
validation for basis sets.

Obviously, Proposition \ref{prop:vari-princi-000} as well as Proposition
\ref{prop:uhm-theo-var} and Corollary \ref{coro:vari-ppp-01} all focus
on eigenvalues of the $\hat{H}_{\mathrm{ele}}$. Indeed, all electronic
structure methods are particularly concerned with the accuracy and
efficiency of electronic energy calculations because, as mentioned
above, chemists consider reaction mechanisms by investigating the
energetic profiles. However, if our goal is the expectation values,
such as geometry, dipole moment, charge distribution, and so on,
attention must be then paid to the quality of the eigenfunctions of
$\hat{H}_{\mathrm{ele}}$, in particular those for excited states.
According to Proposition \ref{prop:uhm-theo-var} and Corollary \ref{coro:vari-ppp-01},
although electronic eigenvectors can be seen as local minima or saddle
points of the approximate energy, the global structure of the exact
electronic energy remains largely unexplored. To overcome this problem,
Burton \cite{bur22:1512} introduced a geometric perspective on the exact
electronic energy landscape, on which electronic states form stationary
points constrained to the surface of a hypersphere, by the $N$-dimensional
representation matrix $\boldsymbol{\phi}\in\mathbb{R}^N$.
\begin{prop}
Each exact eigenstate is represented by two stationary points on the
hypersphere that are related by a sign-change in the wave function.
There are no other stationary points on the exact landscape.
\label{prop:exact-energy-land}
\end{prop}
According to Proposition \ref{prop:ratz-vari-prin-0}, the electronic
energy is given by $E(\boldsymbol{\phi})=\boldsymbol{\phi}^{\mathrm{T}}
\mathbf{H}\boldsymbol{\phi}$, where $\mathbf{H}$ is the representation
matrix of $\hat{H}_{\mathrm{ele}}$ in the orthogonal basis of Slater
determinants. Normalization relation
$\boldsymbol{\phi}^{\mathrm T}\boldsymbol{\phi}=1$ requires
$\boldsymbol{\phi}$ to be constrained to an $(N-1)$-dimensional unit
hypersphere embedded in the $N$-dimensional space. In this way, optimal
ground and excited states are stationary points of the electronic energy
constrained to the surface of this unit hypersphere. A stationary point
requires that the global gradient of the energy in the full hilbert sapce
$\partial_{\boldsymbol{\phi}}E=2\mathbf{H}\boldsymbol{\phi}$ has no
component in the tangent space to the hypersphere. At a point
$\boldsymbol{\phi}$, tangent vectors $\boldsymbol{\varsigma}$ satisfy
condition $\boldsymbol{\varsigma}^{\mathrm{T}}\boldsymbol{\phi}+
\boldsymbol{\varsigma}\boldsymbol{\phi}^{\mathrm{T}}=0$. Similar to
the unitary transformation matrix $\mathbf{U}$, the orthogonal basis
vectors of $(N-1)$-dimensional tangent space form the columns of a
projector into the local tangent basis, denoted by
$\mathbf{U}^{\perp}\in\mathbb{R}^{N\times(N-1)}$.
Noting that $\boldsymbol{\phi}$ and $\mathbf{U}^{\perp}$ are $N\times1$
and $N\times(N-1)$ basis matrices, respectively, one can find the completeness
condition $\boldsymbol{\phi}\boldsymbol{\phi}^{\mathrm{T}}+\mathbf{U}^{\perp}
\mathbf{U}^{\perp}{}^{\mathrm{T}}=\mathbf{I}$.
The constrained energy gradient $\boldsymbol{\nabla}E$ is then obtained
by projecting the global gradient $\partial_{\boldsymbol{\phi}}E$ into
the tangent space leading to
$\boldsymbol{\nabla}E=\mathbf{U}^{\perp}{}^{\mathrm{T}}\partial_{\boldsymbol{\phi}}E
=2\mathbf{U}^{\perp}{}^{\mathrm{T}}\mathbf{H}\boldsymbol{\phi}$ with
contrained stationary points satisfying $\boldsymbol{\nabla}E=\mathbf{0}$.
This requires that $\partial_{\boldsymbol{\phi}}E$ has no component in
the tangent space making $\mathbf{H}\boldsymbol{\phi}$ must parallel
or antiparallel to $\boldsymbol{\phi}$. Then, one can immediately obtain
eigenvalue equation $\mathbf{H}\boldsymbol{\phi}_k=E_k\boldsymbol{\phi}_k$
with $k=1,\cdots,N$ and thus Proposition \ref{prop:exact-energy-land}.
\begin{remark}
Proposition \ref{prop:exact-energy-land} can be compared to the HF theory,
where the occupied orbitals represent the current position on a Grassmann
manifold (approximately say, a set of all linear subspaces) and occupied-virtual
orbital rotations define the tangent basis vectors. Therefore, optimization
of the energy requires the off-diagonal blocks of an effective Hamiltonian
matrix to become zero such that the occupied-virtual coupling between orbitals
vanishes.
\label{remark:hartree-fock-theo}
\end{remark}

%----------------------------------------------------------
\subsection{Principle of Equivalence\label{sec:keo-metric}}

For a general case, we consider the system living in a generally curved
space. To this end, one has to introduce the principle of equivalence
which is another postulate of this work. This postulate is the basis for
deriving the Laplacian operator through the variational principle. By
the Laplacian operaotr, one can obtain the KEO expression. We would like
to emphasize that such KEO expression is applicable to both electronic
and nuclear motions. Since electrons in a molecular system are always
confined to bound states, their KEO is of a fixed form. We therefore
focus this work on the nuclear KEO. As a beginning, let us consider the
geodesic of a space which is supposed to be generally curved.
\begin{defi}
A geodesic is a curve that locally minimizes distance between nearby
points, and whose tangent vector remains parallel to itself along the
curve.
\label{def:geodesic}
\end{defi}
In general, the curvature of a curved space is related to the metric
tensor, which in turn is related to the geometric properties of the
space. The concept of curvature, all of whose manifestations depend on
the so-called connection. The connection connects vectors in the tangent
spaces of neighboring points, allowing us to differentiate, compare
vectors, and define geodesics by Definition \ref{def:geodesic}. Based
on the metric tensor $g_{\mu\nu}$, one can construct a unique connection.
To this end, we introduce the Christoffel symbols
\begin{equation}
\Gamma_{\mu\nu}^{\rho}=
\frac{1}{2}g^{\rho\sigma}\Big(\partial_{\mu}g_{\nu\sigma}+
\partial_{\nu}g_{\sigma\mu}-\partial_{\sigma}g_{\mu\nu}\Big).
\label{eq:christoffel-symbol-000}
\end{equation}
The Christoffel symbols $\Gamma_{\mu\nu}^{\rho}$ have a form similar to
that of a tensor, but in fact $\Gamma_{\mu\nu}^{\rho}$ are not tensors.
This is why they are called ``symbols''. Next, one can define the covariant derivative
which is one manifestation of the connection. For simple, the covariant
derivative of an arbitrary vector field $A^{\nu}$ satisfies
\begin{equation}
\nabla_{\mu}A^{\nu}=\partial_{\mu}A^{\nu}+
\Gamma_{\mu\sigma}^{\nu}A^{\sigma}.
\label{eq:christoffel-symbol-001}
\end{equation}
The covariant derivative of a general tensor can be also written in a
similar manner to Equation \eqref{eq:christoffel-symbol-001}. Simply,
the covariant derivative measures change relative to the parallel transport.
According to Definition \ref{def:geodesic}, the connection allows us
to define a geodesic $x^{\mu}(\lambda)$ through the geodesic equation
\begin{equation}
\frac{\mathrm{d}^2x^{\mu}}{\mathrm{d}\lambda^2}+
\Gamma_{\rho\sigma}^{\mu}\frac{\mathrm{d}x^{\rho}}{\mathrm{d}\lambda}
\frac{\mathrm{d}x^{\sigma}}{\mathrm{d}\lambda}=0,
\label{eq:christoffel-symbol-002}
\end{equation}
where $\lambda$ represents parameter set. We will return to the proof
of Equation \eqref{eq:christoffel-symbol-002} later. Next, Definition \ref{def:geodesic}
implies that the manner in which a family of geodesics spreads or focuses
can be used to define the Riemann tensor. According to Equation
\eqref{eq:christoffel-symbol-002}, curvature can be expressed in terms
of the Riemann tensor $R_{\sigma\mu\nu}^{\rho}$, which is a
$(1,3)$-tensor satisfying
\begin{equation}
R_{\sigma\mu\nu}^{\rho}=\partial_{\mu}\Gamma_{\nu\sigma}^{\rho}-
\partial_{\nu}\Gamma_{\mu\sigma}^{\rho}+\Gamma_{\mu\eta}^{\rho}
\Gamma_{\nu\sigma}^{\eta}-\Gamma_{\nu\eta}^{\rho}
\Gamma_{\mu\sigma}^{\eta}.
\label{eq:christoffel-symbol-003}
\end{equation}
According to $R_{\sigma\mu\nu}^{\rho}$, one can obtain geometric
characteristics of the space. For instance, contracting the indices of
the Riemann tensor, one can obtain the Ricci tensor $R_{\mu\nu}=
R_{\mu\rho\nu}^{\rho}$ and the Ricci scalar $R=g^{\mu\nu}R_{\mu\nu}$.
By the Einstein field equations of general relativity,
\begin{equation}
R_{\mu\nu}-\frac{1}{2}Rg_{\mu\nu}=G_{\mu\nu}=8\pi T_{\mu\nu}, 
\label{eq:christoffel-symbol-004}
\end{equation}
geometric characteristics of the curved sapcetime (the Einstein tensor
$G_{\mu\nu}$) in the left-hand side is related with the momentum-energy
tensor $T_{\mu\nu}$ of matter in the right-hand side.
\begin{remark}
The Einstein field equations describe the relationship between the
gravitational field and the geometry of the spacetime. Typically, the
Einstein equations can be coupled with the EOMs of a system to describe
the coupled motions of gravity and the system. Nevertheless, we neglect
such coupling, thereby assuming that the geometry of the spacetime merely
serve as a background for motions of the system, with the latter not
influencing the former.
\label{rem:coupl-gravity}
\end{remark}

Before giving the Laplacian operator, it is worth noting that a free
particle is moving along a geodesic in the curved spacetime with no
other interaction acting on it. This can be found by the
principle of equivalence, which is another postulate of the present work.
\begin{prop}[weak principle of equivalence]
All objects move in gravity in the same manner. 
\label{prop:wpe}
\end{prop}
According to Proposition \ref{prop:wpe}, the properties of the motion
in a noninertial frame are the same as those in an inertial frame in
the presence of a gravitational field, which implies that a noninertial
reference system is equivalent to a certain gravitational field. 
%In other words, one can obtain the following statement of the principle of equivalence
\begin{post}[principle of equivalence]
In freely falling local reference frames, the laws of physics are the
same as in an inertial frame without gravity.
\label{post:pe}
\end{post}
Postulate \ref{post:pe} indicates that gravity is not a force, but a
consequence of spacetime curvature, meaning that free particle moves
along geodesics in such curved spacetime. Nevertheless, it must be
mention that the field to which a noninertial reference frame is
equivalent, is not completely identical with the gravitational field
which actually occurs in an inertial frame. The essential difference
between them is with respect to their behaviour at infinity. Taking
motion in a uniformly accelerated reference frame as an example, a
particle of arbitrary mass, freely moving in such a reference frame,
has relative to this frame a constant acceleration, equal and opposite
to the frame acceleration itself. The same motion can be observed for
the particle freely moving in a uniform constant gravitational field.
Therefore, a uniformly accelerated reference frame is equivalent to a
constant, uniform external field.
\begin{remark}
The weak principle of equivalence (Proposition \ref{prop:wpe}) states
that gravitational mass is equal to inertial mass. Moreover, gravity
can be locally eliminated by a choice of coordinate system, namely a
freely falling frame. Thus, the strong principle of equivalence (Postulate
\ref{post:pe}) states that, in freely falling local reference frames,
the laws of physics are the same as in an inertial frame without gravity.
\label{rem:wpe-eistein}
\end{remark}

Keeping the above idea in the mind, let us apply the variational principle
to derivation of the Laplacian operator on any Riemannian manifold with
metric $g_{\mu\nu}$. 
\begin{defi}
A local coordinate set is established relative to a particular object,
component, or region, rather than the entire space. A global coordinate
set is a single coordinate system that covers the entire manifold smoothly
and without ambiguity.
\label{def:local-global-coor}
\end{defi}
To this end, we should turn to derivation of the
geodesic equation (see Equation \eqref{eq:christoffel-symbol-002}) by
optimizing the kinetic energy functional of a free particle in parameter
space,
\begin{equation}
E\big[x(\lambda)\big]=\frac{1}{2}\int g_{\mu\nu}\big(x(\lambda)\big)
\frac{\mathrm{d}x^{\mu}}{\mathrm{d}\lambda}
\frac{\mathrm{d}x^{\nu}}{\mathrm{d}\lambda}\mathrm{d}\lambda.
\label{eq:christoffel-symbol-005}
\end{equation}
Obviously, setting $\delta E[x(\lambda)]=0$ gives the geodesic equation.
Similarly, to derive the Laplacian operator, it is worth noting that
the energy functional of a free particle can be given by functional of
a scalar field $f$ on the manifold $M$ in local coordinates $x^{\mu}$,
\begin{equation}
E\big[f\big]=\frac{1}{2}\int_Mg^{\mu\nu}\big(\partial_{\mu}f\big)
\big(\partial_{\nu}f\big)\sqrt{\vert g\vert}\mathrm{d}^4x,
\label{eq:christoffel-symbol-006}
\end{equation}
where $g=\det(g^{\mu\nu})$ and $\sqrt{\vert g\vert}$ accounts for the
4-volume element during variation. Varying $E[f]$ in Equation
\eqref{eq:christoffel-symbol-006} with respect to $f$, one can obtain
\begin{equation}
\delta E=\int_Mg^{\mu\nu}\big(\partial_{\mu}f\big)
\big(\partial_{\nu}\delta f\big)\sqrt{\vert g\vert}\mathrm{d}^4x=
-\int_M\delta f\frac{1}{\sqrt{\vert g\vert}}\partial_{\nu}
\Big(\sqrt{\vert g\vert}g^{\mu\nu}\big(\partial_{\mu}f\big)\Big)
\sqrt{\vert g\vert}\mathrm{d}^4x,
\label{eq:christoffel-symbol-007}
\end{equation}
where the second equality follows from integration by parts and the
divergence
\begin{equation}
\nabla_{\mu}x^{\mu}=\frac{1}{\sqrt{\vert g\vert}}
\partial_{\mu}\left(\sqrt{\vert g\vert}x^{\mu}\right).
\label{eq:div-coord-0999}
\end{equation}
By Equation \eqref{eq:christoffel-symbol-000}, the contracted Christoffel
symbol is given by $\Gamma^{\mu}_{\mu\nu}=\partial_{\nu}\ln(\sqrt{\vert g\vert})$
which can be substituted into Equation \eqref{eq:christoffel-symbol-001}
to obtain the divergence $\nabla_{\mu}x^{\mu}$ in Equation \eqref{eq:div-coord-0999}.
Setting $\delta E=0$ and noting the arbitrariness of $f$, Equation
\eqref{eq:christoffel-symbol-007} directly predicts the Laplacian
operator
\begin{equation}
\nabla_g^2=\frac{1}{\sqrt{\vert g\vert}}\partial_{\nu}
\Big(\sqrt{\vert g\vert}g^{\mu\nu}\partial_{\mu}\Big).
\label{eq:christoffel-symbol-007}
\end{equation}
Therefore, the Laplacian operator is fundamentally the Euler-Lagrange
operator associated with the kinetic energy norm. It measures the
difference between the value of a function at a point and its average
value on an infinitesimal geodesic sphere, hence its deep connection
to geodesic distance.

%----------------------------------------
\section{Nuclear Hamiltonian Operator\label{sec:hamiltonian}}

This section describes construction techinques for nuclear Hamiltonian
operator, as collected in Table \ref{tab:theory-comp}. Remark \ref{rem:hamil-keo-pes}
indicates that the Hamiltonian operator consists of the KEO and PES
terms, which correspond to distinct physical insight and can be therefore
constructed in fundamentally different ways. Section \ref{sec:boa-app}
describes separation of electrons and nuclei through adiabatic
approximation in space and time, such as the Born-Oppenheimer
approximation (BOA). Section \ref{sec:coord-frame} describes coordinates
frame for representing the molecular system, in particular the
molecular reaction system. Section \ref{sec:keo-coord}
gives the geometric framework on deriving the KEO, in particular its
dependence on the geometric characteristics of configuration space.
Section \ref{sec:pes-build-regree} gives geometric consideration in
constructing the PES. FInally, Section \ref{sec:coumb-pot} gives
electromagnetic interaction based on previous discussions on
electrodynamics in curved spacetime.
\begin{remark}
Currently, physical possibility to further unify the KEO and PES terms
of the Hamiltonian operator is still unclear due to the difficulty of
unifying the geometric properties of spacetime and other external field.
Although partial unification (say grand unified theory) was already
established, the unification of geometric characteristics remains an
open problem.
\label{rem:keo-pes-hamil}
\end{remark}

%----------------------------------------
\subsection{Adiabatic Approximation\label{sec:boa-app}}

As given in the end of Section \ref{sec:math-frame}, core task of
chemical dynamics is to solve the molecular Schr{\"o}dinger equation.
Since a molecule system contains electrons and nuclei, solving the
coupled EOMs for electrons and nuclei simultaneously leads to the
so-called diabatic representation. However, it requires a significant
cost in calculations. To overcome this problem, such coupled EOMs can
be separated into the electronic one and the nuclear one to simplify
calculations. This is the so-called adiabatic approximation
\cite{yar96:985,yar96:10456,yar98:971,yar01:6277,jua07:044317,bou08:124322,jua08:211101,alt08:214117,ced04:3,bae06:boa,ced13:224110}.
\begin{defi}
In adiabatic approximation, if the nuclear are moving slowly enough,
the electrons remain in instantaneous eigenstate throughout the
evolution.
\label{def:ad-app-0}
\end{defi}
This is not new because separation of DOFs plays a fundamental role in
the understanding of complex dynamics, where DOFs are typically
partitioned into fast and slow motions. The fast motions are solved
first, and their influence is subsequently incorporated into an
effective potential that governs the effective dynamics of the slow
DOFs. Thus, the molecular system can be naturally partitioned into
electronic and nuclear motions, called the Born-Oppenheimer approximation
(BOA) 
\cite{yar96:985,yar96:10456,yar98:971,yar01:6277,jua07:044317,bou08:124322,jua08:211101,alt08:214117,ced04:3,bae06:boa,ced13:224110},
because the electrons are moving at least $50$ times faster than the
nuclei.
\begin{remark}
The BOA allows us to decouple electronic and nuclear motions, defining
electronic structure and molecular dynamics and providing the nuclear
Hamiltonian as the bridge that connects them.
\label{rem:boa-import-dyna-chem}
\end{remark}
In this approximation, the effective dynamics for the nuclei
is of extraordinary importance in understanding molecular reaction
dynamics, where the nuclei evolve in an effective Hamiltonian
generated by motions of the electrons, while the state of the
electrons instantaneously adjusts to an eigenstate corresponding to
the momentary configuration of the nuclei.
\begin{defi}
The phenomenon that fast DOFs become slaved by slow DOFs that evolve
autonomously is called adiabatic decoupling.
\label{def:decop-adiba}
\end{defi}
Adiabatic approximation (Definition \ref{def:ad-app-0}) and adiabatic
decoupling (Definition \ref{def:decop-adiba}), in particular the BOA,
have been well known by the chemistry community (see Remark \ref{rem:boa-import-dyna-chem}).
However, the adiabatic approximation requires correlations. First,
there exist numerous multi-electronic-state processes that violate
the BOA providing fruitful dynamic and geometric effects. Second, even
for single-electronic-state processes, it remains essential to account
for perturbations associated with the BOA through adiabatic perturbation
theory
\cite{ste71:2231,bra79:199,sok78:339,pan02:250405,teu03:book,pan03:145,rig10:170406,rig08:052508,rig14:022104,cho25:012221}
which was developed not only to yield a rigorous confirmation of adiabatic
approximation, but also to provide new physical insight into adiabatic
problems. Unlike traditional introduction of the adiabatic approximation
by separation of variables, we consider this issue from dynamics because
chemical reaction is essentially a time-evolution process. At the end
of this section, we will briefly re-visit the traditional introduction
of the BOA.

Introducing adiabatic parameter $0<\varepsilon\ll1$ to control the time-scalar
separation between fast and slow DOFs, the samller $\varepsilon$ the
slower is the vriation of $\hat{H}$ on the {\it a priori} fixed fast
time-scalar. Letting $s$ be fast time-scalar, the $t=\varepsilon s$ is
the slow time-scalar. Turning to the propagation given by Equation
\eqref{eq:path-int-002}, one can find that a unitary propagator
$U(t,t_0)=K(X(t),X(t_0))$ is a jointly strongly continuous family of
unitary operators \cite{teu03:book} satisfying transitivity
$U(t,t_0)=U(t,\tau)U(\tau,t_0)$ and identity condition $U(t,t)=1$.
By the Schr{\"o}dinger equation given by Equation \eqref{eq:path-int-004},
the propagator satisfies
\begin{equation}
\mi\frac{\mathrm{d}}{\mathrm{d}s}U(\varepsilon s,\varepsilon s_0)=
\hat{H}(s)U(\varepsilon s,\varepsilon s_0)\quad\mathrm{or}\quad
\mi\varepsilon\frac{\mathrm{d}}{\mathrm{d}t}U(t,t_0)=\hat{H}(t)U(t,t_0).
\label{eq:propagators-00-ad}
\end{equation}
Assuming that the spectrum $\sigma(\hat{H}(t))$ of $\hat{H}(t)$ contains a subset
$\sigma_*(t)\subset\sigma(\hat{H}(t))$, such that there are two bounded continuous
functions $f_{\pm}$ and an interval $I(t)=[f_-(t),f_+(t)]$, we can give
the time-adiabatic theorem \cite{teu03:book,pan03:145,rig10:170406,rig08:052508,rig14:022104}.
\begin{prop}
Assuming existence of the gap condition
\begin{equation}
\sigma_*(t)\subset I(t),\quad
\inf_{t\in\mathbb{R}}\mathrm{dist}\Big(I(t),\sigma\big(\hat{H}(t)\big)
\setminus\sigma_*(t)\Big)>0,
\label{eq:gap-condition-0}
\end{equation}
there exists a constant $C<\infty$ such that
\begin{equation}
\Big\Vert\big(1-P_*(t)\big)U(t,t_0)P_*(t_0)\Big\Vert\leq
C\varepsilon\big(1+\vert t-t_0\vert\big),
\label{eq:propagators-01-ad}
\end{equation}
where $P_*(t)$ is the spectral projection of the Hilbert space $\mathbb{H}$
for $\sigma_*(t)$ and $\hat{H}(t)$ is sufficient regularity.
\label{prop:time-adia-00}
\end{prop}
Mathematically, the gap condition given by Equation \eqref{eq:gap-condition-0}
requires that the quantum states in $\sigma_*(t)$ has a nonzero
separation from the complement $\sigma\setminus\sigma_*$. In this context,
the states under consideration form a relatively independent Hilbert
subspace, such as low-lying electronic states decoupled with high-lying
states in computing photoelectron spectrum.

Usually, Proposition \ref{prop:time-adia-00} can be reorganized in an
easily understood statement, which is precisely the time-adiabatic
theorem \cite{teu03:book,pan03:145,rig10:170406,rig08:052508,rig14:022104}.
\begin{prop}[time-adiabatic theorem]
If $\hat{H}$ and $\sigma_*(t)\subset\sigma(\hat{H}(t))$ satisfy the gap
condition in Equation \eqref{eq:gap-condition-0}, then there is a
constant $C<\infty$ such that
\begin{equation}
\Big\Vert U(t,t_0)-U_{\mathrm{a}}(t,t_0)\Big\Vert\leq
C\varepsilon\big(1+\vert t-t_0\vert\big),\quad
\forall t,t_0\in\mathbb{R},
\label{eq:propagators-02-ad}
\end{equation}
where $U_{\mathrm{a}}(t,t_0)$ is the adiabatic propagator and
intertwines $P_*(t)$ at different times exactly, that is
$P_*(t)U_{\mathrm{a}}(t,t_0)=U_{\mathrm{a}}(t,t_0)P_*(t_0)$
holds.
\label{prop:time-adia-01}
\end{prop}
We refer the reader to Reference \cite{teu03:book} for proof in details.
Propositions \ref{prop:time-adia-00} and \ref{prop:time-adia-01} both
state that, propagating the initial state $\Psi(t_0)\in P_*(t_0)\mathbb{H}$
by solving Equation \eqref{eq:path-int-004}, the time-dependent state
$\Psi(t)$ stays in the subspace $P_*(t)\mathbb{H}$ up to an error of
order $\mathcal{O}(\varepsilon(1+\vert t-t_0\vert)\Vert\Psi(t_0)\Vert)$.
According to Definition \ref{def:decop-adiba}, the way that spectral
subspaces which depend in some sense slowly on some parameter are
approximately invariant under the time-evolution by Equation \eqref{eq:path-int-004}
is adiabatic decoupling.
In practical quantum dynamics calculations, selecting a subspace
$\sigma_*$ is sometimes an art rather than a science. For example,
which electronic states to include in computing photoelectron spectrum
calculations, or which electronic states to consider in photochemical
reaction dynamics, and so on. In general, the gap condition requires
that the separation between $\sigma_*$ and $\sigma\setminus\sigma_*$
remains strictly greater than zero, as shown by Equation \eqref{eq:gap-condition-0}.
This condition is often not straightforward to verify intuitively in
practical calculations. Thus, a variational method becomes an important
tool for establishing the gap condition.

To consider effective dynamics with adiabatic decoupling under reference
space $\mathbb{C}^d$, letting $\{\varphi_{\kappa}(t)\}_{\kappa=1}^d$
be some fixed orthonormal basis of $P_*(t)\mathbb{H}$ and letting
$\{\chi_{\kappa}\}_{\kappa=1}^d$ be an orthonormal basis set of
$\mathbb{C}^d$, one can define a time-dependent unitary operator
$\hat{U}(t)\:P_*(t)\mathbb{H}\to\mathbb{C}^d$ as
$\hat{U}(t)=\sum_{\kappa=1}^d\vert\chi_{\kappa}\rangle\langle\varphi_{\kappa}(t)\vert$.
Then, a unitary propagator on $\mathbb{C}^d$ can be defined as
$U_{\mathrm{eff}}(t,t_0)=\hat{U}(t)U_{\mathrm{a}}(t,t_0)\hat{U}^{\dagger}(t_0)$,
which, according to Equation \eqref{eq:propagators-00-ad}, is the
solution of the equation
\begin{equation}
\mi\varepsilon\frac{\mathrm{d}}{\mathrm{d}t}U_{\mathrm{eff}}(t,t_0)=
\hat{H}_{\mathrm{eff}}(t)U_{\mathrm{eff}}(t,t_0),\quad
U_{\mathrm{eff}}(t,t)=1.
\label{eq:propagators-03-ad}
\end{equation}
According to Equations \eqref{eq:propagators-02-ad} and \eqref{eq:propagators-03-ad},
we directly find that $\hat{H}_{\mathrm{eff}}(t)=\hat{U}(t)\hat{H}(t)\hat{U}^{\dagger}(t)
+\mi\varepsilon\hat{\dot{U}}(t)\hat{U}^{\dagger}(t)$. With the above
results and the gap condition given in Equation \eqref{eq:gap-condition-0},
there exists the following proposition.
\begin{prop}
Assuming that $\sigma_*(t)=\{E(t)\}$ contains a single eigenvalue, the
solution of Equation \eqref{eq:propagators-03-ad} satisfies
\begin{equation}
\Big\Vert\Big(U(t,t_0)-\hat{U}^{\dagger}(t)U_{\mathrm{eff}}(t,t_0)
\hat{U}(t_0)\Big)P_*(t_0)\Big\Vert\leq
C\varepsilon\big(1+\vert t-t_0\vert\big),
\label{eq:propagators-04-ad}
\end{equation}
where $C<\infty$ is a constant.
\label{prop:time-adia-02}
\end{prop}
Under the gap condition (see Equation \eqref{eq:gap-condition-0}),
Propositions \ref{prop:time-adia-01} and \ref{prop:time-adia-02} are
the first-order time-adiabatic theory
\cite{teu03:book,pan03:145,rig10:170406,rig08:052508,rig14:022104}.
By this theory and its generalizations
\cite{ste71:2231,bra79:199,sok78:339,pan02:250405,teu03:book,pan03:145,rig10:170406,rig08:052508,rig14:022104,cho25:012221},
the space-adiabatic problems
\cite{pan02:250405,teu03:book,pan03:145,rig10:170406,rig08:052508,rig14:022104},
such as the BOA
\cite{yar96:985,yar96:10456,yar98:971,yar01:6277,jua07:044317,bou08:124322,jua08:211101,alt08:214117,ced04:3,bae06:boa,ced13:224110}, 
have been derived and deeply discussed. Since the separation between
the fast and slow DOFs in the time-independent fashion, that is the
space adiabatic theorems
\cite{ste71:2231,bra79:199,sok78:339,pan02:250405,teu03:book,pan03:145,rig10:170406,rig08:052508,rig14:022104,cho25:012221}, 
has been well known, we refer the reader to References
\cite{yar96:985,yar96:10456,yar98:971,yar01:6277,jua07:044317,bou08:124322,jua08:211101,alt08:214117,ced04:3,bae06:boa,ced13:224110}
for further details. In summary, by the time-adiabatic theorems, the
slow variation is of dynamical origin, arisen naturally from the
intrinsic properties of the system, rather than artificially imposed
through a Hamiltonian depending slowly on time. Later, we will return
to this issue for the geometric phase effects.

Finally, we should briefly give conventional description on the BOA
\cite{yar96:985,yar96:10456,yar98:971,yar01:6277,jua07:044317,bou08:124322,jua08:211101,alt08:214117,ced04:3,bae06:boa,ced13:224110}
by separating electronic and nuclear motions. Assuming that there are
$N_{\mathrm{ele}}$ electrons and $N_{\mathrm{nuc}}$ nuclei in the
molecular system, the molecular Hamiltonian is geive by
$\hat{H}_{\mathrm{mol}}(X_{\mathrm e},X_{\mathrm n})=\hat{T}_{\mathrm e}(X_{\mathrm e})+
\hat{T}_{\mathrm n}(X_{\mathrm n})+V_{\mathrm{ee}}(X_{\mathrm e})+V_{\mathrm{nn}}(X_{\mathrm n})+
V_{\mathrm{en}}(X_{\mathrm e},X_{\mathrm n})$, where $\hat{T}_{\mathrm e}$
and $\hat{T}_{\mathrm n}$ are electronic and nuclear KEOs, respectively,
while $V_{\mathrm{ee}}$, $V_{\mathrm{nn}}$, and $V_{\mathrm{en}}$ are
electromagnetic interactions for the electron-electron, nuclues-nucleus,
and electorn-nucleus pairs, respectively. Since nuclei are much heavier
than electrons and move more slowly, to a good approximation, the
electrons in the molecular system are consider to be moving in the
electromagnetic field of nuclei that are moving in the effective field
of electrons.
\begin{remark}
By the BOA, the electronic and nuclear motions are completely decoupled,
where the electronic structure parametrically depends on the nuclear
coordinates, while the nuclei are moving in the effective field arisen
from the electrons. The entire wave function of the molecular system is
thus given by product of those for electrons and nuclei.
\label{remark:ele-nucl-decou-000}
\end{remark}
By Remark \ref{remark:ele-nucl-decou-000}, in each electronic-structure
calculation given in Section \ref{sec:elec-struc}, the nuclei are set
to be fixed at some coordinates $X_{\mathrm{n}}$. Within this approximation,
the nuclear KEO can be neglected in each electronic structure calculation
and the $V_{\mathrm{nn}}$ term is constant. Then, one can have the
electronic Hamiltonian $\hat{H}_{\mathrm{ele}}(X_{\mathrm e})=
\hat{T}_{\mathrm e}(X_{\mathrm e})+V_{\mathrm{ee}}(X_{\mathrm e})+
V_{\mathrm{en}}(X_{\mathrm e},X_{\mathrm n})$. Here, the electronic
Hamiltonian $\hat{H}_{\mathrm{ele}}$ parametrically depends on the
nuclear coordinates $X_{\mathrm n}$ that have been already fixed in
electronic structure calculations. It must be noted that any constant
added to an operator only adds to the operator eigenvalues and has no
effect on the operator eigenfunctions. This implies that the BOA is
reasonable for electronic structure theory, and then the total enegry
for fixed nuclei is given by $\epsilon_{\mathrm{tot}}=\epsilon_{\mathrm{ele}}+V_{\mathrm{nn}}$,
where $\epsilon_{\mathrm{ele}}$ is the eigenvalue of $\hat{H}_{\mathrm{ele}}$,
called electronic energy. Solving the eigenvalue problem of
$\hat{H}_{\mathrm{ele}}$ as given by Section \ref{sec:elec-struc}, it
is subsequently possible to solve the nuclear EOM under the BOA.
According to Remark \ref{remark:ele-nucl-decou-000}, it is a reasonable
approximation to replace the electronic coordinates $X_{\mathrm e}$ by
their average values over the electronic wave function. This generates
the nuclear Hamiltonian for the nuclear dynamics in the effective field
of the electrons,
$\hat{H}=\hat{T}_{\mathrm n}+V_{\mathrm{nn}}+\langle\hat{H}_{\mathrm{ele}}\rangle=
\hat{T}_{\mathrm n}+\langle\epsilon_{\mathrm{tot}}\rangle$. Here,
$\hat{T}_{\mathrm n}$ and $\langle\epsilon_{\mathrm{tot}}\rangle$ are
nuclear KEO and PES terms, respectively, which will be denoted by
$\hat{T}$ and $V$, respectively, in the following. For simple denotations,
the nuclear Hamiltonian is directly denoted by $\hat{H}$ in the following
sections.

%---------------------------------------
\subsection{Nuclear Coordinate Frame\label{sec:coord-frame}}

\begin{figure}[h!] 
 \subfigure[\quad Mapping from curved spacetime to configuration space]{%
  \includegraphics[width=18cm]{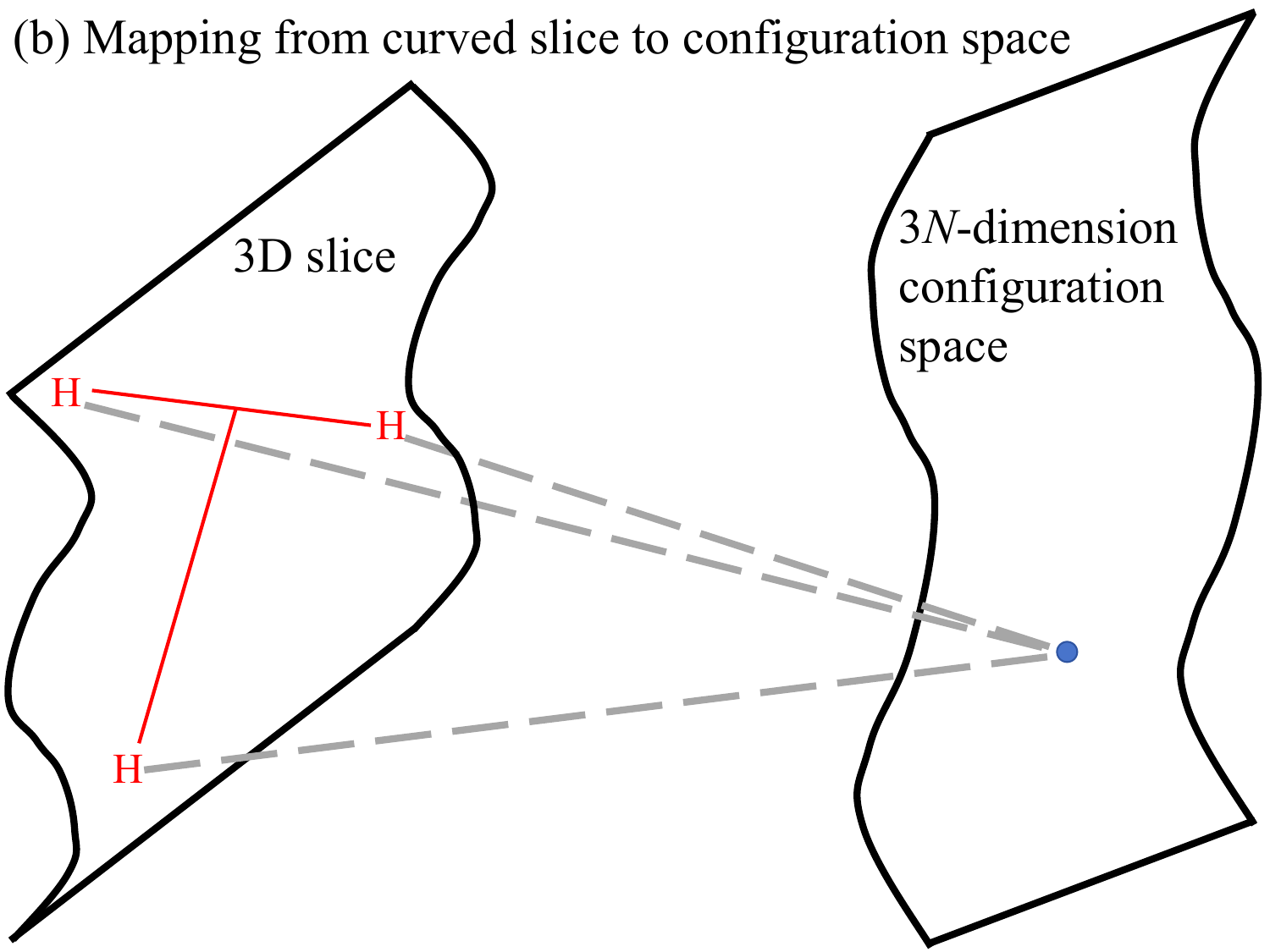}}
   \caption{\figfoot}
    \label{fig:map-slice-conf}
     \end{figure}

As shown in Section \ref{sec:keo-metric}, according to the principle
of equivalence (Postulate \ref{post:pe}) the Laplacian operator given
by Equation \eqref{eq:christoffel-symbol-007} can be obtain by the
variation for the kinetic energy functional of free particle. Before
further deriving the KEO of the molecular system, one has to consider
its coordinate frame. It is worth noting that Equation \eqref{eq:christoffel-symbol-007}
is the Laplacian operator of the local space coordiantes, while the KEO
is essentially the Laplacian operator in the ocnfiguration space. In
addition, one must introduce mass associated with each coordinate. Since
the KEO depends on the metric tensor of configuration space \cite{zha25:20397,zha26:1613},
this section mainly concerns on coordinate frame of the configuration
space, while Section \ref{sec:keo-coord} gives the KEO expression based
on the geometric characteristic of configuration space. The coordinate
frame of the molecular system in the flat space (that is 3D Euclidean
space) has been well described and discussed elsewhere
\cite{pod28:812,lau02:8560,gat09:1,ndo12:034107,sad12:234112,ndo13:204107,mar22:20200388,zha26:1613}.
Here, let us consider a more general case where the molecular system
is living in a general spacetime with nonzero curvature, as shown by
Figure \ref{fig:reax-spacetime}. Usually, one can separate the motion
into its spatial and temporal components (see Figure \ref{fig:reax-spacetime}),
called the $[3+1]$ decomposition \cite{sch12:book,mis17:book,car19:book,zha26:1613}.
\begin{defi}
By the [3 + 1] decomposition, the 4D spacetime is split into 3D spatial
slices and 1D temporal coordinate.
\label{def:3-1-decompo}
\end{defi}
By Definition \ref{def:3-1-decompo}, the Einstein field equations (see
Equation \eqref{eq:christoffel-symbol-004}) can be transferred into a
a time evolution problem suitable for numerical simulations
\cite{sch12:book,mis17:book,car19:book} by the $[3+1]$ decomposition,
where the metric tensor defines the event interval
\cite{sch12:book,mis17:book,car19:book,zha26:1613}
\begin{equation}
\mathrm{d}s^2=g_{\mu\nu}\mathrm{d}x^{\mu}\mathrm{d}x^{\nu}=
-\alpha^2\mathrm{d}t^2
+\gamma_{ij}\big(\mathrm{d}x^i+\beta^i\mathrm{d}t\big)
\big(\mathrm{d}x^j+\beta^j\mathrm{d}t\big),
\label{eq:metric-event-interval}
\end{equation}
where $\alpha$ and $\beta^i$ are freely chosen lapse function and shift
vector, respectively. Determination of an appropriate observer requires
the choice of $\alpha$ and $\beta^i$. In trivial region, a valid gauge
choice of $\alpha=1$ and $\beta^i=0$ corresponds to coordinate time
equal to proper time along Eulerian observers, with spatial grid points
fixed orthogonally to the slices.
\begin{defi}
An Eulerian observer is at rest with respect to a fixed spatial slice,
whereas a Lagrangian observer is comoving with the fluid or matter that
is flowing through space.
\label{def:eul-lag-obs}
\end{defi}
Based on the Eulerian observer, the molecular system living in a 3D
slicing at a given instant is further represented by the configuration
space whose point maps to a specific molecular conformation, as shown
in Figure \ref{fig:map-slice-conf}. Therefore, unless otherwise specified,
we always adopt the Eulerian observer in this work.
\begin{remark}
Different observers correspond to different
ways of decomposing spacetime into space and time. Each observer provides
a distinct physical insight, and the choice depends on the spacetime
symmetry and the system being considered. First, Eulerian observers
(or called fiducial observers) move orthogonal to the spatial slices
and define the absolute space at each instant, implying that they are
natural choice for formulating an initial-value problem. Second, comoving
observers move with a fluid or matter flow and simplify the description of
matter sources. Third, zero-angular-momentum observers are used in rotating
spacetimes and can capture frame-dragging effects. Fourth, static observers
remain at fixed spatial coordinates in static spacetimes. Finally, inertial
observers are freely falling and locally realize special relativity.
\label{rem:observers-other}
\end{remark}
\begin{remark}
In many non-trivial situations, such as dynamical evolution of the
spacetime near black hole, $\alpha=1$ and $\beta^i=0$ lead to
coordinate singularities or numerical instabilities. More flexible
choices are used to avoid singularities and maintain stability.
\label{rem:other-choices}
\end{remark}

Now, turning to the spatial slices shown in Figure \ref{fig:reax-spacetime}
and setting $\alpha=1$ and $\beta^i=0$ in Equation \eqref{eq:metric-event-interval}
for the $N$-atomic system, the Cartesian coordiante set
$\mathbi{x}=\{x^{(i)}\}_{i=1}^{3N}$ satisfies the spatial interval \cite{zha26:1613}
\begin{equation}
\mathrm{d}s^2=
\sum_{n=1}^N\gamma_{i_nj_n}\mathrm{d}x^{(i_n)}\mathrm{d}x^{(j_n)}+
\frac{1}{2}\sum_{n,m=1}^N\gamma_{i_nj_m}\mathrm{d}x^{(i_n)}
\mathrm{d}x^{(j_m)}=\tilde{\gamma}_{ij}\mathrm{d}x^{(i)}
\mathrm{d}x^{(j)},
\label{eq:metric-001-conf-space-mol}
\end{equation}
where $\gamma_{i_nj_n}$ is spatial metric of the 3D slicing local to
the $n$-th atom. According to Equation \eqref{eq:metric-001-conf-space-mol},
one can separate the $3N\times3N$ matrix $\tilde{\gamma}_{ij}$ into
$N^2$ blocks each representing the metric for the body-fixed (BF)
frame at the atom,
\begin{equation}
\tilde{\gamma}_{ij}=\left(\begin{array}{cccc}
\gamma_{i_1j_1} & \gamma_{i_1j_2} & \cdots & \gamma_{i_1j_N} \\
\gamma_{i_2j_1} & \gamma_{i_2j_2} & \cdots & \gamma_{i_2j_N} \\
\cdots & \cdots & \cdots & \cdots \\
\gamma_{i_Nj_1} & \gamma_{i_Nj_2} & \cdots & \gamma_{i_Nj_N} \\
\end{array}\right),
\label{eq:metric-002-conf-space-mol}
\end{equation}
where $\gamma_{i_nj_m}$ is $3\times3$ matrix for the metric connected
coordinates of the $n$-th and $m$-th atoms. In general, the metric
$\tilde{\gamma}_{ij}$ depends on the 4D spacetime
\cite{sch12:book,mis17:book,car19:book,zha26:1613} and thus varies with geometric
characteristics of the slicing. In practical calculations, a set of
generalized coordinates $\mathbi{q}$ must be defined as functions of
$\mathbi{x}$, that is $q^{(\kappa)}=q^{(\kappa)}(x^{(1)},\cdots,x^{(3N)})$
and $x^{(i)}=x^{(i)}(q^{(1)},\cdots,q^{(3N)})$. If they are linear functions,
called rectilinear coordinate; meanwhile, if they are non-linear, called
curvilinear coordinate, the KEO expression becomes complex and has been
thoroughly derived \cite{pod28:812,lau02:8560,gat09:1,ndo12:034107,ndo13:204107,sad12:234112}.
Assuming that one has designed the generalized coordinates $\mathbi{q}$,
the metric given in Equation \eqref{eq:metric-002-conf-space-mol} can
be obtain through the transformation between $\mathbi{x}$ and $\mathbi{q}$.
The existence of such transformation is ensured by Lorentz invariance
of the spacetime interval.
\begin{post}[principle of local Lorentz invariance]
The spacetime interval is invariant under arbitrary local Lorentz
transformations.
\label{post:lorentz-inva}
\end{post}
Under the $[3+1]$ decomposition, the interval invariance of Postulate
\ref{post:lorentz-inva} is fully applicable, but it must be interpreted
locally. Postulate \ref{post:lorentz-inva} can be also simply
expressed in terms of observable quantities that do not depend on the
choice of coordinates. In this work, we only work within the framework of absolute spacetime
and neglect special-relativistic effects. Consequently, the Lorentz
transformation is replaced by its non-relativistic limit, that is the
Galilean transformation, and no boost effects are considered.
\begin{remark}
The inner product between
quantum states defined in the case with the existence of absolute time
(based on Eulerian observer) is not Lorentz invariant \cite{bos25:015003}.
If one wants to describe chemical dynamics in a manner consistent with
relativity, a covariant formalism must be adopted to make the EOMs and
the inner products Lorentz invariant. Quantum field theory (QFT) in
curved spacetime \cite{bir82:book,bar09:book,hac16:book} enables it in
the low-energy regime.
\label{rem:lorentz-0-invari}
\end{remark}

Before deriving the metric of Equation \eqref{eq:metric-002-conf-space-mol}
on the basis of Postulate \ref{post:lorentz-inva}, one should turn to
design of generalized coordinates $\mathbi{q}$ for representing the
molecular system in configuration space, as shown in Figure
\ref{fig:map-slice-conf}. Based on chemical experience, the system is
hierarchically divided into several fragments each containing two or
three atoms. Then, the following three levels of curvilinear coordinates
are usually adopted
\cite{lau02:8560,gat09:1,ndo12:034107,sad12:234112,ndo13:204107,mar22:20200388}
leading to hierarchical representation \cite{zha25:20397}. The first
level is three Cartesian coordinates which describe the position of
the center-of-mass (COM) of the entire system in a space-fixed (SF) or
called laboratory-fixed (LF) frame. The second level is coordinate set
which describes internal motions of one of fragments in a BF frame.
Origin of the BF frame coincides with the fragment COM,
while axes of the BF frame are related to the system in either an
explicit or an implicit way. The third level is three Euler angles
which determine the orientation of a BF frame. We refer the reader to
References \cite{pod28:812,lau02:8560,gat09:1,ndo12:034107,sad12:234112,ndo13:204107,mar22:20200388}
for details of the above three levels of coordinates and to Reference
\cite{zha25:20397} for discussions of the advantages and disadvantages
of such hierarchical representation. In addition, the following ways
were proposed to reduce the system dimensionality. The simplest way
is a rigid-constraint approach which consists freezing some bond
parameters (such as bond lengths, angles, and dihedral angles) or
entire atomic group. This way is too simplistic and discards a
significant amount of dynamics details. Thus, a semi-rigid approach is
often employed instead, where the variations of the inactive coordiantes
are employed to adjust the active coordinates. This approach reduces
the system dimensionality through an effective Hamiltonian in the
active variables.
Designing these coordinate systems requires chemical experience and
insight, whereas the simplest choice is clearly the Cartesian coordinate
system $\mathbi{x}$. In general, Cartesian coordinates offer simplicity
and intuitive geometric interpretation, making them ideal for problems
with rectilinear symmetry, such as quasi-classical molecular dynamics
simulations. However, the situation changes for quantum molecular
dynamics where Cartesian coordinates are rarely emplyed because calculations
often become cumbersome.

%----------------------------------------
\subsection{Kinetics Energy Operator\label{sec:keo-coord}}

At a given moment, where the system is represented by conformation in
one of spatial slices shown in Figure \ref{fig:reax-spacetime}, the
KEO is associated with the Laplacian operator in the general form of
Equation \eqref{eq:christoffel-symbol-007}. For mass-weighted Cartesian
coordinate set $\mathbi{x}=\{x^{(i)}\}_{i=1}^{3N}$, according to Equation
\eqref{eq:christoffel-symbol-007}, the metric tensor $\tilde{\gamma}^{ij}$
and Christoffel symbols $\Gamma_{ij}^k$ of the second kind encode
geometric characteristic of the configuration space and lead to the
KEO,
\begin{equation}
2\hat{T}=-\nabla^2=-\tilde{\gamma}^{ij}\Big(\partial_i\partial_j-
\Gamma_{ij}^k\partial_k\Big)
=-\frac{1}{\sqrt{\vert\tilde{\gamma}\vert}}\partial_i
\Big(\sqrt{\vert\tilde{\gamma}\vert}\tilde{\gamma}^{ij}\partial_j\Big)
=-\left[\tilde{\gamma}^{ij}\partial_i\partial_j+
\Big(\partial_i\tilde{\gamma}^{ij}\Big)\partial_j+
\frac{1}{2}\tilde{\gamma}^{ij}\Big(\partial_i
\ln\tilde{\gamma}\Big)\partial_j\right],
\label{eq:polysph-keo-003}
\end{equation}
where, due to metric compatibility, the Christoffel symbols (see also
Equation \eqref{eq:christoffel-symbol-000}) are uniquely determined by
the components of the metric tensor and their first partial derivatives,
\begin{equation}
\Gamma_{ij}^k=\frac{1}{2}\tilde{\gamma}^{kl}
\Big(\partial_i\tilde{\gamma}_{lj}+\partial_j\tilde{\gamma}_{li}-
\partial_l\tilde{\gamma}_{ij}\Big).
\label{eq:polysph-keo-003-relaca}
\end{equation}
In Equation \eqref{eq:polysph-keo-003}, $\partial_i$ means $\partial/\partial x^{(i)}$
and $\tilde{\gamma}$ is determinant of the metric $\tilde{\gamma}_{ij}$.
In the last expression of Equation \eqref{eq:polysph-keo-003}, the
terms $\tilde{\gamma}^{ij}\partial_i\partial_j$, $(\partial_i\tilde{\gamma}^{ij})\partial_j$,
and $(\tilde{\gamma}^{ij}(\partial_i\ln\tilde{\gamma})\partial_j)/2$
are kinetic term, geometric potential, and first-order derivative term,
respectively. The kinetic term corresponds to the KEO in flat Euclidean
space, where $\tilde{\gamma}^{ij}=\delta^{ij}$ and $\Gamma_{ij}^k=0$
making Equation \eqref{eq:polysph-keo-003} become the simplest form
\begin{equation}
2\hat{T}=-\partial_i\delta^{ij}\partial_j=-\partial_i\partial_i
=-\sum_{i=1}^{3N}\frac{\partial^2}{\partial x^{(i)}{}^2}.
\label{eq:keo-cartesian-f-dofs-000}
\end{equation}
The first-order derivative term can be expressed as an effective gauge
coupling, reflecting the influence of the chosen coordinate frame on
dynamics. It is worth noting that this term is an analogy to the gauge
interaction of the environment with the system, yet the environment is
specially restricted to choosing a coordinate frame. The geometric
potential arises from the curvature of the configuration space represented
by the coordinate frame. Therefore, as long as we have got the metric
tensor of the slice, one can derive the KEO in Cartesian coordinates.
However, the Cartesian coordinates usually cannot intuitively describe
the nuclear time-dependent wave function in breaking and forming of
chemical bond. These limitations prevent the use of the Cartesian
coordinates in quantum dynamics calculations, as shown in the above
Section \ref{sec:coord-frame}.

To obtain the KEO in generalized coordinate set $\mathbi{q}$, the metric
tensor $\varg^{ij}$ should be defined by the transformations between
$\mathbi{q}$ and $\mathbi{x}$, encoding geometric characteristic by
$\mathbi{q}$. Then, based on the principle of local Lorentz invariance
given by Postulate \ref{post:lorentz-inva}, one can obtain the KEO
that depends on $\varg^{\iota\kappa}$ through Equation \eqref{eq:polysph-keo-003}.
To this end, according to Postulate \ref{post:lorentz-inva}, Equation
\eqref{eq:metric-001-conf-space-mol} can be rewritten as
\begin{equation}
\mathrm{d}s^2=\tilde{\gamma}_{ij}\mathrm{d}x^{(i)}\mathrm{d}x^{(j)}=
\varg_{\iota\kappa}\mathrm{d}q^{(\iota)}\mathrm{d}q^{(\kappa)}.
\label{eq:transf-relation-coord-000}
\end{equation}
In orther words, Equation \eqref{eq:transf-relation-coord-000} indicates
that transformations of cooridnates do not change $\mathrm{d}s^2$, such
as the length of a vector in flat space. Thus, the metric tensor
$\varg^{\iota\kappa}$ can be derived in the form
\begin{equation}
\varg^{\iota\kappa}=\tilde{\gamma}^{ij}\partial_{i}q^{(\iota)}
\partial_{j}q^{(\kappa)},\quad
\varg_{\iota\kappa}=(\varg^{\iota\kappa})^{-1},
\label{eq:transf-relation-coord-990}
\end{equation}
while the metric $\tilde{\gamma}^{ij}$ can be given by
\begin{equation}
\tilde{\gamma}^{ij}=\varg^{\iota\kappa}
\partial_{\iota}x^{(i)}\partial_{\kappa}x^{(j)}.
\label{eq:transf-relation-coord-991}
\end{equation}
Now, let define the absolute value of the determinant of $3N\times3N$
Jacobian matrix with elements $\partial_{\kappa}x^{(i)}$ as
\begin{equation}
J=\left\vert\det\left(\frac{\partial x^{(i)}}{\partial q^{(\kappa)}}
\right)\right\vert=
\Big\vert\det\Big(\partial_{\kappa}x^{(i)}\Big)\Big\vert.
\label{eq:transf-relation-coord-992}
\end{equation}
Substituting Equation \eqref{eq:transf-relation-coord-991} into Equation
\eqref{eq:transf-relation-coord-992}, one can directly find that
\begin{equation}
J=\left(\left\vert\frac{\det(\tilde{\gamma}^{ij})}
{\det(\varg^{\iota\kappa})}\right\vert\right)^{\frac{1}{2}}.
\label{eq:transf-relation-coord-993}
\end{equation}
Next, substituting Equation \eqref{eq:transf-relation-coord-991} into
Equation \eqref{eq:polysph-keo-003}, one can obtain the KEO of $\mathbi{q}$
in the form
\cite{pod28:812,lau02:8560,gat09:1,ndo12:034107,sad12:234112,ndo13:204107,mar22:20200388,zha25:20397}
\begin{equation}
2\hat{T}=-J^{-1}\partial_{\iota}J\varg^{\iota\kappa}\partial_{\kappa},
\label{eq:polysph-keo-011}
\end{equation}
As described in Remark \ref{rem:contangent-fiber}, one can define
generalized momentum operators associated with generalized coordinates.
It is then straightforward to eliminate $J$ in Equation \eqref{eq:polysph-keo-011}
and obtain the KEO represented by momentum operators rather than
coordinates \cite{gat09:1},
\begin{equation}
2\hat{T}=\hat{p}_{\iota}^{\dagger}\varg^{\iota\kappa}\hat{p}_{\kappa}.
\label{eq:polysph-keo-014}
\end{equation}
In the coordinate representation, the momentum operators $\hat{p}_{\kappa}$
and $\hat{p}_{\iota}^{\dagger}$ are given by
\begin{equation}
\hat{p}_{\kappa}=-\mi\partial_{\kappa},\quad
\hat{p}_{\iota}^{\dagger}=-\mi J^{-1}\partial_{\iota}J=
\hat{p}_{\iota}+(J^{-1}\hat{p}_{\iota}J)=
\hat{p}_{\iota}+\hat{\Lambda}_{\iota},
\label{eq:polysph-keo-015}
\end{equation}
where $\hat{\Lambda}_{\iota}=(J^{-1}\hat{p}_{\iota}J)$ is a purely
multiplicative operator. The parentheses in the expression of
$\hat{\Lambda}_{\iota}$ means that $\hat{p}_{\iota}$ in it does not
operate beyond the parentheses. In Equations \eqref{eq:polysph-keo-014}
and \eqref{eq:polysph-keo-015}, the momentum operators $\hat{\mathbi{p}}
=\{\hat{p}_{\kappa}\}_{\kappa=1}^{3N}$ are conjugate to the generalized
coordinates $\mathbi{q}=\{q^{(\kappa)}\}_{\kappa=1}^{3N}$, requiring
existence of $J$. Momenta of otherwise kind are often called quasi-momenta,
denoted by $\hat{\boldsymbol{\mathfrak{p}}}=\{\hat{\mathfrak{p}}_{\zeta}\}$.
Similar to Equation \eqref{eq:polysph-keo-014}, the KEO can be written
in terms of quasi-momenta as \cite{gat09:1}
\begin{equation}
2\hat{T}=\hat{\mathfrak{p}}_{\zeta}^{\dagger}\varg^{\zeta\eta}
\hat{\mathfrak{p}}_{\eta},
\quad\varg^{\zeta\eta}=\tilde{\gamma}^{ij}\partial_{i}\mathfrak{q}^{(\zeta)}
\partial_{j}\mathfrak{q}^{(\eta)},
\label{eq:polysph-keo-019}
\end{equation}
where $\boldsymbol{\mathfrak{q}}=\{\mathfrak{q}^{(\zeta)}\}$
means coordinate set associated with quasi-momenta $\hat{\boldsymbol{\mathfrak{p}}}$,
say angular coordinates associated with angular momenta.

Although the KEO expressions are given by Equations \eqref{eq:polysph-keo-011},
\eqref{eq:polysph-keo-014}, and \eqref{eq:polysph-keo-019}, it is
worth noting that motions of the entire system, three translational
motions of the COM and three (or two) rotations, should be considered
differently from the internal motions. Thus, these KEO expressions
derived by the metric in the form of Equation \eqref{eq:metric-002-conf-space-mol}
can be further simplified. To this end, Lauvergnat and Nauts
\cite{lau02:8560} together with Gatti and Iung \cite{gat09:1} have well
discussed in details. Noting that three translations of the COM are
redundant and three (or two) rotations of the entire system are not
only redundant but also coupled with internal motions, Lauvergnat and
Nauts \cite{lau02:8560} introduced the SF frame (or called LF frame)
and E$_2$ frame to represent the molecular system. As given in Section
\ref{sec:coord-frame}, this representation employs Euler angles to
describe the overall orientation of the molecule, while utilizing the
Eckart conditions to define the internal vibrational modes (hence the
name of E$_2$ frame).
\begin{defi}
The E$_2$ coordinate frame can be seen as a combination of Eulerian and
Eckart coordinate frames.
\label{def:e2-coord-frame}
\end{defi}
By separating three translations, one can represent an $N$-atomic
system by a $(d+3)$-dimensional configuration space, where $d=3N-6$
or $d=3N-5$ depending on rotational symmetry of the system. Then, Nauts
and co-workers \cite{nau87:164,lau02:8560} reduced the $3N\times3N$
metric $\tilde{\gamma}_{ij}$ and $\tilde{\gamma}^{ij}$ in the
$(d+3)\times(d+3)$ form
\begin{equation}
\tilde{\gamma}_{ij}=\left(\begin{array}{ccc}
\mathbf{S}_{d\times d} & \mathbf{C}_{d\times3}^{\mathrm{T}} \\
\mathbf{C}_{3\times d} & \mathbf{V}_{3\times3} \\
\end{array}\right),\quad
\tilde{\gamma}^{ij}=\tilde{\gamma}_{ij}^{-1}=\left(\begin{array}{ccc}
\boldsymbol{\upsilon}_{d\times d} & \boldsymbol{\chi}_{d\times3}^{\mathrm{T}} \\
\boldsymbol{\chi}_{3\times d} & \boldsymbol{\varsigma}_{3\times3} \\
\end{array}\right)
\label{eq:metric-red-000}
\end{equation} 
where $\{\mathbf{S}_{d\times d},\boldsymbol{\upsilon}_{d\times d}\}$,
$\{\mathbf{C}_{3\times d},\boldsymbol{\chi}_{3\times d}\}$, and
$\{\mathbf{V}_{3\times3},\boldsymbol{\varsigma}_{3\times3}\}$ are the
internal deformation matrix, the Coriolis coupling matrix, and the
inertia tensor, respectively \cite{nau87:164,lau02:8560}. Having
Equation \eqref{eq:metric-red-000}, the KEO expression can be obtained
in the summation of products of single-mode operators \cite{lau02:8560}.
\begin{remark}
This approach systematically separates the translational, rotational,
and internal vibrational modes of the molecule. In essence, it
represents a methodology for projecting the motion in the SF frame onto
a molecular frame.
\label{rem:e2-frame}
\end{remark}
Having the Jacobian, Lauvergnat and Gatti and co-workers
\cite{lau02:8560,gat09:1,ndo12:034107,sad12:234112,ndo13:204107,mar22:20200388}
developed a sophisticated set of numerical and analytical techniques
to directly obtain the KEO.

Finally, turning to problems mentioned in Remark \ref{rem:lorentz-0-invari},
we must turn to the volume element
$\mathrm{d}\tau\propto\mathrm{d}x^{(1)}\cdots\mathrm{d}x^{(3N)}$. In
solving the SE, the scale factor must be determined. According to
the principle of local Lorentz invariance (Postulate \ref{post:lorentz-inva}),
a direct choice is Jacobian, that is
$\mathrm{d}\tau^{(\mathrm{E})}=J\mathrm{d}x^{(1)}\cdots\mathrm{d}x^{(3N)}$,
called Euclidean volume element. However, in most cases of quantum
molecular calculations, the Euclidean volume element is not adapted
because such choice cannot ensure orthonormality of the multi-dimensional
basis set \cite{gat09:1,ndo12:034107,sad12:234112,ndo13:204107,mar22:20200388}.
To overcome this problem, one should define a volume element
$\mathrm{d}\tau^{(\varrho)}=\varrho\mathrm{d}x^{(1)}\cdots\mathrm{d}x^{(3N)}$
so that the basis set is orthonormal \cite{nau85:1287,cha91:1}. In
this context, according to Equations \eqref{eq:transf-relation-coord-993}
and \eqref{eq:polysph-keo-011}, the KEO adapted to this new volume
element is given by \cite{gat09:1,mar22:20200388,nau85:1287,cha91:1}
\begin{equation}
2\hat{T}^{(\varrho)}=
2\sqrt{\dfrac{J}{\varrho}}\hat{T}\sqrt{\dfrac{\varrho}{J}}=
-\varrho^{-1}\partial_{\iota}\varrho
\varg^{\iota\kappa}\partial_{\kappa}+2V_{\mathrm{ep}},
\label{eq:volumn-ele-keo-09}
\end{equation}
where $V_{\mathrm{ep}}$ is a scalar term, called extra-potential or
pseudo-potential term. This term is arisen from the geometric term
and first-order derivative term. Despite of deep mathematical analogy,
change in the volume element is not a gauge transformation because it
reflects nothing about an internal symmetry of the system. Nevertheless,
when such a change is performed, the extra-potential term depends on
$\varrho$, implies that this term can appear in a form analogous to a
gauge potential. Thus, physical effect of coordinate transformation can
mimic that of a gauge field. For the flat space, Gatti and Iung
\cite{gat09:1} has derived the expression of $V_{\mathrm{ep}}$ in
details for the case of $\varrho=1$ because this choice ensures
orthonormality of the basis set in the flat space. In curved space,
according to Equation \eqref{eq:metric-001-conf-space-mol}, a natural
choice is $\varrho=\tilde{\gamma}$.
\begin{remark}
The extra-potential term $V_{\mathrm{ep}}$ arises because the
configuration space is curved. By choosing coordinates that reflect
the underlying geometry, the curvature effects can be recast as
additional potential terms, which are formally analogous to gravitational
interactions, rather than being encoded in a complicated metric of
configuration sapce. Such a choice is anticipated to eliminate the
need for incorporating various hierarchical fragments and the BF frames
attached to them. These hierarchical constructions bear a striking
resemblance \cite{zha25:20397} to the epicyclic-deferent model
adopted in pre-Copernican astronomy, which analyzed the orbital
motions of our solar system relative to the Earth-based reference
frame.
\label{rem:vep-phys-0}
\end{remark}

%------------------------------------------------------------
\subsection{Potential Energy Surface\label{sec:pes-build-regree}}

Having the KEO, we now turn to the potential term of the nuclear Hamiltonian
operator. Since the PES is a function operator of only coordinates, its
expression in the coordinate representation remains metric invariant
under transformations between frames of nuclear coordinates. In this
context, the task of constructing the PES is to approximate a real-valued
function $f:\mathbb{R}^d\to\mathbb{R}$ through the set of discrete
energies \cite{son22:1983,zha25:20397}. As previously discussed \cite{son22:1983,zha25:20397},
the approximated function parametrically depends on a set of parameters
$\theta\in\mathbb{R}^m$, denoted by $f_{\theta}:\mathbb{R}^d\to\mathbb{R}$.
The GLR-based function approximation \cite{son22:1983} aims to find the
optimal parameters $\theta$ by directly minimizing the distance between
$f$ and $f_{\theta}$. The KMR-based function approximation \cite{son22:1983}
aims to find the optimal superparameters $\theta$ of the kernel function
by maximizing the marginal likelihood of the database. Inspired by GLR
and KMR of entirely different styles, the concept of potenial tangent
bundle (PTB),
\begin{equation}
\mathrm{PTB}\big(f_{\theta},\theta\big)=\mathrm{span}
\left\{\frac{\partial}{\partial\theta_i}f_{\theta}\Big\vert
i=1,\cdots,m\right\},
\label{eq:tangent-bundle-00-ai}
\end{equation}
offers a general approach to function approximation by
\begin{equation}
\mathrm{PTB}\big(f_{\theta},\alpha\big)=\frac{\partial}{\partial\theta}
f_{\theta}\cdot\alpha,\quad\alpha\in\mathbb{R}^m,
\label{eq:tangent-bundle-01-ai}
\end{equation}
where $\alpha$ are the linear combination coefficients and thus the
minimizer in building the PES. To easily derive the present theory,
let us firstly consider the simplest case with an analytical function
$f(\mathbi{q})$ as objective function in function approximation. Then,
we consider the practical case with the discrete database as training
set.

If the analytic form of the objective potential function $f(\mathbi{q})$
has been obtained, the function approximation becomes least-squares problem
\begin{equation}
\alpha=\arg\min_{\alpha\in\mathbb{R}^m}\int\left\Vert
\frac{\partial}{\partial\theta}f_{\theta}(\mathbi{q})\cdot\alpha-
f(\mathbi{q})\right\Vert^2\mathrm{d}^d\mathbi{q}.
\label{eq:tangent-bundle-02-ai}
\end{equation}
This is a semi-convex optimization and has solution expressed in a
transformed form
\begin{equation}
\alpha=\mathscr{G}^{\dagger}_{\theta}P\big(\theta,f\big),\quad
\mathscr{G}_{\theta}\in\mathbb{R}^{m\times m},\quad
P\big(\theta,\cdot\big):\mathcal{L}^2\big(\mathbb{R}^d,\mathbb{R}\big)
\to\mathbb{R}^m,
\label{eq:tangent-bundle-03-ai}
\end{equation}
where
\begin{equation}
\mathscr{G}_{\theta}=\int\frac{\partial}{\partial\theta}
f_{\theta}^{\perp}\big(\mathbi{q}\big)\cdot
\frac{\partial}{\partial\theta}f_{\theta}\big(\mathbi{q}\big)
\mathrm{d}^d\mathbi{q},\quad
P\big(\theta,f\big)=\int\frac{\partial}{\partial\theta}
f_{\theta}^{\perp}\big(\mathbi{q}\big)\cdot
f\big(\mathbi{q}\big)\mathrm{d}^d\mathbi{q}
\label{eq:tangent-bundle-04-ai}
\end{equation}
are metric tensor of the parameter space and projection, respectively,
while $\mathcal{L}^2(\mathbb{R}^d,\mathbb{R})$ is the square-integrable
function space, usually denoted by $L^2$ space. Defining a projection
operator in the $L^2$ space as
\begin{equation}
\mathscr{K}_{\theta}\big[\cdot\big]\big(\mathbi{q}\big)=
\frac{\partial}{\partial\theta}f_{\theta}\big(\mathbi{q}\big)
\mathscr{G}^{\dagger}_{\theta}P\big(\theta,\cdot\big),
\label{eq:tangent-bundle-05-ai}
\end{equation}
one can write the approximation in Equation \eqref{eq:tangent-bundle-01-ai}
of function $f(\mathbi{q})$ as $\mathrm{PTB}(f_{\theta},\alpha)=
\mathscr{K}_{\theta}[f](\mathbi{q})$. Substituting Equation
\eqref{eq:tangent-bundle-04-ai} into Equation \eqref{eq:tangent-bundle-05-ai},
one can finally obtain
\begin{equation}
\mathrm{PTB}\big(f_{\theta},\alpha\big)=
\mathscr{K}_{\theta}\big[f\big]\big(\mathbi{q}\big)=
\frac{\partial}{\partial\theta}f_{\theta}\cdot\left(\int
\frac{\partial}{\partial\theta}f_{\theta}^{\perp}
\big(\mathbi{q}\big)\cdot\frac{\partial}{\partial\theta}
f_{\theta}\big(\mathbi{q}\big)\mathrm{d}^d\mathbi{q}\right)^{\dagger}
\cdot\left(\int\frac{\partial}{\partial\theta}
f_{\theta}^{\perp}\big(\mathbi{q}\big)\cdot
f\big(\mathbi{q}\big)\mathrm{d}^d\mathbi{q}\right).
\label{eq:tangent-bundle-06-ai}
\end{equation}
Comparing to the unified regression model for building the PES \cite{son22:1983},
the PTB approximation of Equation \eqref{eq:tangent-bundle-03-ai}
optimizes the coefficients $\alpha$ rather than the parameters
$\theta$.

Now, we should express the above function approximation in discrete
form leading to the matrix formalism. According to the unified regression
model \cite{son22:1983} the tangent bundle defined by Equation
\eqref{eq:tangent-bundle-00-ai} offers the PES in the expansional form
of Equation \eqref{eq:tangent-bundle-01-ai}. With the training database
$\{\mathbf{X},\mathbf{E}\}=\{\mathbi{X}_i,E_i\vert i=1,\cdots,n\}$, the
least-squares problem of Equation \eqref{eq:tangent-bundle-02-ai} becomes
the matrix form
\begin{equation}
\alpha=\arg\min_{\alpha\in\mathbb{R}^m}\sum_{i=1}^n
\left\Vert\frac{\partial}{\partial\theta}f_{\theta}(\mathbi{X}_i)
\cdot\alpha-E_i\right\Vert^2,
\label{eq:tangent-bundle-902-ai}
\end{equation}
where $\mathbi{X}_i$ and $E_i$ are the $i$-th point in the configuration
space and corresponding energy value, respectively, while $\mathbf{X}$
and $\mathbf{E}$ are $n\times d$ and $n\times1$ matrices, respectively.
Similar to Equation \eqref{eq:tangent-bundle-03-ai}, this optimization
in Equation \eqref{eq:tangent-bundle-902-ai} for $\alpha$ has solution
expressed in a matrix transform
$\alpha=\mathscr{G}^{\dagger}_{\theta}P(\theta,\mathbf{E})$, where
$\mathscr{G}_{\theta}\in\mathbb{R}^{m\times m}$ and
$P(\theta,\cdot):\mathcal{L}^2(\mathbb{R}^d,\mathbb{R})\to\mathbb{R}^m$
are given by
\begin{equation}
\mathscr{G}_{\theta}=\sum_{i=1}^n
\frac{\partial}{\partial\theta}
f_{\theta}^{\perp}\big(\mathbi{X}_i\big)\cdot
\frac{\partial}{\partial\theta}f_{\theta}\big(\mathbi{X}_i\big),\quad
P\big(\theta,\mathbf{E}\big)=\sum_{i=1}^n
\frac{\partial}{\partial\theta}f_{\theta}^{\perp}
\big(\mathbi{X}_i\big)\cdot E_i.
\label{eq:tangent-bundle-904-ai}
\end{equation}
With Equation \eqref{eq:tangent-bundle-05-ai} in the $L^2$ space and
the metric and projection in Equation \eqref{eq:tangent-bundle-904-ai},
one can similarly write the PES prediction $f_{\theta}(\mathbi{q})$ as
\begin{equation}
\mathrm{PTB}\big(f_{\theta},\alpha\big)=
\mathscr{K}_{\theta}\big[\mathbf{E}\big]\big(\mathbi{q}\big)=
\frac{\partial}{\partial\theta}f_{\theta}\cdot\left(\sum_{i=1}^n
\frac{\partial}{\partial\theta}f_{\theta}^{\perp}
\big(\mathbi{X}_i\big)\cdot
\frac{\partial}{\partial\theta}f_{\theta}\big(\mathbi{X}_i\big)
\right)^{\dagger}\cdot\left(\sum_{i=1}^n
\frac{\partial}{\partial\theta}f_{\theta}^{\perp}
\big(\mathbi{X}_i\big)\cdot E_i\right).
\label{eq:tangent-bundle-906-ai}
\end{equation}
Until now, we have not specified the functional form of $f_{\theta}$,
consequently, the PES construction presented above (say Equations
\eqref{eq:tangent-bundle-06-ai} and \eqref{eq:tangent-bundle-906-ai})
is general. We refer the reader to Reference \cite{zha25:20397} for
various functional forms of $f_{\theta}$ in building the PES.

According to the unified regression model \cite{son22:1983}, all of
fitting approaches to build the PES can be seen as linear relation of
the so-called feature space \cite{son22:1983}. This space is a
sufficiently flexible function of coordinates, denoted by
$\tilde{f}_{\tilde{\theta}}(\mathbi{q})$, where
$\tilde{\theta}\in\mathbb{R}^{m-m'}$ and
$\tilde{f}_{\tilde{\theta}}(\mathbi{q})\in\mathrm{R}^{1\times m'}$ with
$1<m'<m$. Thus, the objective potential function $f_{\theta}$ can be
generally written in linear relation,
\begin{equation}
f_{\theta}\big(\mathbi{q}\big)=
\tilde{f}_{\tilde{\theta}}\big(\mathbi{q}\big)\cdot\omega,\quad
\omega\in\mathbb{R}^{m'}.
\label{eq:tangent-bundle-800-ai}
\end{equation}
By Equation \eqref{eq:tangent-bundle-800-ai}, it should be emphasize
that there must exist $f_{\theta}$ in the PTB as deifned by Equation
\eqref{eq:tangent-bundle-00-ai}. This point is easy to understand if
we further rewritten Equation \eqref{eq:tangent-bundle-800-ai} as
\begin{equation}
f_{\theta}\big(\mathbi{q}\big)=
\tilde{f}_{\tilde{\theta}}\big(\mathbi{q}\big)\cdot\omega=
\left(\begin{array}{cr}
\dfrac{\partial}{\partial\tilde{\theta}}
\tilde{f}_{\tilde{\theta}}\cdot\omega &
\quad\tilde{f}_{\tilde{\theta}}
\end{array}\right)\cdot\left(\begin{array}{c}
0 \\
\omega
\end{array}\right)=\frac{\partial}{\partial\theta}f_{\theta}\cdot
\alpha\in\mathrm{span}
\left\{\frac{\partial}{\partial\theta_i}f_{\theta}\Big\vert
i=1,\cdots,m\right\}
=\mathrm{PTB}\big(f_{\theta},\theta\big).
\label{eq:tangent-bundle-801-ai}
\end{equation}
Obviously, Equations \eqref{eq:tangent-bundle-906-ai} and
\eqref{eq:tangent-bundle-801-ai} possess little practical significance
but can achieve reasonable approximation PES, even if the PTB is not
optimal \cite{son22:1983}. In addition, if the feature space
$\tilde{f}_{\tilde{\theta}}$ itself possesses a certain mathematical
structure, say expansion by products basis, then the resulting PES
$f_{\theta}$ will also exhibit the same structure \cite{son22:11128,son24:597}.
This serves as the mathematical foundation for developing a range of
fitting methods \cite{zha25:20397}.

To easily understand the above function approximation, we need to specify
the explicit functional form \cite{son22:1983}. Starting from the original
linear regression (LR), the relation between the conformations $\mathbi{q}$
and and electronic potential energy $f^{(\mathrm{LR})}$ can be
written as $f^{(\mathrm{LR})}(\mathbi{q})=\mathbi{q}^{\mathrm{T}}
\cdot\mathbi{w}$, where $\mathbi{w}$ represets a $d\times1$ parameter
matrix. To optimize $\mathbi{w}$, the
minimization target function weighing the regression accuracy is given
by $J=(\mathbf{E}-\mathbf{X}^{\mathrm{T}}\mathbi{w})^{\mathrm{T}}\cdot
(\mathbf{E}-\mathbf{X}^{\mathrm{T}}\mathbi{w})$. Setting
$\partial_{\mathbi{w}}J\vert_{\mathbi{w}=\overline{\mathbi{w}}}=\mathbf{0}$,
one can directly obtain the resulting prediction
$f^{(\mathrm{LR})}(\mathbi{q})=\mathbi{q}^{\mathrm{T}}
\cdot\overline{\mathbi{w}}=\mathbi{q}^{\mathrm{T}}\cdot
(\mathbf{X}\cdot\mathbf{X}^{\mathrm{T}})^{-1}\cdot\mathbf{X}\cdot\mathbf{E}$,
which clearly shows that $f^{(\mathrm{LR})}(\mathbi{q})$ is
not only a function of $\mathbi{q}$ but also parametrically depends on
the training database $\{\mathbf{X},\mathbf{E}\}$. For complex and
high-dimensional cases, one can transfer $\mathbf{X}$ to another
appropriate matrix $\boldsymbol{\Phi}$, called feature space (or
feature matrix), to characterise the feature of realistic system
elaborately leading to the generalized linear regression (GLR) method.
In other words, an appropriate map of
$\boldsymbol{\mathscr{F}}:\mathbf{X}\to\boldsymbol{\Phi}$ must be
firstly determined for GLR, where $\boldsymbol{\Phi}$ is set to be a
$p\times n$ matrix such that $n$ new $p$-dimensional inputs are used.
Similar to the LR method, one can optimize the parameter matrix and
obtain the GLR prediction $f^{(\mathrm{GLR})}(\mathbi{q})=
\boldsymbol{\phi}^{\mathrm{T}}\cdot\overline{\boldsymbol{\omega}}
=\boldsymbol{\mathscr{F}}(\mathbi{q}^{\mathrm{T}})\cdot
(\boldsymbol{\Phi}\cdot\boldsymbol{\Phi}^{\mathrm{T}})^{-1}
\cdot\boldsymbol{\Phi}\cdot\mathbf{E}$.
Since the target $J$ in either LR or GPR is regression accuracy, both
LR and GLR are concerned on the fitting accuracy rather than the
generalization ability, implying the risk of overfitting in LR and GLR.
To recover this problem, an additional term with a penalized parameter
$\lambda$ should be added into the target function $J$ to shorten the
magnitude of parameters and to balance fitting accuracy and model
generalizability. One of the commonly used improved regressions is
ridge regression with the target 
$J_{\mathrm{ridge}}=(\mathbf{E}-\boldsymbol{\Phi}^{\mathrm{T}}\cdot
\boldsymbol{\omega})^{\mathrm{T}}\cdot(\mathbf{E}-\boldsymbol{\Phi}^{\mathrm{T}}\cdot
\boldsymbol{\omega})+\lambda\boldsymbol{\omega}^{\mathrm{T}}\cdot
\boldsymbol{\omega}$, providing optimized parameter matrix
$\overline{\boldsymbol{\omega}}=(\boldsymbol{\Phi}\cdot\boldsymbol{\Phi}^{\mathrm{T}}
+\lambda\mathbf{I})^{-1}\boldsymbol{\Phi}\cdot\mathbf{E}$. Comparing
the fitting errors of validation set for different values of $\lambda$,
one can determine $\lambda$ and associated $\overline{\boldsymbol{\omega}}$.
Another way to rewrite the parameter matrix and prediction involves
construction of a linear combination
$\boldsymbol{\omega}=-\boldsymbol{\Phi}(\boldsymbol{\Phi}^{\mathrm{T}}
\boldsymbol{\omega}-\mathbf{E})/\lambda=\boldsymbol{\Phi}\cdot\mathbf{a}$,
where $\mathbf{a}=-(\boldsymbol{\Phi}^{\mathrm{T}}\boldsymbol{\omega}-\mathbf{E})/\lambda$
is an $n\times1$ transformation vector. The function model is then given
by $f^{(\mathrm{KMR})}(\mathbi{q})=\boldsymbol{\phi}^{\mathrm{T}}\cdot\boldsymbol{\omega}=
\boldsymbol{\phi}^{\mathrm{T}}\cdot\boldsymbol{\Phi}\cdot\mathbf{a}=
\boldsymbol{\mathscr{F}}(\mathbi{q}^{\mathrm{T}})\cdot\boldsymbol{\Phi}\cdot\mathbf{a}
=\mathbf{K}_*^{\mathrm{T}}\cdot\mathbf{a}$,
where $\mathbf{K}_*^{\mathrm{T}}=\boldsymbol{\phi}^{\mathrm{T}}\cdot\boldsymbol{\Phi}$.
The matrix $\mathbf{K}_*$ only depends on definition of $\boldsymbol{\mathscr{F}}(\cdot)$
and hence has elements of bivariate kernel function $K(\cdot,\cdot)$.
The model $f^{(\mathrm{KMR})}(\mathbi{q})$ is thus named the kernel-model
regression (KMR). Minimizing the target function $J_{\mathrm{ridge}}$,
one can have got the KMR prediction
$f^{(\mathrm{KMR})}(\mathbi{q})=\mathbf{K}_*^{\mathrm{T}}\cdot\overline{\mathbf{a}}=
\mathbf{K}_*^{\mathrm{T}}\cdot(\mathbf{K}+\lambda\mathbf{I})^{-1}\cdot\mathbf{E}$,
where $\mathbf{K}=\boldsymbol{\Phi}^{\mathrm{T}}\cdot\boldsymbol{\Phi}$
and $\mathbf{I}$ is $n\times n$ identity matrix. Therefore, two kinds
of prediction expressions,
\begin{equation}
f^{(\mathrm{GLR})}(\mathbi{q})=
\boldsymbol{\phi}^{\mathrm{T}}\cdot\overline{\boldsymbol{\omega}},\quad
f^{(\mathrm{KMR})}(\mathbi{q})=
\mathbf{K}_*^{\mathrm{T}}\cdot\overline{\mathbf{a}},
\label{eq:glr-kmr-000}
\end{equation}
are found from the original LR method. These expressions are both
consistent with Equations \eqref{eq:tangent-bundle-800-ai} and
\eqref{eq:tangent-bundle-800-ai}. By $f^{(\mathrm{GLR})}$ and
$f^{(\mathrm{KMR})}$, one can derive various methods for building
the PES, such as NN and GPR.

In addition to AI-based techniques, a wide range of function approximation
methods have been developed for building the PES over the past decades.
The practical applicability of these function approximation methods is
determined by (1) accuracy with respect to test data and (2) numerical
efficiency of the resulting function, which must be balanced. Accuracy
is the core requirement that expects the fitted PES to reproduce both
the training data and the test data with high precision. This ensures
that the function approximation faithfully captures the underlying
dynamics of the system. Efficiency, on the other hand, pertains to the
computational cost of evaluating the model during subsequent dynamics
calculations, including quasi-classical and quantum molecular dynamics.
In many practical applications to dynamics calculations, numerical
performance of the PES is just as important as its accuracy. A highly
accurate but computationally expensive PES may become impractical,
whereas a slightly less accurate but significantly faster model can
enable simulations that would otherwise be infeasible. Balancing these
two objectives often requires careful consideration of the intended
application. In practice, the choice of function approximation method
depends on the complexity of the underlying PES, the dimensionality of
the problem, the availability of training data, and the specific demands
of the subsequent dynamics calculations.
\begin{remark}
Tow approaches, force field and deep potential, are typical models for
atomic interactions in quantum molecular dynamics calculations. A force
field is a computationally efficient and well-established physics-based
model that describes interatomic forces using given functions with
well-defined physical insight, such as harmonic-oscillator potential,
Lennard-Jones potentials, and Coulomb interactions. The parameters are
typically fitted to experimental data or a few electronic-structure
calculations. A deep potential, in contrast, is a deep NN potential,
trained on a large set of energy values. Deep potentials retain the
accuracy of electronic-structure calculations. These approaches offer
a spectrum of trade-offs between computational efficiency and accuracy.
\label{rem:ff-deeppo}
\end{remark}

%-----------------------------------------------
\subsection{Electromagnetic Interaction\label{sec:coumb-pot}}

As is well known, molecular systems exemplify electromagnetic confinement,
with their constituents, namely electrons and nuclei, held together by
electromagnetic interaction. Now, we should briefly review electromagnetic
interaction and especially revisit to electronic structure from a geometric
viewpoint. According to Equation \eqref{eq:metric-event-interval}, lapse
function determines the rate at which proper time $\tau$ elapses for
observers moving normally to the slices, relative to coordinate time
$t$, while shift vector describes how the spatial coordinate $x^i$
shifts from one slice to the next. In the $[3+1]$ decomposition, the
choice of lapse function and shift vector is often a matter of gauge
freedom, but not completely arbitrary (see Remark \ref{rem:other-choices}).
\begin{defi}
Gauge freedom is the existence of a family of transformations that
leave a theory unchanged, but modify the mathematical descriptions.
\label{def:gauge-free}
\end{defi}
Thus, one can see $\alpha$ and $\beta^i$ as gauge variables which
should be chosen according to some given condition.
\begin{defi}
Gauge variables are the unobservable, redundant components of a field
that encode gauge freedom and can be fixed by a gauge condition.
\label{def:gauge-var}
\end{defi}
For insatnce, for systems located on Earth, that is in a typical
approximately flat spacetime, the lapse function and shift vector
simply satisfy $\alpha=1$ and $\beta^i=0$. As indicated in Section
\ref{sec:coord-frame} and Remark \ref{rem:other-choices}, the present
work always chooses $\alpha=1$ and $\beta^i=0$.

It should be noted
that the gauge variables $\alpha$ and $\beta^i$ parameterize physically
equivalent configurations of a gravitational theory, where the underlying
physical content is encoded in gauge-invariant quantities, that is the
intrinsic curvature of spatial slices together with the extrinsic
curvature in gravity. Thus, the gauge freedom reflects the redundancy
of description necessary to maintain manifest locality and covariance.
This geometric perspective extends naturally to gauge field theory,
which is formulated as a principal fiber bundle over a manifold with
structure group. The curvature of a connection on this bundle is the
gauge field which is a geometric and gauge-invariant field. The gauge
variable is the connection $1$-form that is a local representation of
the connection on a trivializing open set. It depends on the choice of
a local section ({\it i.e.}, a gauge choice) and encodes how the
internal symmetry group acts along paths. The gauge freedom corresponds
precisely to the freedom to change the local section. For instance, the
electromagnetic field is the gauge field, often described by the gauge
variable, namely the four-potential. The gauge freedom is the redundancy
in this description, where different gauge variables related by the
gauge transformation of the four-potential yield exactly the same
physical field.
\begin{remark}
Geometrically, the above description reflects the fact that both gravity
and electromagnetism admit a description as gauge theories. But, it does
not constitute a unification of the two interactions, as given in
Remark \ref{rem:keo-pes-hamil}.
\label{rem:keo-pes-ha111}
\end{remark}

Based on the Eulerian observer one can accept the adsolute-space and
universal-time viewpoint of Galileo, where the $[3+1]$ formulation of
electrodynamics involves the Maxwell equations in curved spacetime
\cite{tho82:339}. It must be noted that the viewpoint of Galileo,
where all inertial frames share the same space and the same time, is
only available to the Eulerian observers. Considering a charged particle
with rest mass $m$ and charge $q$, such as an electron with mass
$m_{\mathrm{e}}=1$ and charge $q=-1$, its ordinary velocity $\vec{v}$
and momentum $\vec{p}$ are 3-vectors in 3D slice $\Sigma_t$ with $v<1$.
Then, the EOM for the charged particle in 4D spacetime can be written
in the $[3+1]$ decomposed form \cite{tho82:339}
\begin{equation}
\big(\mathcal{D}_{\mathrm{F}}+\vec{v}\cdot\vec{\nabla}\big)\vec{p}=
-\left(\frac{m\vec{a}}{\sqrt{1-v^2}}+\vec{\sigma}\cdot\vec{p}+
\frac{1}{3}\vartheta\vec{p}\right)+q\big(\vec{E}+\vec{v}\times\vec{B}\big),
\label{eq:curved-electro-dyn}
\end{equation}
where $\vec{a}$, $\vec{E}$, and $\vec{B}$ are 3-vectors in $\Sigma_t$
for acceleration, electric field, and magnetic field, respectively. In
Equation \eqref{eq:curved-electro-dyn}, $\vec{\sigma}$ and $\vartheta$
describe shear and expansion of the 4D spacetime, respective, while
$\mathcal{D}_{\mathrm{F}}$ is a time derivative defined by Fermi
transport along the fiducial world lines \cite{tho82:339}, called the
Fermi derivative. Unlike the covariant derivative (see also Equation
\eqref{eq:christoffel-symbol-001}), which describes how a vector
changes due to the curvature of spacetime, the Fermi derivative
describes how a vector changes from the viewpoint of an observer
moving along its own world line. The left-hand side of Equation
\eqref{eq:curved-electro-dyn} is the convective derivative of $\vec{p}$
along the particle world line, that is the rate of momentum change,
where the particle is (1) Fermi-transported from its initial position
on $\Sigma_t$ to the next slice, and then (2) spatially parallel-transported
within the new slice to its final position. The first term of the
right-hand side of Equation \eqref{eq:curved-electro-dyn} is a kind of
inertial force to consider the fact that the fiducial observers have a
relative velocity to inertial observers, while the second term is the
usual Lorentz force. For electrons living in the molecular systems
located on Earth, where the sapcetime is flat, due to $m=m_{\mathrm{e}}$,
$q=-1$, $\vec{\sigma}=0$, $\vartheta=0$, and $v^2\ll1$, Equation
\eqref{eq:curved-electro-dyn} can be rewritten as
\begin{equation}
\big(\mathcal{D}_{\mathrm{F}}+\vec{v}\cdot\vec{\nabla}\big)\vec{p}=
-m_{\mathrm{e}}\vec{a}-\big(\vec{E}+\vec{v}\times\vec{B}\big).
\label{eq:curved-electro-dyn-0}
\end{equation}
Solutions of Equation \eqref{eq:curved-electro-dyn} or
\eqref{eq:curved-electro-dyn-0} predict the motion of charged particle
in the 4D spacetime with general geometric characteristics determined
by the metric, lapse function, and shift vector. However, it is still
difficult to predict electornic structure of the molecular sytems in
the 4D spacetime.

To consider this problem, we should turn to the electromagnetic interaction in
general form with geometric characteristics of the 4D spacetime \cite{nik20:191}.
According to Postulate \ref{post:pls}, on the one hand, if we apply the
principle of least action to the spacetime curvature in a coordinate-invariant
manner, the Einstein field equations can be obtained through a stationarity
condition for the metric tensor. On the other hand, the classical electromagnetic
Lagrangian can be understood through the scalar curvature of a manifold,
implying that the electromagnetic field equations are a special case
of the Einstein field equations. It must be noted that, however, even
though one can derive the electromagnetic field equations by an {\it
ad hoc} action (or Lagrangian) for a special case, there is still no
satisfactory model to link electromagnetic field and the geometric
characteristics of the 4D spacetime. Nevertheless, many attempts to
find such relationship have been reported by Einstein, Eddington, Weyl
and Schr{\"o}dinger, and so on \cite{bal96:book}. Based on Remark
\ref{rem:keo-pes-ha111} and the above considerations, we adopt
Remark \ref{rem:keo-pes-hamil} and accept the following proposition
as the foundation.
\begin{prop}
The molecular Hamiltonian operator is partitioned into the KEO and the
PES. The KEO captures the geometric effects arising from the space in
which the reaction occurs, whereas the PES asymptotically captures the
electromagnetic interactions that govern the reaction in a flat space.
\label{prop:divid-keo-pes}
\end{prop}
The statement regarding the KEO in Proposition \ref{prop:divid-keo-pes}
was already given in detail in Sections \ref{sec:coord-frame} and
\ref{sec:keo-coord}, in particular Remarks \ref{rem:keo-pes-hamil} and
\ref{rem:vep-phys-0}. We should turn to numerically estimate the statement
on the PES. The simplest case is the electromagnetic interaction between
atomic nucleus and an electron in a uniformly curved spacetime
\cite{nob16:032108,que16:102101,agg16:2653} with a constant positive
curvature $R$ but without any charge and rotation,
\begin{equation}
V^{(\mathrm{curved})}_{\mathrm{coul}}\big(r_{\mathrm{e}}\big)
=-\frac{Z}{r_{\mathrm{e}}}\sqrt{1-Rr_{\mathrm{e}}^2}
=V^{(\mathrm{flat})}_{\mathrm{coul}}\big(r_{\mathrm{e}}\big)
\sqrt{1-Rr_{\mathrm{e}}^2},\quad
r<\sqrt{\frac{1}{R}},
\label{eq:coulomb-inter-con-cur}
\end{equation}
where $Z$ is the charge of the atomic nucleus and $r_{\mathrm{e}}$ is
distance between atomic nucleus and electron, while
$V^{(\mathrm{flat})}_{\mathrm{coul}}$ is the electromagnetic interaction
in the flat spacetime. The curvature radius, $\sqrt{1/R}$, can be seen
as apparent radius of the universe of order of magnitude $10^{26}\;
\mathrm{m}$, while $r_{\mathrm{e}}$ is approximately equal to atomic
radius of $10^{-10}\;\mathrm{m}$. Therefore, it is safety to have
approximation $V^{(\mathrm{curved})}_{\mathrm{coul}}\approx
V^{(\mathrm{flat})}_{\mathrm{coul}}$ for electrons moving in the
molecular system. Thus, one can generalize the electronic structure
methods developed under conditions at the flat spacetime to approximate
those under general conditions at the curved spacetime, where it is
only required that the $\sqrt{1-Rr^2_{\mathrm{e}}}$ term is sufficiently
small, that is $R$ is not too large. A similar argument can be made
regarding the interaction among the nuclei leading ultimately to
Proposition \ref{prop:divid-keo-pes}.
\begin{remark}
The unification of gravity and electromagnetism remains a theoretical
aspiration. Although elegant classical and string-theoretic frameworks
exist, a complete and experimentally verified quantum theory of unified
interactions is still far beyond our current capabilities.
\label{rem:uni-geom-elec}
\end{remark}

%-----------------------------------------------
\section{Theory of Wavepacket Propagation\label{sec:dynamics}}

\begin{figure}[h!]
   \centering  
    \includegraphics[width=18cm]{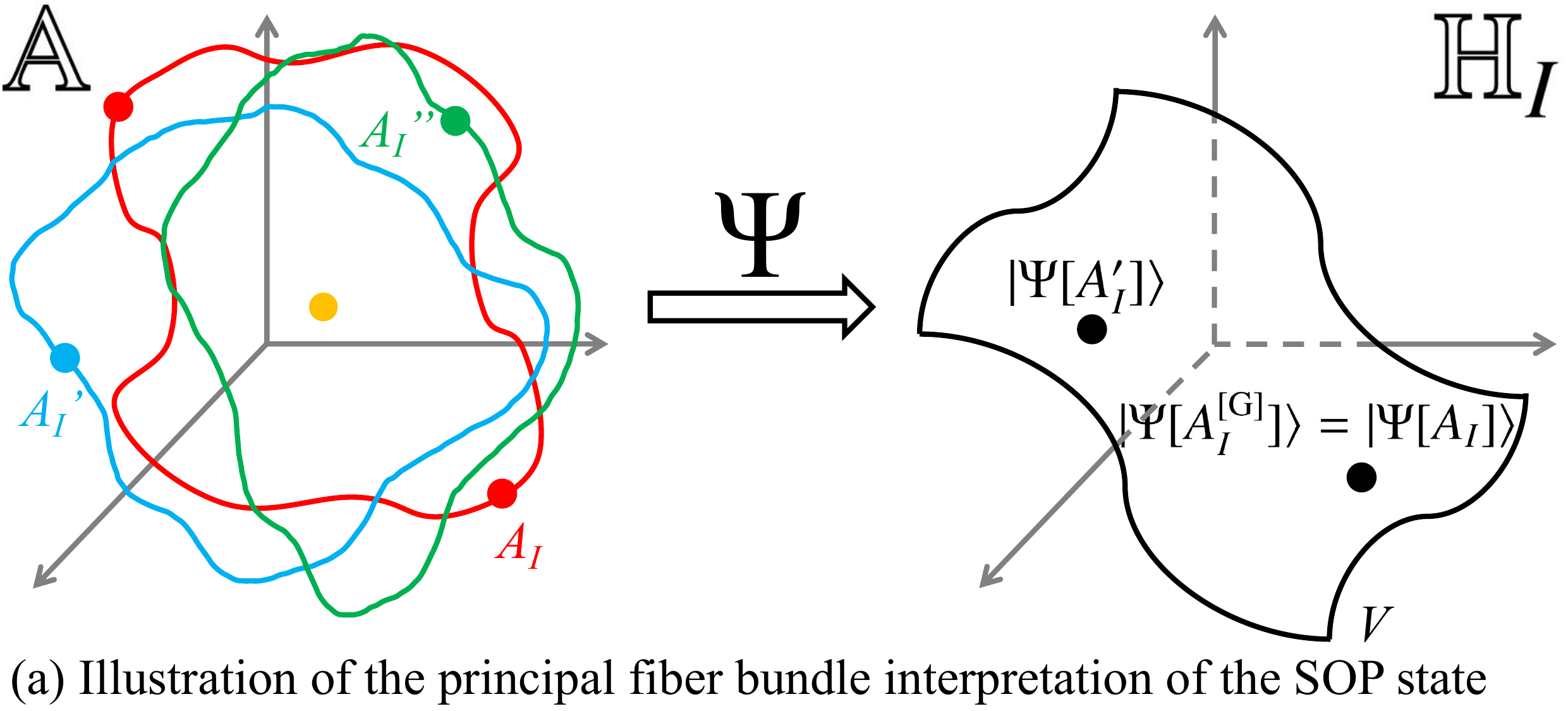}
     \includegraphics[width=18cm]{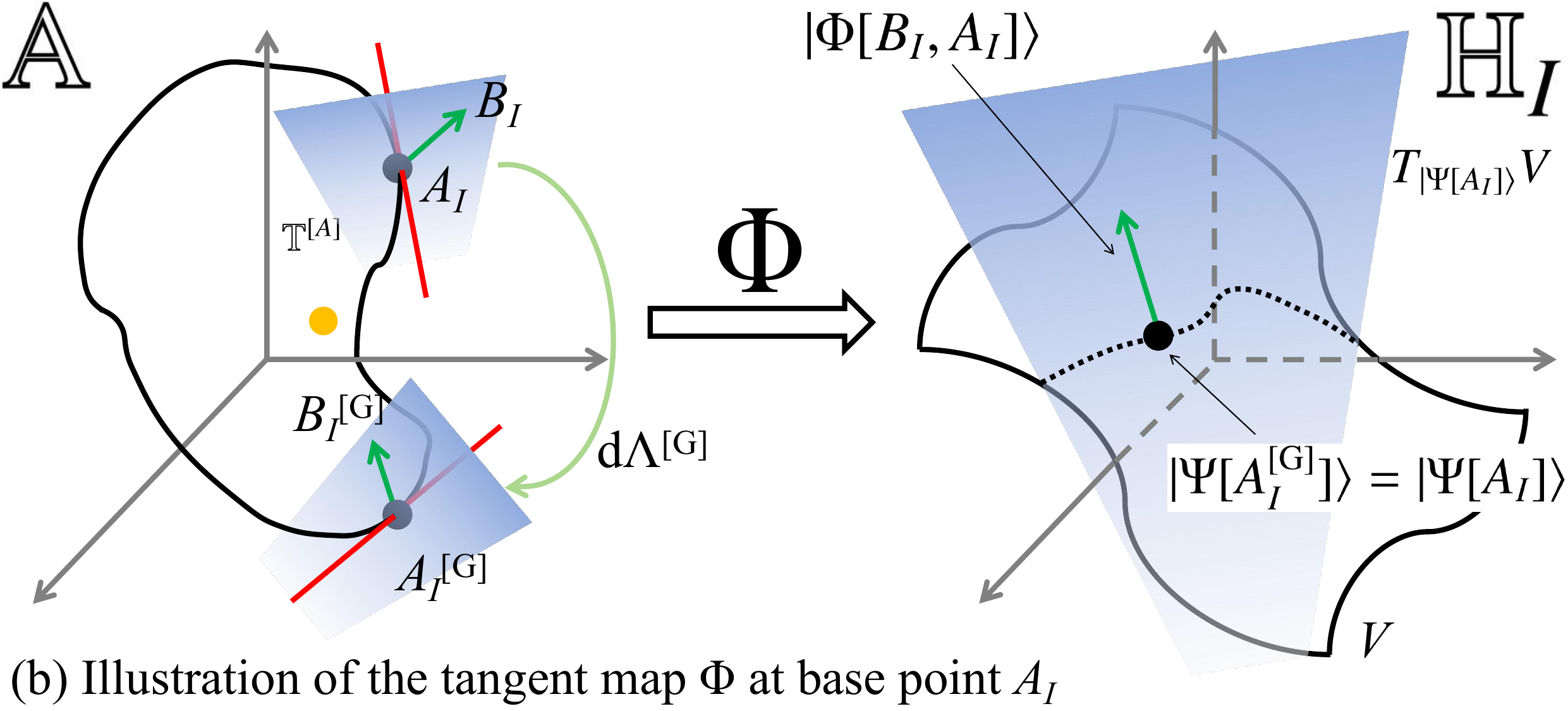}
      \caption{\figfoot}
       \label{fig:spa-sate}
        \end{figure}

This section describes the theory of wavepacket propagation for molecular
reaction dynamics, as collected in Table \ref{tab:theory-dyn}. The most
powerful kind of wavepacket propagation methods is that developed on
the basis of the single-particle approximation (SPA), such as MCTDH,
ML-MCTDH, TD-DMRG, and so on. Such approaches based on the SPA can be
interpreted as a variational framework that selects the best approximation
to the quantum state of the system within the set of configurations.
Section \ref{sec:elec-stru} gives geometric description of a quantum
state assumed in the SPA, as shown in Figure \ref{fig:spa-sate}(a).
Section \ref{sec:math-struc} describes the variational framework in
deriving working equation for solving the single-particle terms, as
shown in Figure \ref{fig:spa-sate}(b). Section \ref{sec:eom-prob} gives
the geometric phase effects during the motions, as explained by Figure
\ref{fig:phase-bundle}. Based on gauge freedom introduced in Section
\ref{sec:eom-prob}, possibility and necessity of quantum simulator for
molecular reaction dynamics is described in Section \ref{sec:qc-simulatior}.
Section \ref{sec:consistency} describes mathematical consistency among
the steps (see also Fgiure \ref{fig:framework}) in wavepacket propagation.

%--------------------------------------
\subsection{Single-Particle Approximation\label{sec:elec-stru}}

In this section, we consider the SPA for the wave function ansatz in
wavepacket propagation. The SPA in conjunction with variational principles
have been widely used in electronic structure and quantum molecular
dynamics \cite{zha25:20397}, where the solution ansatz is expanded by
basis set of products of single-variable functions, such as
single-particle functions (SPFs) and molecular orbitals (MOs). Each
of these terms is the solution of single-variable working equations
derived by substituting the solution ansatz into the EOM. In addition,
the quantum state can be also represented by matrix product state
(MPS). The history of MPSs can date back to the time when they were
referred to as valence bond states or finitely correlated states. Before
discussing the SPA in details, we should briefly descrbe denotations
and some concepts used in this section. As
is well known, quantum states correspond to rays of vectors in affine
Hilbert space and should be identified with the elements of the
projective Hilbert space $P(\mathbb{H})=\mathbb{H}\setminus\mathrm{GL}(1,\mathbb{C})$,
where $\mathrm{GL}(1,\mathbb{C})$ is the multiplicative abelian group
of norm and phase changes which acts on $\mathbb{H}$ as $\Gamma:\mathbb{H}
\times\mathrm{GL}(1,\mathbb{C})\to\mathbb{H}:(\vert\Psi\rangle,c)\mapsto
\Gamma(\vert\Psi\rangle,c)=\vert\Psi\rangle c$. This is because physical
properties $O$ are related to normalized expectations $O:\mathbb{H}\to
\mathbb{C}:\vert\Psi\rangle\mapsto O(\vert\Psi\rangle)=\langle\Psi\vert
\hat{O}\vert\Psi\rangle/\langle\Psi\vert\Psi\rangle$, which is invariant
under the action of $\mathrm{GL}(1,\mathbb{C})$ and thus restricted to
$P(\mathbb{H})$. Although the affine Hilbert space $\mathbb{H}$ is often
chosen to work out physical problem, the projective structure of state
space is required to avoid a number of divergences, and thus the ray of
state $\vert\Psi\rangle\in\mathbb{H}$ is denoted as $[\vert\Psi\rangle]
\in P(\mathbb{H})$. In practical calculations, the elements $[\vert\Psi\rangle]
\in P(\mathbb{H})$ are obtained through any representation of the ray
$[\vert\Psi\rangle]$ and are seen as a set of homogeneous coordinates
for the elements of $P(\mathbb{H})$.
 
Having the above basic concepts of quantum states and Hilbert space, we
turn to the concepts of variational manifold and K{\"a}hler manifold.
Over the past years, Haegeman and co-workers \cite{hae11:070601,hae14:021902,van19:7}
discussed existence of MPS to represent states based on the language
of fiber bundles. We refer the reader to their works and Reference
\cite{zha25:20397} for more discussions on the SPA and comparisons of
various ansatzes. These solution ansatzes form a set of quantum states
depending on variational parameters forming a variational manifold,
which is a smooth manifold embedded in the total Hilbert space
$\mathbb{H}$ of the problem.
\begin{defi}
A variational manifold is a finite-dimensional smooth submanifold of
the Hilbert space or projective Hilbert space parameterized by a set
of variational parameters.
\label{def:variation-manifold} 
\end{defi}
To simplify formulas, state vector $\vert\Psi\rangle$ is normalized to
obtain $[\vert\Psi\rangle]$ in the projective Hilbert space $P(\mathbb{H})$.
The tangent space $T_{[\vert\Psi\rangle]}P(\mathbb{H})$ can be denoted
by $\mathbb{H}/\!\sim$, where ``$\sim$'' means that two vectors
$\forall\vert\Phi_1\rangle,\vert\Phi_2\rangle\in\mathbb{H}$ satisfy
equivalentce relation $\vert\Phi_1\rangle\sim\vert\Phi_2\rangle$ on
$\mathbb{H}$, if and only if $\exists c\in\mathbb{C}$ such that
$\vert\Phi_1\rangle-\vert\Phi_2\rangle=\vert\Psi\rangle c$. In
this context, a unique vector $\vert\Phi\rangle\in T_{[\vert\Psi\rangle]}
P(\mathbb{H})$ can be found by imposing $\langle\Psi\vert\Phi\rangle=0$.
The projective Hilbert space $P(\mathbb{H})$ is a K{\"a}hler manifold
if one endows it with the Fubini-Study metric
\begin{equation}
\tilde{g}_{\vert\Psi\rangle}^{(\mathbb{H})}
\Big(\vert\Phi_1\rangle,\vert\Phi_2\rangle\Big)=
\frac{\big\langle\Phi_2\big\vert\hat{P}^{\perp}_{\vert\Psi\rangle}\big\vert\Phi_1\big\rangle}
{\big\langle\Psi\big\vert\Psi\big\rangle},
\label{eq:spa-metric-000}
\end{equation}
where $\hat{P}^{\perp}_{\vert\Psi\rangle}$ is orthogonal projector onto
$T_{[\vert\Psi\rangle]}P(\mathbb{H})\simeq\mathbb{H}^{\perp}_{\vert\Psi\rangle}$.
K{\"a}hler geometry is the geometric foundations of quantum dynamics.
We refer the reader to mathematical papers, say References
\cite{mor07:book,voi10:85,gua14:297},
for more details of the K{\"a}hler manifold and K{\"a}hler geometry.

Now, since the molecular reaction system is focused, we only consider
the state in the holomorphic tangent space (not to be confused with
Definition \ref{def:hopl-part-dyn-action}). 
\begin{defi}
For a variational manifold $M$ is embedded as a complex submanifold of
the projective Hilbert space, its holomorphic tangent space at a point
$\vert\Psi(z)\rangle$ is given by
\begin{equation}
T_z^{1,0}M=\operatorname{span}_{\mathbb{C}}\left\{
\dfrac{\partial}{\partial z^i}\Big\vert\Psi(z)\Big\rangle-\Big\vert\Psi(z)\Big\rangle
\left\langle\Psi(z)\left\vert\dfrac{\partial}{\partial z^i}
\right\vert\Psi(z)\right\rangle
\right\}=\operatorname{span}_{\mathbb{C}}\Big\{
\partial_i\big\vert\Psi(z)\big\rangle-\Big\vert\Psi(z)\Big\rangle
\big\langle\Psi(z)\big\vert\partial_i\big\vert\Psi(z)\big\rangle\Big\}
\label{eq:spa-metric-99-hopl}
\end{equation}
where $\vert\Psi(z)\rangle$ is restricted by parameters
$z=\{z^i\}_{i=1}^m$.
\label{def:hopl-manifold-99}
\end{defi}
The subscript $1,0$ in Equation \eqref{eq:spa-metric-99-hopl} means
that these vectors are eigenvectors of the complex structure
with eigenvalue $+\mi$, where the complex structure is a linear
map $J:T_{[\vert\Psi\rangle]}P(\mathbb{H})\to T_{[\vert\Psi\rangle]}P(\mathbb{H})$
on the real tangent space satisfying $J^2=-I$.
\begin{remark}
The holomorphic tangent space identifies the directions on the
variational manifold that align with the complex structure of the
projective Hilbert space, ensuring that the effective dynamics of
the parameters follows a Hamiltonian flow and thereby preserving the
symplectic structure inherent to quantum dynamics.
\label{rem:holpo-quantum-state}
\end{remark}
For submanifold $M\subset\mathbb{H}$, the holomorphic tangent space at
$\vert\Psi\rangle\in M$ satisfies $T_{\vert\Psi\rangle}M\subset\mathbb{H}$.
The variational subset $M$ corresponding to a variational ansatz
$\Psi:U\subset\mathbb{C}^m\to\mathbb{H}$ is defined as
$M=\{\vert\Psi(z)\rangle\vert z\in U\}$. Under the injectivity assumption,
a holomorphic inverse map $\Psi^{-1}:M\to U\subset\mathbb{C}^m$ can be
defined making $(M,\Psi^{-1})$ be a global coordinate chart for $M$.
The partial derivatives of $\Psi$ at a point $z\in U$ is denoted by
$\partial_i\Psi:U\to\mathbb{H}:z\mapsto\vert\partial_i\Psi(z)\rangle
=\partial_i\vert\Psi(z)\rangle\vert_z$. In orther words, if the ansatz
$\vert\Psi(z)\rangle$ represents a trial wave function, then
$\partial_i\Psi$ is the direction in $\mathbb{H}$ obtained by varying
the $i$-th parameter. These vectors span the tangent space
$T_{\vert\Psi(z)\rangle}M$. The tangent map $\mathrm{d}\Psi_z$ is then
given by
\begin{equation}
\mathrm{d}\Psi_z:\mathbb{C}^m\to
T_{\vert\Psi(z)\rangle}M\subset\mathbb{H}:\omega^i\partial_i\big\vert_z
\mapsto\omega^i\big\vert\partial_i\Psi(z)\big\rangle=
\big\vert\Phi(\omega,z)\big\rangle,
\label{eq:spa-metric-001}
\end{equation}
which defines map $\Phi:\mathbb{C}^m\times\mathbb{C}^m\to\mathbb{H}$.
This is core idear to variational framework. The pushfroward of $\Psi$
induces a bundle map between the tangent bundles $\mathrm{d}\Psi:
\mathbb{C}^m\times U\to TM:(\omega,z)\mapsto(\vert\Phi(\omega,z)
\rangle,\vert\Psi(z)\rangle)$. In this context, that ansatz defining
a fiber bundle structure, where $U$ or $z$ is base space and tangent
spaces $T_{\vert\Psi(z)\rangle}M$ are fibers. The above geometric
formalism is not only foundational for time-dependent variational
principle because it is inherently a projection onto a tangent space,
but also suitable for time-independent eigenvalue variational principle.
\begin{remark}
Both time-dependent and time-independent variational methods can be
unified by the quantum geometric tensor (comparing with Equation
\eqref{eq:spa-metric-99-hopl}),
$\langle\partial_i\Psi\vert(1-\vert\Psi\rangle\langle\Psi\vert)\vert\partial_j\Psi\rangle$.
Its real part is the Riemannian metric on the projective Hilbert
space, that is Fubini-Study metric given in Equation \eqref{eq:spa-metric-000}.
Its imaginary part is the geometric curvature, or called Berry curvature.
In time-dependent problems, this metric appears in the EOMs. In
time-independent problems, this metric defines the gradient for
optimization, accounting for the curvature of the parameter space.
\label{rem:vari-time-depndent-000}
\end{remark}
\begin{remark}
The above geometric formalism is the foundation of variational quantum
algorithms, such as variational quantum eigensolver (VQE), where parameters
are optimized on a classical computer.
\label{rem:quantum-simulator}
\end{remark}

Having the above basis concepts, we turn to a $d$-dimensional system
that is represented by a set of grids $I=\{i_1,\cdots,i_d\}$ and consider
its ansatz represented in the SOP form, where the $\kappa$-th point of
a grid $I$ contains an $m_{\kappa}$-dimensional variable, so that the
local Hilbert space $\mathbb{H}_{\kappa}$ is spanned by a basis
$\{\vert i_{\kappa}\rangle_{\kappa}\vert i_{\kappa}=1,\cdots,m_{\kappa}\}$.
Then, the total Hilbert space is given by direct product $\mathbb{H}_I
=\otimes_{\kappa=1}^d\mathbb{H}_{\kappa}$ and spanned by the product
basis $\vert i_1\cdots i_d\rangle=\vert i_1\rangle_1\otimes\cdots
\otimes\vert i_d\rangle_d$. The dimension of $\mathbb{H}_I$ is thus
$\prod_{\kappa=1}^dm_{\kappa}$. In this context, an arbitrary state
$\vert\Psi\rangle\in\mathbb{H}_I$ is represented by coefficients $A_I$
corresponding to the basis $\vert i_1\cdots i_d\rangle$. The variational
parameters correspond to a set of $m_{\kappa}$ coefficients for the
$\kappa$-th dimensionality, forming space $\mathbb{A}$. The state is
defined as the holomorphic map $\Psi:\mathbb{A}\to\mathbb{H}_I:A_I
\mapsto\vert\Psi[A_I]\rangle$ and given by prescription
\begin{equation}
\big\vert\Psi[A_I]\big\rangle=\sum_{i_1=1}^{m_1}\cdots
\sum_{i_d=1}^{m_d}A_{i_1\cdots i_d}\big\vert i_1\cdots
i_d\big\rangle=\sum_IA_I\vert I\rangle.
\label{eq:def-state-spa-geom-000}
\end{equation}
By Equation \eqref{eq:def-state-spa-geom-000}, one can have got a
variational set $V=\{\vert\Psi[A_I]\rangle\vert\forall A_I\in\mathbb{A}\}$.
Figure \ref{fig:spa-sate}(a) shows a schematic illustration of the
principal fiber bundle interpretation of the map $\Psi:\mathbb{A}\to
\mathbb{H}_I$. Furthermore, to reduce the multi-dimensional EOMs to
few-particle working equations, we require not only the ansatz but
also the Hamiltonian operator to take the SOP form. To see this point,
let the operator $\hat{O}$ for an observable be a product operator
$\otimes_{\kappa=1}^d\hat{O}_{\kappa}$ of local operators $\hat{O}_{\kappa}$
acting non-trivially only on $\mathbb{H}_{\kappa}$. One can define
superoperator as linear homomorphism,
\begin{equation}
\hat{\mathcal{E}}_{\hat{O}_{\kappa}}^{(\kappa)}=
\sum_{i_\kappa,j_\kappa=1}^{m_\kappa}\big\langle
i_\kappa\big\vert
\hat{O}_\kappa\big\vert j_\kappa\big\rangle
A_I^*\otimes A_{J},\;
\hat{\mathcal{E}}_{\hat{O}_{\kappa}}^{(\kappa)}(x)=
\sum_{i_\kappa,j_\kappa=1}^{m_\kappa}
\big\langle i_\kappa\big\vert\hat{O}_\kappa\big\vert
j_\kappa\big\rangle
A_I^*\otimes x\otimes A_{J},\;
\tilde{\mathcal{E}}_{\hat{O}_{\kappa}}^{(\kappa)}(x)=
\sum_{i_\kappa,j_\kappa=1}^{m_\kappa}
\big\langle i_\kappa\big\vert\hat{O}_\kappa\big\vert
j_\kappa\big\rangle
A_J^*\otimes x\otimes A_I,
\label{eq:def-state-spa-geom-001}
\end{equation}
where $x$ represents virtual operators. Equation \eqref{eq:def-state-spa-geom-001}
further implies the expectation of the product operator $\hat{O}$ in
the form
\begin{equation}
\big\langle\Psi[A_I]\big\vert\hat{O}\big\vert\Psi[A_I]\big\rangle=
\mathrm{tr}\Big(\hat{\mathcal{E}}_{\hat{O}_1}^{(1)}\cdots
\hat{\mathcal{E}}_{\hat{O}_d}^{(d)}\Big)
=\mathrm{tr}\left(\prod_{\kappa=1}^d
\hat{\mathcal{E}}_{\hat{O}_{\kappa}}^{(\kappa)}\right)
\label{eq:def-state-spa-geom-002}
\end{equation}
Since most local operators are trivial on the majority of grids, one
can define the set of virtual densities $l=\{l_\kappa\vert\kappa=1,\cdots,d\}$
and $r=\{r_\kappa\vert\kappa=1,\cdots,d\}$ through the recursive expressions
$l_\kappa=\tilde{\mathcal{E}}_{\hat{1}}^{(\kappa)}(l_{\kappa-1})$ with
$l_0=1$ and $r_{\kappa-1}=\hat{\mathcal{E}}_{\hat{1}}^{(\kappa)}(r_\kappa)$
with $r_d=1$.

In general, the SPA for multi-dimensional quantum state (see Equation
\eqref{eq:def-state-spa-geom-000}) is always feasible as long as the
product basis set $\{\vert i_1\cdots i_d\rangle\vert i_{\kappa}=1,
\cdots,m_{\kappa},\;\kappa=1,\cdots,d\}$ is sufficiently close to
complete. However, complete basis sets are rare and, in practical
calculations, unrealistic because they require an infinite number of
terms, that is $m_{\kappa}\to\infty$ for all dimensions. The fact that
$m_{\kappa}<\infty$ in practical calculations inevitably leads to the
so-called ``correlation'' effects among the single-particle terms
during variation. A typical example is the correlation energy problem
in variational approaches for electronic structure. In addition, the
correlation effects in time propagation lead to convergence problem
and low computational efficiency, particularly when considering the
dynamics of higher-lying states. To overcome such correlation effects,
it is necessary to expand the basis set or to endow it with a certain
physical significance of the system. In this context, as previously
proposed \cite{zha25:20397} the system is firstly separated into
several modes which are lower-dimensional subsystems. Then, these
modes can be further separated into modes in deeper layer. Repeatedly
separating the subsystems until the modes in the deepest layer are
manageable, the system is finally organized within a hierarchical
framework. The quantum state can be accordingly expanded in a multilayer
form, where the modes in the deepest layer are further expanded using
a given basis set. For example, to address electron correlation,
multilayer wave function is adopted by CI, MCSCF, MRCI, quantum
chemistry DMRG, and so on. For time evolution, such multilayer wave
function is employed by ML-MCTDH, time-dependent DMRG, and so on.

%-----------------------------------------------
\subsection{Variational Manifold\label{sec:math-struc}}

Having the SPA, one now define the subset of full-rank states with open
boundary as $A=\{A_I\in\mathbb{A}\vert l_\kappa>0,\;r_n>0,\;\kappa=1,\cdots,d\}$,
as illustrated by Figure \ref{fig:spa-sate}(a). The subset $A$ avoids
pathological cases where the representation collapses or becomes
redundant. To ensure well-behaved geometry, all virtual densities
should be strictly positive definite and thus have full rank on top of
being positive semi-definite. The subsect $A\subset\mathbb{A}$ of
full-rank states is a complex manifold with the same dimensionality of
$\mathbb{A}$. Introducing the group $\mathrm{G}=\{\hat{\mathcal{G}}\}$
of local gauge transformations, the group action of G on $\mathbb{A}$
is defined as the map $\Lambda:\mathbb{A}\times\mathrm{G}\to\mathbb{A}:
(A_I,\hat{\mathcal{G}})\mapsto\Lambda[A_I,\hat{\mathcal{G}}]=
A_I^{[\mathrm{G}]}$ with $\forall\hat{\mathcal{G}}\in\mathrm{G}$. This
implies that the map $\Lambda$ is holomorphic. Then, gauge invariance
of the state means that
$\forall A_I\in\mathbb{A}:\vert\Psi[A_I^{[\mathrm{G}]}]\rangle=
\vert\Psi[A_I]\rangle$ for $\forall\hat{\mathcal{G}}\in\mathrm{G}$.
The stabilizer subgroup of transformations $\hat{\mathcal{G}}$ that
leave $\Lambda[A_I,\hat{\mathcal{G}}]=A_I$ is denoted by
$\mathrm{GL}(1,\mathbb{C})$. With the groups $\mathrm{G}$ and
$\mathrm{GL}(1,\mathbb{C})$, the structure group $\mathrm{S}$ can be
defined as $\mathrm{S}=\mathrm{G}\setminus\mathrm{GL}(1,\mathbb{C})$.
Factoring out the stabilizer removes trivial redundancies leading to a
group $\mathrm{S}$ that acts freely on the parameter manifold.
The group action $\Lambda:A\times\mathrm{S}\to A$ is free, where the
stabilizer subgroup of any $A_I\in A$ is given by the trivial group
containing only the identity. A group action $\Lambda:M\times\mathrm{L}
\to M$ of a real Lie group $\mathrm{L}$ on a complex manifold $M$ is
proper if $\mathrm{L}$ is a closed subgroup of holomorphic automorphisms
on $M$ and preserves a continuous distance on $M$. Thus, the group action
$\Lambda:A\times\mathrm{S}\to A$ is proper. If $M$ is a smooth manifold,
$\mathrm{L}$ is a Lie group, and $\Lambda:M\times\mathrm{L}\to M$ is a
smooth, free, and proper group action, one can then find that the space
$M\setminus\mathrm{L}$ is a smooth manifold and the natural projection
$\pi:M\to M\setminus\mathrm{L}$ is a smooth submersion.

Correspondingly, $\pi:M\to M\setminus\mathrm{L}$ is a principal fiber
bundle with total space $M$, base space $M\setminus\mathrm{L}$ and
structure group $\mathrm{L}$. Since the complex manifold $A$ is in the
first place a smooth manifold, and the holomorphic group action $\Lambda$
is a smooth map, we can find that $A\setminus\mathrm{S}$ is a smooth
manifold. Letting $\Lambda: M\times\mathrm{L}\to M$ be a holomorphic,
free, and proper group action of a complex Lie group $\mathrm{L}$ to a
complex manifold $M$, $M\setminus\mathrm{L}$ is a complex manifold and
the quotient map $\pi:M\to M\setminus\mathrm{L}$ is holomorphic. In
orther words, the subset $A$ is not only a smooth manifold but also a
complex manifold because variation preserves this geometric structure.
In addition, any holomorphic map $\Psi:M\to A$ that is invariant under
the action of $\mathrm{L}$ factorizes as $\Psi=\psi\circ\pi$, where
$\psi$ is a holomorphic map from $M\setminus\mathrm{L}$ to $A$. Thus,
$A\setminus\mathrm{S}$ is a complex manifold. Since $\Psi:A\to\mathbb{H}_I$
invariant under the group action, it has a natural restriction
$\psi:A\setminus\mathrm{S}\to\mathbb{H}_I$. One can define $D$ as the
image of $\psi:A\setminus\mathrm{S}\to\mathbb{H}_I$ or equivalently,
the image of $\Psi:A\to\mathbb{H}_I$. The variational class of $\Psi:A
\to V$ is a principal fiber bundle with structure group $\mathrm{S}$,
base manifold $V$, total manifold $A$ and bundle projection $\Psi$. In
this context, many parameter sets represent the same quantum state and
the redundancy is organized by the gauge group. The variational manifold
$V$ is a complex manifold that is biholomorphic to $A\setminus\mathrm{S}$,
meaning that there is a one-to-one and complex-structure-preserving
correspondence. Therefore, by restricting to the subset of full rank
state (open boudary) or injective state (periodic boundary), we can
identify the SOP representation with a principal fiber bundle. The
variational parameters live in the bundle space. The quantum state
encoded by the variational parameters should be left invariant under a
well-understood set of gauge transformations and therefore be identified
with points in the base space, that is the quotient space of the bundle
space and the structure group (gauge group). This is bijective, and
standard theorems of fiber bundle automatically imply that the set of
the quantum state in the SOP form can be given the structure of a
complex manifold. Since this manifold is embedded in an affine or
projective Hilbert space, that is a K{\"a}hler manifold, such manifold
is also a K{\"a}hler manifold. The corresponding K{\"a}hler metric can
be obtained by inducing the standard metric of Hilbert space.

Turning to Figure \ref{fig:spa-sate}(b), by the pushforward $\mathrm{d}
\Psi$ of the bundle projection $\Psi$, one can deifne a bundle map
$\mathrm{d}\Psi:TA\to TV$ between the holomorphic tangent bundle of
$A$ and the holomorphic tangent bundle of $V$. At any point in $A$ we
have $T_AA\simeq\mathbb{A}$. The holomorphic tangent space
$T_{\vert\Psi[A_I]\rangle}V\subset\mathbb{H}_I$ is biholomorphic to a
subspace of $\mathbb{H}_I$ and defines a variational class to consider
the excited states of a Hamiltonian for which $\vert\Psi[A_I]\rangle$
in Equation \eqref{eq:def-state-spa-geom-000} is a well approximation
of ground state.
Next, a map $\vert\Phi\rangle:TA\to\mathbb{H}_I:(B_I,A_I)\mapsto\vert\Phi[B_I,A_I]\rangle$
can be defined leading to the expression
\begin{align}
\big\vert\Phi[B_I,A_I]\big\rangle=\big(\mathrm{d}\Psi\big)_{A_I}[B_I]&=
\sum_{\kappa=1}^d\left(\sum_{i_{\kappa}=1}^{m_{\kappa}}
B_{i_1\cdots i_\kappa\cdots i_d}\frac{\partial}{\partial
A_{i_1\cdots i_{\kappa}\cdots i_d}}\right)\big\vert\Psi[A_J]\big\rangle
\allowdisplaybreaks[4] \nonumber \\
&=\sum_{\kappa=1}^d\sum_{i_{\kappa}=1}^{m_{\kappa}}\left(\sum_{j_1=1}^{m_1}\cdots
\sum_{j_{\kappa-1}=1}^{m_{\kappa-1}}\sum_{j_{\kappa+1}=1}^{m_{\kappa+1}}
\cdots\sum_{j_d=1}^{m_d}B_{J_i^\kappa}\right)
\big\vert j_1\cdots j_{\kappa-1}i_\kappa j_{\kappa+1}\cdots j_d\big\rangle
\allowdisplaybreaks[4] \nonumber \\
&=\sum_{\kappa=1}^d\left(\sum_{j_{\kappa}=1}^{m_{\kappa}}
\sum_{I_j^{\kappa}}{}^{\kappa}
B_{I_j^\kappa}\right)\big\vert I_j^{\kappa}\big\rangle,
\label{eq:deeq-map-dipara-00}
\end{align}
where $I_j^{\kappa}=\{i_1,\cdots,i_{\kappa-1},j_{\kappa},i_{\kappa+1},\cdots,i_d\}$,
while $\sum_{I_j^{\kappa}}{}^{\kappa}$ means summation over all DOFs
except the $\kappa$-th one. Equation \eqref{eq:deeq-map-dipara-00} is
a general tangent vector $\vert\Phi[B_I,A_I]\rangle$, which is built
from $d$ states and shares identical structures with state space. For
brevity, introducing the notation $T_{\vert\Psi[A_I]\rangle}V=\mathbb{T}^{[A]}$,
the linear homomorphism can be defined as $\Phi[A_I]:T_AA\simeq\mathbb{A}\to
\mathbb{T}^{[A]}:B_I\mapsto\vert\Phi[B_I,A_I]\rangle$. Moreover, one
can omit the explicit notation $[A]$ or $A_I$ of the base point in the
notation of the tangent space $\mathbb{T}$ or its vectors $\vert\Phi[B_I]\rangle$.
The tangent space to the K{\"a}hler manifold for quantum states in the
SOP form has been proven to be a key role in time evolution. Due to the
gauge invariance of the SOP representation, not all partial derivatives
with respect to the variational parameters produce linearly independent
tangent states of the manifold. Within the fiber bundle framework, the
tangent map from the tangent bundle of parameter space to the tangent
bundle of the base manifold has a non-trivial kernel referred to as the
vertical subspace. To obtain unique tangent vector representations, a
principal bundle connection is introduced, defining a complementary
horizontal subspace. This connection provides a canonical representation
for tangent vectors, or called a gauge fixing prescription, and simplifies
physical expectations involving tangent states. Notably, for any given
base point, at least two such canonical representations exist, both of
which render the metric at that point into the unit matrix.

To easily understand the above geometric description on variation
manifold, let us consider the working equations of the MCTDH
method \cite{bec00:1,mey03:251,wor08:569,mey12:351} as the first
example. Beginning with Equation \eqref{eq:def-state-spa-geom-000},
the MCTDH ansatz is given by
\begin{equation}
\Psi\big(\mathbi{q},t\big)=\sum_{i_1=1}^{n_1}\cdots\sum_{i_d=1}^{n_d}A_{i_1\cdots i_d}(t)
\prod_{\kappa=1}^d\varphi_{i_{\kappa}}^{(\kappa)}\big(q^{(\kappa)},t\big)
=\sum_IA_I(t)\Omega_I\big(\mathbi{q},t\big),\quad
\mathbi{q}=\big\{q^{(\kappa)}\big\}_{\kappa=1}^d,
\label{eq:mctdh-wf-000}
\end{equation}
where $\varphi_{i_{\kappa}}^{(\kappa)}(q^{(\kappa)},t)$ is the
single-particle function (SPF) of the $i_{\kappa}$-th grid for the
$\kappa$-th DOF, while $\Omega_I(\mathbi{q},t)=\prod_{\kappa=1}^d
\varphi_{i_{\kappa}}^{(\kappa)}(q^{(\kappa)},t)$ is called configuration.
It is also worth noting inequality of $1\leq n_{\kappa}\leq\prod_{\kappa'=1}^dn_{\kappa'}\leq m_{\kappa}$
for the $\kappa$-th DOF where $\kappa=1,\cdots,d$, otherwise there
will be redundant configurations. Increasing values of
$\{n_{\kappa}\vert\kappa=1,\cdots,m_{\kappa}\}$, the more accurate
the time evolution becomes, and the MCTDH ansatz monotonically converges
towards Equation \eqref{eq:def-state-spa-geom-000}, as $n_{\kappa}$
approaches to $m_{\kappa}$. In time evolution, if the basis states
$\{\vert i_{\kappa}\rangle_{\kappa}\}$ of Equation \eqref{eq:def-state-spa-geom-000}
are chosen to be the discrete variable representation (DVR) or finite
basis representation (FBR) basis, then such ansatz becomes discrete
values on grids $J=\{j_1,\cdots,j_d\}$ in the configuration space and then gives the standard
propagation method \cite{bec00:1,mey03:251,wor08:569,mey12:351}. Next,
substituting Equation \eqref{eq:mctdh-wf-000} into the Dirac-Frenkel
variational principle and employing Equation \eqref{eq:deeq-map-dipara-00},
one can directly obtain two working equations \cite{bec00:1,mey03:251,wor08:569,mey12:351}
by varying with respect to the coefficients and SPFs. These equations
associated with $A_I$ and $\Omega_I$ are strongly coupled together
through the Hamiltonian operator, the metric of the configurations (see
Equation \eqref{eq:spa-metric-000}), and the density defined by virtual
densities. Moreover, these working equations have been already extended
to the multilayer form, serving as the ML-MCTDH working equations for
each layer \cite{wan03:1289,man08:164116,ven11:044135}, where the SPFs
of a layer are further expanded by the SPFs in deeper layers in the
form of Equations \eqref{eq:def-state-spa-geom-000} and \eqref{eq:mctdh-wf-000}.
To implement the MCTDH and ML-MCTDH algorithms, the SPFs in Equation
\eqref{eq:mctdh-wf-000} are finally represented by the basis of the DVR
making the wave function become discrete values
\begin{equation}
\Psi\big(\mathbi{q}_J,t\big)=
\sum_{i_1=1}^{n_1}\cdots\sum_{i_d=1}^{n_d}A_{i_1\cdots i_d}(t)
\prod_{\kappa=1}^d\varphi_{j_\kappa i_{\kappa}}^{(\kappa)}=
\sum_I\Omega_{JI}A_I,\;
\varphi_{j_\kappa i_{\kappa}}^{(\kappa)}=
\varphi_{i_{\kappa}}^{(\kappa)}\big(q_{j_\kappa}^{(\kappa)},t\big),\;
\mathbi{q}_J=\big\{q_{j_\kappa}^{(\kappa)}\big\}_{\kappa=1}^d,\;
\Omega_{JI}=\Omega_I(\mathbi{q}_J,t).
\label{eq:mctdh-wf-999}
\end{equation}
Having Equation \eqref{eq:mctdh-wf-999}, the working equations are represented
by matrix equations and then can be solved through standard time integrators
to propagate an initial state $\vert\Psi(0)\rangle$ providing time-dependent
state $\vert\Psi(t)\rangle$ and then the dynamics. 

Now, taking the MRCI method \cite{sza12:108} for electronic structure
as another example. This method belongs to the CI family which is a
general procedure to approximate solutions to the electronic Schr{\"o}dinger
equation. 
\begin{remark}
Of course, electronic structure is essentially an time-independet
eigen-state problem, rather than wavepacket propagation. We mention
MRCI here just as a comparison with MCTDH.
\label{rem:mrci-mctdh}
\end{remark}
Separating the time and the cooridnates according to the BOA, in
multireferece expansions of the electronic wave function, a set of
reference states $\{\vert\zeta\rangle\vert\zeta=1,\cdots,N_{\mathrm{ref}}\}$
is used to generate the ansatz \cite{sza12:108}
\begin{align}
\big\vert\Psi_{\mathrm{MRCI}}\big\rangle&=\sum_{\zeta=1}^{N_{\mathrm{ref}}}\left[
\big\vert\zeta\big\rangle C\big({\zeta}\big)+
\sum_{ma}\big\vert\zeta_m^a\big\rangle C_m^a\big({\zeta}\big)+
\sum_{m<n}\sum_{a<b}\big\vert\zeta_{mn}^{ab}\big\rangle C_{mn}^{ab}\big({\zeta}\big)+
\cdots\right],
\allowdisplaybreaks[4] \nonumber \\
\big\vert\zeta_m^a\big\rangle&=\hat{a}_a^{\dagger}\hat{a}_m\big\vert\zeta\big\rangle,\quad
\big\vert\zeta_{mn}^{ab}\big\rangle=\hat{a}_a^{\dagger}\hat{a}_b^{\dagger}
\hat{a}_n\hat{a}_m\big\vert\zeta\big\rangle,\quad\cdots,\quad
\zeta=1,\cdots,N_{\mathrm{ref}},
\label{eq:mrci-wf-001}
\end{align}
where $\hat{a}_a^{\dagger}$ and $\hat{a}_m$ are creation and annihilation
operators for the $a$-th virtual orbitals and the $m$-th occupied orbitals,
respectively. The reference states are typically either primitive Slater
determinants of spin-orbitals, that is single-electronic MOs, or configuration
state functions. A Slater determinant is an antisymmetric linear combination
of products of one-electron spin-orbitals (namely configurations). A
configuration state function is linear combination of Slater determinants
which must have the same spatial-orbital occupations and eigenfunctions
of the total spin operator and the spin projection along the $z$-axis.
Therefore, the MRCI ansatz in Equation \eqref{eq:mrci-wf-001} can be
finally written in the summation form of configurations, as given by
Equation \eqref{eq:def-state-spa-geom-000}. Employing Equation \eqref{eq:mrci-wf-001}
in the variational pronciple, one can derive the MRCI working equations
for the coefficients and the one-electron spin-orbitals \cite{sza12:108}.
Solving them iteratively, one can finally obtain the electronic energy
and wave function at the MRCI level. In contrast to the time-dependent
nuclear ansatz of Equation \eqref{eq:mctdh-wf-000}, the electronic ansatz
of Equation \eqref{eq:mrci-wf-001} must satisfy fermionic antisymmetry
and must also be an (approximately) eigenfunction of the total spin-squared
and $z$-component operators.

%----------------------------------------
\subsection{Motion and Phase\label{sec:eom-prob}}

\begin{figure}[h!]
   \centering  
    \includegraphics[width=18cm]{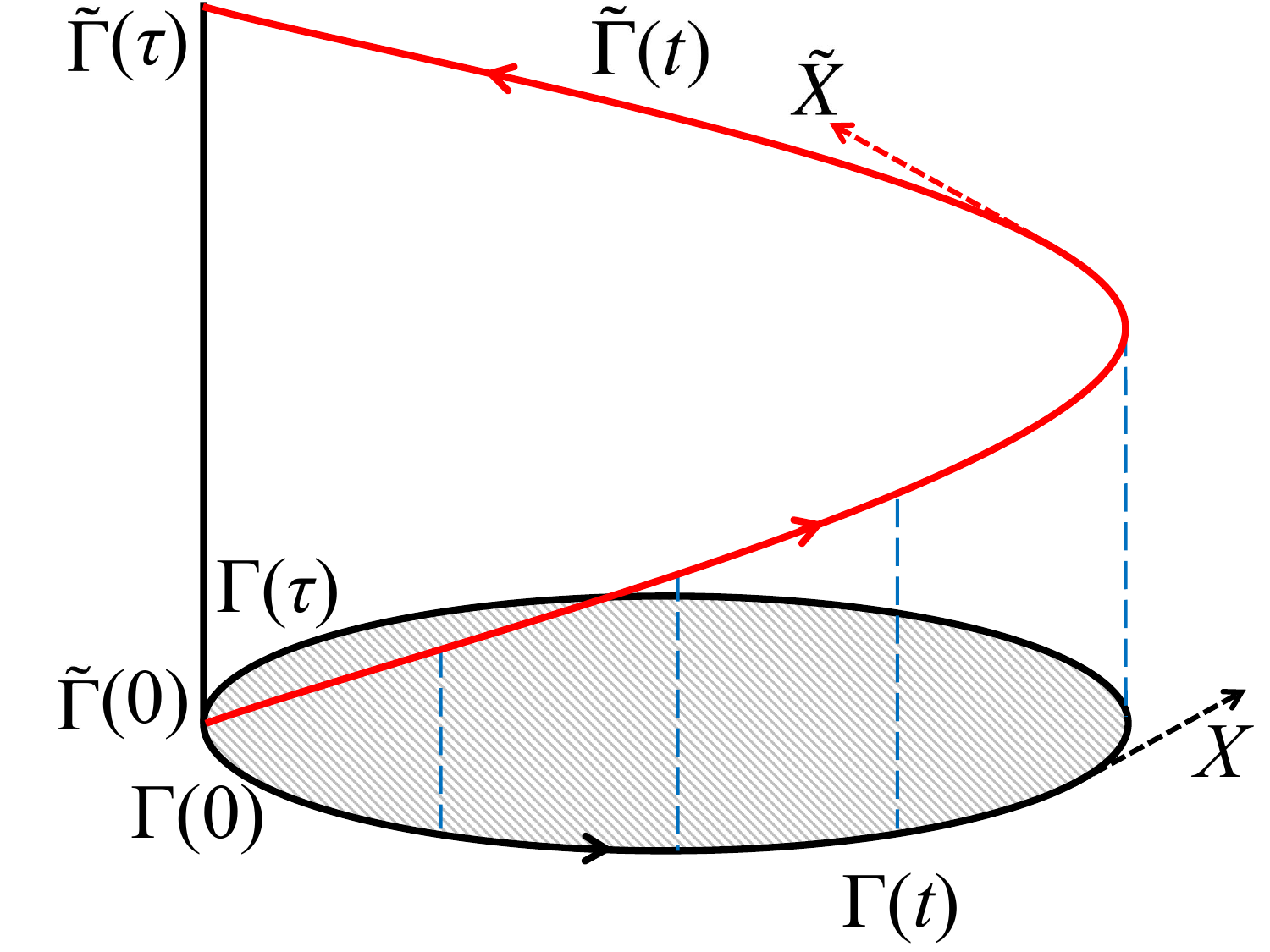}
     \caption{\figfoot}
      \label{fig:phase-bundle}
       \end{figure}

In Section \ref{sec:boa-app}, we have introduced the adiabatic approximation,
based on which the separation of electronic motion from nuclear motion was
discussed. Now, we should turn to concept of the motion and phase for molecular
reaction dynamics and further develop the theory of wavefunction phase
established in our previous work \cite{men22:16415,zha25:20397}. Without
loss of generality, let us consider a system with the Hamiltonian operator
$\hat{H}$ whose normalized eigenstate is denoted by $\vert k\rangle$,
where $k$ is set of quantum numbers. One can then denoted the eigenstate
set by $K=\{\vert k\rangle\vert\langle k\vert k\rangle=1\}$. According
to our previously reported work \cite{men22:16415}, when the state
$\vert k\rangle$ adiabatically evolves along a closed curve $\Gamma$,
the system has got a geometric phase, called the Berry phase. Shortly
after its discovery, Simon \cite{sim83:2167} suggested that the Berry
phase is the holonomy in a Hermitian line bundle since the adiabatic
theorem naturally defines a connection in such a bundle. In this
subsection, based on the geometrically formulated concept of the
Berry phase \cite{sim83:2167}, we will re-examine it from the views of
both electronic and nuclear motions. Simply speaking, each parameter
in the parameter space associates a one-dimensional complex fiber that
is spanned by the instantaneous nondegenerate eigenstate. The adiabatic
evolution defines a parallel transport within this bundle through the
Berry connection. When the parameters traverse a closed loop, the
parallel-transported state acquires a phase factor, that is the Berry
phase. In this context, the Berry phase is a manifestation of the
curvature of the parameter space as captured by the connection arising
from the adiabatic approximation. Figure \ref{fig:phase-bundle}
schamatically illustrates curves in the parameter space.

First, we note that $\vert k\rangle$ and $\vert k\rangle\exp(\mi\varphi)$
cannot be distinguished in the Hilbert space, denoted by $\vert k\rangle
\sim\vert k\rangle\exp(\mi\varphi)$, where $\exp(\mi\varphi)\in\mathrm{U}(1)$.
Here, U(1) means one-dimensional unitary group. Then, the proper quantum
state space of the system is defined as $H=K/\!\sim$ by removing all
similar redundant relationships denoted by the above relation symbol
``$\sim$'' from $K$. Now, letting $H$ be the base manifold, a U(1)-principle
bundle $P(H,\mathrm{U}(1))$ can be defined. The fiber $F_k$ at each
point of $H$ contains the equivalent class of states
\begin{equation}
\pi^{-1}=\Big\{\big\vert k\big\rangle=
\big\vert k\big\rangle g\Big\vert\big\vert k\big\rangle\in K,\;
g\in\mathrm{U}(1)\Big\},
\label{eq:trans-000-fiber-bundle}
\end{equation}
where $\pi$ is a projection with local expression defined using a local
trivialization $\phi$, that is $\pi\circ\phi(\vert k\rangle,\vert k\rangle
g)=\vert k\rangle$. The Berry phase is produced by parallel transport
of a quantum state in $H$ along a loop $\Gamma:[0,1]\to H$, where the
loop $\Gamma$ is supposed to be parameterized by parameter $t$ (may or
may not be time). A section $\sigma:H\to P$ is a smooth map and satisfies
$\pi\circ\sigma=1_H$. Choosing a specific section is equivalent to
locally fixing the phase of $\vert k\rangle$, namely $\sigma(\vert k
\rangle)=\vert k\rangle\exp(\mi\varphi(k))$. If $\pi\circ\tilde{\Gamma}
=\Gamma$, a curve in $P$ given by $\tilde{\Gamma}:[0,1]\to P$ is a
horizontal lift of $\Gamma$, where $\tilde{\Gamma}$ may not be a closed
loop even if $\Gamma$ is. In general, the horizontal lift $\tilde{\Gamma}$
satisfies $\tilde{\Gamma}(1)=\tilde{\Gamma}(0)g_{\Gamma}(1)$, where
$g_{\Gamma}\in\mathrm{U}(1)$ is a transformation on the fiber. The set
of $g_{\Gamma}$ forms a subgroup of the structure group U(1), called
the holonomy group or the Berry holonomy for convenience. Assuming
$\Gamma(0)=\vert k\rangle$ and $\Gamma(t)=\vert k\rangle$, the horizontal
lift $\tilde{\Gamma}(t)$ defines a section $\tilde{\Gamma}(t)=\sigma
(\vert k(t)\rangle)=\vert k(t)\rangle\exp(\mi\varphi(t))$. Figure
\ref{fig:phase-bundle} schamatically illustrates curves of $\Gamma(t)$
and $\tilde{\Gamma}(t)$ and thus the generation of the Berry phase.

Now, we should show the connection on a fiber bundle, in particular
the Ehresmann connection, and then give its application to the phase
in chemical dynamics.
\begin{defi}
Letting $\pi:P\to M$ be a principal fiber bundle with principle bundle
$P$ and structure group G, an Ehresmann connection is a smooth choice
of horizontal subspaces in the tangent bundle of $P$.
\label{def:ehmen-connect-0}
\end{defi}
Since the loop $\Gamma(t)$ is parameterized by $t$ (see Figure
\ref{fig:phase-bundle}), the tangent vector $\mathcal{X}$ to $\Gamma(t)$
can be locally expressed as
\begin{equation}
\mathcal{X}=\mathcal{X}^{\kappa}\frac{\partial}{\partial q^{(\kappa)}}
=\frac{\mathrm{d}q^{(i)}}{\mathrm{d}t}
\frac{\partial}{\partial q^{(i)}}=\frac{\mathrm{d}}{\mathrm{d}t}.
\label{eq:trans-001-fiber-bundle}
\end{equation}
Letting $\tilde{\mathcal{X}}$ be the tangent vector to $\tilde{\Gamma}(t)$,
it forms the tangent bundle associated with $P$, denoted by $TP$ and
thus $\tilde{\mathcal{X}}\in TP$. For instance, Figure \ref{fig:phase-bundle}
shows tangent vectors $\mathcal{X}$ and $\tilde{\mathcal{X}}$ associated
with $\Gamma(t)$ and $\tilde{\Gamma}(t)$, respectively. Noting
$\pi\circ\tilde{\Gamma}=\Gamma$, we have $\pi_*\tilde{\mathcal{X}}=\mathcal{X}$.
The tangent bundle $TP$ can be separated as $TP=HP\oplus VP$, where
$HP$ and $VP$ are horizontal and vertical subspaces, respectively. As
shown in Figure \ref{fig:phase-bundle}, $\tilde{\mathcal{X}}$ is tangent
to the horizontal lift implying that $\tilde{\mathcal{X}}\in HP$. In
the $VP$ subspace, all vectors are tangent to the fiber. At $\vert k
\rangle\in H$, the fiber is given by Equation \eqref{eq:trans-000-fiber-bundle}
and hence a vector in the vertical subspace is given by
\begin{equation}
\left.\frac{\mathrm{d}}{\mathrm{d}s}\Big(\big\vert k\big\rangle\exp\big(\mi\varphi s\big)\Big)
\right\vert_{s=0}
=\big\vert k\big\rangle\mi\varphi,\quad\varphi\in\mathbb{R},\quad\mathrm{and\;thus}\quad
VP_{\vert k\rangle}=\Big\{\big\vert k\big\rangle\mi\varphi\Big\vert\varphi\in\mathbb{R}\Big\},
\label{eq:trans-002-fiber-bundle}
\end{equation}
where $s$ denotes a parameter (coordinate) along the fiber. Equation
\eqref{eq:trans-002-fiber-bundle} indicates that if
$\langle k'\vert k\rangle\neq0$ for $\forall\vert k'\rangle\in H$
then $\mi\cdot\mathrm{arg}\langle\phi\vert k\rangle\in VP_{\vert k\rangle}$
for $\forall\vert\phi\rangle\in\pi^{-1}(\vert k\rangle)$. In other
words, a horizontal vector must not have any component in the vertical
subspace, which implies that the horizontal subspace at $\vert k\rangle$
is given by
\begin{equation}
HP_{\vert k\rangle}=\Big\{\big\vert\psi\big\rangle\Big\vert\big\langle\psi\big\vert
k\big\rangle=0\Big\}.
\label{eq:trans-003-fiber-bundle}
\end{equation}
Letting $\vert\psi(t)\rangle=\vert k\rangle\exp(\mi\varphi(t))$ be a point
on $\tilde{\Gamma}(t)$, the tangent vector at $\vert k\rangle$ is similarly
given by $(\mathrm{d}_P\vert\psi\rangle/\mathrm{d}t)_{t=0}$, where
$\mathrm{d}_P$ is the exterior derivative on $P$ and thus satifies
the horizontal-space condition
\begin{equation}
\big\langle\psi\big\vert\tilde{\mathcal{X}}\big\vert\psi\big\rangle=
\left\langle\psi\left\vert\frac{\mathrm{d}_P}{\mathrm{d}t}\right\vert
\psi\right\rangle=0,\quad\forall t\in\mathbb{R}.
\label{eq:trans-004-fiber-bundle}
\end{equation}
Euation \eqref{eq:trans-004-fiber-bundle} means that one can define a
connection at $\vert\psi\rangle$ on $P$ by
\begin{equation}
\omega_{\vert\psi\rangle}
=\big\langle\psi\big\vert\mathrm{d}_P\big\vert\psi\big\rangle
=\mi\cdot\mathrm{Im}
\big\langle\psi\big\vert\mathrm{d}_P\big\vert\psi\big\rangle,
\label{eq:trans-005-fiber-bundle}
\end{equation}
called Ehresmann connection on $P$ (see Definition \ref{def:ehmen-connect-0}).
The seocnd equality in Equation \eqref{eq:trans-005-fiber-bundle} exists
since $\langle\psi\vert\mathrm{d}_P\vert\psi\rangle$ is imaginary-valued.
With Equation \eqref{eq:trans-005-fiber-bundle}, one can further
express Equation \eqref{eq:trans-004-fiber-bundle} as
$\omega(\tilde{\mathcal{X}})=0$.
\begin{remark}
The Ehresmann connection distinguishes a physical motion (horizontal)
from a gauge transformation which is redundant (vertical). It provides
a geometric realization of the fact that the specification of horizontal
subspaces is the interaction.
\label{rem:insight-ehmen-connet-0}
\end{remark}

The pullback of $\omega$, known as the Berry connection, is defined on
$M$. To this end, if $\mathcal{X}$ is the pushforward of
$\tilde{\mathcal{X}}$, similar to Equation \eqref{eq:trans-005-fiber-bundle},
then one can have the Berry connection
\begin{equation}
A_{\mathrm{B}}\big(\mathcal{X}\big)=\left\langle k\left\vert
\frac{\mathrm{d}}{\mathrm{d}t}\right\vert k\right\rangle.
\label{eq:trans-006-fiber-bundle}
\end{equation}
Due to $\vert\psi(t)\rangle=\vert k\rangle\exp(\mi\varphi(t))$ and
$\omega(\tilde{\mathcal{X}})=0$, by Equation \eqref{eq:trans-004-fiber-bundle}
one can have
\begin{equation}
0=\mi\frac{\mathrm{d}\varphi}{\mathrm{d}t}+\left\langle k\left\vert
\frac{\mathrm{d}}{\mathrm{d}t}\right\vert k\right\rangle=
\mi\frac{\mathrm{d}\varphi}{\mathrm{d}t}+A_{\mathrm{B}}\big(\mathcal{X}\big),
\label{eq:trans-007-fiber-bundle}
\end{equation}
and thus the Berry phase
\begin{equation}
\varphi_{\mathrm{B}}=\mi\oint A_{\mathrm{B}}\big(\mathcal{X}(t)\big)\mathrm{d}t=
\mathrm{arg}\big\langle\psi(0)\big\vert\psi(1)\big\rangle,
\label{eq:trans-008-fiber-bundle}
\end{equation}
where only the closed loops in the paramter space with $\vert k\rangle(0)=\vert
k\rangle(1)$ are considered. Thus, a parallel transport of a vector,
say quantum state, along $\tilde{\Gamma}(t)$ is equivalent to a parallel
transport of the associated fiber element along $\Gamma(t)=\pi(\tilde{\Gamma}(t))$.
It must be noted again that the parameter $t$ of the loop $\Gamma$ is
not necessary to be time. If it is chosen to be time, the above derivations
hold and all results stand. In other words, time evolution of quantum
state along a closed loop in a curved parameter space is what give rise
to the Berry phase, as shown by Equations \eqref{eq:trans-006-fiber-bundle}
and \eqref{eq:trans-008-fiber-bundle}. This is also why the Berry phase
is known as the geometric phase. For instance, parallel transport of a
vector along a 2D M{\"o}bius band results in a reversal of the vector
orientation because the non-trivial monodromy of the band connection
induces a $\pi$-rotation under parallel transport. Similar analysis on
the Klein bottle can be given indicating parallel transport of a vector
in the 3D curved space. Thus, one can introduce the gauge potential into
very fabric of the nuclear motions. This geometric structure introduces
a formulation which reveals how environmental entanglement geometrizes
the underlying motions. It is worth noting that the above idea is not
new in molecular reaction dynamics. For instance, similar to the idea
of the gauge theory, Baer \cite{bae06:boa} deeply discussed on the
concept of molecular field and the theory beyond the BOA, as given in
Table \ref{tab:geom-comp}.
\begin{remark}
According to Equations \eqref{eq:trans-006-fiber-bundle} and \eqref{eq:trans-008-fiber-bundle},
adiabatically transporting a quantum state around a closed loop in flat
sapce typically yields no geometric phase, as the environment is topologically
trivial. This is trivially molecular motion. However, when a molecule
adiabatically encircles a conical intersection, which is a degeneracy
point between two adjacent electronic states, the electronic wave function
acquires a sign change, that is the Berry phase, arising from the underlying
conical topology. It plays a crucial role in nonadiabatic reaction dynamics.
\label{rem:gaugw-field-001}
\end{remark}

To end this section, let us have a look at the gauge theory for molecular
reaction dynamics because we have indicated this theory several times.
In general, there exist two complementary frameworks for the gauge theory,
including physical formulation and geometric formulation. This point
can be clearly found in this section. The physical formulation is
represented in terms of gauge potentials, field strengths, and local
symmetries, while the the geometric formulation is given by concepts
of principal bundles, connections, and curvature. Simply, if the
Lagrangian of a free charged particle remains invariant under a local
$\mathrm{U}(1)$ phase transformation, an additional gauge field must
be introduced. The EOMs of such gauge field are Maxwell equations,
which introduces the electromagnetic interaction. In other words, the
electromagnetic interaction is the static and non-relativistic limit
of quantum electrodynamics which can be obtained by $\mathrm{U}(1)$
gauge symmetry. In this context, one can attribute motions in a molecular
reaction to gauge fields, as summarized in Table \ref{tab:geom-comp},
implying existence of gauge field theory for molecular reaction. This
idea is not surprising because the gauge field theory connects the
interactions in a many-body system (say the molecular system) and its
dynamics properties. Thus, further development of the gauge field theory
for molecular reaction may provide theoretical tools to classify
interactions and the inspiration for experiments, such as quantum
simulators \cite{alb18:015002,cao19:10856,mca20:015003,goo20:1084,oll21:4229,saj22:6475,wei24:102105}.
\begin{remark}
A quantum simulator is a controlled system whose Hamiltonian can be
tuned to match, either exactly or effectively, the Hamiltonian of a
physical system of interest
\cite{alb18:015002,cao19:10856,mca20:015003,goo20:1084,oll21:4229,saj22:6475,wei24:102105}.
By initializing the simulator in a known quantum state, measurements
of its time evolution reveal properties of the target system without
requiring direct experiments. In this context, quantum simulator is a
special-purpose analog quantum computer
\cite{alb18:015002,cao19:10856,mca20:015003,goo20:1084,oll21:4229,saj22:6475,wei24:102105}.
\label{rem:gauge-field-990}
\end{remark}
Due to the non-zero phase effects and entanglement growth, the gauge
field arisen from molecular reaction dynamics might be the most
computationally challenging problems for classical computers. Quantum
simulators might circumvent these problems by directly realizing the
same local gauge symmetry in a controllable system
\cite{alb18:015002,cao19:10856,mca20:015003,goo20:1084,oll21:4229,saj22:6475,wei24:102105,kho24:013032,bes25:052619}.
By engineering the simulator Hamiltonian to match a lattice gauge
theory, one can observe real-time dynamics that are otherwise inaccessible.
For instance, the 1D Fermi-Hubbard model \cite{hub64:401} can be used
to describe electrons in a molecular wire or conjugated polymer. To
simulate electornic structure of such kind of complex systems, Khodaeva
{\it et al.} \cite{kho24:013032} proposed a quantum circuit algorithm
based on the Z$_2$ lattice gauge theory representation of the 1D
Fermi-Hubbard model \cite{hub64:401}. For another example, Bespalova
{\it et al.} \cite{bes25:052619} provided quantum circuits based on the
1D Fermi-Hubbard model to simulate ground and excited states through
the variational quantum eigensolver and the quantum EOM approach,
respectively, as well as product formula decompositions for time
evolution.

%------------------------------------------------------
\subsection{Quantum Simulator for Molecular Reaction\label{sec:qc-simulatior}}

With Remark \ref{rem:gauge-field-990}, in this section let us have a
look at the use of quantum computing
\cite{alb18:015002,cao19:10856,mca20:015003,goo20:1084,oll21:4229,saj22:6475,wei24:102105,kho24:013032,bes25:052619}.
Contrary to the term of quantum, computation based on the traditional
computers is often called classical computing. We refer the reader to
References
\cite{alb18:015002,cao19:10856,mca20:015003,goo20:1084,oll21:4229,saj22:6475,wei24:102105,kho24:013032,bes25:052619}
for details on quantum computing and, in particular, its applications
to optimization and variation approaches, such as ML techniques. In
classical computers binary bits are encoded as $\{0,1\}$ used for all
logical operations. Similar to bit, in a quantum computer qubit (quantum
bit) is commnonly used to encode the primary unit of information. To
define a qubit, two two-dimensional vectors, denoted as $\vert0\rangle$
and $\vert1\rangle$, are referred to as computational basis states. In
physical apparatus, for example, these two basis states may be (1) the
two hyperfine energy levels of an ion like the apparatus of trapped-ion
based quantum computing platforms, or (2) energy levels corresponding
to different number of cooper pairs in a superconducting island created
between a Josephson junction and a capacitor plate or (3) the highly
excited electronic energy levels of a Rydberg atom based cold
atomic-ensembles or (4) polar molecules in pendular states.
Mathematically the two basis states can be represented as matrice
$\vert0\rangle=(1\;0)^{\mathrm{T}}$ and $\vert1\rangle=(0\;1)^{\mathrm{T}}$
and collectively form a basis for the two-dimensional Hilbert space
$\mathbb{H}$ of the system. In general, the state of the qubit in the
basis $\{\vert0\rangle,\vert1\rangle\}$ can be defined by
\begin{equation}
\rho=\big\vert0\big\rangle\left(\frac{1}{2}+n_z\right)
\big\langle0\big\vert+
\big\vert0\big\rangle\big(n_x+in_y\big)\big\langle1\big\vert+
\big\vert1\big\rangle\big(n_x-in_y\big)\big\langle0\big\vert
+\big\vert1\big\rangle\left(\frac{1}{2}-n_z\right)
\big\langle1\big\vert,
\label{eq:compu-000-quantum}
\end{equation}
where $(n_x\;n_y\;n_z)^{\mathrm{T}}\in\mathbb{R}^3$. Equation
\eqref{eq:compu-000-quantum} indicates that $\rho\in\mathscr{L}(\mathbb{H})$
is unit trace positive semi-definite operator, implying that
$(n_x\;n_y\;n_z)^{\mathrm{T}}$ resides within a Bloch sphere.
Parameterizing $(n_x\;n_y\;n_z)^{\mathrm{T}}$ by angles
$\alpha\in[0,\pi]$ and $\beta\in[0,2\pi]$, if $\rho$ is pure then the
associated vector $\vert\psi\rangle\in\mathbb{H}$ reads
$\vert\psi\rangle=\vert0\rangle\cos(\alpha/2)+\vert1\rangle\sin(\alpha/2)\exp(\mi\beta)$.
The angle $\beta$ encodes the relative phase between the $\vert0\rangle$
and $\vert1\rangle$ states. However, $\beta$ itself is not the geometric
phase, but serves as both dynamical and geometric contributions. The
latter plays a fundamental role in geometric quantum computation. Although
$\beta$ is not synonymous with the geometric phase, its controlled variation
along properly designed loop path provides the physical mechanism for
realizing geometrically protected quantum gates.
\begin{remark}
The core idea of geometric quantum computation is to design loop paths
on the Bloch sphere, such that the net change in $\beta$ becomes purely
geometric. The resulting quantum gate depends only on the global topological
feature of the control loop path, making it insensitive to local control
errors, pulse fluctuations, and slow drifts.
\label{remark:geom-comp-quantum}
\end{remark}

Extending to $N$ qubits, the Hilbert space is given by
$\mathbb{H}_A\otimes\mathbb{H}_B\otimes\cdots\otimes\mathbb{H}_N$
where the labels $J=\{A,B,\cdots,N\}$ are physically used to demarcate
the space of each qubit. Similar to Equatoin \eqref{eq:def-state-spa-geom-000},
a computational basis
$\vert I\rangle=\vert i_A\rangle\otimes\vert i_B\rangle\otimes\cdots\otimes\vert i_N\rangle=
\vert i_Ai_B\cdots i_N\rangle$
can be thus defined by Kronecker product, where $I=\{i_A,i_B,\cdots,i_N\}$
represents a set of indices. Unlike the basis of Equatoin \eqref{eq:def-state-spa-geom-000},
since $\{i_K=0,1\vert K\in J\}$, there are $2^N$ basis states for $N$
qubits and hence the basis set is given by $\{\vert I\rangle\}_{i=I}^{2^N-1}$.
Extending Equation \eqref{eq:compu-000-quantum}, a general state
$\rho_{A,B,C,\cdots,N}=\rho_J$ of the multi-qubit system would again
correspond to a positive semi-definite operator with unit trace
$\rho_{A,B,C,\cdots,N}=\rho_J=\sum_{I,K=0}^{2^N-1}\vert I\rangle\rho_{IK}\langle K\vert$,
where $\rho_{IJ}\in\mathbb{C}^2$ represents element of representation
matrix. One can further define a reduced state for each of the qubits
through partial tracing of the state $\rho_J$ as $\rho_{J^K}=\mathrm{Tr}_{J^K}(\rho_J)
\in\mathscr{L}(\mathbb{H}_{J^K})$, where $J^K=\{A,B,\cdots,K-1,K+1,\cdots,N\}$.
This operation for the $K$-th qubit is often called the reduced density
operator. Similar concept is also introduced by the SPA to simply represent
the EOM, as given in Sections \ref{sec:elec-stru} and \ref{sec:math-struc}.
It is completely positive trace preserving map and generates valid state
of the $K$-th qubit. If the general state $\rho_{A,B,C,\cdots,N}=\rho_J$
is pure, then the associated vector $\vert\psi\rangle_J=
\vert\psi\rangle_{A,B,C,\cdots,N}\in\mathbb{H}_A\otimes\mathbb{H}_B
\otimes\cdots\otimes\mathbb{H}_N$ in the $N$-qubit computational basis
$\{\vert I\rangle\}_{I=1}^{2^N-1}$ is given by
\begin{equation}
\big\vert\psi[C_{i_Ai_B\cdots i_N}]\big\rangle_{A,B,C,\cdots,N}
=\sum_{i_A=0}^1\sum_{i_B=0}^1\cdots\sum_{i_N=0}^1
\big\vert i_Ai_B\cdots i_N\big\rangle
C_{i_Ai_B\cdots i_N}\;\mathrm{or}\;
\big\vert\psi[C_I]\big\rangle_J=\sum_{I=0}^{2^N-1}
\big\vert I\big\rangle C_I,
\label{eq:quantum-computing-001}
\end{equation}
where the expansion coefficients $C_{i_Ai_B\cdots i_N}=C_I\in\mathbb{C}^2$
play a similar role as those in Equatoin \eqref{eq:def-state-spa-geom-000}.
For a normalized state, there should exist normalization expression
$\sum_I\vert C_I\vert^2=1$. Since $\{i_K=0,1\vert K\in J\}$, Equation
\eqref{eq:quantum-computing-001} is in stark contrast to Equation
\eqref{eq:def-state-spa-geom-000}, even though their formal identity.
Furthermore, if each coefficient $C_I$ is multiplicatively factorizable
into scalars characterizing the basis states of each qubit, then the
pure state $\vert\psi[C_I]\rangle_J$ given in Equation \eqref{eq:quantum-computing-001}
can be termed separately in the form
$\vert\psi[C_I]\rangle_J=\vert\phi_A\rangle\otimes\vert\phi_B\rangle\otimes\cdots\otimes
\vert\phi_N\rangle=\bigotimes_{K\in J}\vert\phi_K\rangle$, where
$\vert\phi_K\rangle\in\mathbb{H}_K$ and $J=\{A,B,\cdots,N\}$.
If a pure state $\vert\psi\rangle_J$ given in Equation \eqref{eq:quantum-computing-001}
is not separable, we define such state as entangled state \cite{pan12:777}.
\begin{defi}
In entanglement, subsystems are described by a single entangled state,
making their individual properties indeterminate yet perfectly correlated.
\label{def:entangled-state-0}
\end{defi}
Entanglement is a non-classical correlation \cite{pan12:777}. The presence
of entanglement is a feature wherein computation using qubits can be
different from that of the classical bit counterparts. It is a resource
that powers quantum computing and communication.

Having defined states of qubits, let us consider transformation and
manipulation of such states. In the gate-model of quantum computing
paradigm, quantum gates are defined as transformations between states
and often represented by unitary matrices. Due to the unitary property,
the inverse of quantum gate necessarily exists and hence transformations
using quantum gates alone are always reversible. To incorporate
irreversibility, one should make projective measurements as that
disturbs the states irrevocably. These quantum gates might act on
various qubits to implement an operation. Commonly used quantum
gates and their matrix representation in the computational basis have
been reported eslsewhere
\cite{alb18:015002,cao19:10856,mca20:015003,goo20:1084,oll21:4229,saj22:6475,wei24:102105,kho24:013032,bes25:052619}.
Next, a quantum circuit can be defined as an assembly of quantum gates
which transforms the initial multi-qubit state to the final desired
state. In other words, the quantum circuit, the set of quantum gates,
operationally represents a user-defined unitary transformation. The
quantum circuit have been divided into four junctions \cite{saj22:6475}.
At junction I the joint state of the two qubits is the initial computational
basis $\vert00\rangle$. At junction II, the effect of Hadamard on the
first qubit yields a separable state. We need the first qubit in an
equal superposition of the single-qubit basis states and the second
qubit still in $\vert0\rangle$. At junction III, the controlled-not
(CNOT) gate with the first qubit as the control and the second qubit
as the target yields $(\vert00\rangle+\vert11\rangle)/\sqrt{2}$. At
junction IV, the controlled-phase (CPHASE($\alpha$)) gate selectively
creates a phase difference of $\alpha$ between $\vert00\rangle$ and
$\vert11\rangle$. Measurements on the target state on a computational
basis would yield equal probability of observing either the state
$\vert00\rangle$ or $\vert11\rangle$ and zero probability of observing
$\vert01\rangle$ or $\vert10\rangle$. The above quantum computing
paradigm is particularly useful for optimization problems, while there
exist other kinds of paradigms for quantum computing, say those reviewed
in Reference \cite{hau20:054401}.

As shown above, quantum computing for chemistry motions is likely to
develop into a rich subfield of chemistry meaning that the field of
quantum dynamics also needs development of its quantum algorithms,
in particular variational quantum algorithms (VQAs). However, this is
not an easy task \cite{oll21:4229}. As given in Remark \ref{rem:quantum-simulator},
one of solutions is variational quantum eigensolver (VQE) which is
useful and helpful in solving eigenvalue problem, such as electronic
structure (see also Section \ref{sec:elec-struc}). According to Aspuru-Guzik
and co-workers \cite{cao19:10856}, although many state-of-the-arts
electronic-structure tools have been developed on classical computor
and driven much innovation over the past half century, many of these
tools will be supplemented with or even replaced by quantum computing.
Aspuru-Guzik and co-workers \cite{cao19:10856} thus predicted that
quantum computing would have broad prospects for development in chemistry
science. In particular in electronic structure, quantum algorithms have
been currently available as quantum computing is able to naturally
handle electronic wave functions which span the full electronic Hilbert
space. 
Illustrated in Figure \ref{fig:vqe-for-quan-chem} are the procedure
in QVE which consists of following steps. The first step is preparation
of quantum state, where a parametrized state is prepared based on the
quantum device. This step can be finished by applying a carefully
designed parametrized unitary to an easily prepared initial state.
The second step is energy estimation where the energy expectation
is estimated through a Hamiltonian averaging procedure which involves
taking measurements of tensor products of wave function corresponding
to the qubit representation of the target Hamiltonian. The third
step is classical feedback updating the parameters of the quantum
state through a classical nonlinear optimization routine. Repeating
the second and third steps, convergence can be approaches until some
criteria, say energy, are satified.

Tavernelli and co-workers \cite{oll21:4229}
presented a perspective on quantum computational algorithms, aiming to
alleviate the exponential scaling inherent to quantum dynamics calculations.
It is worth noting that, discretizing each of the three coordinates
into $P$ grids, classical computation on propagation of an $N$-body
system requires storing $P^{3N}\times2^N$ complex amplitudes, while
quantum computation only requires $(3\log_2P+1)N$ qubits. Novertheless,
grid-based propagation is largely under-studied compared to VQE for
electronic structure. Currently, quantum dynamics calculation is one
of the most promising applications of quantum computing. A number of
quantum algorithms for this problem have been proposed \cite{oll21:4229}.
%as simply collected in Table \ref{tab:qc-al-dyn}. 
The choice of these algorithms, however, is a delicate question. For
simple, assuming the $d$-dimensional
Hamiltonian of the summation form $\hat{H}=\sum_{i=1}^{d}\hat{h}_i$,
the propagator $\exp(-\mi\hat{H}t)$ can be approximated by the
Lie-Trotter-Suzuki formula, which can be further truncated to first order
$\exp(-\mi\hat{H}t)=(\prod_{i=1}^d\exp(-\mi\hat{h}_it/P))^P+\mathcal{O}(d^2t^2/P)$,
where $t$ must be small and $P$ is number of time grids. This truncated
approximation is implemented by quantum computing when the propagator
$\exp(-\mi\hat{H}t)$ cannot exactly be translated into quantum circuit,
but each individual one-body propagator $\exp(-\mi\hat{h}_it)$ can.

Turning to time-dependent VQAs, we consider a time-dependent variational
trial $\vert\Phi[\boldsymbol{\vartheta}(t)]\rangle$ to approximate the
solution $\vert\Psi(t)\rangle$. To prepare the trial
$\vert\Phi[\boldsymbol{\vartheta}(t)]\rangle$ on a quantum computer, a
parametrized unitary $\hat{U}(\boldsymbol{\vartheta}(t))$, that is a
quantum circuit, is required to act onto a reference qubit state
$\vert\psi\rangle$, namely
$\vert\Phi[\boldsymbol{\vartheta}(t)]\rangle=\hat{U}(\boldsymbol{\vartheta}(t))\vert\psi\rangle$.
Since the parameters enter as qubit-gate angles, all parameters must
be real. Based on complex structure of the McLachlan variational principle,
that is the Dirac-Frenkel variational principle $\langle\delta\Phi\vert
\partial_t-\hat{H}\vert\Phi\rangle=0$, one can obtain the EOM for paramters
$\boldsymbol{\vartheta}(t)$ in the matrix form
$\boldsymbol{\mathcal{M}}\dot{\boldsymbol{\vartheta}}=\boldsymbol{\mathcal{V}}$
with the elements
$\mathcal{M}_{ij}=\mathrm{Re}(\langle\partial_i\Phi\vert\partial_j\Phi\rangle+
\langle\partial_i\Phi\vert\Phi\rangle\langle\partial_j\Phi\vert\Phi\rangle)$
and
$\mathcal{V}_i=\mathrm{Im}(\langle\partial_i\Phi\vert\hat{H}\vert\Phi\rangle-
\langle\partial_i\Phi\vert\Phi\rangle\langle\Phi\vert\hat{H}\vert\Phi\rangle)$,
where $\partial_i=\partial/\partial\vartheta_i$. In principle, this
EOM for $\boldsymbol{\vartheta}(t)$ can be integrated with any numerical
solver. However, the stability of the dynamics simulation depends on
the system and setups, such as errors of numerical integration and
quantum hardware noise on the integration \cite{oll21:4229}. It is worth
noting that the above derivations are obviously valid for either classical
or quantum computer. The distinction is whether the ansatz preparation and
the matrix evaluation adopt quantum circuits or not. The EOM for $\boldsymbol{\vartheta}$
indicates a near-term approach. Other alternative near-term approaches
are the adaptive variational quantum dynamics simulation, the subspace
variational quantum simulation, and the variational fast forwarding.
%as given in Table \ref{tab:qc-al-dyn}. 
Therefore, the first step of
time-dependent VQAs is preparation of quantum state based on the
quantum device. The second step is time evolution in which propagation
of one SPF is simulated by one qubit (see also Equation \eqref{eq:quantum-computing-001})
until the system is propagated by given time. These steps are same as
those in the classical computation. 

%------------------------------------------------------
\subsection{Step Consistency\label{sec:consistency}}

Based on the variational principle, we have described the techniques
for building the Hamiltonian operator and for solving the Schrodinger
equation by the SPA. In practical calculations, these techniques have
been developed independently leading to various methods for different
objectives, some aim for intuitive and simple while some others aim for
accuracy and efficiency. For instance, as illustrated by Figure
\ref{fig:framework} design of coordinates, derivation
of KEO, configurational sampling, electronic-structure calculations,
construction of PES, and choice of solution ansatz are typically
pursued in isolation, giving rise to distinct algorithms, approaches,
and softwares. Nevertheless, practical calculations
ultimately necessitate the integration of the Hamiltonian operator
with the EOM solvers, requiring that these independently developed
algorithms be brought into consistency. For the ansatz in Equation
\eqref{eq:def-state-spa-geom-000} or \eqref{eq:mctdh-wf-000}, the
core idea of this consistency is to reduce the complexity in
solving multi-dimensional Schr{\"o}dinger equation, where one must
evaluate the matrix element $\langle R\vert\hat{H}\vert S\rangle$.
This typically requires writing both the Hamiltonian operator and the
solution ansatz in the same summation of products of single-particle
basis functions making $\langle R\vert\hat{H}\vert S\rangle$
become product of one-dimenional integrals.
%According to Equations \eqref{eq:def-state-spa-geom-000} and \eqref{eq:mctdh-wf-000},
%we require the Hamiltonian operator written as similar form based on
%one-particle operators. 
The KEO is generally in such form, as shown in
Equation \eqref{eq:polysph-keo-003}. However, Equation \eqref{eq:tangent-bundle-801-ai}
indicates that $f_{\theta}$ is smooth without any
special formulation. One way to overcome this problem is to re-fit
the PES in the similar expansion to the solution ansatz. By
another way, the model of function approximation must be specially
designed. Since these two ways have been well discussed elsewhere
\cite{zha25:20397}, only a brief review is given here.

For the first way, the potential function needs to be re-fitted. This
task can be done by various methods. One of typical examples is a
potential re-fitting method associated with MCTDH, called POTFIT
\cite{jae95:5605,jae96:7974,jae98:3772}. By POTFIT, the potential
function is re-fitted in the special form similar to Equations
\eqref{eq:mctdh-wf-000} and \eqref{eq:mctdh-wf-999}. According to
Equation \eqref{eq:mctdh-wf-999}, on the grid $J$ of the configuration
space, an arbitrary smooth function is represented by tensor $V_J=
V_{j_1\cdots j_d}=V(q_{j_1}^{(1)},\cdots,q_{j_d}^{(d)})$, which is
then approximately re-fitted in the sum-of-products (SOP) form \cite{jae95:5605,jae96:7974,jae98:3772}
\begin{equation}
V_J\approx\sum_{i_1=1}^{n_1}\cdots\sum_{i_d=1}^{n_d}
C_{i_1\cdots i_d}\prod_{\kappa=1}^dv_{i_{\kappa}}^{(\kappa)}
\big(q_{j_\kappa}^{(\kappa)}\big)=
\sum_{i_1=1}^{n_1}\cdots\sum_{i_d=1}^{n_d}
C_{i_1\cdots i_d}\prod_{\kappa=1}^dv_{j_{\kappa}i_{\kappa}}^{(\kappa)}=
\sum_I\mathscr{V}_{JI}C_I,\quad
\mathscr{V}_{JI}=\prod_{\kappa=1}^dv_{j_{\kappa}i_{\kappa}}^{(\kappa)},
\label{eq:potfit-002-sop}
\end{equation}
where $v_{j_{\kappa}i_{\kappa}}^{(\kappa)}$ is the $j_\kappa$-th value
of the $i_{\kappa}$-th single-particle potential (SPP) for the $\kappa$-th
coordinate. It is assumed that the SPPs are normalized, but no further
restrictions are imposed, in particular not required that the SPPs are
orthogonal. To determine the SPPs, expansional coefficients and orders
in Equation \eqref{eq:potfit-002-sop}, we define symmetric positive
semi-definite potential density matrix (similar to the case of quantum
state given by Equation \eqref{eq:def-state-spa-geom-002}) for the
$\kappa$-th coordinate, denoted by $\boldsymbol{\varrho}^{(\kappa)}$,
\begin{equation}
\varrho_{ji}^{(\kappa)}=\sum_{i_1=1}^{n_1}\cdots
\sum_{i_{\kappa-1}=1}^{n_{\kappa-1}}\sum_{i_{\kappa+1}=1}^{n_{\kappa+1}}
\cdots\sum_{i_d=1}^{n_d}
V_{i_1,\cdots,i_{\kappa-1},j,i_{\kappa+1},\cdots,i_d}
V_{i_1,\cdots,i_{\kappa-1},i,i_{\kappa+1},\cdots,i_d}
=\sum_{I^\kappa}{}'V_{I_j^\kappa}V_{I_i^\kappa},
\label{eq:potfit-001-sop}
\end{equation}
where $I_{i/j}^\kappa=\{i_1,\cdots,i_{\kappa-1},i/j,i_{\kappa+1},\cdots,i_d\}$
and $I^\kappa=\{i_1,\cdots,i_{\kappa-1},i_{\kappa+1},\cdots,i_d\}$,
while the prime indicates that the summation excludes one coordinate.
The orthonormal eigen-vectors and the corresponding eigen-values of
$\boldsymbol{\varrho}^{(\kappa)}$ are called natural potentials and
natural potential populations, respectively. For simple, we always
assume a decreasing order of the eigen-values. Then, by the natural
potential as the SPPs, one can choose a set of expansion orders
$\{m_\kappa\leq n_\kappa\}_{\kappa=1}^d$ to approximately reduce
Equation \eqref{eq:potfit-002-sop} to the potfitted function by
replacing $\{n_\kappa\}_{\kappa=1}^d$ by $\{m_\kappa\}_{\kappa=1}^d$,
namely $V_J\approx V^{(\mathrm{pf})}_J$. Since the orthonormal
product basis set is complete over the grid $I$ (see Equation
\eqref{eq:potfit-002-sop}), if the expansion orders and the number of
grids are equal, then $V_J$ and $V_J^{(\mathrm{pf})}$ are exactly
identical at the grid $I$. 

According to Equation \eqref{eq:potfit-002-sop},
the expansion coefficients $C_J$ can be determined by the overlaps
between the potential and the natural potentials. Numerical implementation
of POTFIT is thus a typical regression problem to optimize the
expansion coefficients $C_J$, where the optimization target is
error between $V_I$ and $V_I^{(\mathrm{pf})}$. To implement such
optimization, similar to the quantum state cases in Equation
\eqref{eq:deeq-map-dipara-00}, one may define the $\kappa$-th
contracted expansion as
\begin{equation}
D^{(\kappa)}_{I^\kappa}\big(q_{j_{\kappa}}^{(\kappa)}\big)
=D^{(\kappa)}_{j_{\kappa}I^{\kappa}}
=\sum_{i_{\kappa}=1}^{m_{\kappa}}C_Iv_{j_{\kappa}i_{\kappa}}^{(\kappa)},
\label{eq:potfit-007-sop}
\end{equation}
which can be performed over all natural potentials of the $\kappa$-th
logical coordinate by replacing the upper summation limit $m_{\kappa}$
with $n_{\kappa}$. This procedure is called contraction over the
$\kappa$-th coordinate. Employing contracted expansion functions
in Equation \eqref{eq:potfit-007-sop}, the potfitted potential tensor
is re-written as \cite{jae95:5605,jae96:7974,jae98:3772},
\begin{equation}
V^{(\mathrm{pf})}_J=\sum_{I^\kappa}{}'
\Omega_{j_\kappa I_\kappa}^{(\kappa)}D^{(\kappa)}_{j_{\kappa}I^{\kappa}}.
\label{eq:potfit-v-000}
\end{equation}
In general, the so-called contracted mode $\kappa$ can be chosen freely.
This method involved in contracted mode may tightly related with the
variational principle discussed in Section \ref{sec:math-struc}. In
practical calculations, convergence of the potfitted potential must be
carefully inspected by balancing the error of the re-fitting process
and computational cost of subsequent wave function propagation. 
\begin{remark}
In optimization, the parameter space $M$ of a model with
parameter set $\theta$ is naturally seen as a smooth manifold. The
tangent bundle $TM$ associated to each point $\theta\in M$ is the
vector space $T_{\theta}M$ of infinitesimal parameter variations.
Optimization algorithms move the parameters along tangent directions
that reduce the optimization target. The cotangent bundle $T^*M$ is
the dual bundle, whose fibers consist of linear functionals (covectors)
acting on tangent vectors. The differential of the optimization target
is a covector field on $M$. Raising its index with the metric tensor,
one obtain the Riemannian gradient. It is the tangent vector that points
in the steepest descent direction. More generally, the choice of a metric
on $M$ provides an isomorphism between $T_{\theta}M$ and $T_{\theta}^*M$,
converting the covector into a gradient vector. Thus, one can unify
gradient descent, and reveals that the efficiency of an optimization
scheme is governed by the curvature of $M$ and the geodesic flow on
its tangent bundle. Thus, $TM$ encodes where the parameters can move,
$T^*M$ encodes what one aim to minimize, and their interplay through a
metric determines the actual path taken in parameter space.
\label{rem:cotangent-bundle-00}
\end{remark}

In addition to the SOP form in Equation \eqref{eq:potfit-002-sop},
another potential formulation suitable to Equations \eqref{eq:mctdh-wf-000}
and \eqref{eq:mctdh-wf-999} is the canonical polyadic decomposition
(CPD) form, which is also known as parallel factors (PARAFAC) or
canonical decomposition (CANDECOMP) in the literature. Similar to
Equation \eqref{eq:potfit-002-sop}, the CPD form of the $d$-dimensional
function on the grids $J$ is written as \cite{sch20:024108}
\begin{equation}
V_J\approx V^{\mathrm{(cpd)}}_J=\sum_{r=1}^Rc_r
\prod_{\kappa=1}^dv_{j_{\kappa}r}^{(\kappa)}=
\sum_{r=1}^R\mathscr{V}_{Jr}c_r,
\label{eq:potfit-002-sop-cpd}
\end{equation}
where $R$ is the expansion order, called the rank of the CPD expansion.
The remaining task is to find both the expansion functions and the
coefficients. To this end, one starts with a set of functionals for
each coordiante with a positive and coordinate-dependent weight
function which will be integrated out and replaced by unity \cite{sch20:024108}.
A regularization parameter can be also set in the target functional
to square root of machine precision \cite{sch20:024108}. Varying the
target functionals with respect to one SPP and one coefficient of the
coordinate, one can obtain the working equations \cite{sch20:024108}
that can be solved with standard linear algebra tools. The main
bottleneck in solving the working equations is that it contains
multi-dimensional quadratures that in general cannot be separated
into products of lower-dimensional integrals. To overcome this, the
complete sums are replaced by a Monte-Carlo integration leading to
the MCCPD algorithm \cite{sch20:024108}. We refer the reader to
Reference \cite{sch20:024108} for further technical details.

Turning to the second way, the model for function approximation is
specially designed in either the SOP form or the CPD form. To this end,
let us consider prediction of the KMR method $f^{(\mathrm{KMR})}$ given
in Equation \eqref{eq:glr-kmr-000}, based on which one can obtain the
CPD form of the Gaussian process regression (GPR), called CPD-GPR
\cite{son22:11128}. Starting from Equation \eqref{eq:glr-kmr-000}, the
GPR prediction $V^{(\mathrm{gpr})}(\mathbi{q})$ can be expanded as
\cite{son22:11128}
\begin{equation}
V^{(\mathrm{gpr})}\big(\mathbi{q}\big)=
\mathbf{K}_*^{\mathrm{T}}\cdot\overline{\mathbf{a}}
=\sum_{r=1}^n\big(\mathbf{K}_*^{\mathrm{T}}\big)_ra_r
=\sum_{r=1}^na_rK\big(\mathbi{q},\mathbf{X}_r\big),
\label{eq:cpd-gpr-000}
\end{equation}
where $K(\mathbi{q},\mathbf{X}_r)$ and $a_r$ are elements of the
$\mathbf{K}_*^{\mathrm{T}}$ and $\mathbf{a}$ matrices, respectively,
while $\mathbf{X}_r$ is the $r$-th training conformation and $n$ size
of training set, that is $\{\mathbf{X},\mathbf{E}\}=\{\mathbf{X}_r,E_r\}_{r=1}^n$.
Here, we would like to remind reader that $\mathbf{K}_*^{\mathrm{T}}$
and $\mathbf{a}$ are $1\times n$ and $n\times1$ matrices, respectively.
On the grids $J$ in the configuration space, Equation
\eqref{eq:cpd-gpr-000} becomes potential tensor (recall Equation
\eqref{eq:mctdh-wf-999}),
\begin{equation}
V_J^{(\mathrm{gpr})}=V^{(\mathrm{gpr})}\big(\mathbi{q}_J\big)=
\sum_{r=1}^na_rK\big(\mathbi{q}_J,\mathbf{X}_r\big).
\label{eq:cpd-gpr-001}
\end{equation}
For further derivation, we assume that the multi-dimensional kernel
function $K(\cdot,\cdot)$ must have a product form of a series of
kernel functions $\{K^{(\kappa)}(\cdot,\cdot)\}_{\kappa=1}^d$ that
depend on only one coordinates, namely
\begin{equation}
K\big(\mathbi{q},\mathbf{X}_r\big)=\prod_{\kappa=1}^dK^{(\kappa)}
\big(q^{(\kappa)},X^{(\kappa)}_r\big),
\label{eq:cpd-gpr-002}
\end{equation}
where $X^{(\kappa)}_r$ is the $\kappa$-th element of the coordinate
matrix $\mathbf{X}_r$. Substituting Equation \eqref{eq:cpd-gpr-002}
into Equation \eqref{eq:cpd-gpr-001}, we finally obtain the CPD-GPR
potential tensor,
\begin{equation}
V_J^{(\mathrm{cpd-gpr})}=\sum_{r=1}^na_rK\big(\mathbi{q}_J,\mathbf{X}_r\big)
=\sum_{r=1}^na_r\prod_{\kappa=1}^dK^{(\kappa)}\big(q^{(\kappa)}_{j_{\kappa}},X^{(\kappa)}_r\big)
=\sum_{r=1}^na_r\prod_{\kappa=1}^dK^{(\kappa)}_{j_{\kappa}r},
\label{eq:cpd-gpr-003}
\end{equation}
where $K^{(\kappa)}_{j_{\kappa}r}=K^{(\kappa)}(q^{(\kappa)}_{j_{\kappa}},X^{(\kappa)}_r)$.
Comparing Equation \eqref{eq:cpd-gpr-003} with Equation \eqref{eq:potfit-002-sop},
if $c_r=a_r$ and $v^{(\kappa)}_{i_{\kappa}r}=K^{(\kappa)}_{i_{\kappa}r}$,
the GPR potential tensor is equal to the CPD potential tensor, namely
$V_J^{(\mathrm{cpd})}=V_J^{(\mathrm{cpd-gpr})}$. Only restriction for
such expression is the assumption of Equation \eqref{eq:cpd-gpr-002}.
Thus, with the condition of Equation \eqref{eq:cpd-gpr-002}, having
obtained the matrix $\mathbf{a}$ by GPR, one can directly write the PES
in the CPD form by Equation \eqref{eq:cpd-gpr-003}. We would like to
emphasize that CPD-GPR has the same form with CPD but is different from
normal CPD where the $\kappa$-th SPP $v^{(\kappa)}_{i_{\kappa}r}$
predicts energy and $c_r$ is the product of their norms. In CPD-GPR,
however, $a_r$ is related to the energy and $K^{(\kappa)}_{j_{\kappa}r}$
represents the correlation of the grid values for the $\kappa$-th
coordinate and the $r$-th training energy. The latter one is originated
from the linear predictor essence of GPR fitting proposal, providing
another way to broaden the CPD explanation and application. We refer
the reader to References \cite{son22:1983,son22:11128} for details of
GPR and CPD-GPR methods. In addition, there exist other methods to
directly build the PES in the SOP or CPD form, such as the single-layer
NN fitting approach with the SOP form, denoted by SOP-NN
\cite{man06:194105,koc14:021101}.
\begin{remark}
It is worth noting that both SOP-NN and CPD-GPR methods can be seen as
imposing constraints in optimizing the model, which is analogous to
symplectic feature in optimization. The parameter space $M$ is often
equipped with additional geometric structures beyond a metric, such as
a symplectic structure that underpins Hamiltonian optimization, where
the optimization target plays a role of the Hamiltonian. In this context,
the symplectic gradient generates a flow on $M$ that preserves volume
and exhibits long-term stability.
\label{rem:why-cpd-gpr}
\end{remark}

Finally, we must mention a few words about the way where the Hamiltonian
operator and the solution ansatz are inconsistency in formulation,
meanwhile allow us to directly evaluate $\langle R\vert\hat{H}\vert
S\rangle$, in particular $\langle R\vert V\vert S\rangle$. This is
possible because the PES matrix is diagonal under the DVR or FBR
(similar to Equation \eqref{eq:def-state-spa-geom-002}), that is
$\langle R\vert V\vert S\rangle=V_R\delta_{RS}$ if $\{\vert r_{\kappa}\}_{\kappa=1}^d$
and $\{\vert s_{\kappa}\}_{\kappa=1}^d$ are basis sets of the DVR or FBR, where
$\delta_{RS}=\delta_{r_1s_1}\cdots\delta_{r_ds_d}$. Thus, one needs
$\leq N^d$ evalutions to compute the integral $\langle\Psi[A_R]\vert\hat{H}\vert\Psi[A_S]\rangle$,
where $N$ means number of grids. To reduce such expensive computational
costs, a smaller set of time-dependent DVR (TDDVR) grids is used instead of
the DVR grids. These time-dependent grids are obtained by solving the
eigen-value problem of the position operator in the time-dependent
basis of the SPFs. If the PES is also diagonal on the product of
these eigen-value grids, one can then use the TDDVR grids in place of
the primitive DVR grid $J$ (see also Equation \eqref{eq:mctdh-wf-999}).
However, the TDDVR grids result in inaccurate evaluation of
$\langle R\vert V\vert S\rangle$ making introduce of the correlated
DVR (CDVR) scheme necessary. By CDVR, an additional correction term
is added to the potential matrix elements that are arisen from the
TDDVR basis, implying that it is not possible to estimate the error
made in the integration.

%------------------------------------------------
\section{Theory of Quantum Scattering and Resonance\label{sec:scatt-res}}

\begin{figure}[h!]
   \centering
    \includegraphics[width=18cm]{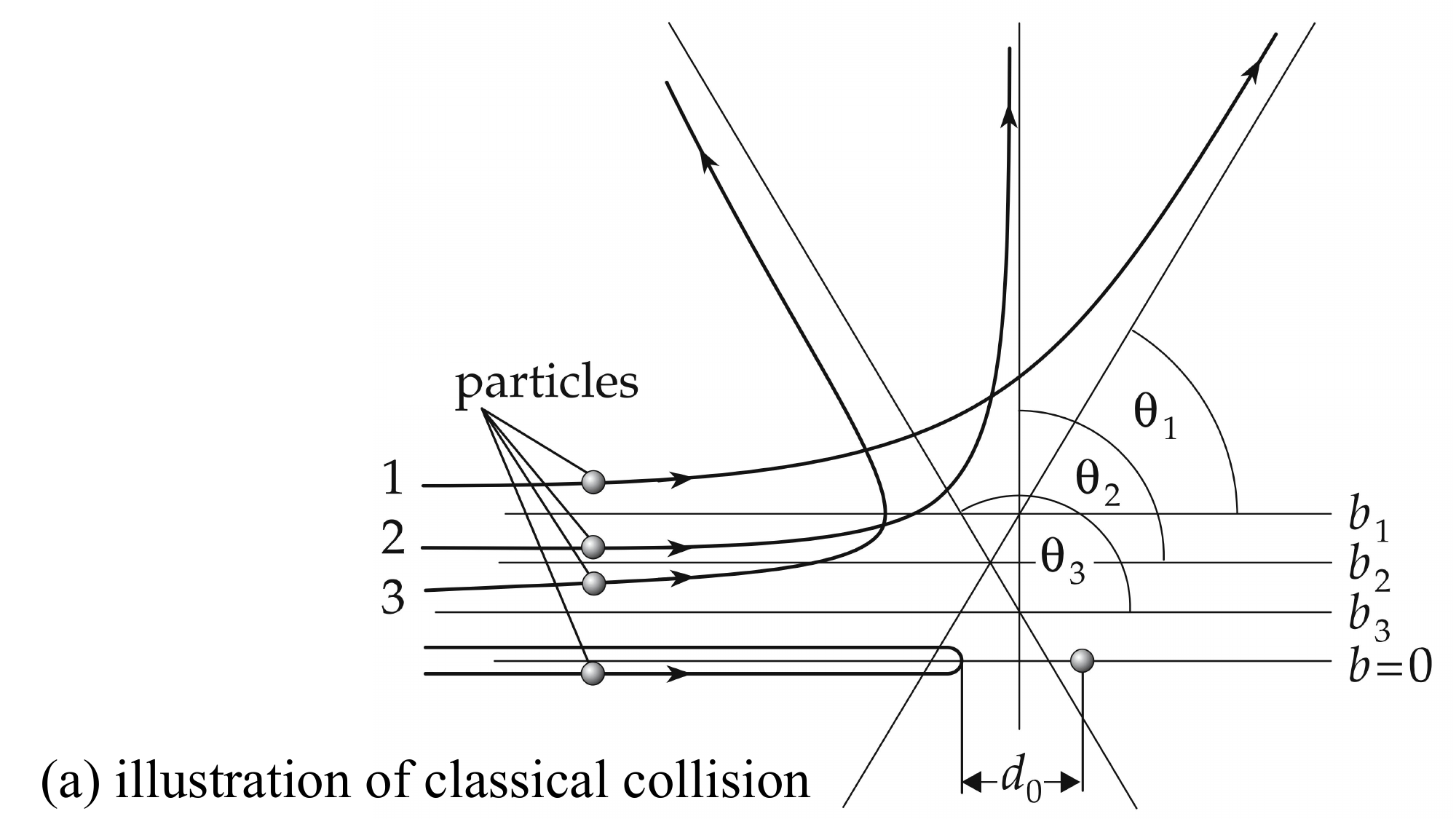}
     \includegraphics[width=18cm]{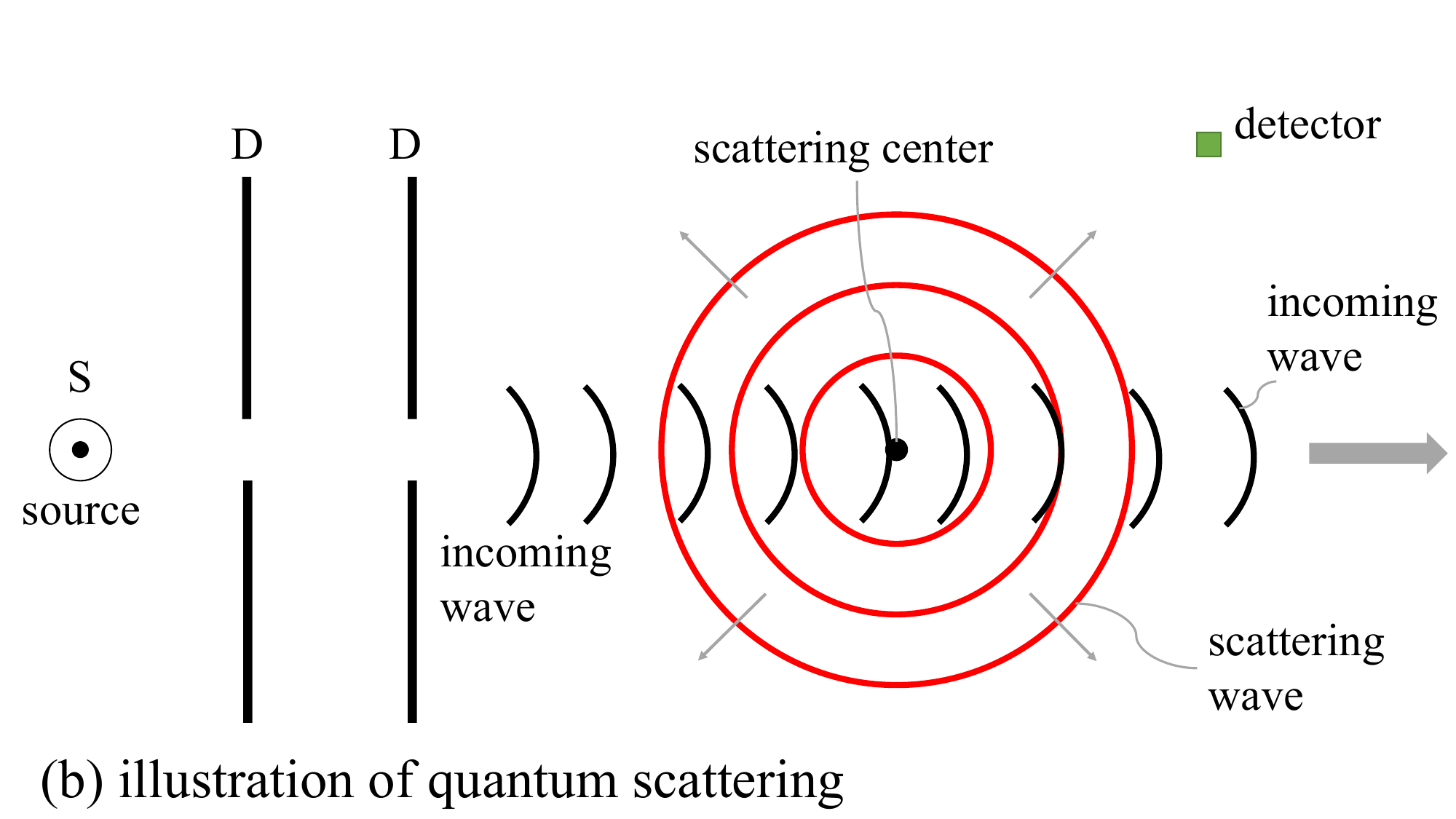}
      \caption{\figfoot}
       \label{fig:scatt-quan}
        \end{figure}

In this section, we turn to the theory of quantum scattering for
state-to-state molecular reaction. Illustrated by Figure \ref{fig:scatt-quan}
are particle trajectories in classical collision (panel (a)) and wave
functions in quantum scattering (panel (b)). Since we are limited to
quantum molecular dynamics, only quantum scattering is considered.
Unlike wavepacket propagation, quantum scattering mainly focuses on
state-to-state reaction and characterizes reactive scattering by the
scattering matrix, which maps incident asymptotic states to outgoing
states, enabling the calculation of reaction probabilities, differential
cross sections (DCSs), and scattering resonances. It is important to
note that, wavepacket propagation and quantum scattering correspond,
respectively, to a time-resolved and a space-resolved perspective on
molecular reaction dynamics. In wavepacket propagation, analyzing the
time-dependent wave function leads to time-resolved properties. Then,
energy-resolved quantities, such as spectra and reaction probabilities,
are extracted by means of Fourier transformation. In contrast, quantum
scattering provides space-resolved properties through the analysis of
the scattering matrix and DCSs, yielding quantities such as the spatial
distribution of quantum states of the products. Section \ref{sec:eigen-state}
describes eigen-state problem and approaches for its solutions which
are asymptotic states. Furthermore, both incoming wavepacket and
scattering wavepacket (see Figure \ref{fig:scatt-quan}(b)) will be
discussed through localizations of the Laplacian eigen-functions.
Section \ref{sec:pot-scatt} describes theory of potential scattering
and reactive scattering, where the PES term is introduced. Section
\ref{sec:sta-2-sta-reax} further describes theory of inverse problem.
Finally, Sections \ref{sec:fesh-res} and \ref{sec:odd-res-hyper}
describe theory of resonance in geometric scattering.

%--------------------------------------------
\subsection{Nuclear Eigen-State Problem\label{sec:eigen-state}}

Theory of quantum scattering predicts predicts transformation between
the reactant and product eigen-states, rather than time-dependent propertiers
during molecular collision. In this section, we consider only single
scattering process, for which the eigenstates can be regarded as stable
and unique. Because quantum scattering is essentially transformation
between the ``in'' and ``out'' quantum states, we should firstly consider
the eigen-state problem. Especially, because the Hamiltonian operator
is herimitian, we focus on the symmetric eigen-state problem which
seeks real eigenvalues and orthogonal eigenvectors. Before removing the
overall translational motion of the molecule, we must give the general
properties of the KEO term and the eigen-states of the KEO.
\begin{prop}
The negative Laplacian operator $-\nabla^2$ with the Sobolev space
$\mathcal{H}^2(\mathbb{R}^d)$ as domain is a self-adjoint operator on
the $L^2$ space and has spectrum $\sigma=[0,\infty)$, where there is
no point spectrum.
\label{prop:scattering-009}
\end{prop}
According to Proposition \ref{prop:scattering-009}, the Fourier transform
can be used to describe the spectral decomposition of $-\nabla^2$ because
$\cos(\lambda X)$ and $\sin(\lambda X)$ are the eigenfunctions of
$-\nabla^2$ for the eigenvalue $\lambda^2\in(0,\infty)$. This implies
that eigen-states of $-\nabla^2$ can be given by the plane wave
$\exp(\mi\lambda X)$. Because plane waves form a complete basis that
diagonalizes the Laplacian operator, any free-particle wave function can
be expressed as a superposition of plane waves. Such wave function is
either a delocalized wave (namely a single plane wave) or a localized
wave (namely a Fourier integral). We will return to the eigen-state
of the Laplacian operator after describing the eigen-states for the
inner DOFs. 

Removing the overall translational motion of the molecule, its internal
vibrations and rotations are bound states. Associated wave functions have
been well known for a long time as localized function. Quantitatively,
a function $u$ defined on a domain $\Omega\subset\mathbb{R}^d$ is called
$L^p$-localized for $p\geq1$ if there exists a bounded subset
$\Omega_0\subset\Omega$ which supports most of the norm of $u$,
\begin{equation}
\frac{\Vert u\Vert_{L^p(\Omega\setminus\Omega_0)}}{\Vert u\Vert_{L^p(\Omega)}}
\ll1,\quad\mathrm{and}\quad\frac{\mu(\Omega_0)}{\mu(\Omega)}\ll1,
\label{eq:local-eigen-00-func}
\end{equation}
where $\mu(\cdot)$ is characteristic length. Qualitatively, a localized
function essentially lives on a small subset of the domain and takes
small values on the remaining part.
Now, we give existence condition of bounded eigen-states which have
negative eigen-values.
\begin{prop}
Assuming there exists $\alpha>0$ and $X_0\in\mathbb{R}^d$ so that
$\int_{\mathbb{R}^d}V(X)\exp(-2\alpha^2(X-X_0)^2)\mathrm{d}^dX<-\alpha\sqrt{\pi/2}$,
then the Hamiltonian $-\nabla^2/2+V(X)$ has a negative eigenvalue.
Moreover, if $V(X)$ is locally square integrable satisfying
$\lim_{\vert X'\vert\to\infty}\int_{X'}^{X'+1}\vert V(X)\vert^2\mathrm{d}^dX=0$
and $\int_{\mathbb{R}^d}\vert X\vert\cdot\vert V(X)\vert\mathrm{d}^dX<\infty$,
then the number of bounded states must not be larger than
$2+\int_{\mathbb{R}^d}\vert X\vert\cdot\vert V(X)\vert\mathrm{d}^dX$.
\label{prop:scattering-109}
\end{prop}
According to Proposition \ref{prop:scattering-109}, one can easily
ensure whether the incoming and outgoing molecules are stable, that is
whether additional dissociation takes place either before or after the
scattering. The simplest example of a bound quantum state in a potential
is the eigen-states of quantum harmonic oscillator, as well described
by any textbook. All these eigen-states are localized in a region around
the minimum of the harmonic potential $m\omega^2X^2/2$ and rapidly decay
outside this region, as expected by Proposition \ref{prop:scattering-109}.
The presence of a strong potential prevents the moleucle from traveling
far from the origin, the size of the localization region being $(m\omega)^{-1/2}$.
This strong localization has been thoroughly investigated in many cases.
In general, a ro-vibrational eigen-state is trapped when it cannot
squeeze outside the localization region through either narrow channels
or branches of the waveguide. This happens if typical spatial variations
of the eigen-state are larger than the size of the narrow part. Thus,
there exists a threshold value $\mu$ which may eventually be zero, such
that the eigen-states with eigen-energy smaller than $\mu$ are localized.
Obviously, $\mu$ can be seen as the upper energy limit for a molecule
to exist stably. It is worth noting that the existence of trapped or
localized eigen-states in bounded domains has attracted less attention.
Even the definition of localization in bounded domains remains
conventional.

Similar to the eigenvalue problem of electronic structure, the eigen-states
of the molecule in the BF coordinate frame must be computed through a given
potential function. Solving such problem is usually called vibrational
structure theory. To do this, the variational principle plays an important
role in computing ro-vibrational structure. Most of such methods
variationally minimises a trial nuclear wave function composed of
Hartree products of one-dimensional vibrational functions, as given
in Equation \eqref{eq:def-state-spa-geom-000}. Over the past
several decades, Bowman, Gerber, Christiansen, and their co-workers
\cite{car98:191,bow03:533,bow06:33,wu06:285,bow07:7317,wan08:071101}
proposed the vibrational self-consistent field (VSCF) method which is
the vibrational-structure counterpart of the HF method. Since then,
Gerber \cite{nor96:11261}, Christiansen \cite{chr03:5773}, Hirata
\cite{yag07:681}, and Hirao \cite{yag01:10647,mil04:5036,oya06:064311}
proposed the vibrational M{\o}ller–Plesset (VMP) method which uses the
Rayleigh–Schr{\"o}dinger perturbation theory (RSPT) with the VSCF wave
function as a reference. Later, Christiansen and co-workers
\cite{chr04:2149,chr05:667,chr05:194105,sei07:204101,sei11:054119}
proposed the vibrational coupled-cluster (VCC) and vibrational
configuration-interaction (VCI) approaches based on the VSCF wave
function again as reference. The VCC method 
\cite{chr04:2149,chr05:667,chr05:194105,sei07:204101,sei11:054119}
is one of the most important post-VSCF methods. 
Neff and Rauhut \cite{nef09:124129,nef14:100} proposed state-specific configuration
selected VCI (ss-VCI) for fast evaluation of state energies. The
multi-configurational treatment of VSCF (namely VMCSCF) was developed
firstly by Culot and Li{\'e}vin \cite{cul94:227}. Later, Heislbetz and
Rauhut \cite{hei10:124102} implemented the VMCSCF method successfully
and showed its accuracy. Moreover, the vibrational multi-reference
coupled cluster (VMRCC) method was developed by Prasad and co-workers
\cite{ban12:114108}. Finally, Christiansen \cite{chr04:2140} proposed
a second-quantization formalism for calculating and analyzing multi-mode
vibrations of the molecular system, where the the nuclear wave function
and Hamiltonian operator are allowed to be rewritten in terms of the
same set of creation and annihilation operators. Similar to the electronic
structure methods, there is size-consistency problem affecting all of
vibrational-structure methods. For instance, VSCF lacks size consistency making the accuracy
VSCF differ for systems of different sizes. Further, any correlation
method based on the VSCF wave function also lacks size consistency.
Another problem of the above methods is the representation of the PES,
because Equation \eqref{eq:def-state-spa-geom-000} is employed as ansatz.
Solutions of this problem have been discussed in Section \ref{sec:consistency},
even though these solutions were designed for wavepacket propagation.

Now, we should turn to the eigen-states of the Laplacian operator to
show its localization, in particular the exponential decay of these
eigen-states in a bounded domain. We would like to show here that, if
an eigenvalue is smaller than a certain threshold defined by the
cross-sections of the domain, then the corresponding eigen-state decays
exponentially in the tail region of the domain. Let us consider
a bound domain $\Omega\subset\mathbb{R}^d$ with a piecewise smooth bound
$\partial\Omega$ and denote $Q(z)=\Omega\cap\{X\in\mathbb{R}^d\vert
X_1=z\}$ the cross section of $\Omega$ at $X_1=z\in\mathbb{R}$ by a
hyperplane perpendicular to the coordinate axis $X_1$. Letting $z_1=
\inf\{z\in\mathbb{R}\vert Q(z)\neq\emptyset\}$ and $z_2=\sup\{z\in
\mathbb{R}\vert Q(z)\neq\emptyset\}$, we fix $z_0$ such that
$z_1<z_0<z_2$ and set $\mu(z)$ be the first eigen-value of the
Laplacian operator in $Q(z)$, with Dirichlet boundary condition on
$\partial Q(z)$ and $\mu=\inf_{z\in(z_0,z_2)}\mu(z)$.
A Dirichlet boundary condition specifies the value of a function at the
boundary of a domain, while a Neumann boundary condition specifies the
slope of a function at the boundary of a domain. Now, letting $u$ be a
eigen-function in $\Omega$ and $\lambda$ the associated eigen-value, if
$\lambda<\mu$ then
\begin{equation}
\Vert u\Vert_{L^2(Q(z))}\leq\Vert u\Vert_{L^2(Q(z_0))}
\exp\Big(-\beta\sqrt{\mu-\lambda}(z-z_0)\Big),\quad
z\geq z_0,
\label{eq:vari-eigen-000-func}
\end{equation}
where $\beta=1/\sqrt{2}$. If $(e_1\cdot n(X))\neq0$ for
$\forall X\in\partial\Omega$ with $X_1>z_0$, then Equation
\eqref{eq:vari-eigen-000-func} holds with $\beta=1$, where $e_1$ is
the unit vector along axis $X_1$ and $n(X)$ is the normal vector at
$X\in\partial\Omega$ directed outwards from the domain. Here, a domain
$\Omega$ is arbitrarily split into two subdomains, $\Omega_1$ with
$X_1<z_0$ and $\Omega_2$ with $X_1>z_0$, by the hyperplane at $X_1=z_0$.
The coordiante axis $X_1$ can be replaced by any straight line. Under
the condition $\lambda<\mu$, the eigen-function $u$ exponentially
decays in the subdomain $\Omega_2$, which is loosely called branch.
It worht noting that choice of $z_0$ determines the threshold $\mu$.
Accoring to Equation \eqref{eq:vari-eigen-000-func}, the norm of an
eigen-function with $\lambda<\mu$ in $\Omega(z)=\Omega\cap\{X\in
\mathbb{R}^d\vert X_1>z\}$ can be made exponentially small providing
that the branch $\Omega_2$ is long enough. Taking $\Omega_0=\Omega\setminus\Omega(z)$,
the first ratio of Equation \eqref{eq:local-eigen-00-func} can be made
arbitrary small, but the second ratio of Equation \eqref{eq:local-eigen-00-func}
may not necessarily be small. Such smallness depends on the domain
$\Omega$. Once again, this is a manifestation of the conventional
character of localization in bounded domains. 

Next, we should turn to high-frequency localization, in particular the
so-called focusing modes, all of which become more and more localized
in a small subdomain when the associated eigenvalue increases. Such
high-frequency or high-energy limit is known as the semi-classical
approximation, by which quantum system can be reduced to the rays in
classical billiards. Jakobson, Nadirashvili, and Toth \cite{jak01:1085},
together with Grebenkov and Nguyen \cite{gre13:601,ngu13:780} gave an overview of
many geometric properties of the Laplacian eigenfunctions on Riemannian
manifolds, with special emphasis on the high-frequency limit \cite{jak01:1085}
and their localizations \cite{gre13:601,ngu13:780}. To begin simply, noting that
spherical coordinates $\{r,\theta,\phi\}$ are suitable to represent
scattering processes in 3D space without any externel potential term,
the Laplacian operator of such 3D domain
$\Omega=\{X\in\mathbb{R}^3\vert R_0<\vert X\vert<R\}$
has explicit representation of eigen-functions, \cite{gre13:601,ngu13:780}
\begin{equation}
\begin{array}{l}
u_{nkl}\big(r,\theta,\phi\big)=\left[j_n\left(\dfrac{\alpha_{nk}r}{R}\right)+
c_{nk}y_n\left(\dfrac{\alpha_{nk}r}{R}\right)\right]
P_n^l\big(\cos\theta\big)\exp\big(\mi l\phi\big),\\
j_n(z)=\sqrt{\dfrac{\pi}{2z}}J_{n+1/2}(z),\quad
y_n(z)=\sqrt{\dfrac{\pi}{2z}}Y_{n+1/2}(z),
\end{array}
\label{eq:lap-spher-coor-eigen-990}
\end{equation}
where $j_n(\cdot)$ and $y_n(\cdot)$ are spherical Bessel functions of
the first and second kind, respectively, while $J_n(\cdot)$ and
$Y_n(\cdot)$ are the Bessel functions of the first and second kind.
In Equation \eqref{eq:lap-spher-coor-eigen-990}, $P_n^l(\cdot)$ are
associated Legendre polynomials, while spherical harmonic functions
are $Y_{nl}(\theta,\phi)\propto P_n^l(\cos\theta)\exp(\mi l\phi)$. The
eigen-functions of free particle in Equation \eqref{eq:lap-spher-coor-eigen-990}
are characterized by three quantum numbers $\{n,k,l\}$, with $n=0,1,2,\cdots$
counting the order of spherical Bessel functions, $k=1,2,\cdots$ counting
zeros (or called nodes), and $l=-n,-n+1,\cdots,n$. The coefficients
$\alpha_{nk}$ and $c_{nk}$ are set by the boundary conditions at $r=R$
and $r=R_0$. 

\begin{figure}[h!]
   \centering    
    \subfigure[\quad Norm of the forcusing modes]{
     \includegraphics[width=18cm]{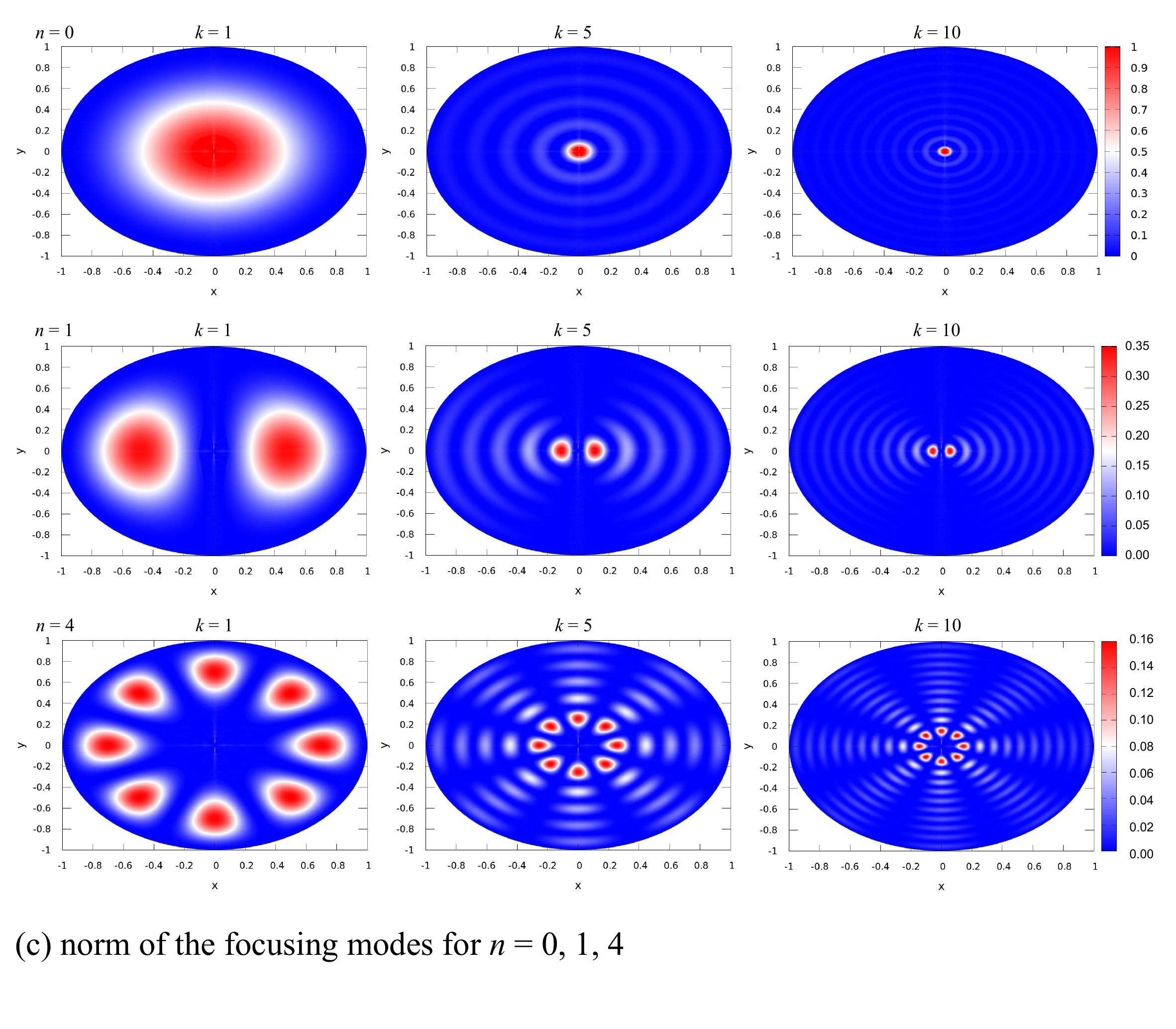}}
      \caption{\figfoot}
       \label{fig:modes-high-freq}
        \end{figure}

Since the crossed molecular beam experiment and the quantum scattering
between molecules occur in a plane, it is necessary to consider the 2D
cases. For the 2D case with $R_0=0$ and $\theta=\pi/2$, the 3D spherical
coordinates become 2D polar coordinates $\{r,\phi\}$ and Equation
\eqref{eq:lap-spher-coor-eigen-990} is re-written as \cite{gre13:601,ngu13:780}
\begin{equation}
u_{nkl}(r,\phi)=\left\{\begin{array}{ll}
J_n\left(\dfrac{\alpha_{nk}r}{R}\right)\cos(n\phi),&l=1 \\
J_n\left(\dfrac{\alpha_{nk}r}{R}\right)\sin(n\phi),&l=2,\quad n\neq0
\end{array}\right.,
\label{eq:lap-spher-coor-eigen-991}
\end{equation}
where $\alpha_{nk}$ are either the positive roots of the Bessel function
$J_n(z)$, or positive roots of its derivative $J_n'(z)$, or the positive
roots of $J_n'(z)+J_n(z)$. When $k$ is fixed and $n$ increases, the
Bessel functions $J_n(\alpha_{nk}r/R)$ are strongly attenuated near
the origin and essentially localized near the boundary, yielding
whispering gallery modes \cite{gre13:601,ngu13:780}. In contrast, when
$n$ is fixed and $k$ increases, the Bessel functions rapidly oscillate
with the amplitude decreasing towards the boundary making the eigen-functions
mainly localized at the origin. This yields focusing modes \cite{gre13:601,ngu13:780}.
Figure \ref{fig:modes-high-freq} illustrates whispering gallery modes
and focusing modes in the unit disk with Dirichlet boundary condition.
Moreover, Grebenkov and Nguyen \cite{gre13:601,ngu13:780} proved that
the eigen-wave-function satisfied Equations \eqref{eq:local-eigen-00-func}
requires $R\to1$ to preserve the localization of free particle
eigen-functions. Simply, this is because the oscillation amplitude of
the focusing modes exhibits a power law decay from the origin toward
the boundary \cite{gre13:601,ngu13:780}, as shown by Figure
\ref{fig:modes-high-freq}. As will be discussed later, the scattering wave
(see Figure \ref{fig:scatt-quan}(b)) in quantum scattering theory for a
central potential can be expanded in partial waves with different angular
quantum numbers, which is analogous to the form of Equation \eqref{eq:lap-spher-coor-eigen-990}.
Consequently, the squared modulus of the eigenfunction shown in Figure
\ref{fig:modes-high-freq}(c) closely resembles the experimental
observations, in particular the case with $n=1$. We are interested in
finding that the localized eigen-states of free particle have already
show many feature of DCSs, implying that semi-classical approximation
plays a fundamental role in theory of quantum scattering.
\begin{remark}
The scattering process is closely associated with both terms of the
Hamiltonian operator. The KEO term accounts for the overall free-particle
character of the scattering wave, whereas the PES term as well as
scattering conditions contribute the finer details.
\label{rem:scatter-00}
\end{remark}

%---------------------------------------------
\subsection{Potential Scattering\label{sec:pot-scatt}}

In this section, we consider the potential scattering and introduce the
concept of scattering matrix, where the potential term exists in the
scattering region in which the impacting molecules interact each other.
Since the Hamiltonian operator $\hat{H}$ is a self-adjoint operator on
the $L^2$ space, the Schr{\"o}dinger equation has a solution
$\Psi(t)=\exp(-\mi t\hat{H})\Psi(0)$ that satisfies $\Vert\Psi(t)\Vert
=\Vert\Psi(0)\Vert$. Similar statement can be given for the Hamiltonian
operator $\hat{H}_0$ of free particle. The potential scattering involves
potential-free
asymptotic states and intermediate states that experience a potential.
The existence of an incoming asymptotic state $\Psi^{(-)}(t)=\exp(-\mi
t\hat{H}_0)\Psi^{(-)}(0)$ for the solution $\Psi(t)$ requires limit
$\lim_{t\to-\infty}\Vert\Psi(t)-\Psi^{(-)}(t)\Vert=0$. Similarly, the
existence of an outgoing asymptotic state
$\Psi^{(+)}(t)=\exp(-\mi t\hat{H}_0)\Psi^{(+)}(0)$ requires limit
$\lim_{t\to+\infty}\Vert\Psi(t)-\Psi^{(+)}(t)\Vert=0$.
\begin{defi}
If $\Psi$ has both incoming and outgoing asymptotic states, then it is
called a scattering state.
\label{def:wave-oper-limi-8}
\end{defi}
To consider the relations between $\Psi(0)$ and $\Psi^{(\pm)}(0)$,
we define the M{\o}ller operator $\hat{\Omega}(t)=\exp(\mi t\hat{H})
\exp(-\mi t\hat{H}_0)$. If $\Psi$ has both the incoming and outgoing
states, then we have $\lim_{t\to\pm\infty}\hat{\Omega}(t)\Psi^{(\pm)}(0)=\Psi(0)$.
By the M{\o}ller operator, one can give the concept of wave operators.
\begin{defi}
The wave operators $\hat{\Omega}_{\pm}$ are defined as the strong
limits of $\hat{\Omega}(t)$ as $t\to\pm\infty$, such that
$\forall\Psi\in\{f\in\mathbb{H}\vert-\infty<\lim_{t\to\pm\infty}\hat{\Omega}(t)f<+\infty\}$
there exists $\hat{\Omega}_{\pm}\Psi=\lim_{t\to\pm\infty}\hat{\Omega}(t)\Psi$.
\label{def:wave-oper-limi-9}
\end{defi}
Now, we should consider the existences and the formulations of the
incoming and outgoing asymptotic states.
\begin{prop}
Assuming that $\hat{H}$ has an eigen-value $\lambda$ for the eigen-state
$\Psi^{(\pm)}$, the solution $\Psi(t)=\exp(-\mi\lambda t)\Psi^{(\pm)}$
has no incoming or outgoing asymptotic states unless $\Psi^{(\pm)}$ is
an eigen-state for $\hat{H}_0$ for the same eigen-value $\lambda$.
\label{prop:pot-scatt-asym-00}
\end{prop}
Proposition \ref{prop:pot-scatt-asym-00} plays a fundamental role in
considering existence of the scattering state.
To understand Proposition \ref{prop:pot-scatt-asym-00}, assuming
$\Psi(t)=\exp(-\mi\lambda t)\Psi^{(-)}$ has an incoming asymptotic
state, there exists a $\Phi^{(-)}$ such that
$\hat{\Omega}_-\Phi^{(-)}=\Psi^{(-)}$ or
$\lim_{t\to-\infty}\hat{\Omega}(t)\Phi^{(-)}=\Psi^{(-)}=\Psi(0)$. As
$t\to-\infty$, this means
$\Vert\exp(-\mi t\hat{H}_0)\Phi^{(-)}-\exp(-\mi t\hat{H})\Psi^{(-)}\Vert
=\Vert\exp(-\mi t\hat{H}_0)\Phi^{(-)}-\exp(-\mi\lambda t)\Psi^{(-)}\Vert
\to0$ which gives $\Vert\Phi^{(-)}-\exp(-\mi t(\lambda-\hat{H}_0))
\Psi^{(-)}\Vert\to0$ as $t\to-\infty$. For fixed $s\in\mathbb{R}$, as
$t\to-\infty$ we have $\Vert\exp(-\mi(t+s)(\lambda-\hat{H}_0))\Psi^{(-)}
-\exp(-\mi t(\lambda-\hat{H}_0))\Psi^{(-)}\Vert\leq\Vert\exp(-\mi(t+s)
(\lambda-\hat{H}_0))\Psi^{(-)}-\Phi^{(-)}\Vert+\Vert\Phi^{(-)}-
\exp(-\mi t(\lambda-\hat{H}_0))\Psi^{(-)}\Vert\to0$ and $\Vert\exp(-\mi
s(\lambda-\hat{H}_0))\Psi^{(-)}-\Psi^{(-)}\Vert\to0$. For arbitrary $s$,
it leads to $\exp(-\mi s(\lambda-\hat{H}_0))\Psi^{(-)}=\Psi^{(-)}$ and
thus $\hat{H}_0\Psi^{(-)}=\lambda\Psi^{(-)}$. A similar statement applies
to the outgoing states. Then, we finally have Proposition \ref{prop:pot-scatt-asym-00}.
For completeness, we directly give a note that the domain and range of
$\hat{\Omega}_{\pm}$ are closed subspace of $\mathbb{H}$.

According to Definition \ref{def:wave-oper-limi-8}, a scattering state
has an incoming state and an outgoing state. According to Proposition
\ref{prop:pot-scatt-asym-00}, their initial states satisfy
$\hat{\Omega}_+\Psi^{(+)}=\Psi(0)=\hat{\Omega}_-\Psi^{(-)}$. Since the
wave operator $\hat{\Omega}_-$ is an isometry from its domain to its
range, it is injective, and we can have expression
$\Psi^{(+)}=\hat{\Omega}_+^{-1}\hat{\Omega}_-\Psi^{(-)}=\hat{S}\Psi^{(-)}$.
\begin{defi}
The scattering operator $\hat{S}=\hat{\Omega}_+^{-1}\hat{\Omega}_-$
relates the incoming asymptotic state with the outgoing asymptotic state
for a scattering sate, that is $\Psi^{(+)}=\hat{S}\Psi^{(-)}$.
\label{def:scatt-oper-000}
\end{defi}
According to Definition \ref{def:scatt-oper-000}, the wave operators
$\hat{\Omega}_{\pm}$ describe possible unitary equivalences of the operators
$\hat{H}$ and $\hat{H}_0$, implying that $\hat{H}$ and $\hat{H}_0$ have
the same spectrum. If this is the case, knowing the wave operators
$\hat{\Omega}_{\pm}$ and the spectral decomposition of one of the
operators gives the spectral decomposition of the other operator.
However, the situation changes because $\hat{\Omega}_{\pm}$
may not be defined on the whole Hilbert space, or because the range of
$\hat{\Omega}_{\pm}$ may not be the whole Hilbert space. In the case
where $\hat{H}_0$ has trivial kernel $\mathrm{ker}(\hat{H}_0)=\{\Psi
\in\mathbb{H}\vert\hat{H}_0\Psi=0\}$, it can be proved that
$\hat{\Omega}_+$ and $\hat{\Omega}_-$ intertwine the self-adjoint
operators. In particular, in case $\hat{\Omega}_{\pm}$ are unitary,
$\hat{H}$ and $\hat{H}_0$ are unitarily equivalent and have completeness
of the wave operators for the Hamiltonian operator in the Hilbert space.
\begin{prop}
Assuming $V(X)\in L^1(\mathbb{R}^d)\cap L^2(\mathbb{R}^d)$ and letting
$\hat{H}_0$ be the Hamiltonian operator of free particle and $\hat{H}=\hat{H}_0+V(X)$,
the wave operators $\hat{\Omega}_{\pm}$ are complete in the Hilbert
space.
\label{prop:compl-wave-oper-00}
\end{prop}
Although the present discussions are physical self-consistency,
mathematical consistency necessitates an examination of the operator
domain.
\begin{prop}
Letting $\hat{H}_0$ be the Hamiltonian operator of free particle and
$\hat{H}=\hat{H}_0+V(X)$ be the Hamiltonian operator with
$\sup_{X'\in\mathbb{R}}\int_{X'}^{X'+1}\vert V(X)\vert^2\mathrm{d}X<\infty$,
if $X\mapsto(1+\vert X\vert)^{\alpha}V(X)$ is an element of $L^2$ for
some $\alpha>1/2$, then domains of $\hat{\Omega}_{\pm}$ are both equal
to $L^2$.
\label{prop:compl-wave-oper-01}
\end{prop}
Proposition \ref{prop:compl-wave-oper-01} holds true, if $V(X)$ is
locally square integrable and for some $\beta>1$ the function
$X\mapsto\vert X\vert^{\beta}V(X)$ is bounded for
$\vert X\vert\to\infty$. This condition is easily met in realistic
scattering.

With the above concepts, we solutions of scattering processes. If the
potential is of sufficient decay at infinity, the eigen-functions of
$\hat{H}$ can be expected to behave as plane wave $\exp(\pm\mi PX)$
for $X\to\pm\infty$, where $P$ is momentum of the plane wave. These
solutions are also known as the Jost solutions. According to Equation
\eqref{eq:path-int-004}, one can derive the Schr{\"o}dinger integral
equations as $t\to\pm\infty$ in the form
\begin{equation}
\Psi\big(X,t\to\pm\infty\big)=\Psi^{(\pm)}\big(X\big)
=\exp\big(\pm\mi PX\big)-\int_{X}^{\pm\infty}
\dfrac{\sin\big(P(X-X')\big)}{P}V\big(X'\big)\Psi^{(\pm)}
\big(X'\big)\mathrm{d}X',
\label{eq:int-sch-eq-00}
\end{equation}
where the sine function is adopted for simple and can be replaced by
plane wave. To understand Equation \eqref{eq:int-sch-eq-00}, it is easy
found that derivatives of $\Psi^{(\pm)}$ satisfy
\begin{align}
\dfrac{\partial}{\partial X}\Psi^{(\pm)}\big(X\big)&=\pm\mi P
\exp\big(\pm\mi PX\big)-\int_{X}^{\pm\infty}
\cos\big(P(X-X')\big)V\big(X'\big)\Psi^{(\pm)}\big(X'\big)\mathrm{d}X',
\label{eq:int-sch-eq-01} \allowdisplaybreaks[4] \\
\dfrac{\partial^2}{\partial X^2}\Psi^{(\pm)}\big(X\big)&
=-P^2\Psi^{(\pm)}\big(X\big)+V\big(X\big)\Psi^{(\pm)}\big(X\big).
\label{eq:int-sch-eq-02}
\end{align}
Obviously, Equation \eqref{eq:int-sch-eq-02} is the eigen-value equation
of the Hamiltonian operator and predicts solutions $\Psi^{(\pm)}_P$.
Nevertheless, Equation \eqref{eq:int-sch-eq-00} is only formal equation for
$\Psi^{(\pm)}_P$ because the integral depends on solutions $\Psi^{(\pm)}_P$
themselves. Hence, it must be recursively substituted into itself an infinite
number of times to arrive at a self-consistent equation for
$\Psi^{(\pm)}_P$. Usually, the resulting integral equation is known as
the Lippmann-Schwinger equation. It can be proved that solutions of the
Lippmann-Schwinger equation are unique and satisfy the estimates as
$X\to\pm\infty$,
\begin{align}
\Big\vert\Psi^{(\pm)}_P\big(X\big)-\exp\big(\pm\mi PX\big)\Big\vert&
\leq\exp\big(\mp XP\big)\left\vert\exp\left(\pm\frac{1}{P}
\int_X^{\pm\infty}\big\vert V\big(X'\big)\big\vert\mathrm{d}X'
\right)-1\right\vert,
\label{eq:int-sch-eq-03} \allowdisplaybreaks[4] \\
\left\vert\frac{\partial}{\partial X}\Psi^{(\pm)}_P\big(X\big)\mp\mi
P\exp\big(\pm\mi PX\big)\right\vert&\leq
P\exp\big(\mp XP\big)
\left\vert\exp\left(\pm\frac{1}{P}\int_X^{\pm\infty}\big\vert
V\big(X'\big)\big\vert\mathrm{d}X'\right)-1\right\vert,
\label{eq:int-sch-eq-04}
\end{align}
where $P\geq0$. The solutions in Equations \eqref{eq:int-sch-eq-03} and
\eqref{eq:int-sch-eq-04} are also known as Jost solutions. By Equations
\eqref{eq:int-sch-eq-03} and \eqref{eq:int-sch-eq-04}, we can directly
find $\Psi^{(\pm)}=\exp(\mp\mi PX)\Psi_P^{(\pm)}(X)$. Furthermore, the
estimates in Equations \eqref{eq:int-sch-eq-03} and \eqref{eq:int-sch-eq-04}
can be improved by imposing more conditions on the potential term, such
as the case that allows $P=0$ and the case that the potential decays
even faster.

Next, we give the transmission and reflection coefficients which produce
dynamics results of quantum scattering. For molecular scattering, we
assume that the potential is a real-valued integrable functin and the
eigen-values of
the Hamiltonian operator is related to $P^2$ with $P\in\mathbb{R}^d
\setminus\{0\}$. It is worth noting that solutions of the eigen-value
equation of the Hamiltonian operator are $\Psi_{\pm P}^{(\pm)}$. To
describe these solutions, we consider the Wronskian
$W(f,g)(X)=f(X)g'(X)-f'(X)g(X)$, which is non-zero if and only if
$f(X)$ and $g(X)$ are linearly independent.
\begin{prop}
The Wronskian of two eigen-states of $\hat{H}$ is constant, and in
particular $W(\Psi_P^{(\pm)},\Psi_{-P}^{(\pm)})=\mp2\mi P$ for
$P\in\mathbb{R}^d\setminus\{0\}$.
\label{prop:wronki-con-999}
\end{prop}
Proposition \ref{prop:wronki-con-999} can be directly proved by
differentiating the Wronskian $W(\Psi_P^{(\pm)},\Psi_{-P}^{(\pm)})$.
According to Proposition \ref{prop:wronki-con-999}, the eigen-states
$\Psi_P^{(\pm)}$ and $\Psi_{-P}^{(\pm)}$ are linearly independent. Now,
since we have four eigen-states $\Psi_{\pm P}^{(\pm)}$, two by two
linearly independent for $P\neq0$, to the Schr{\"o}dinger eigen-value
equation, one can have coefficients $a_P^{(\pm)}$ and $b_P^{(\pm)}$
such that $\Psi_P^{(\pm)}\big(X\big)=a_P^{(\mp)}\Psi_P^{(\mp)}\big(X\big)
+b_P^{(\mp)}\Psi_{-P}^{(\mp)}\big(X\big)$ for $\forall X\in\mathbb{R}^d$
and $P\in\mathbb{R}^d\setminus\{0\}$. The relation $\Psi_P^{(\pm)}{}^*
=\Psi_{-P}^{(\pm)}$ implies $a_P^{(\pm)}{}^*=a_{-P}^{(\pm)}$ and
$b_P^{(\pm)}{}^*=b_{-P}^{(\pm)}$. Using the Wronskian
$W(\Psi_P^{(\pm)},\Psi_{-P}^{(\pm)})=\mp2\mi P$ given by Proposition
\ref{prop:wronki-con-999}, we can further find
$\vert b_P^{(\pm)}\vert^2-\vert a_P^{(\pm)}\vert^2=1$. According to
Proposition \ref{prop:wronki-con-999} and relation $W(f,f)=0$, we can
obtain $2\mi Pa_P^{(\pm)}=\mp W(\Psi_P^{(\mp)},\Psi_{-P}^{(\pm)})$ and
$2\mi Pb_P^{(\pm)}=\pm W(\Psi_P^{(\mp)},\Psi_P^{(\pm)})$. The latter
relation implies $b_P^{(+)}=b_P^{(-)}$. 

According to Equation \eqref{eq:int-sch-eq-00}, the Jost solutions can
be represented by the plane wave. Thus, we define the boundary of the
solution $\Psi_P(X)$ with $P\in\mathbb{R}^d\setminus\{0\}$ as
\begin{equation}
\lim_{X\to-\infty}\Psi_P\big(X\big)=T\big(P\big)\exp\big(-\mi PX\big),
\quad
\lim_{X\to+\infty}\Psi_P\big(X\big)=\exp\big(-\mi PX\big)+R\big(P\big)
\exp\big(\mi PX\big).
\label{eq:int-sch-eq-06}
\end{equation}
The condition at $X\to-\infty$ determines $\Psi_P(X)$ up to a multiplicative
constant, which is determined by the requirement that the coefficient
of $\exp(-\mi PX)$ is 1 as $X\to+\infty$.
\begin{defi}
In Equation \eqref{eq:int-sch-eq-06}, the function $T(P)$ is the
transmission coefficient and the function $R(P)$ is the reflection
coefficient.
\label{def:trans-reflec-coeff}
\end{defi}
According to Proposition \ref{prop:wronki-con-999}, Equation
\eqref{eq:int-sch-eq-06} implies that $\Psi_P(X)=T(P)\Psi_P^{(-)}(X)=
\Psi_{-P}^{(+)}(X)+R(P)\Psi_P^{(+)}(X)$ and thus
\begin{equation}
T\big(P\big)=\frac{1}{b_P^{(+)}}=\frac{1}{b_P^{(-)}}=
\dfrac{2\mi P}{W\big(\Psi_P^{(-)},\Psi_P^{(+)}\big)},\quad
R\big(P\big)=\frac{a_P^{(+)}}{b_P^{(+)}}=\frac{a_P^{(+)}}{b_P^{(-)}}=
-\dfrac{W\big(\Psi_P^{(-)},\Psi_{-P}^{(+)}\big)}{W\big(\Psi_P^{(-)},
\Psi_P^{(+)}\big)}.
\label{eq:int-sch-eq-07}
\end{equation}
By Equation \eqref{eq:int-sch-eq-07}, we have relations
$T^*(P)=T(-P)$, $R^*(P)=R(-P)$, and $\vert T(P)\vert^2+\vert R(P)
\vert^2=1$, that can be interpreted as conservation of particle for
transmitted and reflected waves, as illustrated by Figure
\ref{fig:scatt-quan}(b). Equation \eqref{eq:int-sch-eq-07} indicates
that the reflection coefficient is related to the system travelling
from left to right, denoted by $R^{(+)}(P)$. We can equivalently define
the reflection coefficient $R^{(-)}(P)=a_P^{(-)}/b_P^{(-)}$ for
travelling from right to left. Noting that the transmission coefficient
does not depend on the travelling direction, according to Definition
\ref{def:scatt-oper-000}, the scattering matrix can be given by a
$2\times2$ unitary matrix in U(2),
\begin{equation}
S\big(P\big)=\left(\begin{array}{cc}
T(P) & R^{(-)}(P)  \\
R^{(+)}(P) & T(P)
\end{array}\right).
\label{eq:scatt-matrix-99}
\end{equation}
By Equation \eqref{eq:int-sch-eq-07} one can further find limitations
of $\lim_{P\to\pm\infty}T(P)=1$ and $\lim_{P\to\pm\infty}R^{(\pm)}(P)=0$.
Thus, as the momentum becomes sufficiently large, the scattering exhibits
only transmission and no reflection. It is finally noted that there are
two orientations in defining a scattering process making $S(P)$ in Equation
\eqref{eq:scatt-matrix-99} a $2\times2$ matrix.

Next, we consider the relationship between the potential
term and scattering matrix $S(P)$. To this end, let us consider the
transmission and reflection coefficients that are elements of $S(P)$
in Equation \eqref{eq:scatt-matrix-99}.
\begin{prop}
Assuming $V(X)$ is real-valued integrable and continuous function, the
transmission and reflection coefficients satisfy
\begin{equation}
R\big(P\big)=\frac{1}{2\mi P}\int_{\mathbb{R}}V\big(X\big)
\Psi_P\big(X\big)\exp\big(-\mi PX\big)\mathrm{d}X,\quad
T\big(P\big)=1+\frac{1}{2\mi P}\int_{\mathbb{R}}V\big(X\big)
\Psi_P\big(X\big)\exp\big(\mi PX\big)\mathrm{d}X,
\label{eq:pot-scatt-00-matri}
\end{equation}
and
\begin{equation}
\dfrac{1}{T\big(P\big)}=1-\frac{1}{2\mi P}\int_{\mathbb{R}}V
\big(X\big)\Psi_P^{(+)}\big(X\big)\exp\big(-\mi PX\big)\mathrm{d}X,
\label{eq:pot-scatt-01-matri}
\end{equation}
where $P\in\mathbb{R}^d\setminus\{0\}$.
\label{prop:pot-scatt-matri-991}
\end{prop}
By Equation \eqref{eq:pot-scatt-01-matri}, experimentally measuring
the outgoing state $\Psi_P^{(+)}$, one can precisely measure $T(P)$ in
experiments. To understand Proposition \ref{prop:pot-scatt-matri-991},
we consider the Hamiltonian operator $\hat{H}_0$ for free particle and
$\hat{H}=\hat{H}_0+V(X)$ and eigen-functions $f$ and $g$ of $\hat{H}_0$
and $\hat{H}$, respectively, associated with the eigen-value $P^2$.
The transmission coefficient (also called the transmission probability)
is the probability that an incident particle that travels from one side
will tunnel or pass over the scattering region and be transmitted to
the opposite side. The reflection coefficient (also called reflection
probability) is the probability that the incident particle will be
reflected back from the scattering region to the incident side. Thus,
both transmission and reflection coefficients depend on the potential,
as given by Proposition \ref{prop:pot-scatt-matri-991}.
\begin{remark}
Once the PES term is constructed, the transmission and reflection
coefficients follow directly from Proposition \ref{prop:pot-scatt-matri-991},
allowing us to ultimately determine the reaction probability across the
scattering region.
\label{rem:trans-refle-phys-9}
\end{remark}

Similar to the proof of Proposition \ref{prop:wronki-con-999} by directly
differentiating the Wronskian, due to $\hat{H}_0f=fP^2$ and $\hat{H}g=
\hat{H}_0g+Vg=gP^2$, differentiating the Wronskian $W(f,g)(X)$ gives
\begin{equation}
\frac{\partial}{\partial X}W\big(f,g\big)\big(X\big)=f\big(X\big)
\frac{\partial^2}{\partial X^2}g\big(X\big)-g\big(X\big)
\frac{\partial^2}{\partial X^2}f\big(X\big)
=V\big(X\big)f\big(X\big)g\big(X\big).
\label{eq:pot-scatt-02-matri}
\end{equation}
Integrating Equation \eqref{eq:pot-scatt-02-matri} with $f(x)=\exp(-\mi
PX)$ and $g(X)=\Psi_P(X)$ and taking $X\to\pm\infty$, we have
\begin{align}
\int_{\mathbb{R}}V\big(X\big)\Psi_P\big(X\big)\exp\big(-\mi PX\big)
\mathrm{d}X&=\lim_{X\to+\infty}
W\Big(\exp\big(-\mi PX\big),\Psi_P\Big)\big(X\big)-\lim_{X\to-\infty}
W\Big(\exp\big(-\mi PX\big),\Psi_P\Big)\big(X\big)
\allowdisplaybreaks[4] \nonumber \\
&=R\big(P\big)W\Big(\exp\big(-\mi PX\big),\exp\big(\mi PX\big)\Big)=
2\mi PR\big(P\big),
\label{eq:pot-scatt-03-matri}
\end{align}
where we use Proposition \ref{prop:wronki-con-999}. By Equation
\eqref{eq:pot-scatt-03-matri}, one can directly obtain the first
expression of Equation \eqref{eq:pot-scatt-00-matri}. Similarly,
integrating Equation \eqref{eq:pot-scatt-02-matri} with $f(x)=\exp(\mi
PX)$ and $g(X)=\Psi_P(X)$ and taking $X\to\pm\infty$, we have
\begin{align}
\int_{\mathbb{R}}V\big(X\big)\Psi_P\big(X\big)\exp\big(\mi PX\big)
\mathrm{d}X&=\lim_{X\to+\infty}
W\Big(\exp\big(\mi PX\big),\Psi_P\Big)\big(X\big)-\lim_{X\to-\infty}
W\Big(\exp\big(\mi PX\big),\Psi_P\Big)\big(X\big)
\allowdisplaybreaks[4] \nonumber \\
&=\Big(1-T\big(P\big)\Big)
W\Big(\exp\big(\mi PX\big),\exp\big(-\mi PX\big)\Big)=2\mi
P\Big(T\big(P\big)-1\Big).
\label{eq:pot-scatt-04-matri}
\end{align}
By Equation \eqref{eq:pot-scatt-04-matri}, one can directly obtain the
second expression of Equation \eqref{eq:pot-scatt-00-matri}. To find
Equation \eqref{eq:pot-scatt-01-matri}, one can turn to Equation
\eqref{eq:int-sch-eq-00} and consider relation between $\Psi_P^{(\pm)}(X)$
and $b_P^{(-)}$. Then, according to the first expression of Equation
\eqref{eq:int-sch-eq-07}, one can obtain Equation \eqref{eq:pot-scatt-01-matri}.
Thus, we can have Proposition \ref{prop:pot-scatt-matri-991}. Furthermore,
Equation \eqref{eq:pot-scatt-01-matri} has the asymptotic epxansion for
$P\to\pm\infty$ in the form
\begin{equation}
\dfrac{1}{T\big(P\big)}=1-\frac{1}{2\mi P}
\int_{\mathbb{R}}V\big(X\big)\mathrm{d}X+
\mathcal{O}\left(\frac{1}{P^2}\right).
\label{eq:pot-scatt-05-matri}
\end{equation}
Equation \eqref{eq:pot-scatt-05-matri} describes the simplest relation
between the transmission coefficient and the PES term.

%----------------------------------------
\subsection{Inverse Problem\label{sec:sta-2-sta-reax}}

According to Equation \eqref{eq:pot-scatt-05-matri}, it is possible to
obtain $V(X)$ by measuring $T(P)$. This is the inverse scattering problem,
which is core idea of experimental meansurement for potential interaction.
This is described by the Gelfand-Levitan-Marchenko integral equation.
For simple, we assume that $T(P)$ has no poles in the open upper
half plane, implying that the corresponding translational Hamiltonian
operator has no discrete spectrum. Before continuing with the derivation,
we expand the Jost solution by intergral
\begin{equation}
\Psi_P^{(+)}\big(X\big)=\exp\big(\mi PX\big)+
\int_X^{+\infty}A\big(X,X'\big)\exp\big(\mi PX'\big)\mathrm{d}X',\quad
A\big(X,X'\big)=\frac{1}{2}B\left(X,\frac{X'-X}{2}\right),
\label{eq:four-trans-jost-001}
\end{equation}
where the kernel $B(X,\cdot)\in L^2$ for each $X\in\mathbb{R}^d$ is a
unique real-valued solution of the integral equation
\begin{align}
B\big(X,X'\big)&=\int_{X+X'}^{+\infty}V\big(Y\big)\mathrm{d}Y+
\int_0^{X'}\int_{X+X'-Z}^{+\infty}V\big(Y\big)B\big(Y,Z\big)
\mathrm{d}Z,\quad X'\geq0, 
\allowdisplaybreaks[4] \nonumber \\
\big\vert B\big(X,X'\big)\big\vert&\leq
\exp\left(\int_X^{+\infty}\big(Y-X\big)\big\vert
V\big(Y\big)\big\vert\mathrm{d}Y\right)\int_{X+X'}^{+\infty}\big\vert
V\big(Y\big)\big\vert\mathrm{d}Y,
\label{eq:four-trans-jost-002}
\end{align}
with boundary conditions $B(X,0)=\int_X^{+\infty}V(Y)\mathrm{d}Y$ and
$\lim_{X\to\infty}\Vert B(X,\cdot)\Vert_{\infty}=0$. By Equation
\eqref{eq:four-trans-jost-002}, it was further proved that the PES term
satisfies
\begin{equation}
\frac{\partial}{\partial X}\left(\frac{\partial}{\partial X}B\big(X,X'\big)
-\frac{\partial}{\partial X'}B\big(X,X'\big)\right)=V\big(X\big)
B\big(X,X'\big).
\label{eq:four-trans-jost-992}
\end{equation}
Equation \eqref{eq:four-trans-jost-992} finally gives relation between
the PES term and the kernel of the Jost solution. It is thus possible
to measure the potential interaction by measuring the quantum state.
However, because direct observation of quantum state is difficult,
this remains a challenging task to accomplish directly.

To overcome this problem, because reflection process is observable, let
us consider the kernel of the reflection coefficient by assuming that
the reflection coefficient has a Fourier transform
\begin{equation}
R\big(P\big)=\int_{\mathbb{R}}r\big(X\big)\exp\big(\mi PX\big)
\mathrm{d}X.
\label{eq:four-trans-jost-003}
\end{equation}
According to Equations \eqref{eq:int-sch-eq-06} and \eqref{eq:four-trans-jost-001},
we have intergral expansion
\begin{align}
T\big(P\big)\Psi_P^{(-)}\big(X\big)={}&\Psi_{-P}^{(+)}\big(X\big)+
R\big(P\big)\Psi_P^{(+)}\big(X\big)
\allowdisplaybreaks[4] \nonumber \\
={}&\exp\big(-\mi PX\big)+\int_X^{+\infty}A\big(X,X'\big)
\exp\big(-\mi PX'\big)\mathrm{d}X'+\int_{\mathbb{R}}r\big(X'\big)
\exp\Big(\mi P\big(X+X'\big)\Big)\mathrm{d}X'
\allowdisplaybreaks[4] \nonumber \\
&+\int_{\mathbb{R}}r(Y)\exp\big(\mi PY\big)\mathrm{d}Y\int_X^{+\infty}
A\big(X,X'\big)\exp\big(\mi PX'\big)\mathrm{d}X'
\label{eq:four-trans-jost-004}
\end{align}
By the Paley-Wiener Theorem, together with Equations \eqref{eq:four-trans-jost-003}
and \eqref{eq:four-trans-jost-004}, one can have expression
\begin{equation}
A\big(X,-X'\big)+r\big(X'-X\big)+\int_X^{+\infty}A\big(X,Y\big)
r\big(X'-Y\big)\mathrm{d}Y=0,\quad
X'<-X,
\label{eq:four-trans-jost-005}
\end{equation}
which is a form of the Gelfand-Levitan-Marchenko equation. By Equations
\eqref{eq:four-trans-jost-992} and \eqref{eq:four-trans-jost-005}, one
can obtain the kernel $A(X,X')$ form $r(X)$ and then obtain $V(X)$. 
\begin{remark}
The Paley-Wiener theorem is the mathematical reason why finite spatial
or temporal support forces infinite and analytic frequency support. It
is a rigorous version of the uncertainty principle for Fourier pairs.
\label{remark:pw-th-uncer}
\end{remark}

In case the transmission coefficient has poles, the corresponding
translational Hamiltonian operator has discrete spectrum. The above
approach can be adapted by adding a finite sum to the kernel $r(X)$.
To do this, we consider linear transforms
\begin{equation}
K_c\big(X\big)=\frac{1}{\pi}\int_{\mathbb{R}}R\big(P\big)
\exp\big(2\mi PX\big)\mathrm{d}P\in L^2,\quad
K_d\big(P\big)=2\sum_{n=1}^N\rho_n\exp\big(-2P_nX\big),
\label{eq:four-trans-jost-006}
\end{equation}
where the transmission coefficient is meromorphic in the the open upper
half plane. There are a finite number $N$ of simple poles at $\{\mi P_n
\}_{n=1}^N$ with $P_n>0$, where $\rho_n$ is weight of the $n$-th eigen-state.
Since $K_c$ and $K_d$ correspond to the continuous and discrete spectra,
one can define $K=K_c+K_d$ and give the general form of the
Gelfand-Levitan-Marchenko integral equation.
\begin{prop}
The kernel $B(X,X')$ satisfies the Gelfand-Levitan-Marchenko integral
equation for fixed $X$,
\begin{equation}
K\big(X+X'\big)+\int_0^{+\infty}B\big(X,Y\big)K\big(X+X'+Y\big)
\mathrm{d}Y+B\big(X,X'\big)=0,\quad\forall X'\geq0,
\label{eq:four-trans-jost-007}
\end{equation}
which holds for $B(X,\cdot)\in L^2(0,+\infty)$. Moreover, the Gelfand-Levitan-Marchenko
equation has a unique solution $B(X,\cdot)\in L^2(0,+\infty)$.
\label{prop:glm-inte-equa}
\end{prop}
By Proposition \ref{prop:glm-inte-equa} together with Equation
\eqref{eq:four-trans-jost-992}, one can obtain uniquely the potential
from just the reflection coefficient and the bound states. This is
actually the only part of information related to the transmission
coefficient that is needed.

Another approach to measure the potential function from the quantum
scattering observables is the Fokas-Its Riemann-Hilbert approach, which
solves PDEs by converting them into a problem of finding an analytic
function with a prescribed jump across a curve. Sovling the first
line of Equation \eqref{eq:four-trans-jost-004} gives
\begin{equation}
\Psi_P^{(+)}\big(X\big)=\exp\big(\mi PX\big)+\frac{1}{2\pi\mi}
\int_{\mathbb{R}}\dfrac{R\big(P'\big)\Psi_{P'}^{(+)}\big(X\big)
\exp\big(\mi(P+P')X\big)}{P+P'}\mathrm{d}P',
\label{eq:fi-rie-hil-app-00}
\end{equation}
which determines the Jost solution from the reflection coefficient.
Now, to find the relation between $\Psi_P^{(+)}(X)$ and $V(X)$, we
write $\Psi_P^{(+)}(X)=\exp(\mi PX+u(X,P))$ with $\lim_{X\to+\infty}
u(X,P)=0$ and $\lim_{P\to\pm\infty}u(X,P)=0$ for $P\geq0$. The
Schr{\"o}dinger equation for $\Psi_P^{(+)}(X)$ translates to the
first-order differential equation for $\partial u/\partial X$,
namely $-2\mi P\partial u/\partial X=\partial^2u/\partial X^2+
(\partial u/\partial X)^2-V(X)$. Due to the behaviour of this equation
as $P\to\pm\infty$, one can expand $\partial u/\partial X$ as
$\partial u/\partial X=\sum_{n=1}^{+\infty}v_n(X)(2\mi P)^{-n}$
which gives $v_1(X)=V(X)$. This implies that $2\mi P
\partial u/\partial X=V(X)$ as $P\to\pm\infty$. Moreover, noting
$\exp(u(X,P))\sim1+u(X,P)$, we have finally obtained the PES term
\begin{equation}
V\big(X\big)=\lim_{P\to+\infty}2\mi P\frac{\partial}{\partial X}
\Big(\Psi_P^{(+)}\big(X\big)\exp\big(-\mi PX\big)-1\Big).
\label{eq:fi-rie-hil-app-01}
\end{equation}
By Equation \eqref{eq:fi-rie-hil-app-01}, we can recover the PES term
$V(X)$ from the Jost solution which can be obtained from the reflection
coefficient through Equation \eqref{eq:fi-rie-hil-app-00}.

Finally, before turning to theory of scattering resonance, we must give
some notes on the above theory of reactive scattering for chemical reaction.
we are interested in noting that the entire theory was constructed on
the basis of Jost solutions $\Psi^{(\pm)}_P$ of the Lippmann-Schwinger
equation, whose precise expression is the integral fromulation reformulated
from the time-independent Schr{\"o}dinger equation. In addition to
Equation \eqref{eq:int-sch-eq-00}, the Lippmann-Schwinger equation is
usually written as a more compact expression, $\Psi^{(\pm)}(X)=
\exp(\pm\mi PX)+R_V^{(g)}(P)V(X)\Psi^{(\pm)}(X)$, where $R_V^{(g)}(P)$
is the resolvent discussed later in theory of scattering resonance.
Moreover, by Equation \eqref{eq:int-sch-eq-07}, the Lippmann-Schwinger
equation can be rewritten as $T(P)=V(X)+V(X)R_V^{(g)}(P)T(P)$, which
can be then adopted to solve the inverse problem. In addition to the
above theory for a single Hilbert space, it is worth noting that Chandler
and Gibson
\cite{cha73:1328,cha77:2336,cha78:1610,cha83:80,cha84:1841,cha89:1533,cha92:3477,cha94:1487,cha98:223}
systematically developed $N$-body theory in two Hilbert spaces for
multichannel quantum scattering with short-range potentials. Such theory
\cite{cha73:1328,cha77:2336,cha78:1610,cha83:80,cha84:1841,cha89:1533,cha92:3477,cha94:1487,cha98:223}
is characterized by the sextuple $\mathscr{S}=\{\mathbb{H}_N,\hat{H}_N,
\mathbb{H},\hat{H},\hat{J},\mathfrak{P}_a\}$, where $a$ is index of the
channels. In $\mathscr{S}$, operator $\hat{H}_N$ is the total $N$-body
Hamiltonian acting on the $N$-body Hilbert space $\mathbb{H}_N$, while
$\hat{H}$ is the asymptotic Hamiltonian acting on the asymptotic Hilbert
space $\mathbb{H}$ that has a decomposition $\mathbb{H}=\bigoplus_a\mathbb{H}_a$
of those associated with channels. Moreover, we define $\hat{J}$ as a
bounded identification operator from $\mathbb{H}$ to $\mathbb{H}_N$ and
define $\mathfrak{P}_a$ as the orthogonal projection of $\mathbb{H}$
onto $\mathbb{H}_a$. Obviously, state-to-state processes of a chemical
reaction are typical example. The basic difference between theory in two
Hilbert spaces and the conventional one is that the unkown transition
and resolvent act only on the channel subspaces, but the equations in
two Hilbert spaces are of the Lippmann-Schwinger type, differing only
in the appearance of certain injection operators from one Hilbert space
to the other. By $\mathscr{S}$, the wave operators $\hat{\Omega}_{\pm}:\mathbb{H}\to\mathbb{H}_N$
are given by strong limitations $\hat{\Omega}_{\pm}(\hat{H}_N,\hat{H},\hat{J}\mathfrak{P}_a)
=\lim_{t\to\pm\infty}\exp(\mi\hat{H}_Nt)\hat{J}\mathfrak{P}_a\exp(-\mi\hat{H}t)$.
This definition is equivalent to Definition \ref{def:wave-oper-limi-9}.
But, the M{\o}ller operator $\hat{\Omega}(t)=\exp(\mi t\hat{H})\exp(-\mi t\hat{H}_0)$
is replaced by operators in $\mathscr{S}$. Then, according to Definition
\ref{def:scatt-oper-000}, the scattering operator $\hat{S}$ can be defined.
The sextuple $\mathscr{S}$ is referred to the exact scattering system.
Similar to the derivations given in Section \ref{sec:pot-scatt}, one
can derive the integral equation for scattering state in the Lippmann-Schwinger
type
\cite{cha73:1328,cha77:2336,cha78:1610,cha83:80,cha84:1841,cha89:1533,cha92:3477,cha94:1487,cha98:223}.
According to Tobocman \cite{tob83:1405}, the scatteirng theory in two
Hilbert spaces selects a set of Lippmann-Schwinger equations to serve
as a set of simultaneous coupled integral equations for all the elements
in a row or column of the transition operator matrix. It has the advantage
of coupling the partitions in a symmetrical manner \cite{tob83:1405},
implying that it is less sensitive to truncations on the spectrum of
virtual states (see also the SPA given in Section \ref{sec:elec-stru}).
Moreover, it was shown to be consistent with unitarity provided the
coupling scheme includes all open channels \cite{tob83:1405}. Finally,
Tobocman \cite{tob83:1405} found that the integral equations in two Hilbert
spaces decouple and thus reduce to a single wave operator equation of
particularly simple form.

\subsection{Scattering Resonance\label{sec:fesh-res}}

Having discussed potential scattering and reactive scattering (in Section
\ref{sec:pot-scatt}) and inverse problem (in Section \ref{sec:sta-2-sta-reax}),
let us specially consider the case where scattering matrix (see Equation
\eqref{eq:scatt-matrix-99}) have poles. This phenomenon is called
scattering resonance \cite{dya19:book}. The scattering resonances
generalize eigenvalues or bound states for systems in which energy
can scatter to infinity. A typical state during scattering resonance
has a rate of oscillation (like a bound state does) and a rate of decay
\cite{fer02:67,qiu06:1440,liu12:1,yan15:60,yua18:1289,dya19:book,che20:9202,xie20:767}.
Although the concept of scattering resonance is intrinsically dynamical,
its mathematical formulation comes from considering meromorphic
continuations of Green functions or scattering matrices \cite{dya19:book}.
The poles of these meromorphic continuations capture the physical insight
by identifying the rate of oscillations with the real part of a pole and
the rate of decay with its imaginary part. A typical example is shape
resonance created when the interaction region is separated from the free
region by a potential barrier. This is a simple and typical model for the
potential of an elementary reaction. In this case, the reaction rate is
exponentially small in a way depending on the width of the barrier. We
mainly describe theory of scattering resonance for such case \cite{dya19:book}.
We consider compactly supported possibility that gives rise to resonances.
It is convenient to replace a specific mechanism by an abstractly defined
black box scattering. We denote $\hat{H}_0^{(g)}=-\nabla_g^2/2$ a self-adjoint
operator outside $B_{R_0}$, where $\nabla_g^2$ satisfies the expression
as given in Equation \eqref{eq:polysph-keo-003}. The region $B_{R_0}$
denotes a ball of radius $R_0$ in $\mathbb{R}^d$ centered at the origin.
Moreover, the operator $\hat{H}_0$ acts on a Hilbert space with an
orthogonal decomposition, namely $\mathbb{H}=\mathbb{H}_{R_0}\oplus L^2(\mathbb{R}^d\setminus B_{R_0})$.
The orthogonal projection onto the first component $\mathbb{H}_{R_0}$,
called the back box, is denoted by $\mathscr{P}_{B_{R_0}}$. In this case,
the scattering is called black box scattering. Considering state
$\Psi\in H^2(\mathbb{R}^d)$ which is null outside $B_{R_0}$, we introduce
$h$ by assuming $\hat{H}_0^{(g)}(h)\Psi=-h^2\nabla_g^2\Psi/2$. Now, a
family of unbounded self-adjoint operators $\hat{H}_0^{(g)}(h)$ with
$0<h<1$ on a complex Hilbert space $\mathbb{H}=\mathbb{H}_{R_0}\oplus
L^2(\mathbb{R}^d\setminus B_{R_0})$ is called a semiclassical black box
Hamiltonian. If $h=1$ then $\hat{H}_0^{(g)}(h)$ is called a black box
Hamiltonian. These concepts can be further extended to potential scattering,
where $V\in L_{\mathrm{comp}}^{\infty}(\mathbb{R}^d,\mathbb{R})$ with
$\mathrm{supp}V\subset B_{R_0}$ and $\hat{H}_V^{(g)}(h)=\hat{H}_0^{(g)}(h)+V$.
To consider the case of scattering by compactly supported potentials,
we define the resolvent for a black box Hamiltonian as
$R_V^{(g)}(\lambda)=(\hat{H}_V^{(g)}(h)-\lambda^2)^{-1}$ for
$\mathrm{Im}\lambda>0$, where $\lambda^2$ is not in the spectrum of
$\hat{H}_V^{(g)}(h)$. The resolvent continues meromorphically as an
operator \cite{dya19:book}. Since the meromorphic continuation is purely functional
analytic in nature, we consider $\hat{H}_V^{(g)}=\hat{H}_V^{(g)}(1)$
only. One can prove that spectrum of $\hat{H}_V^{(g)}$ is given by
the union of a continuous (cont) part and a discrete (dis) set \cite{dya19:book},
$\mathrm{spec}\hat{H}_V^{(g)}=\mathrm{spec}_{\mathrm{cont}}\hat{H}_V^{(g)}
\cup\mathrm{spec}_{\mathrm{dis}}\hat{H}_V^{(g)}$ with
$\mathrm{spec}_{\mathrm{cont}}\hat{H}_V^{(g)}=[0,+\infty)$ and
$\mathrm{spec}_{\mathrm{dis}}\hat{H}_V^{(g)}=\{z_j\vert z_j\leq z_{j+1}\}_{j=N_1}^{N_+}$,
where $N_{\pm}$ can take $\pm\infty$, the multiplicities are finite,
and the set $\mathrm{spec}_{\mathrm{dis}}\hat{H}_V^{(g)}$ is
discrete.

With the above concepts, let us turn to the scattering matrix. All the
formulas given in Sections \ref{sec:pot-scatt} and \ref{sec:sta-2-sta-reax}
remain valid for the operator $\hat{H}_V^{(g)}(h)$ with the rescaling
$-h^2\nabla_g^2\rightsquigarrow-\nabla_g^2$, $\hat{H}_V^{(g)}(h)
\rightsquigarrow h^{-2}\hat{H}_V^{(g)}(h)$, and $\lambda\rightsquigarrow
\lambda/h$. The solution of $(\hat{H}_V^{(g)}-\lambda^2)\Psi=0$ is sums
of plane waves away from the black box and outgoing waves. One can define
the incoming and outgoing waves for $\lambda\in\mathbb{R}\setminus\{0\}$
as a solution to $(\hat{H}_V^{(g)}-\lambda^2)\Psi=f\in H_{\mathrm{comp}}$.
\begin{defi}
A solution $\Psi_{\lambda}^{(+)}$ to $(\hat{H}-\lambda^2)\Psi_{\lambda}^{(+)}=f$
is called outgoing if $\exists g\in L_{\mathrm{comp}}^2(\mathbb{R}^d)$
such that $\Psi_{\lambda}^{(+)}=R_0(\lambda)g$, where
$R_0(\lambda)=(\hat{H}_0-\lambda^2)^{-1}:L^2(\mathbb{R}^d)\to L^2(\mathbb{R}^d)$
is the resolvent. A solution $\Psi_{\lambda}^{(-)}$ to $(\hat{H}-\lambda^2)\Psi_{\lambda}^{(-)}=f$
is called incoming if $\Psi_{\lambda}^{(-)}=R_0(-\lambda)g$.
\label{def:001-odd-dimen-mul-reson}
\end{defi}
According to Definition \ref{def:001-odd-dimen-mul-reson}, if and only
if $\exists g\in L_{\mathrm{comp}}^2(\mathbb{R}^d)$ and $R>R_0$ such
that $\Psi=R_0(\lambda)g$ in $\mathbb{R}^d\setminus B_{R}$, the solution
$\Psi$ is called outgoing. A solution $\Psi$ is called incoming if and
only if $\exists g\in L_{\mathrm{comp}}^2(\mathbb{R}^d)$ and $R>0$ such
that $\Psi=R_0(-\lambda)g$ in $\mathbb{R}^d\setminus B_{R}$. The outgoing
solutions to the homogenous equation have to be compactly supported and
that the resonances on $\mathbb{R}\setminus\{0\}$ are given by embedded
eigen-values. For $\lambda\in\mathbb{R}\setminus\{0\}$, we can find that
(1) if $(\hat{H}_V^{(g)}-\lambda^2)\Psi=0$ then the outgoing wave is zero,
and that (2) if $\lambda$ is a pole of $R_V^{(g)}(\lambda)$ then $\lambda^2$
is an embedded eigen-value of $\hat{H}_V^{(g)}$ and $m_R(\lambda)$ is the
multiplicity of that eigen-value \cite{dya19:book}. For obstacle scattering,
$\hat{H}_0^{(g)}$ is the Dirichlet or Neumann Laplacian for a metric $g$
on a connect set $\mathbb{R}^d\setminus\mathscr{O}$, where $\mathscr{O}$
is bounded and has a smooth boundary. If
$g_{ij}-\delta_{ij}\in C_c^{\infty}(\mathbb{R}^d\setminus\mathscr{O})$,
where $\hat{H}_0^{(g)}$ is a compact metric perturbation of the Euclidean
Laplacian, then the meromorphic extension of $R_0^{(g)}(\lambda)=
(\hat{H}_0^{(g)}-\lambda^2)^{-1}:L_{\mathrm{comp}}^2(\mathbb{R}^d
\setminus\mathscr{O})\to H_{\mathrm{loc}}^2(\mathbb{R}^d\setminus\mathscr{O})$
is holomorphic near zero. This means no zero resonance in obstacle
scattering \cite{dya19:book}. Thus, it is ncessary to give a condition
which guarantees dissolution of an embedded eigen-value to a resonance
under such a compact metric perturbation.
\begin{prop}[Fermi Golden Rule for embedded eigen-values]
Assuming that $\lambda^2>0$ is an embedded eigen-value of $\hat{H}_0^{(g)}=\hat{H}_0^{(g)}(0)$
and $(\hat{H}_0^{(g)}-\lambda^2)\Psi=0$ with $\Vert\Psi\Vert=1$, 
where $s\mapsto \hat{H}_0^{(g)}(s)$ satisfies $\hat{H}_0^{(g)}(s)\in
C^{\infty}((-s_0,s_0),\mathcal{L})$, $\hat{H}_0^{(g)}{}^{\dagger}(s)=\hat{H}_0^{(g)}(s)$,
and $\hat{H}_0^{(g)}(s)(u^*)=(\hat{H}_0^{(g)}(s)(u))^*$ with $u\mapsto
u^*$ being an involution, then
\begin{equation}
\mathrm{Im}\ddot{\lambda}=-\frac{\lambda^{d-3}}{4(2\pi)^{d-1}}
\int_{\mathbb{S}^{d-1}}\Big\vert\Big\langle\hat{\dot{H}}_0^{(g)}\Psi
\big\vert\Psi_{\lambda}\big(\omega\big)\Big\rangle\Big\vert^2\mathrm{d}\omega
\label{eq:fermi-gol-000}
\end{equation}
where $\Psi_{\lambda}(\omega)$ is a plane wave traveling in the direction
of $\omega$ with wave number $\lambda$.
\label{prop:fermi-golden-emb-reson}
\end{prop}
Proposition \ref{prop:fermi-golden-emb-reson} is required because the
perturbation theory cannot consider states that lie in the continuous
spectrum. By Equation \eqref{eq:fermi-gol-000} we can compute the decay
rate or lifetime of such a resonance, turning an ill‑defined energy shift
into a physically meaningful width.

According to Definition \ref{def:scatt-oper-000}, we define the scattering
matrix as the map
$S_{\mathrm{abs}}(\lambda):C^{\infty}(\mathbb{S}^{d-1})\to C^{\infty}(\mathbb{S}^{d-1})$
and $S_{\mathrm{abs}}(\lambda):g\mapsto f$ as the absolute scattering
matrix \cite{dya19:book}. It can further extend to a unitary transformation
$S_{\mathrm{abs}}(\lambda):L^2(\mathbb{S}^{d-1})\to L^2(\mathbb{S}^{d-1})$.
With $S_{\mathrm{abs}}(\lambda)$ the scattering matrix can be defined
as $S(\lambda)=\mi^{d-1}S_{\mathrm{abs}}(\lambda)J$ with $Jg(\theta)=g(-\theta)$.
For the black box Hamiltonian $\hat{H}^{(g)}$ and $\rho\in C_c^{\infty}(\mathbb{R}^d)$,
we define $E_{\rho}:L^2(\mathbb{R}^d)\to L^2(\mathbb{S}^{d-1})$ and
$E_{\rho}(\lambda)(X,\omega)=\rho(X)\exp(-\mi\lambda\langle X\vert\omega\rangle)$.
For $\chi_i\in C_c^{\infty}(\mathbb{R}^d,[0,1])$ with $i=1,2,3$, if for
some $R_1>R_0$ we set $\chi_i\vert_{B_{R_1}}=1$ and $\chi_{i+1}\vert_{\mathrm{supp\chi_i}}=1$
with $i=1,2$, then the scattering matrix is given by
\begin{equation}
S\big(\lambda\big)=I+\frac{(2\pi)^{1-d}}{2\mi}\lambda^{d-2}
E_{\chi_3}\big(\lambda\big)\big[\nabla_g^2,\chi_1\big]
R_V^{(g)}\big(\lambda\big)
\big[\nabla_g^2,\chi_1\big]E_{\chi_3}^{\dagger}\big(\lambda^*\big).
\label{eq:fermi-gol-001}
\end{equation}
According to Equation \eqref{eq:fermi-gol-001}, the meromorphy of
$R_V^{(g)}(\lambda)$ shows that $S(\lambda)$ forms a meromorphic family
of operators on $\mathbb{C}$. The unitarity relation is given by
$S^{-1}(\lambda)=S^{\dagger}(\lambda^*)$ for $\lambda\in\mathbb{C}$ and
remains valid at $\lambda=0$ by continuity following from the fact that
$S(\lambda)$ is meromorphic. By Equation \eqref{eq:fermi-gol-001}, one
can find several properties. 
\begin{prop}
For $V\in L_{\mathrm{comp}}^{\infty}(\mathbb{R}^d,\mathbb{C})$, the
scattering matrix $S(\lambda)$ is meromorphic in $\mathbb{C}$ with
poles of finite rank and satisfies $S^{-1}(\lambda)=JS(-\lambda)J$
and $Jf(\theta)=f(-\theta)$ for $\lambda\in\mathbb{C}$. There are
only finitely many poles in the closed upper half-plane. If
$V\in L_{\mathrm{comp}}^{\infty}(\mathbb{R}^d,\mathbb{R})$ then $S^{-1}(\lambda)=S^*(\lambda^*)$.
In particular, $S(\lambda)$ is unitary for $\lambda\in\mathbb{R}$ and
holomorphic on $\mathbb{R}$.
\label{prop:10-odd-dimen-mul-reson}
\end{prop}
For the black box Hamiltonian $\hat{H}^{(g)}$ and the scattering matrix
$S(\lambda)$, the precise relations between the multiplicities of the
poles of $S(\lambda)$ and $R_V^{(g)}(\lambda)$ are given by
\begin{equation}
m_S\big(\lambda\big)=-\frac{1}{2\pi}\oint_{\lambda}
\mathrm{tr}S^{-1}\big(\zeta\big)\partial_{\zeta}
S\big(\zeta\big)\mathrm{d}\zeta=\tilde{m}_R\big(+\lambda\big)
-\tilde{m}_R\big(-\lambda\big),\quad
\tilde{m}_R\big(\lambda\big)=\mathrm{rank}
\mathscr{P}_{\mathbb{R}^d\setminus B_{R_0}}\Pi_{\lambda},
\label{eq:fermi-gol-002}
\end{equation}
where the integral is over a positively oriented circle enclosing
$\lambda$ and no other pole of $S(\lambda)$ or $S^{-1}(\lambda)$, while
$\Pi_{\lambda}$ satisfies
\begin{equation}
R_V\big(\lambda\big)=-\sum_{k=1}^{m_R(\lambda_0)}
\dfrac{\big(\hat{H}^{(g)}-\lambda_0^2\big)^{k-1}\Pi_{\lambda_0}}
{\big(\lambda^2-\lambda_0^2\big)^k}+A\big(\lambda,\lambda_0\big),\quad
\Pi_{\lambda_0}=-\frac{1}{2\mi\pi}\oint_{\lambda_0}R_V\big(\lambda\big)
2\lambda\mathrm{d}\lambda,\quad
\big(\hat{H}-\lambda_0^2\big)^{m_R(\lambda_0)}\Pi_{\lambda_0}=0.
\label{eq:resona-sca-04}
\end{equation}
In Equation \eqref{eq:resona-sca-04}, $\lambda\mapsto A(\lambda,\lambda_0)$
is holomorphic near $\lambda_0$.
%while $\mathrm{Ran}\Pi_{\lambda_0}=\mathrm{span}\{u_1,\cdots,u_{m_R(\lambda_0)}\}$.
Due to $\tilde{m}_R(\lambda)=m_R(\lambda)$ for $\lambda^2\notin\mathbb{R}$,
any compactly supported resonant states have to be eigen-states, and
hence $m_S(\lambda)=-\mathrm{sgn}(\mathrm{Im}\lambda)m_R(\lambda)$ for
$\lambda^2\notin\mathbb{R}$. The difference is relevant only on the
imaginary axis. So far, resonances for the black box Hamiltonian was
defined as poles of the meromorphic continuation of the resolvent.
The structure of that continuation near the poles shows that the
meromorphic continuation behaves like a resolvent of a non-self-adjoint
operator.

Now, we consider an abstract non-trapping condition which guarantees
the presence of resonance-free regions for black box Hamiltonians. This
provides an expansion of scattered waves for non-trapping black boxes.
Assuming that $V\in C_c^{\infty}(\mathbb{R}^d,\mathbb{R})$ and
$R_V^{(g)}(\lambda)=(\hat{H}^{(g)}-\lambda^2)^{-1}$ with $\mathrm{Im}\lambda>0$,
for any $M>0$ there exists $C_0$ such that $R_V^{(g)}(\lambda):L_{\mathrm{comp}}^2\to L_{\mathrm{loc}}^2$
continues holomorphically to
$\Omega_M=\{\lambda\in\mathbb{C}\vert\mathrm{Im}\lambda>-M\ln\vert\lambda\vert,\;\vert\lambda\vert>C_0\}$.
Moreover, $\forall\chi\in C_c^{\infty}(\mathbb{R}^d)$ there exist $C_1$
and $T$ such that $\Vert\chi R_V^{(g)}(\lambda)\chi\Vert_{L^2\to L^2}
\leq C_1\vert\lambda\vert^{-1}\exp(T\max(0,-\mathrm{Im}\lambda))$ with
$\vert\lambda\vert\in\Omega_M$. Turning to the general black box case,
since the balck box Hamiltonian $\hat{H}^{(g)}$ is a self-adjoint
operator, we define $U(t)=\sin t\sqrt{\hat{H}^{(g)}}/\sqrt{\hat{H}^{(g)}}$.
If $\forall a>R_0$, $\exists T_a$, $\forall\chi\in C_c^{\infty}(B_a)$,
$\chi\vert_{B_{R_0+\epsilon}}=1$, and $\chi U(t)\chi\vert_{t>T_a}\in
C^{\infty}((T_a,\infty),\mathcal{L})$, then the self-adjoint operator
$\hat{H}^{(g)}$ is non-trapping. To understand this concept, we consider
$\hat{H}^{(g)}_0$ with $g^{ij}$ a Riemannian metric on $\mathbb{R}^d$
such that $g^{ij}-\delta^{ij}\in C_c^{\infty}(B_{R_0})$ and suppose that
the metric is classically non-trapping, namely $\forall(X,P)\in T^*\mathbb{R}^d\setminus\{0\}$
as $t\to\pm\infty$ and $\pi(\exp(tH_p(X,P)))\to\infty$ there exists
$p(X,P)=\sum_{i,j=1}^dg^{ij}P_iP_j$, where $\pi:T^*\mathbb{R}^d\to\mathbb{R}^d$
is the natural projection and $H_p$ is the Hamiltonian vector field of
$p$. In particular, this implies that for any $a>0$ there exists $T_a$
such that $\vert X\vert<a$, $\vert t\vert>T_a$ and $p(X,P)=1$ lead to
$\vert\pi(\exp(tH_p(X,P)))\vert>a$. Hence, the result on propagation of
singularities gives that $\forall\chi\in C_c^{\infty}(B_a)$ and $\exists N>0$
such that $\chi(\sin t\sqrt{\hat{H}_0^{(g)}}/\sqrt{\hat{H}_0^{(g)}})\chi\in C^{\infty}((T_a,\infty),\mathcal{L})$. 
This means that classical non-trapping for the metric implies non-trapping
for the propagator in the sense of non-trapping black box Hamiltonian.
For the non-trapping black box Hamiltonian $\hat{H}^{(g)}$ and
$R_V^{(g)}(\lambda):\mathcal{H}_{\mathrm{comp}}\to\mathcal{D}_{\mathrm{loc}}$
is the meromorphically continued resolvent, there exists $C_0$ such that
$\Omega_M=\{\lambda\in\mathbb{C}\vert\mathrm{Im}\lambda>-M\ln\vert\lambda\vert,\;\vert\lambda\vert>C_0\}$
for any $M$. Moreover, $\forall\chi\in C_c^{\infty}(\mathbb{R}^d)$ for
$\chi=1$ near $B_{R_0}$ there exist $C_1$ and $T$ such that
$\Vert\chi R_V^{(g)}(\lambda)\chi\Vert_{\mathcal{H}\to\mathcal{H}}\leq
C_1\vert\lambda\vert^{-1}\exp(T\max(0,-\mathrm{Im}\lambda))$ with
$\lambda\in\Omega_M$.

Now, it is possible to consider the geometric theory in quantum scattering
by showing meromorphic continuation of the resolvent of the Laplacian operator
on complete asymptotically hyperbolic Riemannian manifolds. The hyperbolic
space $\mathbb{H}^d$ is a basic example and can be seen as the open unit
ball in $\mathbb{R}^d$ with the metric $g=4\mathrm{d}w^2/(1-\vert w\vert^2)^2$
and coordinates $w\in B_{\mathbb{R}^d}(0,1)$. On an asymptotically hyperbolic
manifold of dimension $d$, the scattering resolvent is the meromorphic
continuation of $R_0^{(g)}(\lambda)=(-\nabla_g^2-\lambda^2-(d-1)^2/4)^{-1}:L^2(M)\to L^2(M)$
to the entire complex plane for $\mathrm{Im}\lambda>0$. Compared to
Euclidean scattering, we have an additional $(d-1)^2/4$ term, which is
related to the fact that the essential spectrum of $-\nabla^2_g$ is
$[(d-1)^2/4,\infty)$. In the Euclidean case, the method of complex
scaling provides meromorphic continuation of the resolvent by relating
it to a resolvent of a non-self-adjoint operator $\hat{H}_{\theta}^{(g)}$
such that $\hat{H}_{\theta}^{(g)}-\lambda^2$ is a Fredholm operator for
$\mathrm{arg}\lambda>-\theta$. In the case of asymptotically hyperbolic
manifolds the resolvent $R_0^{(g)}(\lambda)$ is related to the inverse
of $\hat{H}_0^{(g)}(\lambda)$ for a family of Fredholm operators
$\hat{H}_0^{(g)}(\lambda)$ on suitably defined spaces. Unlike the
method of complex scaling, where resonances with $\mathrm{arg}\lambda>-\theta$
are identified with eigen-values $\lambda^2$ of $\hat{H}_{\theta}^{(g)}$,
$\hat{H}_0^{(g)}(\lambda)$ has the Fredholm property in $\{\mathrm{Im}
\lambda>-C\}$ on spaces that depends on contsant $C\in\mathbb{R}$.
Extending the manifold across the boundary at infinity, the operators
$\hat{H}_0^{(g)}(\lambda)$ are constructed and then the family
$\hat{H}_0^{(g)}(\lambda)$ has the property of translating the outgoing
condition for silutions $(-\nabla^2_g-\lambda^2-(d-1)^2/4)u=f$ with
$f\in C_c^{\infty}(M)$ to smoothness of solutions to $\hat{H}_0^{(g)}(\lambda)
\tilde{u}=\tilde{f}$. To do this, $\hat{H}_0^{(g)}(\lambda)$ transitions
from elliptic to hyperbolic have to behavior as one moves across the
boundary at infinity. Now, let us give details of these technically
complicated parts. We refer the reader to Reference \cite{dya19:book}
for more details.

%%-------------------------------------------
\subsection{Scattering on Hyperbolic Manifolds\label{sec:odd-res-hyper}}

First, let us consider an asymptotically hyperbolic manifold. It is a
complete Riemannian manifold whose infinite ends are locally asymptotic
to the infinity of the hyperbolic space. A standard way to study these
manifolds is to define them as interiors of compact manifolds with boundary.
Letting $\overline{M}$ be a compact manifold with boundary $\partial M$
and interior $M$, a boundary defining function on $\overline{M}$ is a
$C^{\infty}$ function $y_1:\overline{M}\to[0,\infty)$ such that $y_1=0$
and $\mathrm{d}y_1\neq0$ on $\partial M$, and $y_1>0$ on $M$. Such
functions always exist and any two boundary defining $y_1$ and $\tilde{y}_1$
are multiples of each other, $\tilde{y}_1=\exp(f)y_1$, for some
$f\in C^{\infty}(\overline{M},\mathbb{R})$. Moreover, for any boundary
defining function $y_1$ and $\epsilon_1>0$ small enough, there exists
a diffeomorphism $y_1^{-1}([0,\epsilon_1))\to[0,\epsilon_1)\times\partial
M$, where $X\mapsto(y_1(X),y_2(X),\cdots,y_d(X))=(y_1(X),y'(X))$ and
$y'\vert_{\partial M}=I$, where $y'$ is a set of $d-1$ complementary
functions of $y_1$. Now we can define an asymptotically hyperbolic manifold
as a complete Riemannian manifold $(M,g)$, where (1) $M$ is the interior
of a compact manifold with boundary, denoted by $\overline{M}$; (2) for
a boundary defining function $y_1$, the metric $y_1^2g$ extends to a smooth
Riemannian metric on $\overline{M}$; and (3) $\vert\mathrm{d}y_1\vert_{y_1^2g}=1$
on $\partial\overline{M}$. Manifolds with properties (1) and (2) are
called conformally compact, while property (3) has the additional effect
that the sectional curvatures on $M$ converge to $-1$ at infinity. The
boundary metric $(y_1^2g)\vert_{\partial M}$ depends on the choice of
the boundary defining function $y_1$. However, a different choice of
$y_1$ multiplies the boundary metric by a conformal factor. Therefore,
to each asymptotically hyperbolic manifold corresponds a conformal class
of Riemannian metrics on $\partial M$, denoted by $[g]_{\partial M}=\{(y_1^2g)\vert_{\partial M}\}$,
where $y_1$ is a boundary defining function.
Letting $(M,g)$ be an asmptotically hyperbolic manifold, a boundary
defining function $y_1$ is called canonical if $\vert\mathrm{d}y_1\vert_{y_1^2g}=1$
in a neighborhood of $\partial M$. For such a function $y_1$, a product
structure $(y_1,y')$ is called canonical if the pushforward of the
metric $g$ has the form $y_1^2g=\mathrm{d}y_1^2+g_1(y_1,y',\mathrm{d}y')$
in a neighborhood of $\partial M$, where $g_1(y_1,y',\mathrm{d}y')$ is
a family of Riemannian metrics on $\partial M$ depending on
$y_1\in[0,\epsilon)$. For fixed $g_0\in[g]_{\partial M}$, there exists
a canonical product structure $(y_1,y')$ on $\overline{M}$ such that
$g_0=(y_1^2g)\vert_{\partial M}$. The asmptotically hyperbolic manifold
$(M,g)$ is called even if there exists a canonical
product structure $(y_1,y')$ such that the corresponding metric $g_1$
satisfies $\partial_{y_1}^{2k+1}g_1(0,y',\mathrm{d}y')=0$ for $k\in\mathbb{N}\oplus\{0\}$.
For a fixed canonical product structure
$(y_1,y')\in[0,\epsilon_1)\times\partial M$ and the diffeomorphism
$M\cap\{y_1<\epsilon_1\}\to(0,\epsilon_1^2)\times\partial M$, one can
define (1) the even compactification $\overline{M}_{\mathrm{even}}$ of
$M$ to be the manifold with boundary obtained by gluing $M$ with $[0,\epsilon^2_1)\times\partial M$
and (2) the even extension $\overline{X}=\overline{X}_{\epsilon}$ of
$M$ to be the manifold with boundary obtained by gluing $M$ with
$[-\epsilon,\epsilon_1^2)\times\partial M$, where $\epsilon>0$ is a
given constant.
\begin{remark}
Here, we consider the scattering in a bounded black box and formalize
the concept of escape at infinity as the outgoing condition.
\label{rem:hybloc-scatt-1}
\end{remark}

Second, we should introduce the modified spectral family of the Laplacian
whose Fredholm property on appropriately chosen spaces will ultimately
give meromorphic continuation of the resolvent. We denote by $X$ the
interior of the even extension $\overline{X}\supset\overline{M}_{\mathrm{even}}$.
Letting $x_1=y_1^2$ so that $x_1$ is a boundary defining function of
$\overline{M}_{\mathrm{even}}$ and $x_1+\epsilon$ is a boundary defining
function of $\overline{M}$. One can have $\overline{X}=M\cup\overline{Y}$,
$M=\{x_1>0\}$, and $\overline{Y}=\{-\epsilon<x_1<\epsilon_1^2\}$, where
$\{y_1<\epsilon_1\}$ is the domain of the product structure $(y_1,y')$.
On $\overline{Y}$ we have the product structure $(x_1,x')\in[-\epsilon,\epsilon_1^2)\times\partial M$
for $x'=y'$. Since $(M,g)$ is an even metric we can write on $M\cap Y$
the expression $g=\mathrm{d}x_1^2/(4x_1^2)+g_1(x_1,x',\mathrm{d}x')/x_1$,
where $g_1$ is smooth in $x_1\in[0,\epsilon_1^2)$. We fix an extension
of $g_1$ to $x_1\in[-\epsilon,\epsilon_1^2)$ as a smooth family of Riemannian
metrics on $\partial M$, namely $g_1\in C^{\infty}([-\epsilon,\epsilon_1^2)
\times\partial M,\otimes^2T^*\partial M)$, where $\otimes^2T^*\partial M$
is referred to as the second tensor product of the cotangent bundle of
the boundary $\partial M$. On $M\cap Y=\{0<x_1<\epsilon_1^2\}$, the
differential operator on $M$ satisfies
\begin{equation}
x_1^{\mi\lambda/2-(d+3)/4}\left(-\nabla_g^2-\lambda^2-\frac{(d-1)^2}{4}
\right)x_1^{(d-1)/4-\mi\lambda/2}=
-4x_1\partial_{x_1}^2+4\big(\mi\lambda-1\big)\partial_{x_1}-\gamma
\big(4x_1\partial_{x_1}+d-1-2\mi\lambda\big)-\nabla^2_{g_1},
\label{eq:extension-even-000}
\end{equation}
where $\nabla_{g_1}^2$ is the laplace-Beltrami operator of $g_1$ in
the $x'$ variables and $\gamma(x_1,x')=J^{-1}\partial_{x_1}J\in
C^{\infty}(\overline{Y},\mathbb{R})$ with $J=\sqrt{\vert\det g_1\vert}$
is the logarithmic derivative of the Jacobian of the metric $g_1$.
Here, the metric $g_1$ is independent of the choice of local
coordinates on $\partial M$. Similar to Equation \eqref{eq:extension-even-000}
on $M$, one can define the extended modified Laplacian as the second
order differential operator $\hat{H}^{(g)}(\lambda)$ on $\overline{X}$.
On $\overline{Y}$, the similar operator is given by
\begin{equation}
\hat{H}^{(g)}\big(\lambda\big)=-4x_1\partial_{x_1}^2+4\big(\mi\lambda-
1\big)\partial_{x_1}-\gamma\big(x_1,x'\big)
\big(4x_1\partial_{x_1}+d-1-2\mi\lambda\big)-\nabla^2_{g_1}.
\label{eq:extension-even-001}
\end{equation}
Now, we introduce the semiclassically rescaled version of $\hat{H}^{(g)}(\lambda)$
in the form of $\hat{H}^{(g)}_h(\omega)=h^2\hat{H}^{(g)}(h^{-1}\omega)$
with $0<h\ll1$ and $\omega\in\mathbb{C}$. It is essential for the high-frequency
limit $\mathrm{Re}\lambda\to\infty$. By Equation \eqref{eq:extension-even-001},
on $M$ one can find
\begin{equation}
\hat{H}^{(g)}_h\big(\omega\big)=x_1^{\mi\omega/(2h)-(d+3)/4}\left(-h^2
\nabla_g^2-\omega^2-\frac{(d-1)^2}{4}h^2\right)x_1^{(d-1)/4-\mi\omega/(2h)},
\label{eq:extension-even-002}
\end{equation}
while on $\overline{Y}$ one can have
\begin{equation}
\hat{H}^{(g)}_h\big(\omega\big)=4x_1\big(h\mathscr{D}_{x_1}\big)^2-
4\big(\omega+\mi h\big)h\mathscr{D}_{x_1}-h^2\nabla_{g_1}^2-\mi h
\gamma\big(x_1,x'\big)\Big(4x_1h\mathscr{D}_{x_1}+\mi(1-d)h-2\omega
\Big),\quad\mathscr{D}_{x_1}=-\mi\partial_{x_1}.
\label{eq:extension-even-003}
\end{equation}
Equations \eqref{eq:extension-even-002} and \eqref{eq:extension-even-003}
indicate that $\hat{H}^{(g)}_h(\omega)$ is a second-order semiclassical
differential operator on $X$ with coefficients smooth up to the boundary,
depending holomorphically on the parameter $\omega$. The semiclassical
principal symbol of $\hat{H}^{(g)}_h(\omega)$ is given by
$p=p(X,P,\omega)=\sigma_h(\hat{H}^{(g)}_h(\omega))\in\mathrm{Poly}^2(T^*X)$,
where $\mathrm{Poly}^2(T^*X)$ means the space of smooth functions on
$T^*X$ which are polynomials of degree at most two in the momentum $P$.
It is worth noting that the symbol $p$ depends on the choice of $\omega$.

Third, we simplify denotation of $\hat{H}^{(g)}(\lambda)$ as $\hat{H}(\lambda)$
to enshort expressions. Noting that the even extension $\overline{X}$ is a compact manifold
with boundary $\partial X=\{x_1=-\epsilon\}$ and interior $X=\{x_1>-\epsilon\}$,
we consider the semiclassical Sobolev space with boundary $\bar{\mathcal{H}}_h^s(X)$
and the homogeneous one $\dot{\mathcal{H}}_h^s(\overline{X})$ both for
$s\in\mathbb{R}$. The sapces $\bar{\mathcal{H}}_h^s(X)$ and
$\dot{\mathcal{H}}_h^{-s}(\overline{X})$ are dual to each other with
respect to the standard $L^2$ pairing. The norms of $\bar{\mathcal{H}}_h^s(X)$
and $\dot{\mathcal{H}}_h^s(\overline{X})$ depend on $h$, but the
underlying Hilbert spaces $\bar{\mathcal{H}}^s(X)$ and $\dot{\mathcal{H}}^s(X)$
are $h$-independent. Since $\hat{H}^{(g)}_h(\omega)$ is a second-order
semiclassical differential operator with coefficients smooth up to the
boundary of $X$, it defines a bounded operator $\bar{\mathcal{H}}^s(X)\to\bar{\mathcal{H}}^{s-2}(X)$.
As proved in Reference \cite{dya19:book}, the Fredholm
property of $\hat{H}(\lambda)$ follows from the semiclassical propagation
estimates for the rescaled operator $\hat{H}_h^{(g)}(\omega)$, whose
principal symbol generates a Hamiltonian flow with radial sources and
sinks at the boundary. Since the family of operators $\hat{H}(\lambda)$
has a meromorphic inverse $\hat{H}^{-1}(\lambda)$, we use $\hat{H}^{-1}(\lambda)$
to show meromorphic continuation of the scattering resolvent. We have
estimated $\Vert u\Vert_{\bar{\mathcal{H}}_h^s(X)}$ in term of
$\Vert\hat{H}_h^{(g)}(\omega)u\Vert_{\bar{\mathcal{H}}_h^{s-1}(X)}$,
where $\hat{H}_h^{(g)}(\omega)$ is the semiclassical rescaling of
$\hat{H}(\lambda)$. However, the second-order differential operator
$\hat{H}(\lambda)$ is not bounded. To resolve this issue, we consider
its domain $\mathcal{X}^s=\{u\in\bar{\mathcal{H}}^s(X)\vert\hat{H}(0)u
\in\bar{\mathcal{H}}^{s-1}(X)\}$, endowed with the semiclassical norm
$\Vert u\Vert_{\mathcal{X}^s}=(\Vert u\Vert^2_{\bar{\mathcal{H}}^s_h(X)}
+\Vert\hat{H}(0)u\Vert^2_{\bar{\mathcal{H}}^{s-1}_h(X)})^{1/2}$. It is
easy to find that $\mathcal{X}^s$ is a Hilbert space, by identifying it
with the closed subspace $\{(u,f)\vert\hat{H}(0)u=f\}\subset\bar{\mathcal{H}}^s(X)
\oplus\bar{\mathcal{H}}^{s-1}(X)$. Moreover, the norms
$\Vert\cdot\Vert_{\mathcal{X}^s_h}$ for different $h$ are equivalent,
with constants depending on $h$. Since $\hat{H}(\lambda)-\hat{H}(0)$
is a first-order differential operator, it is bounded
$\bar{\mathcal{H}}^s(X)\to\bar{\mathcal{H}}^{s-1}(X)$ and therefore
$\hat{H}(\lambda):\mathcal{X}^s\to\bar{\mathcal{H}}^{s-1}(X)$ is a
family bounded operators holomorphic in $\lambda\in\mathbb{C}$. To
consider meromorphy of $\hat{H}^{-1}(\lambda)$, we introduce notations
for the kernel, cokernel, and kernel of the adjoint of $\hat{H}(\lambda)$
as
\begin{equation}
\begin{array}{l}
\mathrm{ker}^s\hat{H}(\lambda)=\Big\{u\in\bar{\mathcal{H}}^s(X)\big\vert
\hat{H}(\lambda)u=0\Big\},\quad
\mathrm{ker}^{1-s}\hat{H}^{\dagger}(\lambda)=
\Big\{v\in\dot{\mathcal{H}}^{1-s}(\overline{X})\big\vert
\hat{H}^{\dagger}(\lambda)v=0\Big\},  \\
\mathrm{coker}^s\hat{H}(\lambda)=\Big\{
v\in\dot{\mathcal{H}}^{1-s}(\overline{X})\big\vert
\big\langle\hat{H}(\lambda)u\vert v\big\rangle_{L^2}=0,\;\forall u\in\mathcal{X}^s\Big\},
\end{array}
\label{eq:sca-estima-hyp-001}
\end{equation}
where $\dot{\mathcal{H}}^{1-s}(\overline{X})$ is the dual to
$\bar{\mathcal{H}}^{s-1}(X)$ with respect to the $L^2$ inner product.
For all $s$ and $\lambda$, we have $\mathrm{coker}^s\hat{H}(\lambda)=
\mathrm{ker}^{1-s}\hat{H}^{\dagger}(\lambda)$. Now, fixing
$s\in\mathbb{R}$ the operator $\hat{H}(\lambda):\mathcal{X}^s\to
\bar{\mathcal{H}}^{s-1}(X)$ is a Fredholm operator of index zero for
$\mathrm{Im}\lambda>1/2-s$ and has a meromorphic inverse with poles of
finite rank $\hat{H}^{-1}(\lambda):\bar{\mathcal{H}}^{s-1}(X)\to
\mathcal{X}^s$ with $\mathrm{Im}\lambda>1/2-s$. For each $\lambda$,
the inverse $\hat{H}^{-1}(\lambda)$ does not depend on the choice of
$s$, and moreover, maps $\bar{\mathcal{H}}^{s-1}(X)\to\bar{\mathcal{H}}^{s}(X)$
for all $s>1/2-\mathrm{Im}\lambda$. Thus, one can have a meromorphic
family of continuous operators $\hat{H}^{-1}(\lambda):C^{\infty}(\overline{X})
\to C^{\infty}(\overline{X})$ with $\lambda\in\mathbb{C}$. Assuming that
$s<t$ and letting $\hat{H}^{-1}(\lambda,s)$ and $\hat{H}^{-1}(\lambda,t)$
be the inverses of $\hat{H}(\lambda)$ as operators
$\mathcal{X}^s\to\bar{\mathcal{H}}^{s-1}(X)$ and $\mathcal{X}^t\to\bar{\mathcal{H}}^{t-1}(X)$,
respectively, we have $\hat{H}^{-1}(\lambda,s)f=\hat{H}^{-1}(\lambda,t)f$
with $f\in\bar{\mathcal{H}}^{t-1}(X)$ and $\mathrm{Im}\lambda>1/2-s$.
\begin{remark}
In the black box, an operator transforms the complex scattering problem
in curved space into a familiar problem.
\label{rem:scatt-hybloc-2}
\end{remark}

Finally, turning to the pole, if $\lambda\in\mathbb{C}$ is a pole of
$\hat{H}^{-1}(\lambda):\bar{\mathcal{H}}^{s-1}(X)\to\mathcal{X}^s$ for
$s>1/2-\mathrm{Im}\lambda$ then $\lambda$ is an extended resonance of
$(M,g)$. In this case, elements of $\mathrm{ker}^s\hat{H}(\lambda)$
are extended resonant states and elments of $\mathrm{coker}^s\hat{H}(\lambda)$
are extended coresonant states. Since $\hat{H}(\lambda)$ is a Fredholm
operator of index zero, $\lambda$ is an extended resonance if and only
if $\dim\ker^s\hat{H}(\lambda)>0$. Moreover, $\dim\ker^s\hat{H}(\lambda)
=\dim\mathrm{coker}^s\hat{H}(\lambda)$. For all $t>s>1/2-\mathrm{Im}\lambda$,
we have $\ker^t\hat{H}(\lambda)\subset\ker^s\hat{H}(\lambda)$ and
$\mathrm{coker}^s\hat{H}(\lambda)\subset\mathrm{coker}^t\hat{H}(\lambda)$.
These inequalities indicate that the spaces of extended resonant and
coresonant states do not depend on $s$. By these results, we have
$\ker^s\hat{H}(\lambda)=\{u\in C^{\infty}(\overline{X})\vert\hat{H}(\lambda)u=0\}$
and $\mathrm{coker}^s\hat{H}(\lambda)=\{v\in\mathfrak{D}'(X)\vert\mathrm{supp}v
\subset\overline{X}_{\mathrm{even}},\;\hat{H}^{\dagger}(\lambda)v=0\}$
for $s>1/2-\mathrm{Im}\lambda$, where $\mathfrak{D}'(X)$ is the linear
space of all distributions (Schwartz distributions). Therefore, one can
drop the superscript $s$ in $\ker^s\hat{H}(\lambda)$ and $\mathrm{coker}^s\hat{H}(\lambda)$.
Returning to the even asymptotically hyperbolic manifold $(M,g)$. Since
it is complete, the Lapacian operator $\nabla^2_g$ is essentially
self-adjoint on $C_c^{\infty}(M)$. The domain of the self-adjoint
extension is denoted by $\mathcal{H}^2(M)=\{u\in L^2(M)\vert\nabla_g^2u\in L^2(M)\}$,
where one can understand $\nabla_g^2u$ in the sense of distributions
and use the volume form of $g$ to define the space $L^2(M)$. For
$\mathrm{Im}\lambda>0$ with $\lambda\notin\mi[0,(d-1)/2]$, the holomorphic
$L^2$ resolvent is given by $(-\nabla_g^2-\lambda^2-(d-1)^2/4)^{-1}:
L^2(M)\to\mathcal{H}^2(M)$. This family admits a meromorphic continuation
with poles of finite rank $R(\lambda):L^2_{\mathrm{comp}}(M)\to
H_{\mathrm{loc}}^2(M)$ with $\lambda\in\mathbb{C}$. Moreover, for each
$s>1/2-\mathrm{Im}\lambda$ one can extend $R(\lambda)$ to $R(\lambda):
x_1^{(d+3)/4-\mi\lambda/2}\bar{\mathcal{H}}^{s-1}(M_{\mathrm{even}})\to
x_1^{(d-1)/4-\mi\lambda/2}C^{\infty}(\overline{M}_{\mathrm{even}})$,
where we should note that $x_1=y_1^2$ is a boundary defining function
of the even compactification $\overline{M}_{\mathrm{even}}$. In
particular, if $f\in C_c^{\infty}(M)$ then we have $R(\lambda)f\in
x_1^{(d-1)/4-\mi\lambda/2}C^{\infty}(\overline{M}_{\mathrm{even}})$. A
solution $u\in C^{\infty}(M)$ to $(-\nabla_g^2-\lambda^2-(d-1)^2/4)u=f
\in C_c^{\infty}(M)$ is outgoing if $u\in x_1^{(d-1)/4-\mi\lambda/2}
C^{\infty}(\overline{M}_{\mathrm{even}})$. We can define resonances of
$(M,g)$ as the poles of $R(\lambda)$. The set of resonances is contained
in the set of extended resonanaces. However, there could be extended
resonances which are not resonances. This means that the singular part
of $\hat{H}^{-1}(\lambda)$ is annihilated by restricting to $M$. Thus,
we could not define resonant states for the asymptotically hyperbolic
case, as there could also exist an outgoing solution with $f=0$ for
some $\lambda$ which is not a resonance.

%-----------------------------------------------
% DISCUSSIONS
%-----------------------------------------------
\section{Optimization Insight\label{sec:perspec}}

Now, we should discuss the present theoretical framework from optimization
insight. Section \ref{sec:disscusions} gives discussions on chemical
insight of the principle of least action. Section \ref{sec:outlook}
discusses the reaction dynamics in general spacetime. Sections
\ref{sec:qc4qd} and \ref{sec:gener-ai} give perspectives on potential applies
of quantum computation and generative artificial intelligence (GenAI) to
molecular reaction dynamics, respectively. In implememntations of
optimization within the variational
principle, the stochastic characteristics of practical optimization
might be significant because of stochastic algorithms, inherent
randomness in some algorithms, and hardware issues. This implies
necessity to discuss the numerical accuracy of optimization from
the perspective of stochastic process dynamics.
Sections \ref{sec:kinetics} gives precision of optimizations
from the viewpoint of statistical mechanics by introducing the
thermodynamics uncertainty relation. Section \ref{sec:markov-proc}
simply describes the Markov process because optimization result depends
only on the current setup, not on the sequence of past optimizations.

%---------------------------------------
\subsection{Principle of Least Action in Geometry\label{sec:disscusions}}

First, we should turn to Section \ref{sec:math-frame} providing further
mathematical interpretations on the principle of least action (Postulate
\ref{post:pls}) that plays a central role in introducing the requirement
of the variational principles. In particular, we should give more discussions
on Definition \ref{def:hopl-part-dyn-action} and Remark \ref{rem:qm-path-in}.
As is well known, the principle of least action in Equation
\eqref{eq:000-least-action} states that solutions extremize the action
with fixed end points on each finite time interval. Until now, we show
nothing about the end points, in particular elliptic fixed point, that
is fundamentally stable. It is the case for chemical reaction, where
all reactants and products are in motion around their equilibria, but
stable. Therefore, a deep connection between the local behavior near
fixed points and the minimal action should be discussed through the
Birkhoff normal form.
\begin{defi}
The Birkhoff normal form is a simplified canonical representation of
a Hamiltonian system near its equilibrium, obtained by a sequence of
near-identity coordinate transformations that remove as many nonlinear
terms as possible.
\label{def:birk-normal-form-00}
\end{defi}
By Definition \ref{def:birk-normal-form-00}, the Birkhoff normal form
is essentially a geometric formulation of perturbation theory for the
dynamics of a Hamiltonian system.
\begin{defi}
Perturbation theory considers the effects of small disturbances (or
perturbations) on the behaviour of a system, irrespective of the
source of the disturbance to find approximate solutions to problems
that cannot be solved exactly.
\label{def:perb-theo}
\end{defi}
For a general elliptic fixed point, the symplectic character of the map
(that connects the elliptic fixed points) permits a transformation to
symplectic coordinates that puts it into the particularly simple Birkhoff
normal form. In other words, due to the freedom to change coordinates
in a way that preserves volume in configuration space, one can simplify
the map near the point through coordinate transformation to capture the
essence of the dynamics leading to the Birkhoff normal form. This is
similar to perturbation theory, as indicated by Definition \ref{def:birk-normal-form-00}.
For instance, let $H=H_0+V$ be Hamiltonian of a nonharmonic oscillator
as function of coordinate $q$ with $q_0$ the equilibrium point, where
$H_0$ is Hamiltonian of a harmonic oscillator and $V$ a real-valued
function due to the Taylor expansion
\begin{equation}
H\big(q\big)=H\big(q_0\big)+\sum_{k=1}^r
\left.\frac{\mathrm{d}^kH}{\mathrm{d}q^k}\right\vert_{q=q_0}
\cdot\big(q-q_0\big)^k+\mathcal{O}(q^{r+1}).
\label{eq:birkho-norm-def-0}
\end{equation}
Thus, there exists a real analytical canonical transformation making
$H$ in the form $H_0+\mathcal{Z}+\mathcal{O}(q^{r+1})$, where
$\mathcal{Z}$ is a polynomial of order $r\geq3$ in normal form.

Second, noting that the Birkhoff normal form is a simplified polynomial
expression, the coefficients of this normal form are symplectically
invariant and hence called Birkhoff invariants. In this context, as
long as the coordinates are transferred symplectically, the Birkhoff
invariants remain the same and thus a kind of fingerprint of the fixed
point. From the Birkhoff normal form, one obtains an asymptotic
approximation whcih coincides with the original map only up to a term
that vanishes asymptotically when one approaches the fixed point. It
was shown in Section \ref{sec:math-frame} that there is symplectically
invariant minimal action $\alpha$ given by Equation \eqref{eq:003-least-action}
associated with a volume-preserving map near a general elliptic fixed
point. It is a local invariant and contains dynamics information near
the fixed point. Moreover, the Taylor coefficients of the convex
conjugate $\alpha^*$ (Legendre transform of $\alpha$) are the Birkhoff
invariants. Next, the germ of the minimal action can be associated with
a general elliptic closed geodesic on a two-dimensional Riemannian
manifold. The minimal action carries information about the geodesic
flow near the closed geodesic. In higher dimensions, considering a
symplectic diffeomorphism $\phi$ in a neighbourhood of an invariant
torus $\Lambda$, if the dynamics on $\Lambda$ satisfy a certain
non-resonance condition, one can transform $\phi$ into Birkhoff normal
form. If it is positive definite, the map $\phi$ determines the germ
of the minimal action $\alpha$ and the minimal action contains the
Birkhoff invariants as Taylor coefficients of $\alpha^*$. Therefore,
for elliptic and non-resonant Hamiltonian systems, the minimal action
contains all local dynamical information, which can be seen as the
coefficients of the Birkhoff normal form. In this context, unlike
the elliptic and non-resonant case, near a hyperbolic transition state
the minimal action germ does not encode all local dynamics, and the
Birkhoff normal form series diverges, implying that some perturbation
theory inapplicable and some others applicable.
\begin{remark}
A stable molecule without symmetry-imposed degeneracy or accidental
resonances is a prototypical elliptic and non-resonant Hamiltonian
system. A symmetric molecule is a typical resonant Hamiltonian system,
but may or may not be elliptic, where degeneracies or resonances (say
Fermi resonance) are present.
\label{rem:elliptic-00}
\end{remark}
\begin{remark}
In molecular reaction dynamics, the transition state is a hyperbolic
invariant set, which is either a hyperbolic equilibrium (namely saddle
point) on the PES or a normally hyperbolic invariant manifold forming
a dividing surface between reactants and products. Thus, the dynamics
near a transition state is hyperbolic, making the transition state a
prototypical example of a hyperbolic structure in the Hamiltonian system.
\label{rem:elliptic-01}
\end{remark}

Finally, instead of single system shown in the above sections, let us
simply discuss all systems on a symplectic manifold $(M,\omega_0)$ at
once, collected in the Hamiltonian diffeomorphism group Ham($M,\omega_0$)
which carries the Hofer metric. Here, it should be recalled that
$\omega_0$ is the symplectic form. 
\begin{remark}
Formally, Hofer geometry considers the space of Hamiltonian diffeomorphisms
equipped with the Hofer metric. It measures ``distance'' between two
Hamiltonian systems in terms of the energy cost needed to deform one
into the other.
\label{rem:hofer-geom}
\end{remark}
In general, the Hofer metric on
Ham($M,\omega_0$) is defined by $d(\phi,\psi)=\Vert\phi^{-1}\cdot\psi\Vert$
with $\forall\phi,\psi\in\mathrm{Ham}(M,\omega_0)$. The group Ham($M,\omega_0$)
can be seen as infinite-dimensional Lie group with Lie algebra composed
of all smooth, compactly supported functions $H:M\to\mathbb{R}$ with
mean value of zero. If any function norm $\Vert\cdot\Vert$ is invariant
under the adjoint action, the Hofer distance of a diffeomorphism $\phi$
from the identity is the infimum of the lengths of all paths in
Ham($M,\omega_0$) that connect $\phi$ to the identity. Since the Hamiltonian
system is determined by the first derivatives of the Hamiltonian, which
is variant under the adjoint action, the appropriate choice may be the
oscillation norm $\Vert\cdot\Vert=\max-\min$. Despite its simple definition,
the Hofer distance is hard to obtain making one know vary little about
the relation between the Hofer geometry and dynamics of a Hamiltonian
diffeomorphism. For Hamiltonians on the cotangent bundle $T^*\mathbb{T}^d$
satisfying a Legendre condition, one can define the minimal action
$\alpha$ for convex Lagrangians on $T^*\mathbb{T}^d$. As given in Section
\ref{sec:math-frame}, $\mathbb{T}^d$ is $d$-dimensional torus which is
a periodic, compact, smooth manifold. For a moelcular reaction system,
however, the Hamiltonian is unbounded and cannot fit into the framework
of the Hofer metric. To consider the stable states of reactants and
products, the Hamiltonian has to be restricted to a compact part of
$T^*\mathbb{T}^d$, such as the unit ball cotangent bundle $B^*\mathbb{T}^d$,
allowing us to stay in the range of Mather theory that provides a
variational framework for understanding the complex behavior of
dynamical systems, particularly the persistence of minimal invariant
sets (Mather sets) under perturbations. Consequently, there exists a
relation between the Hofer geometry of convex Hamiltonian mappings and
their dynamical behavior, that is, the minimal action bounds the Hofer
distance, linking variational dynamics to symplectic topology.
\begin{remark}
Restricting the reaction Hamiltonian to the unit ball $B^*\mathbb{T}^d$
is an {\it ad hoc} truncation for considering states of reactants and
products. It is well-known that the molecular reaction requires unbounded
momenta in the phase space to move over barrier. Such truncation is
a mathematical convenience to consider dynamics of the reactants and
products using perturbation theory, but not a physical approximation.
This is thus consistent with Remark \ref{rem:elliptic-01}.
\label{rem:perturb-000}
\end{remark}

%--------------------------------
\subsection{Reaction Dynamics in Geometry\label{sec:outlook}}

Now, we should turn to the mountain pass theorem (Proposition \ref{theo:mpt})
given in Section \ref{sec:mpt-math}, which predicts existence of the
saddle points between two separated minima. There are two aspects of
Proposition \ref{theo:mpt} worth paying attention to. On one hand,
Proposition \ref{theo:mpt} is widely applied to functional variation
and predicts existence of a saddle point between two local minima in
space, such as parameter set $\theta$ in defining the approximated
function $f_{\theta}$ (see Equation \eqref{eq:tangent-bundle-00-ai}).
Proposition \ref{theo:mpt} and Corollary \ref{coro:mpt-coro} indicate
that, when optimizing a complex functional, it is quite common for the
optimization to converge to a local minimum rather than the global
minimum. The presence of saddle point can hinder optimization iterations
from escaping one local minimum to reach another, often resulting in
outcomes that are either stochastic in nature or exhibit significant
fluctuations. This highlights the inherent challenges in non-convex
optimization, where the landscape of the objective function is rugged
and the path to the true global extremum is obstructed by numerous
critical points. Consequently, the choice of initial guess and
optimization algorithm plays a crucial role in determining the final
result. On the other hand, Proposition \ref{theo:mpt} indicates that
between two local minima (namely intermediates)
on the PES, there must exist at least one first-order saddle point
(namely transition state), which corresponds to a pathway of elementary
reaction. For special cases, there is no saddle point on the PES if one
of the intermediates is not an energy minimum, such as in reactions
involving two radicals. However, when statistical effects are taken
into account, it becomes necessary to consider the damping effect of
non-reactive motions on the reaction, meaning that the change of the
reaction must be examined on the free energy surface (see also Remark
\ref{rem:pass-mountain-th}). On such surface, a transition state can
usually be identified. To understand this point, it is worth noting
that the free energy barrier of elementary reaction combines the
potential-energetic resistance from the PES and the and the entropic
resistance from statistics and disorder. For reactions without potential
energy barrier, the latter term arises the transition state.
\begin{remark}
As noted in the beginning of Section \ref{sec:boa-app}, the separation
of DOFs inevitably induces gauge interactions. For instance, the separation
of electronic and nuclear motions leads to gauge effects on nuclear dynamics.
Consequently, the decoupling of subsystem and bath DOFs can also produce
gauge interactions under suitable conditions, and such gauge interactions
may be interpreted as the source of free energy. In this context, free
energy reflects the impact of the external environment (namely the bath
of reaction subsystem) on molecular reactions, including statistical effects
(such as temperature and pressure), as well as the influence of non-reactive
ro-vibrational motions on the reaction.
\label{rem:pass-mount-990}
\end{remark}

Here, we would like discuss free energy in more detail. As is well known,
the motion of a molecular system in configuration space is confined to
the PES. Even if the initial condition of the reaction is changed, the
system can never ``leave'' the PES. Geometrically, this represents a
form of boundary rigidity phenomenon in geometry. Nevertheless, if we
consider mode-/bond-specific reactions and state-to-state reactions,
the PES can only roughly determine each individual channel, but cannot
encompass all the dynamical information. However, the initial state, as
an eigenstate, is computed from the PES, indicating the existence of
some invariant connection. Moreover, when statistical effects are taken
into account, the molecular system is alternatively constrained to the
free energy surface. In this case, boundary rigidity is lost. However,
one must note that at least entropy still serves as a connecting link
between the free energy and the potential energy. That is, the dynamical
features on the PES influence the motion of the system on the free energy
surface and leave indelible traces. This is related with the concept of
the non-removable intersection in geometry. Next, let us turn to boundary
rigidity and non-removable intersection.
Letting $\Sigma$ be a hepersurface bounding a domain $U_{\Sigma}$ and
containing Lagrangian submanifold $\Lambda$, one can summarize existence
of boundary rigidity as the question whether can $\Lambda$ be pushed
inside $U_{\Sigma}$ by an exact Lagrangian deformation that can preserve
the Liouville class $a_{\Lambda}$ or not. If one cannot even move
$\Lambda$ at all, then we call this phenomenon boundary rigidity.
Setting $\mathcal{L}$ to be the class of all Lagrangian submanifolds
of $T^*X$, which are Lagrangian isotopic to the zero section
$\mathfrak{O}$, if $\Lambda,\Upsilon\in\mathcal{L}$ are Lagrangian
submanifolds both lying in an optical hypersurface $\Sigma$ and in
$U_{\Sigma}$ with the same Liouville class $a_{\Upsilon}=a_{\Lambda}$,
then $\Lambda$ and $\Upsilon$ can be found to be identical, that is
$\Upsilon=\Lambda$. This result is applicable to a wider class of
hypersurfaces and does not require strict convexity. Even if boundary
rigidity may fail for a given Lagrangian submanifold $\Lambda$, in
many cases the intersection $\Lambda\cap\Sigma$ cannot be empty set,
called non-removable intersection. The intersection always contains an
invariant set of the characteristic foliation of $\Sigma$.
\begin{remark}
By boundary rigidity in symplectic geometry, a semiclassical quantum
state cannot be completely confined below the PES without changing its
topological phase and must intersect the PES at turning points that
are connected by classical trajectories. Thus, we give a rigorous
formulation of the relationship between Hamiltonian flow on the PES
and the geometry of Lagrangian submanifolds that represent semiclassical
quantum states.
\label{rem:bou-rigi-non-rem-0}
\end{remark}

Finally, the present geometric theory of molecular reaction dynamics
might open a way to consider reaction in curved spacetime. As is well
known, the $[3+1]$ decomposition given by Definition \ref{def:3-1-decompo}
has been widely applied to develop numerical general relativity through
methods of hydrodynamics. Algorithms of hydrodynamics in conjunction with
general relativity become the foundation for simulating astrophysical
phenomena where strong gravity interacts with matter. In essence, the
$[3+1]$ decomposition enables the evolution of curved spacetime, and
the hydrodynamics describes how the matter moves and interacts on the
spacetime. Moreover, choosing appropriate ensemble as boundary conditions,
the reaction kinetics can be simulated through hydrodynamics in phase
space if the ensemble of the system is considered as a continuous fluid.
That is, the motion of the ensemble in phase space and the motion of a
fluid in configuration space share the same mathematical structures.
Therefore, simulation of reaction kinetics in phase space can be extended
to a general spacetime framework. In this context, since the configuration
space becomes curved, the corresponding phase space can accordingly be
constructed in a generally curved one. As a result, the algorithms
developed for relativistic hydrodynamics within the framework of
numerical general relativity can be naturally applied to simulate
reaction kinetics.
\begin{remark}
To introduce statistical-mechanical effects, one constructs an
ensemble of the system of interest and treats it as a continuous
fluid in phase space with suitable boundary conditions. The reaction
kinetics then emerges from hydrodynamic simulations of this phase-space
fluid, linking molecular reaction dynamics to reaction kinetics. For
instance, sampling initial states of reactants ensemble in phase space,
integrating EOMs to obtain trajectories, and counting how fast and how
many cross a dividing surface into products, obtaining rate constants,
product distributions, and recrossing coefficients.
\label{rem:kinetics-reaction}
\end{remark}
Of course, there exist many theoretical problems on
this issue. For example, how to write the EOM in curved spacetime, how
to construct the Lagrangian or Hamiltonian of the molecular system in
curved spacetime, and how to develop hydrodynamics EOM in curved phase
space for reaction kinetics. A more interesting theoretical issue is
what would happen if the measurements were made by another observer,
that is a non-Eulerian observer. In Sections \ref{sec:theory} and
\ref{sec:hamiltonian}, we have provided ways to find the EOM and to
construct the Hamiltonian. Nevertheless, it is still an open question
to finally solve the above problems. As recently discussed \cite{zha25:20397},
it may be possible to extend the SPA to its field-theory counterpart
through similar pathway in condensed matter many-body theory. If
molecular reaction dynamics were fully understood within the quantum
field theory, it would then be possible to further generalize it to
couple with theory of curved spacetime.

%----------------------------------------
\subsection{Towarding to Quantum Computing\label{sec:qc4qd}}

Now, we would like to mention the fact that, although
all of current theories and algorithms for quantum molecular dynamics
were based on real numbers, satndard quantum theory was formulated in
terms of operators acting on complex Hilbert spaces. Now, we have a
look at the quantum mechanics based on real numbers
\cite{wei20:060406,ren21:625,wu22:140401,li22:040402,bar26:240202},
which might play an important role in quantum simulation of chemical
reaction given in Section \ref{sec:qc-simulatior}. In 2020, Weilenmann
and Colbeck \cite{wei20:060406} analyzed a range of generalized probabilistic
theories (GPTs) and found that none of them perform better than quantum
theory.
\begin{defi}
Generalized probabilistic theories (GPTs) are a class of theories that
include classical and quantum theory as well as theories with more nonlocal
correlations still, where the state space can be expressed as a compact
convex set in a real vector space and measurements are collections of
effects, linear maps from states to probabilities, while joint states
defined for multipartite systems lead to nonsignaling correlations.
\label{def:gpts-multipa-sys}
\end{defi}
Weilenmann and Colbeck \cite{wei20:060406} proposed a new way to axiomatize
quantum theory, enabling an experimental test that simultaneously rules out
all nonquantum GPTs. Starting from the above theoretical results \cite{wei20:060406},
Navascu{\'e}s and co-workers \cite{ren21:625} experimentally observed
that complex quantum theory outperforms real quantum theory when a
non-local game is played in the entanglement-swapping scenario. Later,
two independent experiments \cite{wu22:140401,li22:040402} demonstrated
that a complex formulation of quantum mechanics rather than the real one
is necessary to reproduce experimental results. However, the conclusion
that complex quantum theory is indispensable to describe microscopic
motion true only if we accept the postulates of standard quantum mechanics.
As Navascu{\'e}s and co-workers \cite{ren21:625} point out, this conclusion
would not be applicable based on different postulates (say Bohmian mechanics).
Therefore, these experiments \cite{ren21:625,wu22:140401,li22:040402}
could stimulate attempts to go beyond the standard quantum mechanics,
which is sometimes considered inadequate \cite{nor17:book}.

In contrast to the conclusions by above experiments \cite{ren21:625,wu22:140401,li22:040402}
that real-valued quantum mechanics can be falsified, Barrios Hita and
co-workers \cite{bar26:240202} theoretically argued that real-valued
quantum mechanics cannot be falsified. To figure out the conclusion that
the use of complex numbers in quantum mechanics is a matter of convenience,
Barrios Hita and co-workers \cite{bar26:240202} proposed four postulates
for both complex and real quantum mechanics. First, there is a complex
(real) Hilbert space $\mathbb{H}_{\mathrm A}$ in which the state of
system A is represented by a normalized vector $\vert\phi\rangle$. Second,
A measurement $\Pi$ of system A corresponds to an ensemble of positive
semi-definite operator $\{\hat\Pi_r\}$ indexed by the measurement result
$r$, acting on $\mathbb{H}_{\mathrm A}$ and obeying $\sum_r\hat\Pi_r=I$,
where $I$ is the identity. Third, the probability of the $r$-th measuring
result is $P(r)=\langle\phi\vert\hat\Pi_r\vert\phi\rangle$. Fourth, the
operators described measurements of system A act trivially on other system
B, and vice versa. Given a ficed basis, a state vector of a $d$-dimensional
system can be represent by matrix $\vert\psi\rangle=(c_1\;c_2\;\cdots\;c_d)^{\mathrm T}$,
which can also be given by $(\mathrm{Re}c_1\;\mathrm{Re}c_2\;\cdots\;
\mathrm{Re}c_d,\mathrm{Im}c_1\;\mathrm{Im}c_2\;\cdots\;\mathrm{Im}c_d)^{\mathrm T}$.
Denoting $\mathrm{Re}\vert\psi\rangle=(\mathrm{Re}c_1\;\mathrm{Re}c_2\;
\cdots\;\mathrm{Re}c_d)^{\mathrm T}$ and $\mathrm{Im}\vert\psi\rangle=
(\mathrm{Im}c_1\;\mathrm{Im}c_2\;\cdots\;\mathrm{Im}c_d)^{\mathrm T}$,
a map from the complex to the real representation of a pure state can
be given by
\begin{equation}
\mathfrak{S}:\mathbb{C}^d\to\mathbb{R}^{2d}:
\vert\psi\rangle\mapsto\mathrm{Re}\vert\psi\rangle\otimes\vert0\rangle_F
+\mathrm{Im}\vert\psi\rangle\otimes\vert1\rangle_F=
\vert\tilde{\psi}\rangle,\quad\vert\psi\rangle\in\mathbb{C}^d,
\quad\vert\tilde{\psi}\rangle\in\mathbb{R}^{2d},
\label{eq:phase-real-qm-00}
\end{equation}
where the subscript stands for flag whose two real DOFs indicate the
real and imaginary part of $\vert\psi\rangle$. Although $\mathfrak{S}:
\vert\psi\rangle\mapsto\vert\tilde{\psi}\rangle$ formally resembles
the structure of an entangled state, $\vert\tilde{\psi}\rangle$ is a
single state and thus is not entangled.

As given in Section \ref{sec:eom-prob}, a normalized state $\vert\psi\rangle$
is equilvalent to $\vert\psi\rangle\exp(\mi\varphi)$ with $\varphi\in\mathbb{R}$.
With the global phase $\varphi$, the map $\mathfrak{S}$ satisfies
\cite{bar26:240202}
\begin{equation}
\mathfrak{S}:\vert\psi\rangle\exp\big(\mi\varphi\big)\mapsto
\mathrm{Re}\vert\psi\rangle\otimes
\Big(\vert0\rangle_F\cos\big(\varphi)+\vert1\rangle_F
\sin\big(\varphi)\Big)+\mathrm{Im}\vert\psi\rangle\otimes
\Big(-\vert0\rangle_F\sin\big(\varphi)+\vert1\rangle_F
\cos\big(\varphi)\Big)=R_F\big(\varphi\big)\mathfrak{S}
\big(\vert\psi\rangle\big),
\label{eq:phase-real-qm-01}
\end{equation}
which means that $\varphi$ or U(1) ambiguity in a complex state vector
corresponds to an SO(2) ambiguity in the real formulation. The physical
indistinguishability of two real state vectors $\vert\tilde{\psi}\rangle$
and $\vert\tilde{\phi}\rangle=\vert\tilde{\psi}\rangle R_F(\varphi)$
is guaranteed if and only if the real measurement operators lead to
identical outcome probabilities. If the map from a complex operator to
its real representation is denoted by $\mathfrak{T}$, while the real
image of a complex operator $\hat{\Pi}$ is correspondingly
$\widehat{\tilde{\Pi}}=\mathfrak{T}(\hat{\Pi})$, then such
indistinguishability requires $\langle\tilde{\psi}\vert
\widehat{\tilde{\Pi}}\vert\tilde{\psi}\rangle=
\langle\tilde{\phi}\vert\widehat{\tilde{\Pi}}\vert\tilde{\phi}\rangle=
\langle\tilde{\psi}\vert R_F^{\dagger}(\varphi)\widehat{\tilde{\Pi}}
R_F(\varphi)\vert\tilde{\psi}\rangle$. This indicates that
$\widehat{\tilde{\Pi}}$ must commute with $R_F(\varphi)$ for
$\forall\varphi\in\mathbb{R}$. Due to $\mathfrak{S}(\hat\Pi\vert\psi\rangle)
=\mathfrak{T}(\hat\Pi)\mathfrak{S}(\vert\psi\rangle)$, real operators
are given by \cite{bar26:240202}
\begin{equation}
\mathfrak{T}(\hat{\Pi})=\mathrm{Re}(\hat{\Pi})\otimes I_F+
\mathrm{Im}(\hat{\Pi})\otimes J_F,\quad
I_F=\left(\begin{array}{cc}
1 & 0 \\
0 & 1
\end{array}\right)\quad
J_F=\left(\begin{array}{cc}
0 & -1 \\
1 & 0
\end{array}\right),
\label{eq:phase-real-qm-02}
\end{equation}
where $I_F$ and $J_F$ are identity and symplectic matrices, respectively,
while $\mathrm{Re}(\hat{\Pi})$ and $\mathrm{Im}(\hat{\Pi})$ are real and
imaginary parts of the elements of $\hat{\Pi}$ in the basis for defining
$\mathrm{Re}\vert\psi\rangle$ and $\mathrm{Im}\vert\psi\rangle$,
respectively. The real representation in Equation \eqref{eq:phase-real-qm-02}
preserves the algebraic structure $\mathfrak{T}(\hat{A}_1\hat{A}_2)=
\mathfrak{T}(\hat{A}_1)\mathfrak{T}(\hat{A}_2)$ for any observables
$A_1$ and $A_2$ and satisfies $\mathfrak{S}(\vert\psi\rangle)^{\mathrm{T}}
\mathfrak{T}(\hat{A})\mathfrak{S}(\vert\psi\rangle)=\langle\psi\vert\hat{A}
\vert\psi\rangle$. This expression means that the maps $\mathfrak{S}$
and $\mathfrak{T}$ leave all expectation values invariant \cite{bar26:240202}.

By extending $\mathfrak{S}$ in Equation \eqref{eq:phase-real-qm-00} to
$N$-body system whose $N$ subsystsems are simply supposed to be all
represented by $d$-dimensional vectors $\{\vert\psi_i\rangle\in\mathbb{C}^d\}_{i=1}^d$,
one can find that the real and complex quantum theories are isomorphic
\cite{bar26:240202}. However, a natural extension of Equation \eqref{eq:phase-real-qm-00}
to $\mathfrak{S}^{\otimes N}:\vert\psi_1\rangle\otimes\cdots\otimes
\vert\psi_N\rangle\mapsto\mathfrak{S}(\vert\psi_1\rangle)\otimes\cdots
\otimes\mathfrak{S}(\vert\psi_N\rangle)$ is not well defined. This is
because a vector in $\bigotimes_{i=1}^N\mathbb{C}_i^d$ has $2Nd$ real
parameters, while a vector in $\bigotimes_{i=1}^N\mathbb{R}_i^{2d}$
has $(2d)^N$ parameters (see also Equation \eqref{eq:phase-real-qm-00}).
To overcome this problem, we must note that each of independent subsystems
can have an undetectable phase. Due to linearity in each subsystem, a
global phase is equal to the product of the individual phases and hence
can be split up between the subsystems in an arbitrary way (also denoted
by phase kickback). Since $\mathfrak{S}$ in Equation \eqref{eq:phase-real-qm-00}
is invertible and its extension $(\mathfrak{S}^{-1})^{\otimes N}$ is
well defined, we further consider the invertible map
\begin{equation}
\mathfrak{Q}:\mathbb{C}^d\otimes\cdots\otimes\mathbb{C}^d\to
\mathbb{R}^{2d}\otimes\cdots\otimes\mathbb{R}^{2d}/
\ker\Big[\big(\mathfrak{S}^{-1}\big)^{\otimes N}\Big],
\label{eq:phase-real-qm-03}
\end{equation}
where the kernel of $(\mathfrak{S}^{-1})^{\otimes N}$ is employed, as
two elements in the domain of $(\mathfrak{S}^{-1})^{\otimes N}$ have
the same image. Each element in the quotient space in the right-hand
side of Equation \eqref{eq:phase-real-qm-03} is an equivalence class
that is $\vert v\rangle\sim\vert w\rangle$ if $\vert v\rangle-\vert
w\rangle\in\ker[(\mathfrak{S}^{-1})^{\otimes N}]$. Thus, one can
distinguish $2d^N$ linearly independent equivalence classes $[\vert
r_i\rangle\otimes\vert0_1\cdots0_i\cdots0_N\rangle_F]$ and $[\vert
r_i\rangle\otimes\vert0_1\cdots1_i\cdots0_N\rangle_F]$ for $\{\vert
r_i\rangle\}_{i=1}^{d^N}$ a basis of $\mathbb{R}^N$ and can define
\cite{bar26:240202}
\begin{equation}
\big(\mathbb{R}^d\big)^{\otimes N}\otimes
\big(\mathbb{R}^2\big)^{\otimes N}/
\ker\Big[\big(\mathfrak{S}^{-1}\big)^{\otimes N}\Big]=
\mathrm{span}\left(\bigcup_{i=1}^{d^N}
\left\{\Big[\vert r_i\rangle\otimes\vert0_1
\cdots0_i\cdots0_N\rangle_F\Big],
\Big[\vert r_i\rangle\otimes\vert0_1\cdots1_i
\cdots0_N\rangle_F\Big]
\right\}\right).
\label{eq:phase-real-qm-04}
\end{equation}
For any $\vert v\rangle\in(\mathbb{R}^d)^{\otimes N}\otimes
(\mathbb{R}^2)^{\otimes N}$, the canonical form of a representative of
the equivalence class $[\vert v\rangle]$ is $\hat{P}^{\perp}\vert v\rangle$,
where $\hat{P}^{\perp}$ is the projector onto
$\ker[(\mathfrak{S}^{-1})^{\otimes N}]^{\perp}$. Then, one can have the
projection \cite{bar26:240202}
\begin{equation}
\mathfrak{P}:\big(\mathbb{R}^d\big)^{\otimes N}\otimes
\big(\mathbb{R}^2\big)^{\otimes N}/
\ker\Big[\big(\mathfrak{S}^{-1}\big)^{\otimes N}\Big]\to
\ker\Big[\big(\mathfrak{S}^{-1}\big)^{\otimes N}\Big]^{\perp}
\subset\big(\mathbb{R}^d\big)^{\otimes N}:\big[\vert v\rangle\big]
\mapsto\vert v\rangle^{\perp}=\hat{P}^{\perp}\vert v\rangle.
\label{eq:phase-real-qm-05}
\end{equation}
It should be noted that $\vert v\rangle^{\perp}$ is the canonical
representative of the equivalence class and that $\vert v\rangle^{\perp}$
is unique and independent of the choice of $\vert v\rangle$ from the
equivalence class. In this context, one can define a pure state in
real quantum mechanics as separable if and only if it is either a
product state or equivalent to a product state \cite{bar26:240202},
as given in Equation \eqref{eq:def-state-spa-geom-000}. Otherwise,
it is entangled. Moreover, the embedding of the local actions on
subsystems into the operators for the composite system is well
defined \cite{bar26:240202}. The reduced density operator in the
real-valued formalism is still given via the (real) partial trace.

%---------------------------------------------------------------
\subsection{Applications of Generative Artificial Intelligence\label{sec:gener-ai}}

Usually, GenAI refers to AI models that can generate new content based
on patterns learned from training data and can serve as a creative assistant
that helps us explore alternative possibilities, as indicated in Remarks
\ref{rem:general-ai} and \ref{rem:general-ai-llm}. In particular, GenAI
provides data-driven solutions to PDEs by learning from data based on
the minimax methods in mathematics (see also Section \ref{sec:mpt-math}).
Over the past years, it was well discussed on the relationship between
the mountain pass theorem and the minimax methods, as indicated in the
end of Section \ref{sec:mpt-math}, together with their applications to
solving PDEs. For instance, to solve semilinear elliptic problem in
configuration space $X\in\mathbb{R}^d$,
\begin{equation}
-\nabla^2u+\lambda u=-f\big(u\big),\;\lambda>0,\; d\neq3,\;
u\in\mathcal{H}^1\big(\mathbb{R}^d\big)=\Big\{u\in
\mathcal{L}^2\big(\mathbb{R}^d,\mathbb{R}\big)\Big\vert\boldsymbol{\nabla}u\in
\mathcal{L}^2\big(\mathbb{R}^d,\mathbb{R}\big)\Big\},
\label{eq:semi-lin-ell-999}
\end{equation}
where $\mathcal{H}^1(\mathbb{R}^d)$ is Sobolev space, the ground state
solution can be obtained through the minimax of the associated functional
\begin{equation}
I\big[u(X)\big]=\frac{1}{2}\int_{\mathbb{R}^d}\Big(\big\vert\boldsymbol{\nabla}
u\big\vert^2+\lambda u^2\Big)\mathrm{d}^dX-\int_{\mathbb{R}^d}F\big(u\big)
\mathrm{d}^dX,\quad
F\big(u\big)=\int_0^uf\big(t\big)\mathrm{d}t.
\label{eq:semi-lin-ell-998}
\end{equation}
Maia and co-workers \cite{mai21:642} transferred this solution to the
minimum of the functional $I[u]$ constrained to the Poho\v{z}aev manifold
\begin{equation}
P_O=\Big\{u\in\mathcal{H}^1\big(\mathbb{R}^d\big)\setminus\big\{0\big\}
\Big\vert J(u)=0\Big\},\quad
J(u)=\big(d-2\big)\int_{\mathbb{R}^d}\big\vert\boldsymbol{\nabla}u\big\vert^2
\mathrm{d}^dX-2d\int_{\mathbb{R}^d}\left(-\frac{\lambda}{2}u^2+F(u)\right)
\mathrm{d}^dX.
\label{eq:pohozaev-mani-999}
\end{equation}
Therefore, Maia and co-workers \cite{mai21:642} found that the ground
state solution of Equation \eqref{eq:semi-lin-ell-999} can be given by
$\min_{u\in P_O}I[u]$. This result implies that one can solve special
type of PDEs through an variational method. Moreover, noting that
problem of Equation \eqref{eq:semi-lin-ell-999} is a type of the SE,
it is possible to develope optimization methods for solving the SE
based on the minimax methods making the GenAI applicable in solving
the SE.

To implement the GenAI model for solving the SE, the wave function is
parameterized as an AI model (such as NN) that takes the coordinates
as input and outputs a complex number as the probability amplitude and
phase factor. The model should be designed to respect physical symmetries.
Then, the model parameters are optimized to minimize some target functional
based on the variational principle. This is obviously similar to the
framework based on the ansatz in Equation \eqref{eq:def-state-spa-geom-000}
and the variational principle. For solving electronic-structure problem,
notable successes have achieved highly accurate ground-state energies
as well as lower-lying excited-states energies for small molecules.
Furthermore, while still computationally intensive, the AI model can
be trained from additional information obtained by enforcing the physical
laws leading to physics-informed mechine learning \cite{kar21:422}.
Such physics-informed model integrates noisy data and mathematical
models, and implements them through NNs or KMR methods. It may be
possible to design specialized architectures that automatically satisfy
some of the physical laws for better accuracy, faster training and
improved generalization \cite{kar21:422}. For instance, operator
learning \cite{azi24:320,sub26:123} may offer a new pathway for
solving the SE by learning the mapping from initial state to future
states, bypassing traditional step-by-step numerical time integration.
For time-independent Hamiltonian, the solution operator, that is the
time evoluation operator, is unitary, implying that a more data-efficient
learning approach may involve parameterizing a prespecified operator
class such that every operator in it is unitary \cite{sub26:123}.
Extending existing parameterizations of unitary matrices could provide
a useful starting point. Once trained, the wave function at any future
time can be predicted. Althrough challenges remain in ensuring unitarity
and handling high-dimensional systems, operator learning \cite{azi24:320,sub26:123}
is emerging as a powerful tool for accelerating quantum dynamics calculations,
complementing traditional methods rather than replacing them.

For another example, a chemistry-informed genAI model was proposed by some of
us \cite{mia24:532} providing the chemistry-informed generative adversarial
network (CI-GAN) approach. Answering the question of how to sample the
training data in the configurational space, the original CI-GAN approach
was developed \cite{mia24:532}. To easily build database for complex
molecular systems, an image-input intersection was also implemented \cite{mia24:532}
making CI-GAN have capability to directly recognize the molecular image.
Therefore, the CI-GAN approach can use the training data in either a
numerical fashion or in image. Calculations and analysis \cite{mia24:532}
on typical examples, H + H$_2$, OH + HO$_2$, and H$_2$O/TiO$_2$(110),
indicated that CI-GAN can generate distributions of geometry and
associated energy making it have the capability to replace classical
dynamics simulations as well as electronic structure calculations. On
the basis of these calculations, deep discussions on the power of
CI-GAN were given. Due to its power, systematical and comprehensive
studies on the predictions of the dynamics through GenAI are necessary.
Nevertheless, it is worth noting that GenAI is not new in the fields
of chemistry. Many numerical developments should be considered in next
studies on molecular reaction dynamics. First, one should compare the
generated geometries and energies with a realistic target and then
systematically consider the accuracy of the GenAI models. Second, the
computational efficiency of GenAI should be further systematically
inspected, in particular, if we were to apply it to constructing the
PES. Third, for CI-GAN, influences of the input training data and the
chemistry inspiration contained in the training data should be carefully
considered.

As previously discussed \cite{mia24:532}, calculations for
benchmarks implied that CI-GAN has the potential capability to predict
classical trajectories and electronic energy values avoiding expensive
solving of equations for them, {\it i.e.} Newton equation for trajectory
and electronic SE for electronic structure. Chemistry insight of the
CI-GAN generated trajectory depends on that contained in the training
data. If the training data were computed for fitting the PES, then the
distribution of the energy values in the configuration space can be
obtained by training the CI-GAN model \cite{mia24:532}. With such a
model, one can generate the energy values rather than solving the
electronic SE. If the training data were computed in time-series order,
the generated geometries should, in principle, in time-series order.
Thus, it was expected \cite{mia24:532} that the CI-GAN model has
capability to generate time-dependent geometries and thus obtain
trajectories. Nevertheless, applications of time series analysis to
molecular reaction dynamics through GenAI is not a simple and direct
extension of the current GenCI approaches (such as CI-GAN) because the
time-dependent wave function is certainly different from geometries,
associated energies, and classical trajectories. In generating the wave
function propagation, the target is given by the tensor form, say
Equation \eqref{eq:def-state-spa-geom-000}, instead of those in
function (say the PES) or in scalar (say the classical trajectory).
Turning to Section \ref{sec:mpt-math} and above discussions by Maia and
co-workers \cite{mai21:642}, either the mountain pass theorem together
with the minimax methods or the variational methods for functional were
developed for scalar. The applicability of the minimax approaches to
the tensor still requires further investigation from the viewpoint of
either mathematics or implementation. As given in the end of Section
\ref{sec:mpt-math}, similar existence of critical point for vector-valued
functional was proved by Molho and co-workers \cite{bed11:569}. The
critical point of vector-valued functions was found to be a solution
of a minimax problem consisting of an inner vector maximization problem
and of an outer set-valued minimization problem.

The above PDE-oriented bottom-top framework, where solving PDE is the
core goal and GenAI a powerful tool, is common in the computational
chemistry community. There exist alternative perspectives in data-driven
computing community. As given in above sections, solving the SE (one
kind of PDEs) has been the main tool for describing molecular reaction
dynamics. Consequent simulations on the reaction kinetics based on the
solutions of the SE have provided many discoveries, but facing major
limitations for complex reaction systems, such as combusion kinetics,
atmospheric chemistry, chemical biology, and so on. The main problem
in simulating these complex systems with bottom-top framework is
prohibitively expensive. Nevertheless, data-driven methodologies
present new opportunities and frontiers. For instance, Koumoutsakos
\cite{kou25:e29} suggested that developing algorithmic alloys between
AI and scientific computing offers new tools to tackle complex systems.
Furthermore, understanding the world through AI models that interact
with and challenge those built around PDEs offers new views and
scientific frontiers. A key limitation of data-driven methods is the
lack of interpretability of the resulting model. Unlike the bottom-top
framework (for example, see the end of Section \ref{sec:pes-build-regree}),
they are focused on prediction, but do not provide governing equations
or clearly interpretable relationship in terms of the original dataset.
An alternative kind of data-driven approaches employs symbolic regression
to identify directly the structure of a dynamical system from database.
Over the past years, such interpretable data-driven approaches work
remarkably well for discovering dynamics models, implying the diversity
of strategies employed to extract meaningful dynamics from database. It
has been increasingly recognized that AI models designed by either sparse
regression or neural networks may be capable of learning physics and
chemistry principles and constraints, if from database alone the AI
model can enact (i) the discovery of first-principles models, (ii)
the identification of physical constraints and conservation laws,
and (iii) improved models using known physics. Data-driven methods
allow us to develop black box and gray box modeling strategies for
complex dynamics where basic principle is only partially known. In
addition, data-driven methods are required not only to allow us to
impose physics constraints, or bake-in physics, but also to discover
physical constraints that need to be baked-in. Thus, we require that
these models can constrain learning and can learn constraints.

%------------------------------------------
% kinetics
%------------------------------------------
\subsection{Precision of Optimization\label{sec:kinetics}}

\begin{figure}[h!]
  \centering
   \includegraphics[width=17cm]{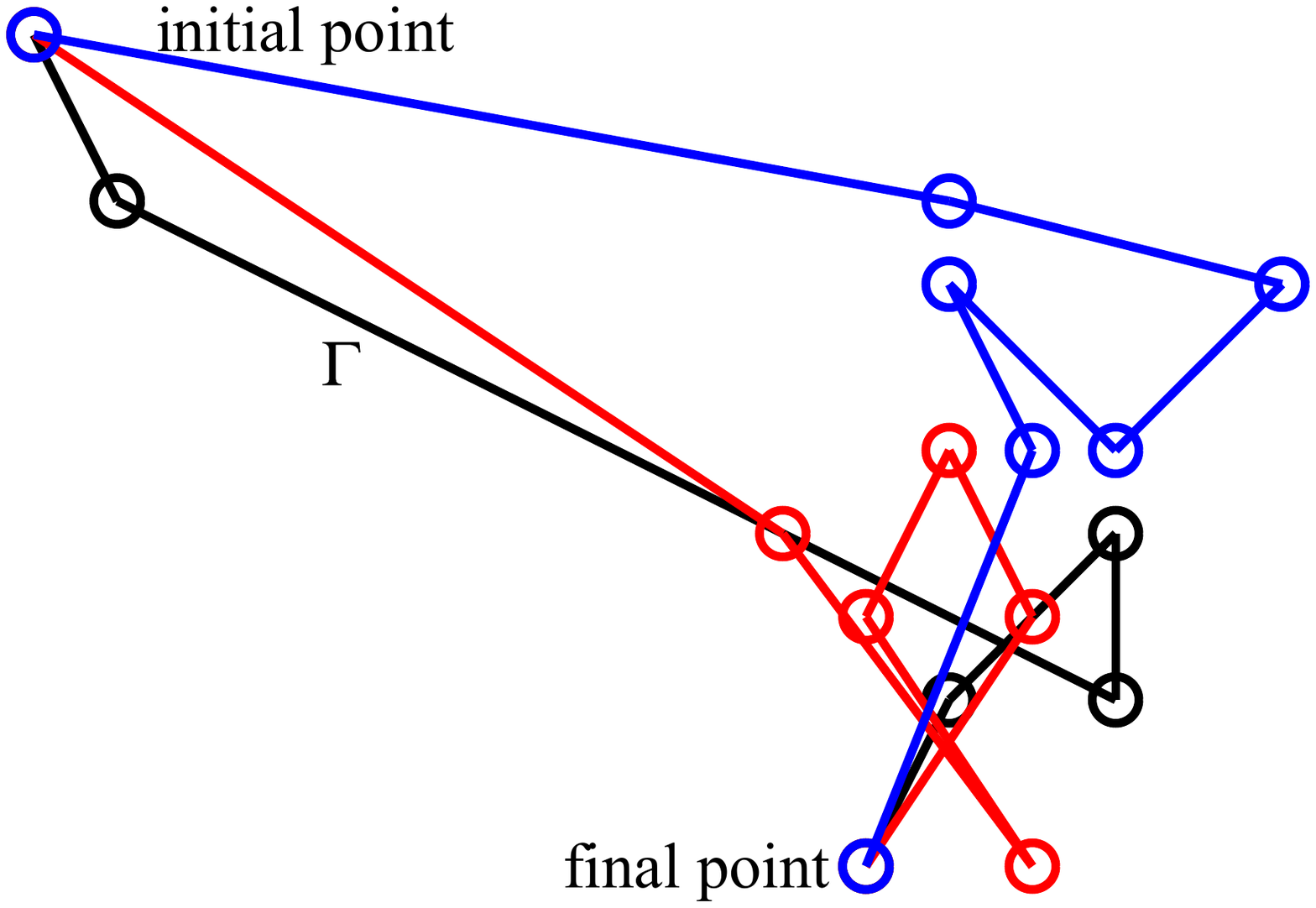}
    \caption{\figfoot}
     \label{fig:lear-net}
      \end{figure}

As given in Section \ref{sec:dynamics}, the present framework of molecular
reaction dynamics was devloped on the basis of the variational principle,
in which a specified functional should be optimized. In practical optimization,
stochastic algorithms may be employed to improve initialization sampling,
escape local minima, enhance convergence efficiency, and so on. In addition,
the inherent randomness in some algorithms and hardware, such as stochastic
gradient descent, random initialization, and floating-point rounding, make
the stochastic characteristics of practical optimization become significant.
In this context, practical optimization of a functional can be interpreted
as a stochastic process unfolding in parameter space, which implies that
a kinetic description provides a natural and powerful framework for its
analysis (see Figure \ref{fig:lear-net} for schematic illustration). By
optimization, one finds a way that contains the largest amount of the
information in the functional with specified restrictions and/or conditions,
as illustrated by Figure \ref{fig:mountain-pass}. By the mountain pass
theorem (Proposition \ref{theo:mpt} and Corollary \ref{coro:mpt-coro}),
there exists critical point on the manifold making such way exist. This
is a typical constrained optimization problem, where the marginal
likelihood $\ln\mathcal{P}(\theta,\cdot)$ of a ``landscape'' can be
chosen to be the optimization target. According to Bayesian inference,
extremizing $\ln\mathcal{P}(\theta,\cdot)$ yields the final parameters
$\theta$. At the first glance, one may analogize $\mathcal{P}(\theta,\cdot)$
with probability distribution of microstates of the parameter set
$\theta$ making $\ln\mathcal{P}(\theta,\cdot)$ a kind of entropy. For
a given macrostate of the parameter set $\theta$, to determine the
entropy is equilvalent to finding the minimum number
of questions required to be answered to fully specify the microstate
of $\theta$. The final way (the red dash line in Figure \ref{fig:mountain-pass})
typically centers around the mean value but also on examining its
behavior up to its tails, that is probability distribution. This
consists in assessing the likelihood of a stochastic process
fluctuating from its usual state to a particular rare value. This is
theory of large deviations (LD) in statistical mechanics, where one
can precisely and systematically estimate the probability of observing
a particular rare event relative to its most likely value. Now,
determining the way that contains the largest amount of information at
a given preciseness is such rare event and hence needs evaluation of
its observing probability. It is worth noting that molecular reaction
is also a kind of rare event in the configuration space, implying
statistical similarities between molecular reaction and stochastic
deviations. Indeed, a large deviation framework has already well
known to understand the mechanism of rare transitions between
metastable states.

As illustrated in Figure \ref{fig:lear-net}, an optimization process
can be decomposed into a series of transformations $\mathfrak{Z}'
\leftarrow\mathfrak{Z}$ forming a path $\Gamma$ in the space of
intermediate parameter values represented by $\mathfrak{z}$. There are
various possibilities of the optimization path, even though the initial
and finial states are fixed. Thus, the quality of the optimization can
be represented by an additive property,
\begin{equation}
A(t)=\int_0^t\mathrm{d}\tau\alpha(\tau),
\label{eq:time-additive-00}
\end{equation}
where $\alpha(\tau)$ depends on an underlying stochastic dynamics at
$\tau$. Moreover, we would like to mention that parameters $t$ and
$\tau$ may be time or may not time. Equation \eqref{eq:time-additive-00}
is very similar to the expression of action given in Equation \eqref{eq:000-least-action}.
The former represents an intermediate state in the optimization process,
while the latter represents a possible state of motion. In this context,
quantity $A(t)$ and $\alpha(\tau)$ in Equation \eqref{eq:time-additive-00}
are analogous to the action and Lagrangian, respectively. Due to the
stochastic feature, values of $A(t)$ fluctuate around a steady value
that represents its more likely outcome. This is similar to the principle
of least action because both optimization property $A(t)$ and action
describe a certain intermediate state during the change. Characterizing
the probability distribution $\mathcal{P}(A,t)$ of $A(t)$ at $t$ to
describe events of interest, quantity $A(t)$ is approximately given by
$A(t)=at$ in large enough $t$ with $t\to\infty$. By the LD theory, we
can generally have
\begin{equation}
\mathcal{P}\big(A=at,t\big)\asymp\exp\big(-tI(a)\big),\quad
\lim_{t\to\infty}\frac{1}{t}\ln\mathcal{P}\big(A=at,t\big)=-I(a),
\label{eq:time-additive-01}
\end{equation}
where $I(a)$ is a rate function. It quantifies the rate at which the
probability of observing a value $A$ deviates from that of its typical
value. If $I(a)$ reaches its minimum in $\overline{a}$, then $\overline{a}$
is the typical value of $A/t$. Equation \eqref{eq:time-additive-01}
means that convergence rate of the stochastic optimization process
satisfies a relationship similar to that of the reaction rate, further
demonstrating the similarity between them.

Next, we should consider the rate function $I(a)$ to show its insight.
According to the LD theory, the optimization process is equivalent to
working in a conditional probability distribution
\begin{equation}
\mathcal{P}\big(\mathfrak{z}\big\vert A/t=a\big)=
\int\mathcal{D}\big(\mathfrak{Z}\big)
p\big(\mathfrak{Z}=\mathfrak{z}\big)
\delta\big(A(\mathfrak{Z})/t-a\big),
\label{eq:time-additive-02}
\end{equation}
where $\mathcal{D}(\mathfrak{Z})$ represents path integral over all
trajectories depending on $\mathfrak{Z}$ (see also Figure
\ref{fig:lear-net}), while $p(\cdot)$ represents distribution function.
Equation \eqref{eq:time-additive-02} implies that conditional probability
$\mathcal{P}(\mathfrak{z}\vert A/t=a)$ is the dynamical analogy to a
microcanonical ensemble in equilibrium systems where $A(t)$ is arbitrary
and plays a role of constraining variable. Following the G{\"a}rtner-Ellis
theorem \cite{gar77:24,ell84:1}, which is a fundamental result in the LD
theory, one can obtain the rate function as
\begin{equation}
I(a)=\max_{\lambda}\lambda a-\lim_{t\to\infty}\frac{1}{t}
\ln\left\langle\exp\Big(\lambda A\Big)\right\rangle
=\max_{\lambda}\lambda a-\psi(\lambda).
\label{eq:time-additive-05}
\end{equation}
The extremalization expression in Equation \eqref{eq:time-additive-05}
gives precisely $\lambda$ conjugated to $a$ and further indicates that
the likelihood of fluctuations of $A(t)$ can be obtained from the
limiting calculation of $\psi(\lambda)$ alone. In Equation
\eqref{eq:time-additive-05}, we introduce rate function and thus
averaged dynamics in optimization process where the averaging process
sums over all possible optimization paths. As illustrated in Figure
\ref{fig:lear-net}, we denote an optimization path in the intermediate
space by $\Gamma$. An operation of parameter reversal $\Gamma\to\tilde{\Gamma}$
can be defined as that changes the order of the intermediate states
and reverses the $t$-dependence contained in the $t$-dependent driving
protocols. In optimization process, such parameter reversal operation
is realizable because it is exactly the process of continuously adjusting
values in the parameter space.
\begin{remark}
If one is aware that $\psi(\lambda)$ plays the role of the pressure and
the rate function $I(a)$ is the entropy, the G{\"a}rtner-Ellis theorem
provides a rigorous basis for computing free energy and entropy in
non-equilibrium system.
\label{rem:ge-theorem}
\end{remark}
\begin{remark}
In a driven system with time-reversible dynamics, the probability of
observing a given amount of entropy production $\mathcal{S}$ is
exponentially larger than the probability of observing its opposite by
a factor of $\exp(\mathcal{S})$. This is the detailed fluctuation theorem.
By this theorem, the second law of thermodynamics is violated in a
quantitative manner, demonstrating that it is not an absolute prohibition.
\label{rem:dft-theorem}
\end{remark}

With concept of path and the detailed fluctuation theorem, the path level
entropy production can be further defined as
\begin{equation}
\mathcal{S}\big(\Gamma\big)=k_{\mathrm{B}}\ln\frac{\mathcal{P}(\Gamma)}{\tilde{\mathcal{P}}(\Gamma)},
\label{eq:fluctuation-data-000}
\end{equation}
where $k_{\mathrm{B}}$ is the Boltzmann constant. Rearranging Equation
\eqref{eq:fluctuation-data-000} and averaging the results, one can
find the integral fluctuation theorem
\begin{equation}
\left\langle\exp\left(-\frac{\mathcal{S}}{k_{\mathrm{B}}}\right)\right\rangle=
\sum_{\Gamma}\mathcal{P}(\Gamma)\frac{\tilde{\mathcal{P}}(\tilde{\Gamma})}{\mathcal{P}(\Gamma)}=
\sum_{\Gamma}\tilde{\mathcal{P}}(\tilde{\Gamma})=
\sum_{\tilde{\Gamma}}\tilde{\mathcal{P}}(\tilde{\Gamma})=1.
\label{eq:fluctuation-data-001}
\end{equation}
To obtain the third equality of Equation \eqref{eq:fluctuation-data-001},
one should recognize that the sum over all paths can be enumerated
equivalently using the forward or the reversed paths. Averaging the
entropy production defined by Equation \eqref{eq:fluctuation-data-000},
one can find the inequality
\begin{equation}
\big\langle\mathcal{S}\big\rangle=k_{\mathrm{B}}\sum_{\Gamma}\mathcal{P}(\Gamma)
\ln\frac{\mathcal{P}(\Gamma)}{\tilde{\mathcal{P}}(\tilde{\Gamma})}\geq0.
\label{eq:fluctuation-data-002}
\end{equation}
Combining Equations \eqref{eq:fluctuation-data-001} and
\eqref{eq:fluctuation-data-002}, one directly has got so-called Jensen
inequality
\begin{equation}
1=\left\langle\exp\left(-\frac{\mathcal{S}}{k_{\mathrm{B}}}\right)\right\rangle
\geq\exp\left(\frac{\langle\mathcal{S}\rangle}{k_{\mathrm{B}}}\right)\geq
1-\frac{\langle\mathcal{S}\rangle}{k_{\mathrm{B}}},
\label{eq:fluctuation-data-003}
\end{equation}
which is essentially an expression of the thermodynamic uncertainty
relation (TUR) \cite{bar15:158101}. The TUR expression bounds the
signal-to-noise ratio of a current $\mathcal{J}$, such as measure of
the precision during the optimization process, in terms of the entropy
production needed to achieve it \cite{bar15:158101},
\begin{equation}
\mathcal{P}\big(\mathcal{J}\big)=\frac{\langle\mathcal{J}\rangle^2}{\langle(\mathcal{J}
-\langle\mathcal{J}\rangle)^2\rangle}
\leq\frac{\langle\mathcal{S}\rangle}{2k_{\mathrm{B}}},
\quad2\mathcal{P}\big(\mathcal{J}\big)k_{\mathrm{B}}\leq\langle\mathcal{S}\rangle.
\label{eq:fluctuation-data-004}
\end{equation}
Obviously, Equation \eqref{eq:fluctuation-data-004} provides a bound
relation on the precision in terms of an observable and further implies
that such observable can be exploited either to optimize the precision
or to infer the entropy production. Therefore, by the entropy production
of the optimization process it is possible to estimate the optimization
precision.

%------------------------------------------------
\subsection{Markov Process in Optimization\label{sec:markov-proc}}

As is well known, model of Markov state \cite{hus18:2386} is a powerful
framework for analyzing stochastic processes and has gained widespread
use over the past several decades. Now, let us consider its application
to determining the distribution $p(\mathfrak{z},t)$ which is useful to
obtain $\mathcal{P}$ by Equation \eqref{eq:time-additive-02}. Dynamics
of an optimization process is determined by transition rates
$r(\mathfrak{z}\leftarrow\mathfrak{z}')$ and satisfies the master
equation for the probability distribution of states,
\begin{equation}
\frac{\partial}{\partial t}p(\mathfrak{z},t)=
\sum_{\mathfrak{z}'\neq\mathfrak{z}}
\Big[r(\mathfrak{z}\leftarrow\mathfrak{z}')p(\mathfrak{z}',t)
-r(\mathfrak{z}'\leftarrow\mathfrak{z})p(\mathfrak{z},t)\Big].
\label{eq:master-equation-000}
\end{equation}
The additive observable $A$ defined by Equation \eqref{eq:time-additive-00}
is written as a summation of $K$ jumps,
\begin{equation}
A=\sum_{0\leq k\leq K-1}\alpha_{\mathfrak{z}_k\mathfrak{z}_{k+1}},
\label{eq:master-equation-001}
\end{equation}
where $\alpha_{\mathfrak{z}_k\mathfrak{z}_{k+1}}$ are individual
contributions along the path of successively states $\{\mathfrak{z}_k\}_{k=0}^K$.
To obtain the likelihood of observing fluctuations of $A$, according to
Equation \eqref{eq:master-equation-000} we consider a joint probability
distribution satisfying relation
\begin{equation}
\frac{\partial}{\partial t}p(\mathfrak{z},A,t)=
\sum_{\mathfrak{z}'\neq\mathfrak{z}}
\Big[r(\mathfrak{z}\leftarrow\mathfrak{z}')
p(\mathfrak{z}',A-\alpha_{\mathfrak{z}'\mathfrak{z}},t)-
r(\mathfrak{z}'\leftarrow\mathfrak{z})p(\mathfrak{z},A,t)\Big],
\label{eq:master-equation-002}
\end{equation}
where $\alpha_{\mathfrak{z}\mathfrak{z}'}$ represents the amount by
which $A$ is increased during a jump from $\mathfrak{z}$ to $\mathfrak{z}'$
(see also Equation \eqref{eq:master-equation-001}). Transferring the
microcanonical probability distribution $p(\mathfrak{z},A,t)$ to the
biased canonical one
\begin{equation}
\tilde{p}\big(\mathfrak{z},\lambda,t\big)=\sum_A\exp\big(\lambda A\big)
p\big(\mathfrak{z},A,t\big),
\label{eq:master-equation-003}
\end{equation}
one can transfer Equation \eqref{eq:master-equation-002} into joint
probability distribution that satisfies
\begin{equation}
\frac{\partial}{\partial t}\tilde{p}\big(\mathfrak{z},\lambda,t\big)=
\sum_{\mathfrak{z}'\neq\mathfrak{z}}
\Big[\exp\Big(-\lambda\alpha_{\mathfrak{z}'\mathfrak{z}}\Big)
r(\mathfrak{z}\leftarrow\mathfrak{z}')
\tilde{p}\big(\mathfrak{z}',\lambda,t\big)-
r(\mathfrak{z}'\leftarrow\mathfrak{z})
\tilde{p}\big(\mathfrak{z},\lambda,t\big)\Big],
\label{eq:master-equation-004}
\end{equation}
where $\lambda$ is the conjugate variable with $A$. Employing Equation
\eqref{eq:time-additive-05} we have got
\begin{equation}
\sum_{\mathfrak{z}}\tilde{p}\big(\mathfrak{z},\lambda,t\big)
=\big\langle\exp\big(\lambda A\big)\big\rangle
\underset{t\to\infty}{\asymp}
\exp\big(t\psi(\lambda)\big).
\label{eq:master-equation-005}
\end{equation}
By Equations \eqref{eq:master-equation-004} and \eqref{eq:master-equation-005},
one can obtain expectation values with respect to $\tilde{p}(\mathfrak{z},\lambda,t)$
and hence the likelihood of fluctuations of $A(t)$.

Here, the optimization process has been transferred into the Markov
process composed by a series of intermediate states. Dynamics of this
process is thus required two ways to focus on these interacting states.
In one way, we consider these states on a fixed network and increase
the number of intermediate states until there is a large number of
states on each site. In other way, we consider intermediate states on
a lattice or in continuous space and rescale space until the density of
states $\rho(\cdot)$ at every point is finite. Obviously, the first way
generally represents dynamics of population models. A typical example
of population model is chemical reaction networks. The second way is
dynamics of diffusion models of $\rho(\cdot)$ even if the resulting
dynamics is sometimes determined trajectory, that is ballistic. In
either way, the Markov process can be seen as an optimization process
on the action that is a functional of path with physical insight of
interest. For a finite step $\delta t$, the probability of a path is
a simple product
\begin{equation}
P_t\Big[\big\{\rho_k\big\}\Big]=\prod_{k=1}^K
P_{\delta t}\big(\rho_{k+1}\big\vert\rho_k\big).
\label{eq:path-integral-000}
\end{equation}
Keeping only the initial and final conditions fixed (see also Figure
\ref{fig:lear-net}), a finite transition probability in terms of these
paths is obtained by Equation \eqref{eq:path-integral-000} as summation
of probabilities associared with all of probable paths,
\begin{equation}
P_t\big(\rho_K\big\vert\rho_0\big)=\sum_{\{\rho_k\}}P_t\Big[\big\{\rho_k\big\}\Big]=
\sum_{\{\rho_k\}}\prod_{k=1}^{K-1}
P_{\delta t}\big(\rho_{k+1}\big\vert\rho_k\big)
=\sum_{\{\lambda_k\}}\prod_{k=0}^{K-1}
P_{\delta t}\big(\lambda_k\big\vert\rho_k\big)
\delta\big(\rho_{k+1}-\rho_k-\nabla\cdot\lambda\delta t\big).
\label{eq:path-integral-001}
\end{equation}
Limitation of $\delta t\to0$ turns Equation \eqref{eq:path-integral-001}
into path integral where the variable is either displacement rate
$\dot{\rho}$ or current $\lambda$. With an extra continuity constraint
implemented by delta functions, Equation \eqref{eq:path-integral-001}
is further given by
\begin{align}
P_t\big(\rho_t\big\vert\rho_0\big)&\asymp
\int\exp\left(-\int_0^t\mathfrak{L}(\dot{\rho},\rho)\mathrm{d}\tau\right)
\delta\big(\rho(t)-\rho_t\big)\delta\big(\rho(0)-\rho_0\big)
\mathrm{d}\big[\rho(\tau)\big]
\allowdisplaybreaks[4] \nonumber \\
&\asymp\int\exp\left(-\int_0^t\mathscr{L}(\lambda,\rho)\mathrm{d}\tau\right)
\delta\big[\dot{\rho}+\nabla\cdot\lambda\big]
\delta\big(\rho(t)-\rho_t\big)\delta\big(\rho(0)-\rho_0\big)
\mathrm{d}\big[\lambda(\tau)\big],
\label{eq:path-integral-002}
\end{align}
where $\mathscr{L}(\lambda,\rho)$ and $\mathfrak{L}(\dot{\rho},\rho)$ are detailed
Lagrangian and standard Lagrangian, respectively. The Lagrangians
$\mathscr{L}(\lambda,\rho)$ and $\mathfrak{L}(\dot{\rho},\rho)$ are related with
$P_{\delta t}(\lambda\vert\rho)$ and $P_{\delta t}(\dot{\rho}\vert\rho)$,
respectively. The two exponentials in Equation \eqref{eq:path-integral-002}
are thus actions giving, respectively, the detailed action and standard
action.

Therefore, the optimization process corresponds to one path in the
intermediate parameter space (say path $\Gamma$ in Figure \ref{fig:lear-net}).
This path connects the initial setup and the finally optimized result.
However, it is necessary to further consider the optimization paths
with various initial setups, that is EOMs of intermediate states. To
this end, one might note that we have got Lagrangian of intermediate
sates by Equation \eqref{eq:path-integral-002}. By Lagrangian, the
principle of least action can be used to obtain the reuslting
Euler-Lagrange equation,
\begin{equation}
\partial_{\rho}\mathfrak{L}-\frac{\mathrm{d}}{\mathrm{d}t}
\partial_{\dot{\rho}}\mathfrak{L}=0,\quad
\nabla\partial_{\rho}\mathscr{L}+\frac{\mathrm{d}}{\mathrm{d}t}
\partial_{\lambda}\mathscr{L}=0.
\label{eq:path-integral-003}
\end{equation}
Moreover, one can define Hamiltonian of the optimization Markov process
through Legendre transformations,
\begin{equation}
\mathfrak{H}(u,\rho)=\sup_{u}\Big[u\cdot\dot{\rho}-
\mathfrak{L}(\dot{\rho},\rho)\Big],\quad
\mathscr{H}(f,\rho)=\sup_{\lambda}\Big[f\cdot\lambda-\mathscr{L}(\lambda,\rho)\Big],
\label{eq:path-integral-004}
\end{equation}
where $u=\partial_{\dot{\rho}}\mathfrak{L}$ and $f=\partial_{\lambda}\mathscr{L}$.
By Equation \eqref{eq:path-integral-004}, one has got
\begin{equation}
\dot{u}=-\partial_{\rho}\mathfrak{H},\quad
\dot{\rho}=\partial_u\mathfrak{H};\quad
\dot{f}=\nabla\cdot\partial_{\rho}\mathscr{H},\quad
\dot{\rho}=-\nabla\cdot\partial_f\mathscr{H}.
\label{eq:path-integral-005}
\end{equation}
Equatoins \eqref{eq:path-integral-003} and \eqref{eq:path-integral-005}
both are EOMs for the optimization process which has been seen as the
Markov process. One can directly find that these EOMs have identical
formalism to those of classical mechanics. This is not surprising
because principle of least action is adopted.

Now, we would like to take back from the mathematics for Markov process
to its physical-insight that one might typically want to consider. To
this end, the intermediate states in optimization process are virtually
supposed to be particles randomly walking in optimizing flux. The
restrictions on the optimization process are specific source of the
randomness, such as implementation of regression model and quality of
training database in building the PES (that is a typical optimization
process). At every point in optimization, these restrictions exert an
effective force on the randomly optimizing process whose distribution
is independent of the intermediate states. According to Equation
\eqref{eq:path-integral-005}, the dynamical variable $f$ of the
detailed Hamiltonian $\mathscr{H}(f,\rho)$ can be interpreted as the
random force produced by the restrictions. When there are fixed external
forces, they will appear additively with $f$. This is particularly
clear for the process with detailed balance. In general, this variable
can instead be taken as the definition of a force being the quantity
conjugate to flux through Legendre transformations. The variable $u$
of the standard Hamiltonian $\mathfrak{H}(u,\rho)$ can be interpreted
as a random potential, in situations where all random forces derive
from potentials. All of observables of optimization process can be
analysed or expressed in terms of solutions to EOMs. It is often
complicated to explicitly perform. In order to make things simpler,
one usually observe the optimization process at long times where it
converges to characteristic and interpretable behaviours. Such long-time
behaviours for stochastic process is mostly characterised by the
attractors of the deterministic dynamics.

%-----------------------------------------------------------------------------
% Conclusions
%-----------------------------------------------------------------------------
\section{Conclusions\label{sec:con}}

In this work, based on previous perspectives \cite{zha25:20397} on
wavepacket propagation, we try to provide a unified geometric theory
of moleuclar reaction dynamics. In general, the previously reported
hierarchical framework \cite{zha25:20397} centers on the concept of
modes that combine several coordinates along with their hierarchical
separations and revolves around the variational methods. From a unified
perspective \cite{zha25:20397}, all of steps in this framework are
optimizations performed in an appropriately defined space employing
variational methods for special functional. As the beginning, the
present theory provides a description on the SPA that provides a
powful ansatz to solve the Schr{\"o}dinger equation of either
electronic structure or wavepacket propagation by the variational
principle. Its core idea centers on the concept of modes that combine
several particles or DOFs \cite{zha25:20397}. In this context, one can
formulate almost all of variation-based electronic-structure and wavepacket
propagation approaches. In addition, the hierarchical framework
\cite{zha25:20397} can be rearranged by TN or TTN. Previously comparing
the function-based methods with those in the TN or TTN form, the
present theory provides a unified framework for the ansatz. Moreover,
construction of the nuclear Hamiltonian operator that is a summation
of the KEO and the PES can be also unified in the variation-based
framework. The present unified theory provides general expression of
the KEO from the viewpoint of geometric characteristics of configuration
space which is the space of nuclear conformation. Using the curvilinear
cooridnates, configuration space is curved. Its curvature depends on the
coordinate frame and the curvature of the spacetime where the reaction
occurs. For the PES, based on the unified regression model, we unify the
PES construction into a geometric optimization by which one can suggest
probable pathways to develop new regression methods. Next, separation
of the fast and slow DOFs leads to the Berry phase effects of the fast
motions. The present theory provides a geometric interpretation of the
Berry phase effects and clarifies its role in reaction dynamics. Based
on the present theory, perspectives on gauge field theory of chemical
reaction are given. Finally, practical optimization often employs stochastic
algorithms to improve initialization sampling, escape local minima, and
enhance convergence efficiency. In addition, stochastic characteristics
of practical optimization can be significant due to the inherent randomness
in some algorithms, such as stochastic gradient descent and random
initialization, and hardware effects, such as floating-point rounding.
Thus, we also discuss the numerical accuracy of optimization from the
perspective of stochastic process dynamics, aiming to deepen understanding
of the practical accuracy limits.

%--------------------------------------------------------------------------
% Supplementary Material
%-------------------------------------------------------------------------
\section*{Supplementary Material}

The Supporting Information documents are available free of charge at
https://www.doi.org/X. In the Supporting Information file, we give
details of abbreviations, notations, and operators used in this
work.

%--------------------------------------------
% Declaration of interests
%----------------------------------------------
\section*{Interest Statement}

The authors declare that they have no known competing financial interests
or personal relationships that could have appeared to influence the work
reported in this paper.

%----------------------------------------------------------------
% Data availability
%-----------------------------------------------------------------
\section*{Data Availability}

All data have been reported in this work.

%--------------------------------------------------------------------
% Acknowledgements
%--------------------------------------------------------------------
\section*{Acknowledgments}

The financial supports of National Natural Science Foundation of China
(Grant No. 22273074) and Fundamental Research Funds for the Central
Universities (Grant Nos. 2025JGZY34 and 2025KCW017) are gratefully
acknowledged. The authors are also grateful to anonymous reviewers
for their thoughtful suggestions.

%------------------------------------------------
% Tables
%-------------------------------------------------
\section*{Tables}

%-------------------------------------
\clearpage
\begin{table}
\caption{Roles of the variational principles in different issues of
molecular reaction dynamics. The second coulumn gives the issues that
need to be solved. The third coulumn gives target functional to be
optimized. The fourth coulumn gives physical insight corresponding to
the target functional and its variation. The rightmost coulumn gives
section of the main text in which the corresponding issue is given.
}
\begin{tabular}{lclclclcr}
\hline
No. &~~& Issue &~~& Functional &~~& Description &~~& Section \\
\hline
1 && the principle && action && It summarizes dynamics of a system 
&& \ref{sec:math-frame} \\
 && of least action && && by providing EOM. && \\
2 && the principle  && kineitc energy &&
 It provides the Laplacian operator for && \ref{sec:keo-metric} \\
  && of equivalence && of free particle && the spacetime with a given metric.  &&  \\
3 && adiabatically  && distance between the && Infimum of the distance remains 
 && \ref{sec:boa-app} \\ 
  && decoupled subset && subset and its complement && strictly greater than zero. &&  \\
4 && coordinate && kinetic energy of &&
It provides coordinates frame and corresponding && \ref{sec:coord-frame} \\  
  && frame       && many-body system && Laplacian operator of compound reaction system && \\
5 && function && loss function && 
It provides the potential function of  && \ref{sec:pes-build-regree} \\
  && approximation && of a model && the configuration space. &&  \\
6 && eigenvalue && $\langle\Psi[A_I]\vert\hat{H}\vert\Psi[A_I]\rangle$ && 
It provides working equations for single-partical && \ref{sec:elec-struc} \\
 && && && terms and expansion coefficients. && \ref{sec:eigen-state} \\
7 && time evolution && $\langle\Psi[A_I]\vert\mi\partial_t-\hat{H}\vert\Psi[A_I]\rangle$ &&
It provides working equations for single-partical  && \ref{sec:elec-stru} \\
 && && && terms and expansion coefficients. && \\
\hline
\end{tabular}
\label{tab:summ-var-gauge}
\end{table}

%--------------------------------------
\clearpage
\begin{table}
\caption{Preliminaries of the present unified geometric framework of
molecular reaction dynamics, as given in Sections \ref{sec:theory} and
\ref{sec:hamiltonian}. These theoretical preliminaries contain
(1) the principle of least action, (2) the mountain pass theorem,
(3) the principle of equivalence, (4) the principle of local
Lorentz invariance, (4) coordinate frame of the molecular system, and
(5) constructions of the KEO and PES. The principle of least action,
the principle of equivalence, and the principle of local Lorentz
invariance are three postulates of the present theory. The mountain
pass theorem predicts
existence of the saddle point during the variational optimization of
the functional, that is existence of transition state which is key to
determining the reaction mechanism. The Hamiltonian constructions play
a foundation role in dynamical theory. The first column gives the numbers
of these objectives. The second column gives object of the present
theoretical framework. The third column describes the details of the
corresponding object. The fourth column gives remarks. The rightmost
column indicates section of the main text in which the corresponding
object is discussed.
}
\label{tab:theory-comp}
\end{table}
%%%%%%%%%%%%%%%%%%%%%
\clearpage
\begin{sidewaystable}
\begin{tabular}{lclclclcr}
\hline
No. &~~& Object &~~& Description &~~& Remark &~~& Section \\
\hline
1 && the principle of
&& It provides the working equations through variation of different
&& It is the only postulate of  && \ref{sec:math-frame} \\
&& least action
\footnote{The principle of least action states that action must take
its least value when the working equations are obeyed. It is the foundation of
dynamical theory.}
&& functionals enabling variational principles formulated on
&& the present framework. &&  \\
&& && distinct manifolds to serve as the mathematical foundation. && && \\
2 && the mountain
&& It states that moving from one locally optimized result to  &&
It is the foundation of  && \ref{sec:mpt-math} \\
&& pass theorem
\footnote{The mountain pass theorem states existence of a saddle point
between two local minima for a functional under suitable geometric and
compactness conditions.}
&& another result requires a certain amount of effort implying &&
mechanisms of an  &&  \\
&& && that optimization requires iteration. It also predicts that &&
elementary reaction. &&  \\
&& && there must exist a transition state between two intermediates. && && \\
3 && the principle  && It states that a free particle moves along geodesics in curved
&& Gravity is related with && \ref{sec:keo-metric} \\
  && of equivalence && spacetime, implying that gravity is not a force.
  && spacetime curvature. \footnote{Such relationship satisfies the Einstein field equations.}
  && \\
4 && the principle of local && It provides relation of metric tensors between &&
It is a key in deriving the KEO. && \ref{sec:coord-frame} \\
  && Lorentz invariance  && spatial slice and configuration space && && and \ref{sec:keo-coord} \\ 
5 && coordinate frame && It can be designed according to the system motion &&
It may depend on variation && \ref{sec:coord-frame} \\
&& && && of kinetic energy functional. && \\
6 && the KEO && It can be derived along geodesic of the configuration space. &&
It depends on metric of  && \ref{sec:keo-coord}  \\
&& && && the spacetime. && \\
7 && the PES && It can be constructed by optimizing potential tangent bundle. &&
It depends on metric of && \ref{sec:pes-build-regree} \\
&& && && the training data space.
\footnote{The training data are computed through extensive electronic-structure
energy calculations.} && \\
\hline
\end{tabular}
%\label{tab:theory-comp}
\end{sidewaystable}

%--------------------------------------
\clearpage
\begin{table}
\caption{Dynamical theory of the present unified geometric framework
of molecular reaction dynamics given in Sections \ref{sec:hamiltonian} and \ref{sec:dynamics},
including (1) the electromagnetic interaction in general spacetime, (2) the
single-particle approximation, (2) the variational principle, (3) time
evolution and geometric phase, and (4) gauge theory in molecular reaction.
The electromagnetic interaction plays a foundation role in dynamical theory
for molecular reaction.
The single-particle approximation gives ansatz for solving the Schr{\"o}dinger
equation in the sum-of-products form. The variational principle gives
working equations when ansatz is determined. Solving the working equation,
time evolution of the system can be ``seen'', while the geometric phase
might be arisen due to the system partition. The gauge field thoery may
thus be helpful to consider the gauge freedom. The first column gives
the numbers of these objectives. The second column gives object of the
present framework. The third column describes the details of the
corresponding object. The fourth column gives remarks. The rightmost
column indicates section of the main text in which the corresponding
object is discussed.
}
\label{tab:theory-dyn}
\end{table}
%%%%%%%%%%%%%%%%%%%%%
\clearpage
\begin{sidewaystable}
\begin{tabular}{lclclclcr}
\hline
No. &~~& Object &~~& Description &~~& Remark &~~& Section \\
\hline
1 && the electromagnetic && It provides elementary interactions in molecular system &&
It is especially important && \ref{sec:coumb-pot} \\
&& interaction && and a foundation of reaction dynamics in general spacetime. &&
for electronic structure. && \\
2 && the single-particle && It provides ansatz in the summation of products 
&& It partitions the many-body problem into && \ref{sec:elec-stru} \\
&& approximation && of single-particle terms, such as SPFs and MOs. && a set of coupled few-body problems. && \\
&& && It also introduces a gauge freedom into the system.
\footnote{The gauge freedom refers to the arbitrariness how to partition
the system into subsystems.}  && && \\
3 && the variational && It provides working equations by optimizing
&& The ansatz is simply substituted into && \ref{sec:math-struc}  \\
&& principle 
\footnote{The variational principle is arisen from the fact that the
principle of least action is adopted, as indicated in Table \ref{tab:theory-comp}.}
&& the target functional that depends on ansatz. 
\footnote{The functional of electronic energy is the target in electronic structure,
while the functional of action is the target in propagation.}
&& the variational
expression of the principle.
\footnote{The ansatz for electronic structure is substituted
into the linear variational principle, while the ansatz for propagation
is substituted into the time-dependent one.} && \\
4 && time evolution && Solving the working equations for SPFs, one can 
&& It is different from quantum scattering, && \ref{sec:eom-prob} \\
  && && obtain propagated wave function based on && which compares the initial and final states && \\
  && && which analysis can be made. && rather than propagating the initial state. && \\
5 && geometric phase && It is arisen from the single-particle approximation &&
If the phase undergoes a change of $\pi$, && \ref{sec:eom-prob} \\
&&  && that partitions the system into many subsystems. && it gives rise to quantum interference.
\footnote{A typical example is that when a molecular system encircles a
conical intersection, the electronic wave function undergoes a phase
change of $\pi$.} && \\
6 && gauge theory && It introduce a vector gauge field $A_{\mu}$ which &&
It provides the final unified geometric  && \ref{sec:eom-prob} \\
&& in reaction
\footnote{It is a mathematical framework that describes optimization
processes for molecular reaction dynamics through local symmetry.}
&& transforms and mediates the gradients in variations.
&& theory for molecular reaction dynamics. &&  \\
\hline
\end{tabular}
%\label{tab:theory-dyn}
\end{sidewaystable}

%----------------Compare geom-----------------------------------------
\clearpage
\begin{sidewaystable}
 \caption{%
Comparisons of the concepts in the geometric phase effects on the nuclear
and electronic properties, together with these concepts associated with
electromagnetic (EM) field. The second column provides geometric concepts
that will be compared to various fields. These concepts are the Berry
connection, Berry curvature, and geometric phase. The third and fourth
columns list physical insight of these concepts for the nuclear and
electronic properties. The fifth column lists the electromagnetic
quantities or properties associated with the geometric features. The
rightmost column gives remarks of these physical quantities or geometric
properties.
}%
 \begin{tabular}{lcllllllllr}
  \hline
No. &~~& Feature \footnote{For clarity, the term ``Berry'' is omitted for the Berry
 connection, Berry curvature, and Berry phase.}  
 &~~& Nuclear &~~& Electron &~~& EM field &~~& Remarks \\
\hline
1 && connection &~~& $A_{\mathrm{B}}$ in Equation \eqref{eq:trans-006-fiber-bundle}
           &~~& diagonal element of 
           &~~& analogous to the vector potential
\footnote{It is a concept in differential geometry but has similar
expression to the corresponding concepts in the classical EM field
leading to analogous phenomena.\label{foot:geom-em-field}}
           &~~& a gauge-dependent object   \\
&& && && electronic NACM % $\boldsymbol{\tau}$
\footnote{If the $j$-th electronic state $\upsilon_j(\mathbi{x},\mathbi{y})$
depends on electronic cooridnates $\mathbi{x}$ and parameters $\mathbi{y}$,
the elements of the electronic NACM satisfy
$\tau_{ij}^{\mathrm{elec}}(\mathbi{y})=\langle\upsilon_i\vert\nabla_{\mathbi{y}}\vert\upsilon_j
\rangle_{\mathbi{x}}$.\label{foot:elec-connection}} 
 &&  &~~& encoding system's dependence  \\
&& && && &&  &~~& on the parameters space  \\
2&&curvature  &~~& curl of $A_{\mathrm{B}}$
           &~~& curl of $\tau_{nn}^{\mathrm{elec}}(\mathbi{y})$
\footnote{The Berry curvature of the $n$-th electronic state is
$\Omega_n(\mathbi{y})=\nabla_{\mathbi{y}}\times\tau_{nn}^{\mathrm{elec}}(\mathbi{y})$.
See also Footnote \textsuperscript{\ref{foot:elec-connection}}.}
           &~~& effective magnetic field in 
           &~~& a local geometric property of \\
&& && && && parameter space \textsuperscript{\ref{foot:geom-em-field}}
           &~~& parameter space encoding the  \\
&& && && &&  &~~& states' anholonomy under  \\
&& && && &&  &~~& adiabatic parameter variations  \\ 
3&&phase      &~~& nuclear wave function &~~& electronic wave function 
           &~~& geometric propertities in both &~~& geometric properties of \\ 
&& && && &&  classical polarization dynamics &~~& wave eigenstates   \\
&& && && &&  and quantum photonic system    &~~&  \\
4&&Chern number&& number of critical points &~~& number of CI points &~~& global properties of EM modes
\footnote{In general, the EM modes are defined for photons, or more pricisely bosons,
such as photon bands in periodic structures.}
           &~~& a topological invariant classifing \\
&& && && &&  &~~& the global topology of the system’s \\
&& && && &&  &~~& eigenstate bundle  \\ 
\hline
\end{tabular}
\label{tab:geom-comp}
\end{sidewaystable}

%%-----------------------------------
%\clearpage
%\begin{table}
%\caption{???????}
%\begin{tabular}{lr}
%\hline
%No. & \\
%\hline
%\end{tabular}
%\label{tab:qc-al-dyn}
%\end{table}

%--------------------------------------------------
% FIGURES
%---------------------------------------------------
\clearpage
\section*{Figures and Captions}

\figcaption{fig:framework}{
Systematic illustration of the quantum dynamics methods. The black
words represent computational goals or results, while the red words
represent computational techniques. The arrows represent the direction
from the input to the computational results. Designing a set of
coordinates, the kenetic energy operator (KEO) can be derived through
the polyspherical approach. Then, the database is constructed through
{\it ab initio} energy calculations on a set of sampled geometries.
The potential energy surface (PES) is then constructed in an appropriate
form. With the Hamiltonian operator, extensive wavepacket propagations
can be launched on the basis of the rovibrational eigenstate of the
reactant which can be computed by relaxation of a guess funciton. Then,
the initial wave function is computed by assigning a given value of
momentum to the initial eigenstate. Having time-dependent wave function,
flux analysis and expectation calculations are launched to compute
reaction probability and time-dependent expectations. Reprinted with
permission from Reference \cite{zha25:20397}. Copyright 2025 Royal
Society of Chemistry.
}

\figcaption{fig:mountain-pass}{
Illustration of the mountain pass theorem taking a 2D smooth function
in configuration space $X$ as an example, where solid lines are contour
lines. Two distinct circle symbols represent two minima. The symbol
$\times$ represents the saddle point. The straight red dotted line
represents the shortest distance between two minima. The curved blue
line $I$ represents a general path.
}

\figcaption{fig:reax-spacetime}{%
Decomposition of the 4D spacetime with metric tensor
$g_{\mu\nu}$, which can be foliated into a family of non-intersecting
spacelike 3D surfaces $\Sigma_t$, called time slices. Subfigures (a)
and (b) illustrate the decompositions of flat and curved spacetime,
respectively. The vertical axis represents the time axis. For easily
understanding, a reaction process of H + H$_2$ is shown in each slice
as a typical example. These 3D surfaces arise, at least locally, as
the level surfaces of a scalar function $t$ which can be interpreted
as a global time function. From $t$, one can define a 1-form as
$\Omega_{\mu}=\nabla_{\mu}t$ and then define the lapse function
$\alpha$. By the unit normal vector $n^\mu$ to the slice $\Sigma_t$,
one can introduce the spatial metric tensor $\gamma_{\mu\nu}$. To
introduce $t^{\mu}$ that connects the spatial coordinates of adjacent
slices, the spatial shift vector $\beta^{\mu}$ is also defined to represent
the offset of spatial coordinates relative to the normal vector. With
$n^{\mu}$ and $\gamma^{\mu\nu}$, the extrinsic curvature $K_{\mu\nu}$
is defined by $K_{\mu\nu}=-\gamma_{\mu}^{\rho}\gamma_{\nu}^{\sigma}\nabla_{\rho}n_{\sigma}$.
Having the time basis vector, the geometric characteristics of each
slice are given by $(\gamma_{ij},K_{ij})$.
}%

\begin{figure}[h!]
 \centering
  \subfigure[\quad Reaction in curved spacetime]{%
   \includegraphics[height=22cm]{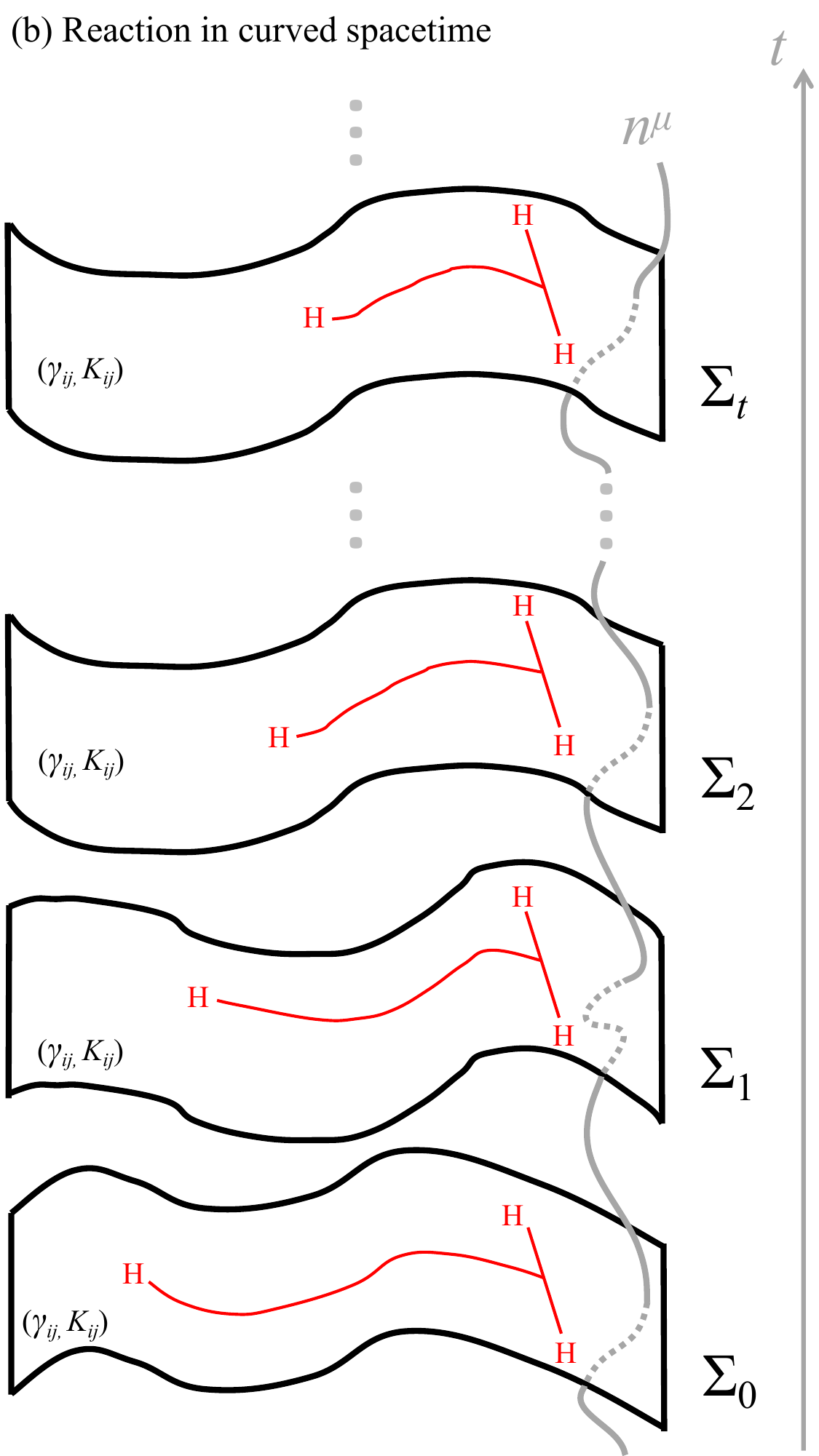}}    
    \caption{\figfoot}
     \label{fig:reax-spacetime}
      \end{figure}

\figcaption{fig:map-slice-conf}{%
Mapping relation between 3D slice and $3N$-dimension configuration
space, where subfigures (a) and (b) give the cases of flat and curved
slices, respectively. Similar to Figure \ref{fig:reax-spacetime}, the
reaction process of H + H$_2$ is shown as a typical example. The left
panel represents either flat or curved slice, while the right panel
represents $3N$-dimensional configuration space. In the configuration
space, the coordinates of $N$-atomic molecular system as given in the
left panel by red words are represented by one point, as shown by the
blue point in the right panel. The gray dashed lines connects atoms (in
red) and the representation point (in blue).
}%

\figcaption{fig:spa-sate}{
Subfigure (a) is a schematic illustration of the principal fiber bundle interpretation of
the map $\Psi:\mathbb{A}\to\mathbb{H}_I$. The left and right panels are
parameter space $\mathbb{A}$ and Hilbert space $\mathbb{H}_I$, respectively.
The middle arrow means the map $\Psi:\mathbb{A}\to\mathbb{H}_I$. The closed
colored lines in parameter space $\mathbb{A}$ are gauge orbits that are
mapped to identical states in Hilbert space $\mathbb{H}$. The yellow dot
in the middle of $\mathbb{A}$ corresponds to a state that does not have
full rank. The gauge orbit looks fundamentally different and this point
has to be excluded from the set $A$ in order to define a principal fiber
bundle. The right panel shows base manifold $V$ of the variational class
of $\Psi:A\to V$ which is a principal fiber bundle with structure group
$\mathrm{S}$, base manifold $V$, total manifold $A$ and bundle projection
$\Psi$, as given in the main text. The variational manifold $V$ is a
complex manifold that is biholomorphic to $A\setminus\mathrm{S}$.
Subfigure (b) is a schematic illustration of the tangent map $\Phi$ at
base point $A_I$. Similar to (a), the left and right panels are parameter
space $\mathbb{A}$ and Hilbert space $\mathbb{H}_I$, respectively; meanwhile,
the back line in $\mathbb{A}$ and surface $V$ in $\mathbb{H}_I$ are
gauge orbit and variational manifold, respectively. In parameter space
(left panel), there is a vertical subspace (green plane) of vectors
tangent to the gauge orbits (red lines). A unique parameterization of
vectors $\vert\Phi[B_I,A_I]\rangle$ in the tangent space (right panel)
requires the definition of a complementary horizontal subspace (green
plane). If this horizontal subspace is defined as the kernel of a principal
bundle connection, then it transforms equivariantly according to the
adjoint representation.
}

\figcaption{fig:phase-bundle}{
Schematic illustration of concepts of the curves $\Gamma(t)$ (the black
loop with arrow) and $\tilde{\Gamma}(t)$ (the red curve with arrow),
together with tangent vectors $X$ (the black dashed arrow) and $\tilde{X}$
(the red dashed arrow). The Berry phase is produced by parallel transport
of a state in the proper quantum phase space $H$ along the loop
$\Gamma:[0,1]\to H$. For the U(1)-principle bundle $P(H,\mathrm{U}(1))$,
the curve $\tilde{\Gamma}:[0,1]\to P$ is the horizontal lift of $\Gamma$
if $\pi\circ\tilde{\Gamma}$.
}

\figcaption{fig:vqe-for-quan-chem}{
Simple illustration of the VQE algorithm for electronic-structure theory
in quantum computing. In VQE, quantum computer is used to prepare a set
of parametrized quantum states as shown in green box. The classical computer 
then takes the individual estimates of the wave function parameters and
averages them to compute a single value of each parameter as shown in
yellow box. This cost
function value is fed into an optimization routine, which produces an
updated set of parameters as input for the quantum circuit in the next
optimization loop. This procedure is repeated until the energy converges.
Recent and future efforts are to improve each VQE component as well as
replace certain ``classical'' subroutines with quantum counterparts to
further leverage the capabilities of quantum computers. Reprinted with
permission from Reference \cite{cao19:10856}. Copyright 2019 American
Chemical Society.
}

\figcaption{fig:scatt-quan}{
Illustartions of (a) trajectories in classical collision of particles
by a fixed target and (b) wavepacket in quantum scattering of incoming
molecular beam by a fixed potential scattering center. In subfigure (a),
four trajectories with various impacting parameters $b$ are given, where
$d_0$ means minimum distance in a central collision. The angle between
the asymptotes of the trajectories at infinity before and after collision
is called the scattering angle, denoted by $\theta$, which depends on
impacting parameter. In subfigure (b), a beam of molecules hits a target
fixed at the scatteirng center that is located at the origin of the
interacting potential. The incident beam is towards to the scattering
center from left to right and is represented by the income plane wave
packet. The scattering wave function is represented by a spherical
wavepacket.
}

\figcaption{fig:modes-high-freq}{
Illustrations of (a) whispering gallery modes and (b) focusing modes
in the unit disk with Dirichlet boundary condition, together with the
$L^2$-norm of the focusing modes. In subfigure (a), the upper and lower
panels give those for $k=0$ and $k=1$, respectively, where an increase
of the index $n$ leads to stronger localization of the eigen-function
near the boundary. Reprinted with permission from Reference \cite{ngu13:780}.
Copyright 2013 Society for Industrial and Applied Mathematics.
In subfigure (b), the upper and lower panels give those for $n=0$ and $n=1$,
respectively, where an increase of the index $k$ leads to stronger localization
of the eigen-function at the origin. Reprinted with permission from Reference
\cite{ngu13:780}. Copyright 2013 Society for Industrial and Applied Mathematics.
In subfigure (c), the upper, middle, and lower panels gives $L^2$-norm
of the focusing modes for $n=0$, $n=1$, and $n=4$, respectively, while
the left, middle, and right panels gives those for $k=1$, $k=5$, and
$k=10$, respectively.
}

\figcaption{fig:lear-net}{%
Diagrammatic sketch of the optimization processes in the hyperparameters
set $\boldsymbol{\theta}$, where colored paths represent different
optimization processes. These processes share the same initial and
finial states. Each circle represents an intermediate state of
$\boldsymbol{\theta}$ during the optimizations, where $\Gamma$ represents
a path during optimization in regression. Coordinates frame
$\mathfrak{z}$ is defined to represent space of the hyperparameters
and the intermediate state during optimization (represented by symbols
``$\circ$'') is represented by $\mathfrak{Z}$.
}%

%--fig. 1------------------------------------
\clearpage
     
%--fig. 2----------------------------------------
 \clearpage

%-fig. 3----------------------------------------------
\clearpage
\begin{figure}[h!]
 \centering
  \subfigure[\quad Reaction in flat spacetime]{%
   \includegraphics[height=22cm]{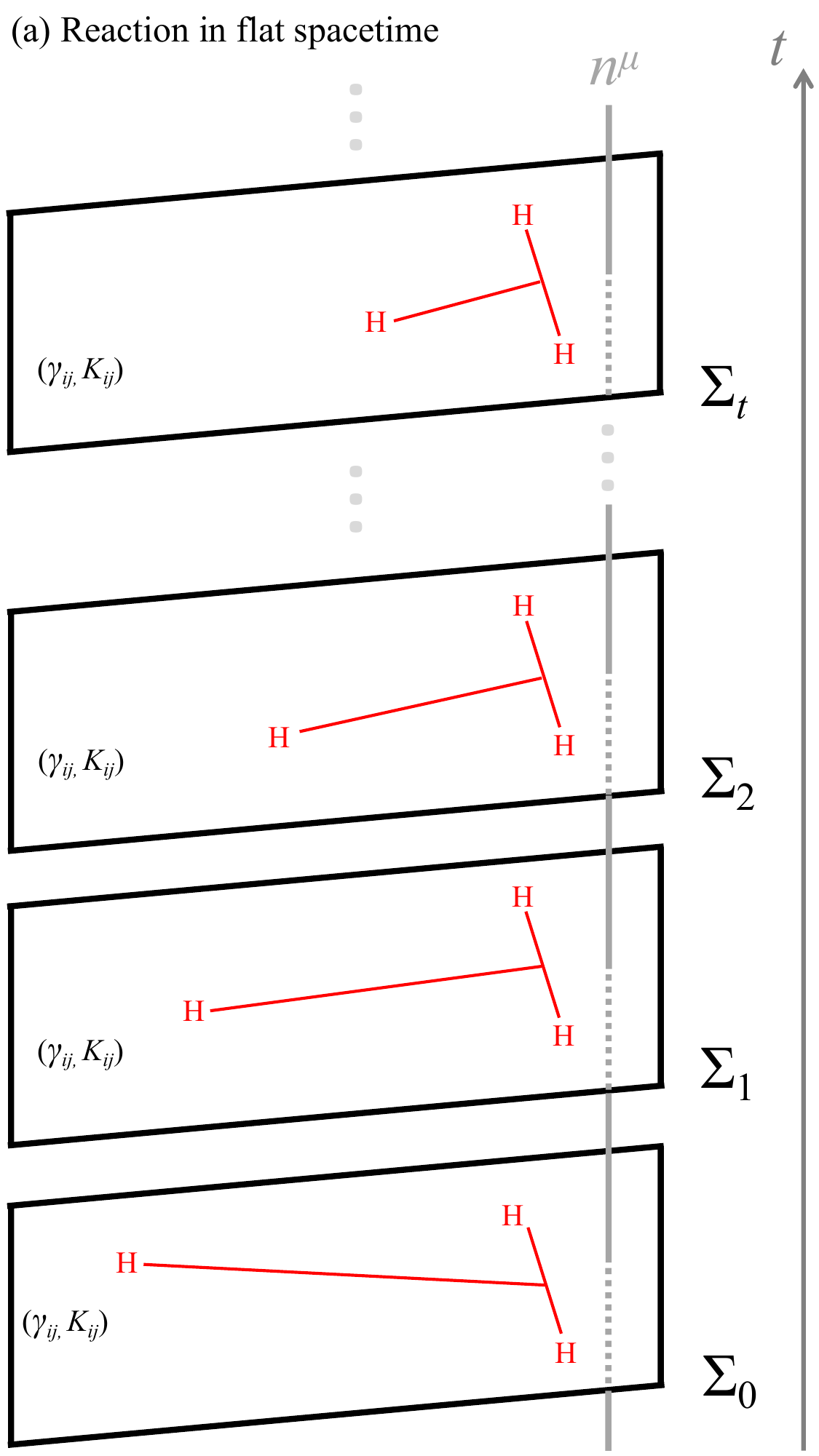}}
    \end{figure}
%-----------------------------------------------
\clearpage

%-fig. 4-----------------------------------------------
\clearpage
\begin{figure}[h!]
 \subfigure[\quad Mapping from flat spacetime to configuration space]{%
  \includegraphics[width=18cm]{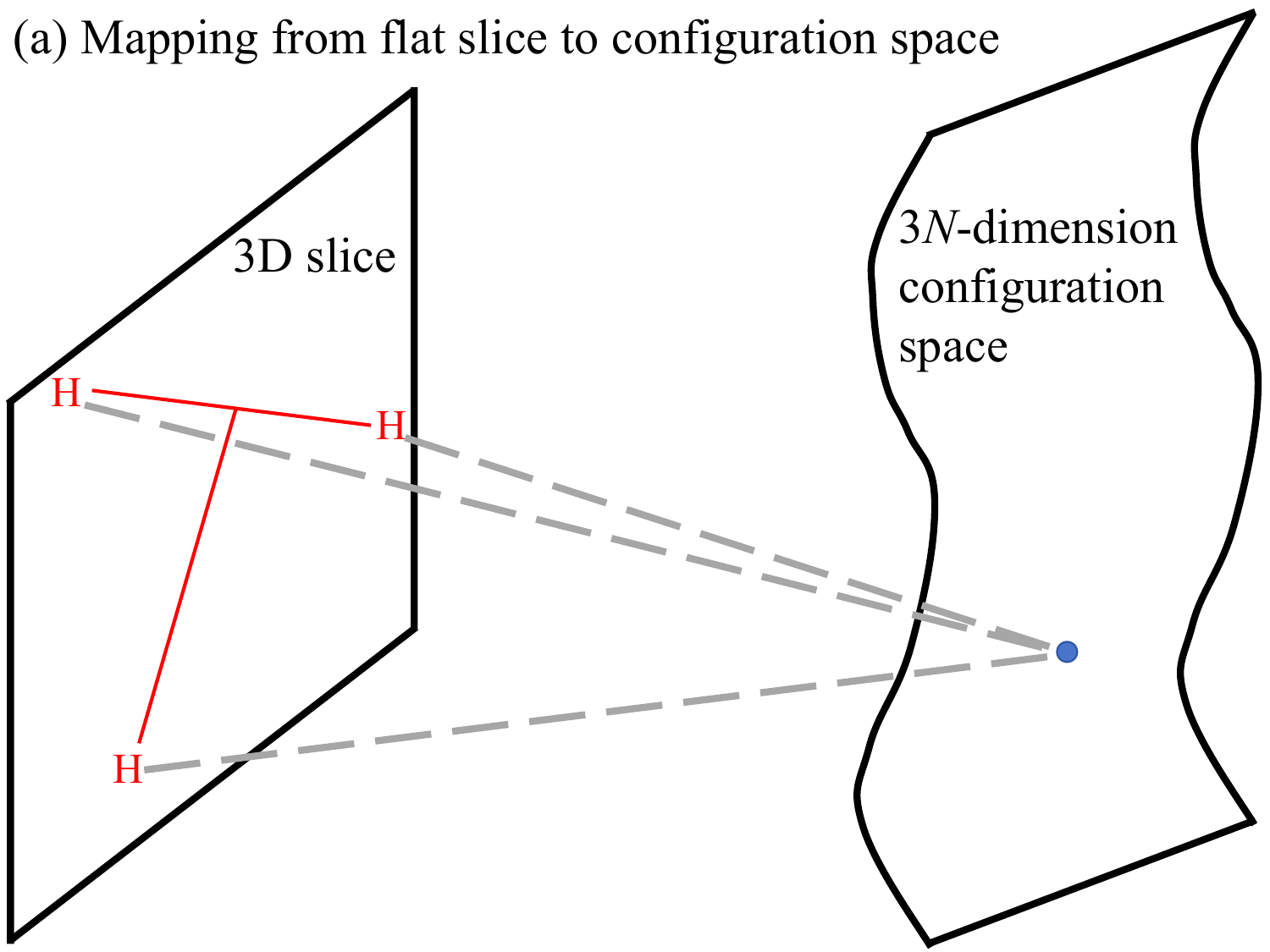}}
   \end{figure}
%------------------------------------------------
\clearpage

%--fig. 5------------------------------------------
 \clearpage

%--fig. 6-------------------------------------------
 \clearpage

%----------fig61---------------
 \begin{figure}[h!]
  \centering
   \includegraphics[width=18cm]{./figure/fig61-vqe-for-quan-chem.pdf}
    \caption{\figfoot}
     \label{fig:vqe-for-quan-chem}
      \end{figure}

%-fig. 7------------------------
 \clearpage

%-fig. 8--------------------------------------
\clearpage
 \begin{figure}[h!]
  \centering
   \subfigure[\quad Whispering gallery modes]{  
    \includegraphics[width=18cm]{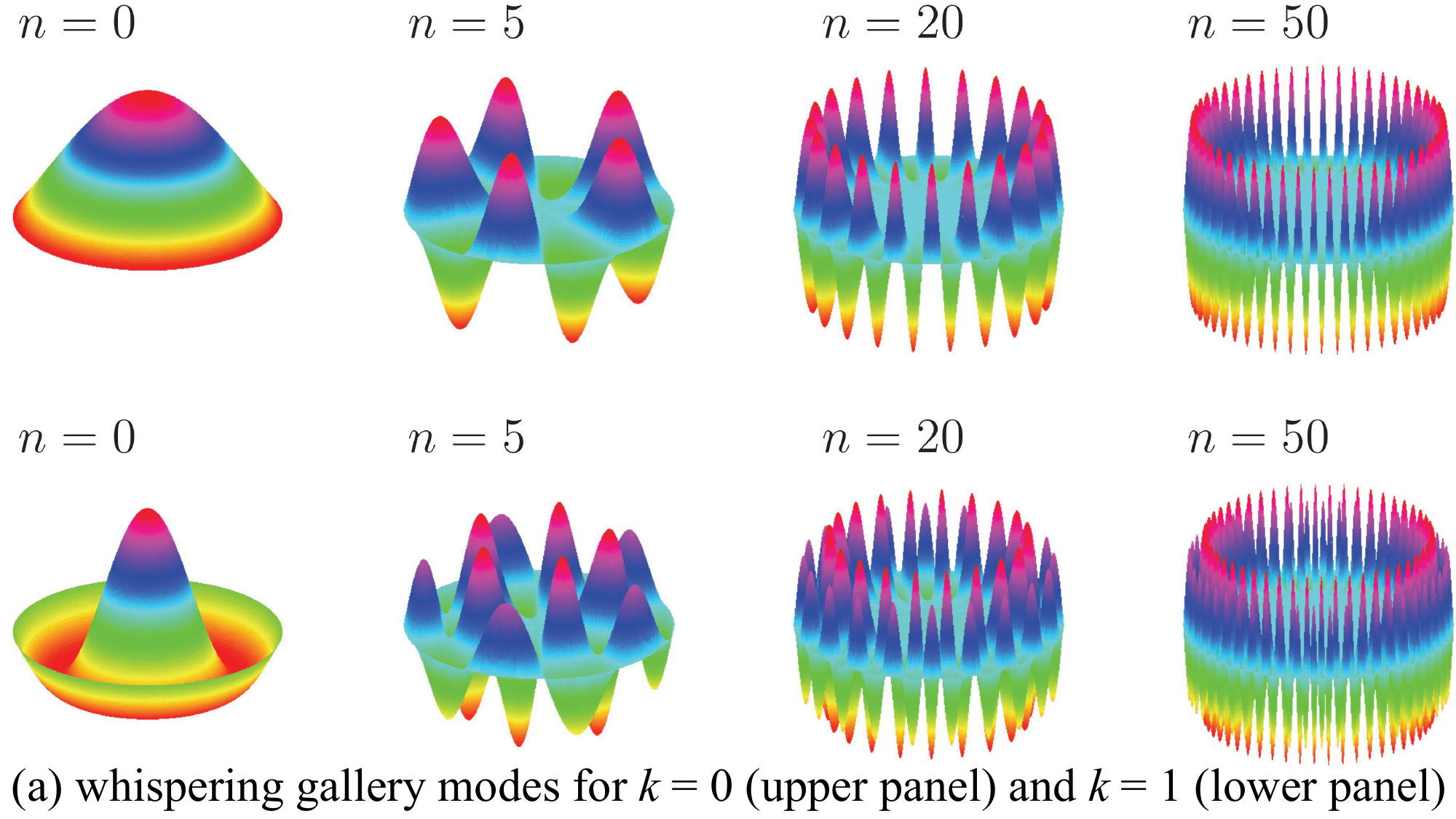}}
     \end{figure}
%-----------------     
  \begin{figure}[h!]
   \centering
    \subfigure[\quad Focusing modes]{
     \includegraphics[width=18cm]{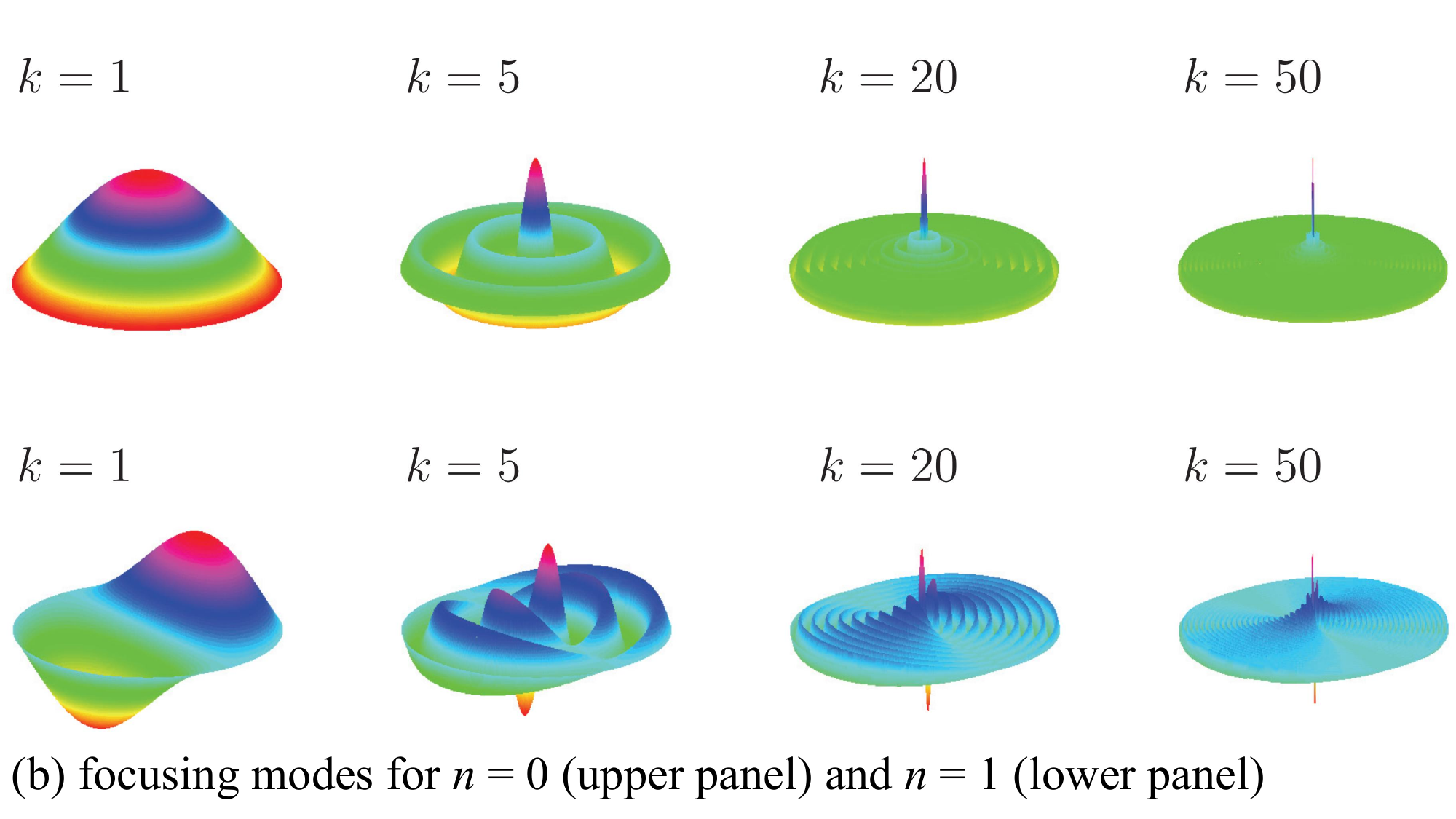}}
      \end{figure}
%------------------     
       
%-fig. 9-------------------------------
\clearpage
       
%-----------------------------------------------------
% References
%------------------------------------------------------
\clearpage
%\bibliographystyle{../bibtex/mybibu}
%\bibliography{../bibtex/refs,../bibtex/refs-1,../bibtex/refs-2} %,../bibtex/refs-3}

\begin{thebibliography}{100}
\ifnum\language=1 \def\biband{und}\else\def\biband{and}\fi

\bibitem{zha25:20397}
{\rm X.~Zhang \biband~ Q.~Meng}.
\newblock A hierarchical wavepacket propagation framework via ml-mctdh for
  molecular reaction dynamics.
\newblock {\em Phys.\ Chem.\ Chem.\ Phys. \bf 27\/} (2025), 20397--20420.

\bibitem{son24:597}
{\rm Q.~Song, X.~Zhang, Z.~Miao,  \biband~ Q.~Meng}.
\newblock Construction of mode-combination hamiltonian under the gridbased
  representation for the quantum dynamics of oh + ho$_2$ $\to$ o$_2$ + h$_2$o.
\newblock {\em J.~Chem.\ Theory Comput. \bf 20\/} (2024), 597--613.

\bibitem{son22:11128}
{\rm Q.~Song, X.~Zhang, D.~Pel{\'a}ez,  \biband~ Q.~Meng}.
\newblock Direct canonical-polyadic-decomposition of the potential energy
  surface from discrete data by decoupled gaussian process regression.
\newblock {\em J.~Phys.\ Chem.\ Lett. \bf 13\/} (2022), 11128--11135.

\bibitem{men22:16415}
{\rm Q.~Meng, J.~Chen, J.~Ma, X.~Zhang,  \biband~ J.~Chen}.
\newblock Adiabatic models for the quantum dynamics of surface scattering with
  lattice effects.
\newblock {\em Phys.\ Chem.\ Chem.\ Phys. \bf 24\/} (2022), 16415--16436.

\bibitem{men21:2702}
{\rm Q.~Meng, M.~Schr{\"o}der,  \biband~ H.-D. Meyer}.
\newblock High-dimensional quantum dynamics study on excitation-specific
  surface scattering including lattice effects of a five-atom surface cell.
\newblock {\em J.~Chem.\ Theory Comput. \bf 17\/} (2021), 2702--2713.

\bibitem{lar24:e2306881}
{\rm H.~R. Larsson}.
\newblock A tensor network view of multilayer multiconfiguration time-dependent
  hartree methods.
\newblock {\em Mol.\ Phys. \bf 122\/} (2024), e2306881.

\bibitem{sza12:108}
{\rm P.~G. Szalay, T.~M{\"u}ller, G.~Gidofalvi, H.~Lischka,  \biband~
  R.~Shepard}.
\newblock Multiconfiguration self-consistent field and multireference
  configuration interaction methods and applications.
\newblock {\em Chem. Rev. \bf 112\/} (2012), 108--181.

\bibitem{goi18:e1341}
{\rm J.~J. Goings, P.~J. Lestrange,  \biband~ X.~Li}.
\newblock Real-time time-dependent electronic structure theory.
\newblock {\em WIREs:\ Comput.\ Mol.\ Sci. \bf 8\/} (2018), e1341.

\bibitem{li20:9951}
{\rm X.~Li, N.~Govind, C.~Isborn, A.~E. {DePrince III},  \biband~ K.~Lopata}.
\newblock Real-time time-dependent electronic structure theory.
\newblock {\em Chem.\ Rev. \bf 120\/} (2020), 9951--9993.

\bibitem{sve23:e1666}
{\rm B.~{Sverdrup Ofstad}, E.~Aurbakken, {\O}.~{Sigmundson Schøyen}, H.~E.
  Kristiansen, S.~Kvaal,  \biband~ T.~B. Pedersen}.
\newblock Time-dependent coupled-cluster theory.
\newblock {\em WIREs:\ Comput.\ Mol.\ Sci. \bf 13\/} (2023), e1666.

\bibitem{yar96:985}
{\rm D.~R. Yarkony}.
\newblock Diabolical conical intersections.
\newblock {\em Rev.\ Mod.\ Phys. \bf 68\/} (1996), 985--1013.

\bibitem{yar96:10456}
{\rm D.~R. Yarkony}.
\newblock On the consequence of nonremovable derivative couplings. {I.} {T}he
  geometric phase and quasidiabatic states: {A} numerical study.
\newblock {\em J.~Chem.\ Phys. \bf 105\/} (1996), 10456--10461.

\bibitem{yar98:971}
{\rm D.~R. Yarkony}.
\newblock {On the description of potential energy surfaces exhibiting conical
  intersections: a compact representation of the energies and derivative
  couplings and locally diabatic bases for the HOH and OHH portions of the 1
  $^1A'-2^1A'$ seam of conical intersection in water}.
\newblock {\em Mol.\ Phys. \bf 93\/} (1998), 971.

\bibitem{yar01:6277}
{\rm D.~R. Yarkony}.
\newblock Conical intersections: the new conventional wisdom.
\newblock {\em J.~Chem.\ Phys. \bf 105\/} (2001), 6277--6293.

\bibitem{jua07:044317}
{\rm J.~C. Juanes-Marcos, S.~C. Althorpe,  \biband~ E.~Wrede}.
\newblock Effect of the geometric phase on the dynamics of the
  hydrogen-exchange reaction.
\newblock {\em J.~Chem.\ Phys. \bf 126\/} (2007), 044317.

\bibitem{bou08:124322}
{\rm F.~Bouakline, S.~C. Althorpe,  \biband~ D.~Pel{\'a}ez}.
\newblock Strong geometric-phase effects in the hydrogen-exchange reaction at
  high collision energies.
\newblock {\em J.~Chem.\ Phys.}, 12 (2008), 124322.

\bibitem{jua08:211101}
{\rm J.~C. Juanes-Marcos, A.~J.~C. Varandas,  \biband~ S.~C. Althorpe}.
\newblock Geometric phase effects in resonance-mediated scattering: H+h2+ on
  its lowest triplet electronic state.
\newblock {\em J.~Chem.\ Phys. \bf 128\/} (2008), 211101.

\bibitem{alt08:214117}
{\rm S.~C. Althorpe, T.~Stecher,  \biband~ F.~Bouakline}.
\newblock Effect of the geometric phase on nuclear dynamics at a conical
  intersection: Extension of a recent topological approach from one to two
  coupled surfaces.
\newblock {\em J.~Chem.\ Phys. \bf 129\/} (2008), 214117.

\bibitem{ced04:3}
{\rm L.~S. Cederbaum}.
\newblock Born-oppenheimer approximation and beyond.
\newblock In {\em Conical {I}ntersections\/} (Singapore, 2004), W.~Domcke,
  D.~R. Yarkony, \biband~ H.~K{\"o}ppel, Eds., World Scientific Co., pp.~3--40.

\bibitem{bae06:boa}
{\rm M.~Baer}.
\newblock {\em Beyond {Born-Oppenheimer}: {E}lectronic Nonadiabatic Coupling
  Terms and Conical Intersections}.
\newblock Wiley, Hoboken, {NJ}, 2006.

\bibitem{ced13:224110}
{\rm L.~S. Cederbaum}.
\newblock The exact molecular wavefunction as a product of an electronic and a
  nuclear wavefunction.
\newblock {\em J.~Chem.\ Phys. \bf 138\/} (2013), 224110.

\bibitem{wan03:1289}
{\rm H.~Wang \biband~ M.~Thoss}.
\newblock Multilayer formulation of the multiconfiguration time-dependent
  {H}artree theory.
\newblock {\em J.~Chem.\ Phys. \bf 119\/} (2003), 1289--1299.

\bibitem{man08:164116}
{\rm U.~Manthe}.
\newblock A multilayer multiconfigurational time-dependent {H}artree approach
  for quantum dynamics on general potential energy surfaces.
\newblock {\em J.~Chem.\ Phys. \bf 128\/} (2008), 164116.

\bibitem{ven11:044135}
{\rm O.~Vendrell \biband~ H.-D. Meyer}.
\newblock {Multilayer multiconfiguration time-dependent Hartree method:
  Implementation and applications to a Henon-Heiles Hamiltonian and to
  pyrazine}.
\newblock {\em J.~Chem.\ Phys. \bf 134\/} (2011), 044135.

\bibitem{bec00:1}
{\rm M.~H. Beck, A.~J{\"a}ckle, G.~A. Worth,  \biband~ H.-D. Meyer}.
\newblock The multi-configuration time-dependent {H}artree ({MCTDH}) method:
  {A} highly efficient algorithm for propagating wave packets.
\newblock {\em Phys.~Rep \bf 324\/} (2000), 1--105.

\bibitem{mey03:251}
{\rm H.-D. Meyer \biband~ G.~A. Worth}.
\newblock {Q}uantum molecular dynamics: {P}ropagating wavepackets and density
  operators using the multiconfiguration time-dependent {H}artree ({MCTDH})
  method.
\newblock {\em Theor.\ Chem.\ Acc. \bf 109\/} (2003), 251--267.

\bibitem{wor08:569}
{\rm G.~A. Worth, H.-D. Meyer, H.~K\"oppel, L.~S. Cederbaum,  \biband~
  I.~Burghardt}.
\newblock Using the {MCTDH} wavepacket propagation method to describe multimode
  non-adiabatic dynamics.
\newblock {\em Int.~Rev.~Phys.~Chem. \bf 27\/} (2008), 569--606.

\bibitem{mey12:351}
{\rm H.-D. Meyer}.
\newblock Studying molecular quantum dynamics with the multiconfiguration
  time-dependent {H}artree method.
\newblock {\em WIREs:~Comput.~Mol.~Sci. \bf 2\/} (2012), 351--374.

\bibitem{hae14:021902}
{\rm J.~Haegeman, M.~Mari{\"e}n, T.~J. Osborne,  \biband~ F.~Verstraete}.
\newblock Geometry of matrix product states: Metric, parallel transport, and
  curvature.
\newblock {\em J. Math. Phys. \bf 55\/} (2014), 021902.

\bibitem{hae11:070601}
{\rm J.~Haegeman, J.~I. Cirac, T.~J. Osborne, I.~Pi\ifmmode~\check{z}\else
  \v{z}\fi{}orn, H.~Verschelde,  \biband~ F.~Verstraete}.
\newblock Time-dependent variational principle for quantum lattices.
\newblock {\em Phys.\ Rev.\ Lett. \bf 107\/} (2011), 070601.

\bibitem{van19:7}
{\rm L.~Vanderstraeten, J.~Haegeman,  \biband~ F.~Verstraete}.
\newblock {Tangent-space methods for uniform matrix product states}.
\newblock {\em SciPost Phys. Lect. Notes\/} (2019), 7.

\bibitem{dor01:19}
{\rm D.~C. Brody \biband~ L.~P. Hughston}.
\newblock Geometric quantum mechanics.
\newblock {\em J. Geom. Phys. \bf 38\/} (2001), 19--53.

\bibitem{ber84:45}
{\rm M.~V. Berry}.
\newblock Quantal phase factors accompanying adiabatic changes.
\newblock {\em Proc.\ R.\ Soc.\ A \bf 392\/} (1984), 45.

\bibitem{son22:1983}
{\rm Q.~Song, X.~Zhang, Z.~Miao, Q.~Zhang,  \biband~ Q.~Meng}.
\newblock Unified regression models in fitting potential energy surfaces for
  quantum dynamics.
\newblock {\em J.~Math.\ Chem. \bf 60\/} (2022), 1983--2012.

\bibitem{fei25:123}
{\rm Y.~Fei, Y.~Liu, C.~Jia, Z.~Li, X.~Wei,  \biband~ M.~Chen}.
\newblock A survey of geometric optimization for deep learning: From euclidean
  space to riemannian manifold.
\newblock {\em ACM Comput. Surv. \bf 57\/} (2025).

\bibitem{zha26:1613}
{\rm X.~Zhang, J.~Yu,  \biband~ Q.~Meng}.
\newblock The [3+1] formulation of chemical dynamics in curved spacetime under
  an eulerian observer.
\newblock {\em ACS Earth Space Chem. \bf 10\/} (2026), 1613--1625.

\bibitem{her60:1870}
{\rm D.~R. Herschbach}.
\newblock Analysis of reactive scattering in crossed molecular beams.
\newblock {\em J.~Chem.\ Phys. \bf 33\/} (1960), 1870--1871.

\bibitem{lee69:1402}
{\rm Y.~T. Lee, J.~D. McDonald, P.~R. LeBreton,  \biband~ D.~R. Herschbach}.
\newblock Molecular beam reactive scattering apparatus with electron
  bombardment detector.
\newblock {\em Rev.\ Sci.\ Instrum. \bf 40\/} (1969), 1402--1408.

\bibitem{bus79:34}
{\rm R.~J. Buss \biband~ Y.~T. Lee}.
\newblock Crossed beam studies related to gas kinetics.
\newblock {\em J.~Phys.\ Chem. \bf 83\/} (1979), 34--36.

\bibitem{pol81:1837}
{\rm J.~E. Pollard, D.~J. Trevor, Y.~T. Lee,  \biband~ D.~A. Shirley}.
\newblock Photoelectron spectroscopy of supersonic molecular beams.
\newblock {\em Rev.\ Sci.\ Instrum. \bf 52\/} (1981), 1837--1846.

\bibitem{lee87:793}
{\rm Y.~T. Lee}.
\newblock Molecular beam studies of elementary chemical processes.
\newblock {\em Science \bf 236\/} (1987), 793--798.

\bibitem{sch94:1345}
{\rm W.~Sch{\"o}llkopf \biband~ J.~P. Toennies}.
\newblock {\em Science \bf 266\/} (1994), 1345.

\bibitem{qui06:1440}
{\rm M.~Qui, Z.~Ren, L.~Che, D.~Dai, S.~A. Harich, X.~Wang, X.~Yang, C.~Xu,
  D.~Xie, M.~Gustafsson, R.~T. Skodje, Z.~Sun,  \biband~ D.~H. Zhang}.
\newblock Observation of {F}eshbach resonances in the {F}+{H}$_2$ $\rightarrow$
  {HF}+{H} reaction.
\newblock {\em Science \bf 311\/} (2006), 1440.

\bibitem{yan07:433}
{\rm X.~M. Yang}.
\newblock State-to-state dynamics of elementary bimolecular reactions.
\newblock {\em Ann.\ Rev.\ Phys.\ Chem. \bf 58\/} (2007), 433--459.

\bibitem{zha10:12782}
{\rm W.~Zhang, Y.~Zhou, G.~Wu, Y.~Lu, H.~Pan, B.~Fu, Q.~Shuai, L.~liu, S.~Liu,
  L.~Zhang, B.~Jiang, D.~Dai, S.-Y. Lee, Z.~Xie, B.~J. Braams, J.~M. Bowman,
  M.~A. Collins, D.~H. Zhang,  \biband~ X.~Yang}.
\newblock Depression of reactivity by the collision energy in the single
  barrier {H+CD$_4$} {$\rightarrow$} {HD} + {CD}$_3$ reaction.
\newblock {\em Proc.\ Natl.\ Acad.\ Sci.\ U.S.A. \bf 107\/} (2010), 12782.

\bibitem{don10:1501}
{\rm W.~Dong, C.~Xiao, T.~Wang, D.~Dai, X.~M. Yang,  \biband~ D.~H. Zhang}.
\newblock Transition-state spectroscopy of partial wave resonances in the
  {F}+{HD}.
\newblock {\em Science \bf 327\/} (2010), 1501.

\bibitem{xia11:440}
{\rm C.~Xiao, X.~Xu, S.~Liu, T.~Wang, W.~Dong, T.~Yang, Z.~Sun, D.~Dai, X.~Xu,
  D.~H. Zhang,  \biband~ X.~Yang}.
\newblock Experimental and theoretical differential cross sections for a
  four-atom reaction: {HD + OH $\to$ H$_2$O + D}.
\newblock {\em Science \bf 333\/} (2011), 440--442.

\bibitem{yan12:1650}
{\rm X.~M. Yang, T.~K. Minton,  \biband~ D.-H. Zhang}.
\newblock Rethinking chemical reactions at hyperthermal energies.
\newblock {\em Science \bf 336\/} (2012), 1650.

\bibitem{yan14:3106}
{\rm J.~Yang, K.~Shao, D.~Zhang, Q.~Shuai, B.~Fu, D.~H. Zhang,  \biband~
  X.~Yang}.
\newblock Trapped abstraction in the o(1d) + chd3 $\to$ oh + cd3 reaction.
\newblock {\em J.~Phys.\ Chem.\ Lett. \bf 5\/} (2014), 3106--3111.

\bibitem{yan15:60}
{\rm T.~Yang, J.~Chen, L.~Huang, T.~Wang, C.~Xiao, Z.~Sun, D.~Dai, X.~Yang,
  \biband~ D.~H. Zhang}.
\newblock {Extremely short-lived reaction resonances in Cl + HD (v = 1) $\to$
  DCl + H due to chemical bond softening}.
\newblock {\em Science \bf 347\/} (2015), 60--63.

\bibitem{yua18:1289}
{\rm D.~Yuan, Y.~Guan, W.~Chen, H.~Zhao, S.~Yu, C.~Luo, Y.~Tan, T.~Xie,
  X.~Wang, Z.~Sun, D.~H. Zhang,  \biband~ X.~Yang}.
\newblock {Observation of the geometric phase effect in the H plus HD -> H-2 +
  D reaction}.
\newblock {\em Science \bf {362}\/} ({2018}), {1289}.

\bibitem{che20:9202}
{\rm Z.~Chen, J.~Chen, R.~Chen, T.~Xie, X.~Wang, S.~Liu, G.~Wu, D.~Dai,
  X.~Yang,  \biband~ D.~H. Zhang}.
\newblock {Reactivity oscillation in the heavy-light-heavy Cl + CH4 reaction}.
\newblock {\em Proc.\ Natl.\ Acad.\ Sci.\ U.S.A. \bf {117}\/} ({2020}),
  {9202--9207}.

\bibitem{xie20:767}
{\rm Y.~Xie, H.~Zhao, Y.~Wang, Y.~Huang, T.~Wang, X.~Xu, C.~Xiao, Z.~Sun, D.~H.
  Zhang,  \biband~ X.~Yang}.
\newblock {Quantum interference in H plus HD -> H-2 + D between direct
  abstraction and roaming insertion pathways}.
\newblock {\em Science \bf {368}\/} ({2020}), {767}.

\bibitem{hac20:048}
{\rm L.~Hackl, T.~Guaita, T.~Shi, J.~Haegeman, E.~Demler,  \biband~ J.~I.
  Cirac}.
\newblock {Geometry of variational methods: dynamics of closed quantum
  systems}.
\newblock {\em SciPost Phys. \bf 9\/} (2020), 048.

\bibitem{amb73:349}
{\rm A.~Ambrosetti \biband~ P.~H. Rabinowitz}.
\newblock Dual variational methods in critical point theory and applications.
\newblock {\em J. Funct. Anal. \bf 14\/} (1973), 349--381.

\bibitem{puc85:142}
{\rm P.~Pucci \biband~ J.~Serrin}.
\newblock A mountain pass theorem.
\newblock {\em J. Differ. Equations \bf 60\/} (1985), 142--149.

\bibitem{puc87:115}
{\rm P.~Pucci \biband~ J.~Serrin}.
\newblock The structure of the critical set in the mountain pass theorem.
\newblock {\em Trans. Amer. Math. Soc. \bf 299\/} (1987), 115--132.

\bibitem{cho93:417}
{\rm Y.~S. Choi \biband~ P.~J. McKenna}.
\newblock A mountain pass method for the numerical solution of semilinear
  elliptic problems.
\newblock {\em Nonlinear Anal. \bf 20\/} (1993), 417--437.

\bibitem{kat94:189}
{\rm G.~Katriel}.
\newblock Mountain pass theorems and global homeomorphism theorems.
\newblock {\em Ann. Inst. Henri Poincaré C \bf 11\/} (1994), 189--209.

\bibitem{hil00:731}
{\rm S.~Hill \biband~ L.~D. Humphreys}.
\newblock Mountain pass solutions for a system of partial dierential equations:
  An existence theorem with computational results.
\newblock {\em Nonlinear Anal. \bf 39\/} (2000), 731--743.

\bibitem{rup16:89}
{\rm H.-J. Ruppen}.
\newblock A generalized mountain-pass theorem: the existence of multiple
  critical points.
\newblock {\em Calc. Var. \bf 55\/} (2016), 89.

\bibitem{bed11:569}
{\rm E.~M. Bednarczuk, E.~Miglierina,  \biband~ E.~Molho}.
\newblock A mountain pass-type theorem for vector-valued functions.
\newblock {\em Set-Valued Anal. \bf 19\/} (2011), 569--587.

\bibitem{bur22:1512}
{\rm H.~G.~A. Burton}.
\newblock Energy landscape of state-specific electronic structure theory.
\newblock {\em J.~Chem.\ Theory Comput. \bf 18\/} (2022), 1512--1526.

\bibitem{ste71:2231}
{\rm D.~P. Stern}.
\newblock Classical adiabatic perturbation theory.
\newblock {\em J. Math. Phys. \bf 12\/} (1971), 2231--2242.

\bibitem{bra79:199}
{\rm M.~A. Braun \biband~ A.~D. Gurchumeliya}.
\newblock Relativistic adiabatic perturbation theory for degenerate levels.
\newblock {\em Teor. Mat. Fiz. \bf 45\/} (1979), 199--209.

\bibitem{sok78:339}
{\rm V.~V. Sokolov}.
\newblock Adiabatic perturbation theory for quasilevels.
\newblock {\em Teor. Mat. Fiz. \bf 35\/} (1978), 339--351.

\bibitem{pan02:250405}
{\rm G.~Panati, H.~Spohn,  \biband~ S.~Teufel}.
\newblock Space-adiabatic perturbation theory in quantum dynamics.
\newblock {\em Phys. Rev. Lett. \bf 88\/} (2002), 250405.

\bibitem{teu03:book}
{\rm S.~Teufel}.
\newblock {\em Adiabatic Perturbation Theory in Quantum Dynamics}.
\newblock Springer, Heidelberg, 2003.

\bibitem{pan03:145}
{\rm G.~Panti, H.~Spohn,  \biband~ S.~Teufel}.
\newblock Space-adiabatic perturbation theory.
\newblock {\em Adv. Theor. Math. Phys. \bf 7\/} (2003), 145--204.

\bibitem{rig10:170406}
{\rm G.~Rigolin \biband~ G.~Ortiz}.
\newblock Adiabatic perturbation theory and geometric phases for degenerate
  systems.
\newblock {\em Phys. Rev. Lett. \bf 104\/} (2010), 170406.

\bibitem{rig08:052508}
{\rm G.~Rigolin, G.~Ortiz,  \biband~ V.~H. Ponce}.
\newblock Beyond the quantum adiabatic approximation: Adiabatic perturbation
  theory.
\newblock {\em Phys. Rev. A \bf 78\/} (2008), 052508.

\bibitem{rig14:022104}
{\rm G.~Rigolin \biband~ G.~Ortiz}.
\newblock Degenerate adiabatic perturbation theory: Foundations and
  applications.
\newblock {\em Phys. Rev. A \bf 90\/} (2014), 022104.

\bibitem{cho25:012221}
{\rm H.-S. Chou}.
\newblock Adiabatic perturbation theory invariant under the unitary
  transformations.
\newblock {\em Phys. Rev. A \bf 112\/} (2025), 012221.

\bibitem{pod28:812}
{\rm B.~Podolsky}.
\newblock Quantum-mechanically correct form of hamiltonian function for
  conservative systems.
\newblock {\em Phys.\ Rev. \bf 32\/} (1928), 812.

\bibitem{lau02:8560}
{\rm D.~Lauvergnat \biband~ A.~Nauts}.
\newblock Exact numerical computation of a kinetic energy operator in
  curvilinear coordinates.
\newblock {\em J.~Chem.\ Phys. \bf 116\/} (2002), 8560.

\bibitem{gat09:1}
{\rm F.~Gatti \biband~ C.~Iung}.
\newblock Exact and constrained kinetic energy operators for polyatomic
  molecules: The polyspherical approach.
\newblock {\em Phys.\ Rep. \bf 484\/} (2009), 1--69.

\bibitem{ndo12:034107}
{\rm M.~Ndong, L.~{Joubert Doriol}, H.-D. Meyer, A.~Nauts, F.~Gatti,  \biband~
  D.~Lauvergnat}.
\newblock Automatic computer procedure for generating exact and analytical
  kinetic energy operators based on the polyspherical approach.
\newblock {\em J.~Chem.\ Phys. \bf 136\/} (2012), 034107.

\bibitem{sad12:234112}
{\rm K.~Sadri, D.~Lauvergnat, F.~Gatti,  \biband~ H.-D. Meyer}.
\newblock Numeric kinetic energy operators for molecules in polyspherical
  coordinates.
\newblock {\em J.~Chem.\ Phys. \bf 136\/} (2012), 234112.

\bibitem{ndo13:204107}
{\rm M.~Ndong, A.~Nauts, L.~{Joubert-Doriol}, H.-D. Meyer, F.~Gatti,  \biband~
  D.~Lauvergnat}.
\newblock Automatic computer procedure for generating exact and analytical
  kinetic energy operators based on the polyspherical approach: general
  formulation and removal of singularities.
\newblock {\em J.~Chem.\ Phys. \bf 139\/} (2013), 204107.

\bibitem{mar22:20200388}
{\rm E.~Marsili, F.~Agostini, A.~Nauts,  \biband~ D.~Lauvergnat}.
\newblock Quantum dynamics with curvilinear coordinates: models and kinetic
  energy operator.
\newblock {\em Phil. Trans. R. Soc. A \bf 380\/} (2022), 20200388.

\bibitem{sch12:book}
{\rm B.~Schutz}.
\newblock {\em A First Course in General Relativity}.
\newblock Cambridge University Press, 2012.

\bibitem{mis17:book}
{\rm C.~W. Misner, K.~S. Thorne,  \biband~ J.~A. Wheeler}.
\newblock {\em Gravitation}.
\newblock Princeton University Press, 2017.

\bibitem{car19:book}
{\rm S.~M. Carroll}.
\newblock {\em Spacetime and Geometry: An Introduction to General Relativity}.
\newblock Cambridge University Press, 2019.

\bibitem{bos25:015003}
{\rm S.~Bose, I.~Fuentes, A.~A. Geraci, S.~M. Khan, S.~Qvarfort, M.~Rademacher,
  M.~Rashid, M.~Toro\ifmmode~\check{s}\else \v{s}\fi{}, H.~Ulbricht,  \biband~
  C.~C. Wanjura}.
\newblock Massive quantum systems as interfaces of quantum mechanics and
  gravity.
\newblock {\em Rev. Mod. Phys. \bf 97\/} (2025), 015003.

\bibitem{bir82:book}
{\rm N.~D. Birrell \biband~ P.~C.~W. Davies}.
\newblock {\em Quantum Fields in Curved Space}.
\newblock Cambridge University Press, 1982.

\bibitem{bar09:book}
{\rm C.~B{\"a}r \biband~ K.~Fredenhagen}.
\newblock {\em Quantum Field Theory on Curved Spacetimes}.
\newblock Springer, 2009.

\bibitem{hac16:book}
{\rm T.-P. Hack}.
\newblock {\em Cosmological Applications of Algebraic Quantum Field Theory in
  Curved Spacetimes}.
\newblock Springer, 2016.

\bibitem{nau87:164}
{\rm A.~Nauts \biband~ X.~Chapuisat}.
\newblock Hamiltonians for constrained n-particle systems.
\newblock {\em Chem.\ Phys.\ Lett. \bf 136\/} (1987), 164.

\bibitem{nau85:1287}
{\rm A.~Nauts \biband~ X.~Chapuisat}.
\newblock Momentum, quasi-momentum and hamiltonian operators in terms of
  arbitrary curvilinear coordinates, with special emphasis on molecular
  hamiltonians.
\newblock {\em Mol.\ Phys. \bf 55\/} (1985), 1287.

\bibitem{cha91:1}
{\rm X.~Chapuisat, A.~Nauts,  \biband~ J.-P. Brunet}.
\newblock Exact quantum molecular hamiltonians: {P}art {I}. {A}pplication to
  the dynamics of three particles.
\newblock {\em Mol.\ Phys. \bf 72\/} (1991), 1--31.

\bibitem{tho82:339}
{\rm K.~S. Thorne \biband~ D.~MacDonald}.
\newblock Electrodynamics in curved spacetime: 3 + 1 formulation.
\newblock {\em Mon. Not. R. Astr. Soc. \bf 198\/} (1982), 339--343.

\bibitem{nik20:191}
{\rm N.~N. Nikolaev \biband~ S.~N. Vergele}.
\newblock Maxwell equations in curved space-time: non-vanishing magnetic field
  in pure electrostatic systems.
\newblock {\em J. High Energy Phys. \bf 2020\/} (2020), 191.

\bibitem{bal96:book}
{\rm D.~Baldomir \biband~ P.~Hammond}.
\newblock {\em Geometry of Electromagnetic Systems}.
\newblock Oxford University Press, Oxford, 1996.

\bibitem{nob16:032108}
{\rm J.~H. Noble \biband~ U.~D. Jentschura}.
\newblock Dirac hamiltonian and reissner-nordstr\"om metric: Coulomb
  interaction in curved space-time.
\newblock {\em Phys. Rev. A \bf 93\/} (2016), 032108.

\bibitem{que16:102101}
{\rm C.~Quesne}.
\newblock Quantum oscillator and kepler–coulomb problems in curved spaces:
  Deformed shape invariance, point canonical transformations, and rational
  extensions.
\newblock {\em J. Math. Phys. \bf 57\/} (2016), 102101.

\bibitem{agg16:2653}
{\rm L.~Aggoun, N.~Bounouioua, F.~Benamira,  \biband~ L.~Guechi}.
\newblock Path integral solution for the coulomb potential in a curved space of
  constant positive curvature.
\newblock {\em Int. J. Theor. Phys. \bf 55\/} (2016), 2653--2667.

\bibitem{mor07:book}
{\rm A.~Moroianu}.
\newblock {\em Lectures on K{\"a}hler Geometry}.
\newblock London Mathematical Society Student Texts. Cambridge University
  Press, 2007.

\bibitem{voi10:85}
{\rm C.~Voisin}.
\newblock Algebraic geometry versus k{\"a}hler geometry.
\newblock {\em Milan J. Math. \bf 78\/} (2010), 85--116.

\bibitem{gua14:297}
{\rm M.~Gualtieri}.
\newblock Generalized k{\"a}hler geometry.
\newblock {\em Commun. Math. Phys. \bf 331\/} (2014), 297--331.

\bibitem{sim83:2167}
{\rm B.~Simon}.
\newblock Holonomy, the quantum adiabatic theorem, and berry's phase.
\newblock {\em Phys.\ Rev.\ Lett. \bf 51\/} (1983), 2167--2170.

\bibitem{alb18:015002}
{\rm T.~Albash \biband~ D.~A. Lidar}.
\newblock Adiabatic quantum computation.
\newblock {\em Rev. Mod. Phys. \bf 90\/} (2018), 015002.

\bibitem{cao19:10856}
{\rm Y.~Cao, J.~Romero, J.~P. Olson, M.~Degroote, P.~D. Johnson,
  M.~Kieferov{\'a}, I.~D. Kivlichan, T.~Menke, B.~Peropadre, N.~P.~D. Sawaya,
  S.~Sim, L.~Veis,  \biband~ A.~Aspuru-Guzik}.
\newblock Quantum chemistry in the age of quantum computing.
\newblock {\em Chem.\ Rev. \bf 119\/} (2019), 10856--10915.

\bibitem{mca20:015003}
{\rm S.~{McArdle}, S.~Endo, A.~Aspuru-Guzik, S.~C. Benjamin,  \biband~
  X.~Yuan}.
\newblock Quantum computational chemistry.
\newblock {\em Rev. Mod. Phys. \bf 92\/} (2020), 015003.

\bibitem{goo20:1084}
{\rm Google}.
\newblock Hartree-fock on a superconducting qubit quantum computer.
\newblock {\em Science \bf 369}, 6507 (2020), 1084--1089.

\bibitem{oll21:4229}
{\rm P.~J. Ollitrault, A.~Miessen,  \biband~ I.~Tavernelli}.
\newblock Molecular quantum dynamics: A quantum computing perspective.
\newblock {\em Acc.\ Chem.\ Res. \bf 54\/} (2021), 4229--4238.

\bibitem{saj22:6475}
{\rm M.~Sajjan, J.~Li, R.~Selvarajan, S.~H. Sureshbabu, S.~S. Kale, R.~Gupta,
  V.~Singh,  \biband~ S.~Kais}.
\newblock Quantum machine learning for chemistry and physics.
\newblock {\em Chem. Soc. Rev. \bf 51\/} (2022), 6475--6573.

\bibitem{wei24:102105}
{\rm J.~D. Weidman, M.~Sajjan, C.~Mikolas, Z.~J. Stewart, J.~Pollanen, S.~Kais,
   \biband~ A.~K. Wilson}.
\newblock Quantum computing and chemistry.
\newblock {\em Cell Rep. Phys. Sci. \bf 5\/} (2024), 102105.

\bibitem{kho24:013032}
{\rm U.~E. Khodaeva, D.~L. Kovrizhin,  \biband~ J.~Knolle}.
\newblock Quantum simulation of the one-dimensional fermi-hubbard model as a
  {${\mathrm{Z}}_{2}$} lattice-gauge theory.
\newblock {\em Phys. Rev. Res. \bf 6\/} (2024), 013032.

\bibitem{bes25:052619}
{\rm T.~A. Bespalova, K.~Deli\ifmmode~\acute{c}\else \'{c}\fi{}, G.~Pupillo,
  F.~Tacchino,  \biband~ I.~Tavernelli}.
\newblock Simulating the fermi-hubbard model with long-range hopping on a
  quantum computer.
\newblock {\em Phys. Rev. A \bf 111\/} (2025), 052619.

\bibitem{hub64:401}
{\rm J.~Hubbard}.
\newblock Electron correlations in narrow energy bands iii. an improved
  solution.
\newblock {\em Proc. R. Soc. A. Math. Phys. Sci. \bf 281}, 1386 (1964),
  401--419.

\bibitem{pan12:777}
{\rm J.-W. Pan, Z.-B. Chen, C.-Y. Lu, H.~Weinfurter, A.~Zeilinger,  \biband~
  M.~$\dot{\mathrm{Z}}$ukowski}.
\newblock Multiphoton entanglement and interferometry.
\newblock {\em Rev.\ Mod.\ Phys. \bf 84\/} (2012), 777--838.

\bibitem{hau20:054401}
{\rm P.~Hauke, H.~G. Katzgraber, W.~Lechner, H.~Nishimori,  \biband~ W.~D.
  Oliver}.
\newblock Perspectives of quantum annealing: methods and implementations.
\newblock {\em Rep.\ Prog.\ Phys. \bf 83\/} (2020), 054401.

\bibitem{jae95:5605}
{\rm A.~J{\"a}ckle \biband~ H.-D. Meyer}.
\newblock Reactive scattering using the multiconfiguration time-dependent
  {H}artree approximation: {G}eneral aspects and application to the collinear
  {H+H$_2\rightarrow$ H$_2$+H} reaction.
\newblock {\em J.~Chem.\ Phys. \bf 102\/} (1995), 5605.

\bibitem{jae96:7974}
{\rm A.~J{\"a}ckle \biband~ H.-D. Meyer}.
\newblock Product representation of potential energy surfaces.
\newblock {\em J.~Chem.\ Phys. \bf 104\/} (1996), 7974.

\bibitem{jae98:3772}
{\rm A.~J{\"a}ckle \biband~ H.-D. Meyer}.
\newblock Product representation of potential energy surfaces~{II}.
\newblock {\em J.~Chem.\ Phys. \bf 109\/} (1998), 3772.

\bibitem{sch20:024108}
{\rm M.~Schr\"oder}.
\newblock {Transforming high-dimensional potential energy surfaces into a
  canonical polyadic decomposition using Monte Carlo methods}.
\newblock {\em J.~Chem.\ Phys. \bf 152\/} (2020), 024108.

\bibitem{man06:194105}
{\rm S.~Manzhos \biband~ T.~{Carrington, Jr.}}
\newblock Using neural networks to represent potential surfaces as sum of
  products.
\newblock {\em J.~Chem.\ Phys. \bf 125\/} (2006), 194105.

\bibitem{koc14:021101}
{\rm W.~Koch \biband~ D.~H. Zhang}.
\newblock Communication: Separable potential energy surfaces from
  multiplicative artificial neural networks.
\newblock {\em J.~Chem.\ Phys. \bf 141\/} (2014), 021101.

\bibitem{car98:191}
{\rm S.~Carter, J.~M. Bowman,  \biband~ N.~C. Handy}.
\newblock Extensions and tests of "multimodes': a code to obtain accurate
  vibration/rotation energies of many-mode molecules.
\newblock {\em Theor.\ Chem.\ Acc. \bf 100\/} (1998), 191.

\bibitem{bow03:533}
{\rm J.~M. Bowman, S.~Carter,  \biband~ X.~Huang}.
\newblock {MULTIMODE}: a code to calculate rovibrational energies of polyatomic
  molecules.
\newblock {\em Int.~Rev.~Phys.~Chem. \bf 22\/} (2003), 533--549.

\bibitem{bow06:33}
{\rm J.~Bowman, X.~Huang, L.~Harding,  \biband~ S.~Carter}.
\newblock The determination of molecular properties from {MULTIMODE} with an
  application to the calculation of franck-condon factors for photoionization
  of {CF}$_3$ to {CF}$_3$.
\newblock {\em Mol.\ Phys. \bf 104\/} (2006), 33.

\bibitem{wu06:285}
{\rm X.~H. J.~Wu, S.~Carter,  \biband~ J.~M. Bowman}.
\newblock Tests of multimode calculations of rovibrational energies of
  {CH}$_4$.
\newblock {\em Chem.\ Phys.\ Lett. \bf 426\/} (2006), 285.

\bibitem{bow07:7317}
{\rm J.~M. Bowman, X.~Huang, N.~C. Handy,  \biband~ S.~Carter}.
\newblock Vibrational levels of methanol calculated by the reaction path
  version of {MULTIMODE}, using an \emph{ab initio}, full-dimensional
  potential.
\newblock {\em J.~Phys.\ Chem.~A \bf 111\/} (2007), 7317.

\bibitem{wan08:071101}
{\rm Y.~Wang, S.~Carter, B.~J. Braams,  \biband~ J.~M. Bowman}.
\newblock Multimode quantum calculations of intramolecular vibrational energies
  of the water dimer and trimer using ab initio-based potential energy
  surfaces.
\newblock {\em J.~Chem.\ Phys. \bf 128\/} (2008), 071101.

\bibitem{nor96:11261}
{\rm L.~S. Norris, M.~A. Ratner, A.~E. Roitberg,  \biband~ R.~B. Gerber}.
\newblock Moller-plesset perturbation theory applied to vibrational problems.
\newblock {\em J.~Chem.\ Phys. \bf 105\/} (1996), 11261.

\bibitem{chr03:5773}
{\rm O.~Christiansen}.
\newblock Moller-plesset perturbation theory for vibrational wave functions.
\newblock {\em J.~Chem.\ Phys. \bf 119\/} (2003), 5773.

\bibitem{yag07:681}
{\rm K.~Yagi, S.~Hirata,  \biband~ K.~Hirao}.
\newblock Multiresolution potential energy surfaces for vibrational state
  calculations.
\newblock {\em Theor.\ Chem.\ Acc. \bf 118\/} (2007), 681--691.

\bibitem{yag01:10647}
{\rm K.~Yagi, T.~Taketsugu,  \biband~ K.~Hirao}.
\newblock Generation of full-dimensional potential energy surface of
  intramolecular hydrogen atom transfer in malonaldehyde and tunneling
  dynamics.
\newblock {\em J.~Chem.\ Phys. \bf 115\/} (2001), 10647.

\bibitem{mil04:5036}
{\rm G.~V. Mil'nikov, K.~Yagi, T.~Taketsugu, H.~Nakamura,  \biband~ K.~Hirao}.
\newblock Simple and accurate method to evaluate tunneling splitting in
  polyatomic molecules.
\newblock {\em J.~Chem.\ Phys. \bf 120\/} (2004), 5036.

\bibitem{oya06:064311}
{\rm C.~Oyanagi, K.~Yagi, T.~Taketsugu,  \biband~ K.~Hirao}.
\newblock {\em J.~Chem.\ Phys. \bf 124\/} (2006), 064311.

\bibitem{chr04:2149}
{\rm O.~Christiansen}.
\newblock Vibrational coupled cluster theory.
\newblock {\em J.~Chem.\ Phys. \bf 120\/} (2004), 2149.

\bibitem{chr05:667}
{\rm O.~Christiansen \biband~ J.~Luis}.
\newblock Beyond vibrational self-consistent-field methods: {B}enchmark
  calculations for the fundamental vibrations of ethylene.
\newblock {\em Int.\ J.\ Quant.\ Chem. \bf 104\/} (2005), 667.

\bibitem{chr05:194105}
{\rm O.Christiansen}.
\newblock Response theory for vibrational wave functions.
\newblock {\em J.~Chem.\ Phys. \bf 122\/} (2005), 194105.

\bibitem{sei07:204101}
{\rm P.~Seidler \biband~ O.~Christiansen}.
\newblock Vibrational excitation energies from vibrational coupled cluster
  response theory.
\newblock {\em J.~Chem.\ Phys. \bf 126\/} (2007), 204101.

\bibitem{sei11:054119}
{\rm P.~Seidler, M.~Sparta,  \biband~ O.~Christiansen}.
\newblock {Vibrational coupled cluster response theory: A general
  implementation}.
\newblock {\em J.~Chem.\ Phys. \bf 134\/} (2011), 054119.

\bibitem{nef09:124129}
{\rm M.~Neff \biband~ G.~Rauhut}.
\newblock Toward large scale vibrational configuration interaction
  calculations.
\newblock {\em J.~Chem.\ Phys. \bf 131}, 12 (2009), 124129.

\bibitem{nef14:100}
{\rm M.~Neff \biband~ G.~Rauhut}.
\newblock Towards black-box calculations of tunneling splittings obtained from
  vibrational structure methods based on normal coordinates.
\newblock {\em Spectrochimica Acta part A \bf 119\/} (2014), 100--106.

\bibitem{cul94:227}
{\rm F.~Culot \biband~ J.~Li\'evin}.
\newblock A multiconfigurational {SCF} computational method for the resolution
  of the vibrational {S}chr\"odinger equation in polyatomic molecules.
\newblock {\em Theor.\ Chem.\ Acc. \bf 89\/} (1994), 227.

\bibitem{hei10:124102}
{\rm S.~Heislbetz \biband~ G.~Rauhut}.
\newblock Vibrational multiconfiguration self-consistent field theory:
  Implementation and test calculations.
\newblock {\em J.~Chem.\ Phys. \bf 132\/} (2010), 124102.

\bibitem{ban12:114108}
{\rm S.~Banik, S.~Pal,  \biband~ M.~D. Prasad}.
\newblock Vibrational multi-reference coupled cluster theory in bosonic
  representation.
\newblock {\em J.~Chem.\ Phys. \bf 137\/} (2012), 114108.

\bibitem{chr04:2140}
{\rm O.~Christiansen}.
\newblock A second quantization formulation of multimode dynamics.
\newblock {\em J.~Chem.\ Phys. \bf 120\/} (2004), 2140--2148.

\bibitem{jak01:1085}
{\rm D.~Jakobson, N.~S. Nadirashvili,  \biband~ J.~Toth}.
\newblock Geometric properties of eigenfunctions.
\newblock {\em Russian Math. Surveys \bf 56\/} (2001), 1085--1105.

\bibitem{gre13:601}
{\rm D.~S. Grebenkov \biband~ B.-T. Nguyen}.
\newblock Geometrical structure of laplacian eigenfunctions.
\newblock {\em SIAM Rev. \bf 55\/} (2013), 601--667.

\bibitem{ngu13:780}
{\rm B.-T. Nguyen \biband~ D.~S. Grebenkov}.
\newblock Localization of laplacian eigenfunctions in circular, spherical, and
  elliptical domains.
\newblock {\em SIAM J. Appl. Math. \bf 73\/} (2013), 780--803.

\bibitem{cha73:1328}
{\rm C.~Chandler \biband~ A.~G. Gibson}.
\newblock {T}ransition from time‐dependent to time‐independent multichannel
  quantum scattering theory.
\newblock {\em J. Math. Phys. \bf 14\/} (1973), 1328--1335.

\bibitem{cha77:2336}
{\rm C.~Chandler \biband~ A.~G. Gibson}.
\newblock {$N$‐body quantum scattering theory in two Hilbert spaces. I. The
  basic equations}.
\newblock {\em J. Math. Phys. \bf 18\/} (1977), 2336--2347.

\bibitem{cha78:1610}
{\rm C.~Chandler \biband~ A.~G. Gibson}.
\newblock {$N$‐body quantum scattering theory in two Hilbert spaces. II. Some
  asymptotic limits}.
\newblock {\em J. Math. Phys. \bf 19\/} (1978), 1610--1616.

\bibitem{cha83:80}
{\rm C.~Chandler \biband~ A.~Gibson}.
\newblock {$N$-body quantum scattering theory in two Hilbert spaces. III.
  Theory of approximations}.
\newblock {\em J. Func. Anal. \bf 52\/} (1983), 80--105.

\bibitem{cha84:1841}
{\rm C.~Chandler \biband~ A.~G. Gibson}.
\newblock {$N$‐body quantum scattering theory in two Hilbert spaces. IV.
  Approximate equations}.
\newblock {\em J. Math. Phys. \bf 25}, 6 (1984), 1841--1856.

\bibitem{cha89:1533}
{\rm C.~Chandler \biband~ A.~G. Gibson}.
\newblock {$N$‐body quantum scattering theory in two Hilbert spaces. V.
  Computation strategy}.
\newblock {\em J. Math. Phys. \bf 30\/} (1989), 1533--1544.

\bibitem{cha92:3477}
{\rm C.~Chandler \biband~ A.~G. Gibson}.
\newblock {$N$‐body quantum scattering theory in two Hilbert spaces. VI.
  Compactness conditions}.
\newblock {\em J. Math. Phys. \bf 33\/} (1992), 3477--3492.

\bibitem{cha94:1487}
{\rm C.~Chandler \biband~ A.~G. Gibson}.
\newblock {$N$‐body quantum scattering theory in two Hilbert spaces. VII.
  Real‐energy limits}.
\newblock {\em J. Math. Phys. \bf 35\/} (1994), 1487--1512.

\bibitem{cha98:223}
{\rm C.~Chandler \biband~ A.~G. Gibson}.
\newblock {$N$-body quantum scattering theory in two Hilbert spaces: $N$-body
  integral equations}.
\newblock {\em Few-Body Sys. \bf 23\/} (1998), 223--258.

\bibitem{tob83:1405}
{\rm W.~Tobocman}.
\newblock Comparison of the chandler-gibson and the baer, kouri, levin, and
  tobocman many-body scattering formalisms.
\newblock {\em Phys. Rev. C \bf 27\/} (1983), 1405--1414.

\bibitem{dya19:book}
{\rm S.~Dyatlov \biband~ M.~Zworsk}.
\newblock {\em Mathematical Theory of Scattering Resonances}.
\newblock American Mathematical Society, Providence, 2019.

\bibitem{fer02:67}
{\rm F.~Fernandez-Alonso \biband~ R.~N. Zare}.
\newblock Scattering resonances in the simplest chemical reaction.
\newblock {\em Ann.\ Rev.\ Phys.\ Chem. \bf 53\/} (2002), 67--99.

\bibitem{qiu06:1440}
{\rm M.~Qiu, Z.~Ren, L.~Che, D.~Dai, S.~A. Harich, X.~Wang, X.~M. Yang, C.~Xu,
  D.~Xie, M.~Gustafsson, R.~T. Skoedje, Z.~Sun,  \biband~ D.-H. Zhang}.
\newblock Observation of {F}eshbach resonances in the {F+H$_2$ $\rightarrow$
  HF+H} reaction.
\newblock {\em Science \bf 311\/} (2006), 1440.

\bibitem{liu12:1}
{\rm K.~Liu}.
\newblock Quantum dynamical resonances in chemical reactions: From a + bc to
  polyatomic systems.
\newblock {\em Adv.\ Chem.\ Phys. \bf 149\/} (2012), 1--46.

\bibitem{wei20:060406}
{\rm M.~Weilenmann \biband~ R.~Colbeck}.
\newblock Self-testing of physical theories, or, is quantum theory optimal with
  respect to some information-processing task?
\newblock {\em Phys. Rev. Lett. \bf 125\/} (2020), 060406.

\bibitem{ren21:625}
{\rm M.-O. Renou, D.~Trillo, M.~Weilenmann, T.~P. Le, A.~Tavakoli, N.~Gisin,
  A.~Ac{\'i}n,  \biband~ M.~Navascu{\'e}s}.
\newblock Quantum theory based on real numbers can be experimentally falsified.
\newblock {\em Nature \bf 600\/} (2021), 625--629.

\bibitem{wu22:140401}
{\rm D.~Wu, Y.-F. Jiang, X.-M. Gu, L.~Huang, B.~Bai, Q.-C. Sun, X.~Zhang, S.-Q.
  Gong, Y.~Mao, H.-S. Zhong, M.-C. Chen, J.~Zhang, Q.~Zhang, C.-Y. Lu,
  \biband~ J.-W. Pan}.
\newblock Experimental refutation of real-valued quantum mechanics under strict
  locality conditions.
\newblock {\em Phys. Rev. Lett. \bf 129\/} (2022), 140401.

\bibitem{li22:040402}
{\rm Z.-D. Li, Y.-L. Mao, M.~Weilenmann, A.~Tavakoli, H.~Chen, L.~Feng, S.-J.
  Yang, M.-O. Renou, D.~Trillo, T.~P. Le, N.~Gisin, A.~Ac\'{\i}n,
  M.~Navascu\'es, Z.~Wang,  \biband~ J.~Fan}.
\newblock Testing real quantum theory in an optical quantum network.
\newblock {\em Phys. Rev. Lett. \bf 128\/} (2022), 040402.

\bibitem{bar26:240202}
{\rm P.~{Barrios Hita}, A.~Trushechkin, H.~Kampermann, M.~Epping,  \biband~
  D.~Bru{\ss}}.
\newblock Quantum mechanics based on real numbers: A consistent description.
\newblock {\em Phys. Rev. Lett. \bf 136\/} (2026), 240202.

\bibitem{nor17:book}
{\rm T.~Norsen}.
\newblock {\em Foundations of Quantum Mechanics - An Exploration of the
  Physical Meaning of Quantum Theory}.
\newblock Springer, 2017.

\bibitem{mai21:642}
{\rm L.~A. Maia, D.~Raom, R.~Ruviaro,  \biband~ Y.~D. Sobral}.
\newblock Mini-max algorithm via pohozaev manifold.
\newblock {\em Nonlinearity \bf 34\/} (2021), 642.

\bibitem{kar21:422}
{\rm G.~E. Karniadakis, I.~G. Kevrekidis, L.~Lu, P.~Perdikaris, S.~Wang,
  \biband~ L.~Yang}.
\newblock Physics-informed machine learning.
\newblock {\em Nat. Rev. Phys. \bf 3\/} (2021), 422--440.

\bibitem{azi24:320}
{\rm K.~Azizzadenesheli, N.~Kovachki, Z.~Li, M.~Liu-Schiaffini, J.~Kossaifi,
  \biband~ A.~Anandkumar}.
\newblock Neural operators for accelerating scientific simulations and design.
\newblock {\em Nat. Rev. Phys. \bf 6\/} (2024), 320--328.

\bibitem{sub26:123}
{\rm U.~Subedi \biband~ A.~Tewari}.
\newblock Operator learning: A statistical perspective.
\newblock {\em Ann. Rev. Stat. App. \bf 13\/} (2026), 123--148.

\bibitem{mia24:532}
{\rm Z.~Miao, X.~Zhang, Y.~Zhang, L.~Wang,  \biband~ Q.~Meng}.
\newblock Chemistry-informed generative model for classical dynamics
  simulations.
\newblock {\em J.~Phys.\ Chem.\ Lett. \bf 15\/} (2024), 532--539.

\bibitem{kou25:e29}
{\rm P.~Koumoutsakos}.
\newblock Machine learning and partial differential equations: benchmark,
  simplify, and discover.
\newblock {\em Data-Centric Eng. \bf 6\/} (2025), e29.

\bibitem{gar77:24}
{\rm J.~G\"{a}rtner}.
\newblock On large deviations from the invariant measure.
\newblock {\em Theory Probab. Appl. \bf 22\/} (1977), 24--39.

\bibitem{ell84:1}
{\rm R.~S. Ellis}.
\newblock Large deviations for a general class of random vectors.
\newblock {\em Ann. Probab. \bf 12\/} (1984), 1--12.

\bibitem{bar15:158101}
{\rm A.~C. Barato \biband~ U.~Seifert}.
\newblock Thermodynamic uncertainty relation for biomolecular processes.
\newblock {\em Phys.\ Rev.\ Lett. \bf 114\/} (2015), 158101.

\bibitem{hus18:2386}
{\rm B.~E. Husic \biband~ V.~S. Pande}.
\newblock Markov state models: From an art to a science.
\newblock {\em J.~Am.\ Chem.\ Soc. \bf 140\/} (2018), 2386--2396.

\end{thebibliography}

\end{document}